\documentclass[12pt, a4paper, notitlepage, twoside]{report}

\RequirePackage[l2tabu, orthodox]{nag}

\usepackage[top=20mm,bottom=20mm,left=25mm,right=25mm]{geometry}

\usepackage[T1]{fontenc}
\usepackage[utf8]{inputenc}

\usepackage{amsmath,amssymb,amsfonts}

\usepackage{centernot}

\usepackage{multirow}

\usepackage{underbracket}
\usepackage{pgf}
\usepackage{tikz}
\usetikzlibrary{calc}
\usetikzlibrary{arrows}
\usetikzlibrary{decorations.markings}

\usepackage{ifthen}

\newcommand*{\colorboxed}{}
\def\colorboxed#1#{%
  \colorboxedAux{#1}%
}
\newcommand*{\colorboxedAux}[3]{%
  \begingroup
    \colorlet{cb@saved}{.}%
    \color#1{#2}%
    \boxed{%
      \color{cb@saved}%
      #3%
    }%
  \endgroup
}

\numberwithin{equation}{section}

\usepackage{makeidx}
\makeindex
\usepackage{tocbibind}

\usepackage{amsthm}
\newtheoremstyle{break}{9pt}{9pt}{\itshape}{}{\bfseries}{}{\newline}{}
\theoremstyle{break}    
\newtheorem{exo}{Exercise}[chapter]
\newtheorem{hyp}{Axiom}[chapter]

\usepackage[colorlinks=true, linktoc=all, linkcolor=black, citecolor=red, urlcolor=blue, backref=page]{hyperref}

\usepackage{etoolbox}
\makeatletter
\patchcmd{\BR@backref}{\newblock}{\newblock(}{}{}
\patchcmd{\BR@backref}{\par}{)\par}{}{}
\makeatother

\title{\bfseries Conformal field theory on the plane}

\author{Sylvain Ribault \vspace{2mm}
\\
{\normalsize CEA Saclay, Institut de Physique Th\'eorique}
 \\
 {\footnotesize \ttfamily sylvain.ribault@ipht.fr }
}

\date{}

\begin{document}

\maketitle

\begin{abstract}
We review conformal field theory on the plane in the conformal bootstrap approach.
We introduce the main ideas of the bootstrap approach to quantum field theory, and how they apply to two-dimensional theories with local conformal symmetry.
We describe the mathematical structures that appear in such theories, from the Virasoro algebra and its representations, to BPZ equations and conformal blocks.
Examples include Liouville theory, (generalized) minimal models, free bosonic theories, the $H_3^+$ model, and the $SU_2$ and $\widetilde{SL}_2(\mathbb{R})$ WZW models. 
We also discuss relations between some of these models, and limits of these models when the central charge and/or conformal dimensions tend to particular values.
\end{abstract}

\vspace{1cm}
\begin{center}
 \textbf{Keywords}
\end{center}
\noindent conformal field theory, operator product expansion, conformal bootstrap, Virasoro algebra, BPZ equation, conformal block, crossing symmetry, Liouville theory, DOZZ formula, minimal models, affine Lie algebra, free boson, Sugawara construction,
KZ equations, KZ-BPZ relation, $H_3^+$ model, WZW model

\vspace{1cm}
\begin{center}
\textbf{Public domain notice}
\end{center}
\noindent To the extent possible under law, Sylvain Ribault has waived all copyright and related or neighboring rights to this text.

\tableofcontents
\hypersetup{linkcolor=blue}

\addtocounter{chapter}{-1}

\chapter{Preliminaries \label{secprel}}

\section{Context and approach}

Just like free field theories, conformal field theories are relatively simple and tractable quantum field theories, and can serve as starting points for perturbative computations in more generic quantum field theories. However, conformal field theories are less trivial than free field theories, and are often better approximations of realistic quantum field theories:
\begin{align}
 \begin{tikzpicture}[scale = .8, thick,decoration={
    markings,
    mark=at position 0.5 with {\arrow{>}}}
    ] 
  \draw (-3, .6) node[draw, fill = red!5] (f) {Free QFT};
  \draw (0, -.6) node[draw, fill = red!15] (c) {CFT};
  \draw (4, 0) node[draw, fill = red!25] (r) {Realistic QFT};
  \draw[postaction={decorate}] (f.east) to (r.west);
  \draw[postaction={decorate}] (c.east) to (r.west);
  \draw[->] (-4,-1.3) node[below right]{Trivial} -- (5.5,-1.3) node[below left]{Complicated};
 \end{tikzpicture}
\end{align}
Many conformal field theories
are also of independent interest, in particular for describing critical phenomenons. In this text we will focus on building and solving conformal field theories: for the applications, see some of the works cited in Section \ref{sec:womt}.

In two dimensions, unlike in higher dimensions, the algebra of conformal transformations is infinite-dimensional. As a result, it has been possible to exactly solve certain nontrivial two-dimensional conformal field theories. Therefore, the subject of two-dimensional conformal field theory deserves a separate treatment. Nevertheless, the two-dimensional case may hold valuable lessons for
the higher-dimensional case, for example that unitarity only plays a minor role in the structure and classification of conformal field theories.

As an introduction to two-dimensional conformal field theory, this text has the particularity of using exclusively the bootstrap approach, which consists in systematically exploiting symmetry and consistency assumptions.
While this approach is widely recognized as very powerful, much of the literature uses it alongside the Lagrangian approach, which can be quite useful for heuristics and for some computations.
Renouncing Lagrangians however has the advantages of simplicity and consistency, and we will try to show that we do not lose much in terms of heuristics and computational power -- we will mainly have to deplore the loss of a simple definition for WZW models. Moreover, we need not assume any previous knowledge of quantum field theory:
the prerequisites are limited to 
elementary complex analysis, and some familiarity with Lie algebras and their representations.

Another advantage of the bootstrap approach is that by making assumptions explicit, we can better understand them, and if necessary lift them.
For example, results such as the C-theorem, and the fact that scale invariance implies conformal invariance in two-dimensional unitary quantum field theories, rely on the strong assumption that there exists an energy-momentum tensor -- a conserved field with spin two and conformal dimension two that generates conformal transformations.
We will not lift this particular assumption, as we will study theories with local (i.e.
not just global) conformal invariance, which always have an energy-momentum tensor.

We will however lift a number of other common assumptions, in particular the existence of a vacuum state, and the existence of theories on any Riemann surface (and not just on the plane).
It would be interesting to investigate how this affects the set of possible models, and in particular the classification of minimal models.
It is however mainly for the sake of generality and simplicity that
we will lift these assumptions.
We will indeed be studying not only rational, but also non-rational theories such as Liouville theory.
Non-rational theories do not necessarily have a vacuum state, and assuming their consistency on the torus is of little help for determining their spectrums \cite{rib14c}. 

By definition, rational conformal field theories have spectrums made of finitely many representations.
However, in order to ensure closure under fusion, these representations must have complicated structures.
In Liouville theory, the spectrum is made of a continuum of representations, but these representations are merely Verma modules, so we will study this theory first. The fundamental nature of Liouville theory is also apparent in the relation between its structure constants, and the fusing matrix of the Virasoro algebra.
Then we will study minimal models, which are rational and therefore less generic and more complicated.
Even later will come free bosonic theories, which do not play any fundamental role in the bootstrap approach.

\section{Plan of the text \label{secplan}}

We will give a systematic exposition of the ideas and techniques of the conformal bootstrap approach in Chapters \ref{secintr} and \ref{secccs}, before studying particular models in Chapters \ref{seccbm} and \ref{secaff}.

In Chapter \ref{secintr}, we introduce the bootstrap approach to quantum field theory, before focusing on the particular case of conformal field theories with their Virasoro symmetry algebra.
We then give a preview of some of the simplest models.

In Chapter \ref{secccs}, we work out the technical consequences of these ideas for the spectrum and correlation functions.
We study the highest-weight representations of the Virasoro algebra, the fields that correspond to states in these representations, and the correlation functions of these fields. 
In particular we derive Ward identities for correlation functions of primary fields, and BPZ equations for correlation functions that involve degenerate fields. 
We decompose correlation functions into conformal blocks, and write crossing symmetry equations.

In Chapter \ref{seccbm}, we introduce and solve some of the simplest nontrivial two-dimensional conformal field theories: Liouville theory, generalized minimal models, A-series minimal models, and Runkel--Watts-type theories.
We define these theories by a few simple assumptions on their spectrums and correlation functions.
We pay particular attention to the three-point structure constants, whose explicit expressions we derive. 
We study the analytic properties of correlation functions, and find relations between these theories by taking limits in the central charge and/or conformal dimensions.

In Chapter \ref{secaff}, we introduce extended symmetry algebras, starting with the affine $\hat{\mathfrak{u}}_1$ algebra.
This not only allows us to study free bosonic theories, but also provides alternative perspectives on Liouville theory.
We then consider nonabelian affine symmetry algebras, and in particular $\widehat{\mathfrak{sl}}_2$.
The KZ-BPZ relation between differential equations satisfied by $\widehat{\mathfrak{sl}}_2$-symmetric and Virasoro-symmetric correlation functions leads us to define the $H_3^+$ model from Liouville theory, and allows us to solve it.
We then consider WZW models, and in particular the $SU_2$ and $\widetilde{SL}_2(\mathbb{R})$ WZW models. 

In each chapter, the last section is devoted to exercises.
Some exercises are intended to test and improve one's comprehension of the material; such exercises become scarcer in later chapters.
Some exercises sketch proofs of results from the main text, in cases when hints and guidance seem warranted.
And some exercises provide supplementary material.

This text comes with an index, which points to the definition of each listed term. 
These terms appear in bold where they are defined.

\section{Why one more text on conformal field theory?}\label{sec:womt}

Let us argue that this text fills an available niche, by considering which neighbouring niches are already filled: 
\begin{itemize}
\item 
The book \cite{zz90} by A. and Al. Zamoldochikov is in spirit quite close to the present text.
That book is a concise exposition of conformal field theory, from the basic principles to advanced results, mostly in the conformal bootstrap approach.
However, the depth and breadth of the ideas may be too much for beginning students.
Non-rational theories such as Liouville theory are not covered.
And misprints are frequent.
\item
Again in the spirit of the present text, Teschner's review article \cite{tes17} is both more concise and more advanced, and provides a guide to the literature.
\item
Non-rational theories are covered in a review by Schomerus \cite{sch05}.
That article uses multiple approaches, and pays particular attention to boundary conformal field theory, with applications to string theory in mind. 
\item
The Big Yellow Book \cite{fms97} by di Francesco, Mathieu and Sénéchal is a useful reference, especially on rational conformal field theories.
As its epic length suggests, it is rather encyclopedic. 
\item
The encyclopedic approach was extended in the direction of Liouville theory by Nakayama's review \cite{nak04}, which includes systematic guides to the literature.
\item 
On the more pedagogical side, Gaberdiel's review \cite{gab99} provides a consistent exposition using vertex operators, which mathematically formalize the conformal bootstrap approach.
That text deals more with the symmetry algebras and their representations, than with correlation functions.
Examples include logarithmic theories, but no non-rational theories.
\item
Motivations and applications are mostly absent from the present text.
On the statistical mechanics side, they are given in Cardy's lecture notes \cite{car08}, which also provide a concise introduction to the formalism.
\item 
$W$-symmetry is reviewed in depth in Bouwknegt and Schoutens's article \cite{bs92}, where $W$-symmetric conformal field theories are however not solved systematically.
An exposition of the $W$-symmetric minimal models and conformal Toda theories would be a natural extension of the present text, although these theories have not been fully solved so far. 
\end{itemize}
A much abridged version of this text is available \cite{rib16}, which includes a number of corrected exercises. An even briefer treatment is found in Wikipedia, with in particular the articles on \href{https://en.wikipedia.org/wiki/Two-dimensional_conformal_field_theory}{Two-dimensional conformal field theory}, \href{https://en.wikipedia.org/wiki/Liouville_field_theory}{Liouville theory},
\href{https://en.wikipedia.org/wiki/Minimal_models}{minimal models}, and \href{https://en.wikipedia.org/wiki/Virasoro_conformal_block}{Virasoro conformal blocks}.
(As of 2022, these Wikipedia articles were mostly written by the present author.) Moreover, the video recordings of 7 lectures at YRISW 2019 are available online.

While we mostly review known results, the following aspects may be original:
\begin{itemize}
\item in Section \ref{secaco}, the formulation of Al. Zamolodchikov's recursive representation of conformal blocks as an explicit formula,
\item in Section \ref{secliou}, a complete and correct solution of Liouville theory in the bootstrap approach, 
\item in Section \ref{secvmm}, the logical sequence ``Liouville theory $\rightarrow$ generalized minimal models $\rightarrow$ minimal models'',
\item in Section \ref{secltf}, the systematic discussion of the limits of Liouville theory, (generalized) minimal models, and Runkel--Watts-type theories,
\item in Section \ref{secsacf}, the study of compactified free bosons for any value of $c$ and not just $c=1$,
\item in Section \ref{seclld}, the derivation of the light and heavy asymptotic limits of Liouville theory, without using the Lagrangian definition,
\item in Section \ref{seckzbpz}, the derivation of $\widehat{\mathfrak{sl}}_2$ degenerate representations and fusion rules from the KZ-BPZ relation,
\item in Section \ref{secsu}, the definition of the generalized $SU_2$ WZW model,
\item in Section \ref{secslr}, a concise derivation of the spectrum and fusion rules of the $\widetilde{SL}_2(\mathbb{R})$ WZW model.
\end{itemize}

\section{User's manual}

This text is intended as a tool for learning, and as a collection of technical results.
It provides neither a history of the subject, nor a guide to the literature.
Accordingly, citations to the existing literature are not meant to distribute credit to researchers in the field.
Citations are solely meant to be 
\href{http://researchpracticesandtools.blogspot.fr/2013/08/write-for-humans-not-for-robots.html}
{helpful to the reader}, and are therefore used sparingly.
Each cited work comes with a hopefully clear and precise indication of what could be useful therein.
In choosing cited works, the criterions have been
\begin{enumerate}
\item ease of access (only freely and if possible legally available texts are cited), 
\item clarity and completeness (this often favours review articles over original works),
\item my familiarity with the cited works (admittedly this favours my own works). 
\end{enumerate}
Citations are only one tool for finding more information on the subject. 
The reader is assumed to have access to other tools such as Google and Wikipedia. 
To facilitate searches, an effort has been made to use standard terminology and notations. 

Traditional scientific articles, with their frozen texts and authors, are obviously a poor way of conveying the ever-evolving knowledge of a community of researchers.
They may become obsolete, but it is not clear what will replace them.
Meanwhile, it seems prudent to distribute the present text so that it can easily be 
\href{http://researchpracticesandtools.blogspot.fr/2014/02/the-case-for-emancipating-articles-from.html}
{reused and modified by others}.
For this purpose:
\begin{itemize}
 \item This text is in the public domain, in order to eliminate legal restrictions to its reuse. 
\item This article is distributed not only on Arxiv, but also on \href{https://github.com/ribault/CFT-Review}{GitHub}, so that it can be collaboratively modified. 
\item In the Latex file, there are no user-defined global macros.
This makes it easier to reuse excerpts of that file. 
\end{itemize}

I have tried to follow some stylistic good practices, such as: providing a clickable table of contents, not clogging the bibliography with superfluous information such as journal data, \href{http://researchpracticesandtools.blogspot.fr/2018/01/will-no-one-rid-me-of-these-tiresome.html}{avoiding Latin plurals} when regular plurals are available, and numbering all equations while boxing the important ones.

\section{Acknowledgements}

I am grateful to my colleagues at IPhT Saclay, and in particular to Antoine Duval, Riccardo Guida, Santiago Migliaccio, Vincent Pasquier, and Pierre Ronceray, for the opportunity to give lectures on this subject. I am also grateful to the organizers and participants of the 10th APCTP Focus Program on Liouville, Integrability and Branes, and in particular to Soojong Rey, for including my lectures in an otherwise very advanced workshop. Moreover, I am grateful to the organizers and participants of the 2016 Carg\`ese school on Quantum integrable systems, conformal field theory and stochastic processes, for challenging me to introduce the subject in about four hours. (See \cite{rib16}.) 

I wish to thank Philippe Di Francesco, Sheer El-Showk, Matthias Gaberdiel, Christoph Keller, Miguel Paulos, R\'emi Rhodes, Slava Rychkov, Hubert Saleur, Joerg Teschner, Vincent Vargas, and G\'erard Watts, for useful discussions and correspondence.

I am grateful to Aditya Bawane, Connor Behan, Fran\c{c}ois David, Quang-Dien Duong, Bruno Le Floch, Omar Foda, Santiago Migliaccio, Nikita Nemkov, Paul Roux, Raoul Santachiara, and Vincent Vargas, for helpful comments on this text.

A version of this text served as my habilitation thesis, defended on 21 December 2018. I wish to thank the jury members Denis Bernard, Matthias Gaberdiel, Jesper Jacobsen, Vyacheslav Rychkov, V\'eronique Terras, G\'erard Watts and Jean-Bernard Zuber for a lively defense, and for valuable feedback. Special thanks to Matthias Gaberdiel and G\'erard Watts for their specific suggestions and comments.

\chapter{Introduction \label{secintr}}

\section{Quantum field theory \label{secqft} }

Conformal field theory is a special type of \textbf{\boldmath quantum field theory}\index{quantum field theory}, and we start with a brief reminder on quantum field theory.

\subsection{Definition} 

A \textbf{\boldmath theory}\index{theory} can be either a general framework such as quantum field theory or general relativity, or a more specific \textbf{\boldmath model}\index{model} such as the standard model of particle physics or Liouville theory.
The standard model is formulated in the framework of quantum field theory, so we can call it a model of quantum field theory or simply a quantum field theory.
Similarly, Liouville theory is a conformal field theory.
A model need not be directly related to a physical system: a given physical system can be described by several models, and a given model can describe a number (possibly zero) of physical systems.
This holds not only for physical systems, but also for what could be called theoretical systems, such as quantum gravity in $d$ dimensions.
These notions are illustrated by the following examples:
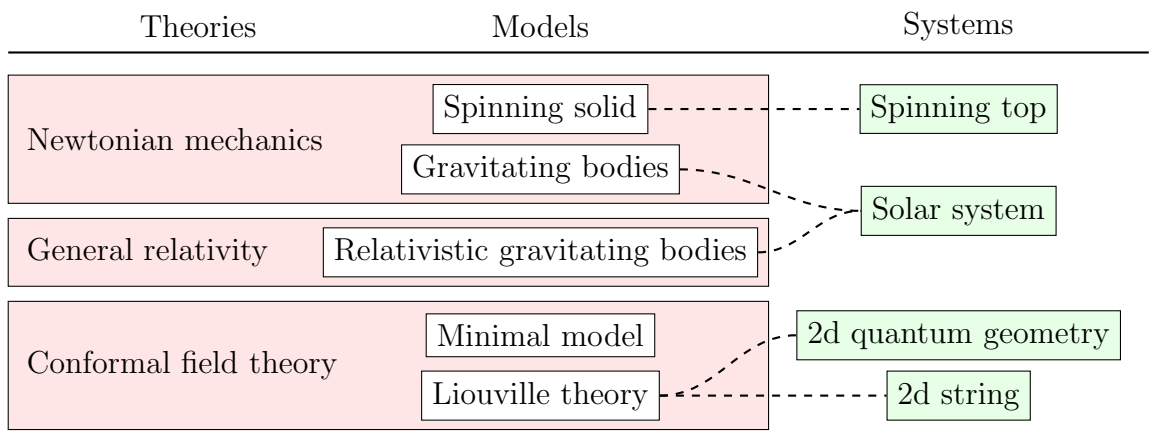
\begin{align}
 \begin{tikzpicture}[scale = .5]
  \node at (5,11){Theories};
  \node at (14,11){Models};
  \node at (25,11){Systems};
  \draw[thick] (0,10.3) -- (30,10.3);
  \draw[fill = red!10] (0, 6.3) rectangle (20, 9.7);
  \draw[fill = red!10] (0, 4.1) rectangle (20, 5.9);
  \draw[fill = red!10] (0, .3) rectangle (20, 3.7);
  \node at (.2, 8)[right] {Newtonian mechanics};
  \node at (.2, 5)[right] {General relativity};
  \node at (.2, 2)[right] {Conformal field theory};
  \draw (14, 8.8) node[draw, fill = white] (ss) {Spinning solid};
  \draw (14, 7.2) node[draw, fill = white] (gb) {Gravitating bodies};
  \draw (14, 5) node[draw, fill = white] (rgb) {Relativistic gravitating bodies};
  \draw (14, 2.8) node[draw, fill = white] (mm) {Minimal model};
  \draw (14, 1.2) node[draw, fill = white] (lt) {Liouville theory};
  \draw (25, 8.8) node[draw, fill = green!10] (st) {Spinning top};
  \draw (25, 6.1) node[draw, fill = green!10] (ssys) {Solar system};
  \draw (25, 2.8) node[draw, fill = green!10] (qg) {2d quantum geometry};
  \draw (25, 1.2) node[draw, fill = green!10] (str) {2d string};
  \draw[thick, dashed] (ss.east) to [out = 0, in = 180] (st.west);
  \draw[thick, dashed] (gb.east) to [out = 0, in = 180] (ssys.west);
  \draw[thick, dashed] (rgb.east) to [out = 0, in = 180] (ssys.west);
  \draw[thick, dashed] (lt.east) to [out = 0, in = 180] (qg.west);
  \draw[thick, dashed] (lt.east) to [out = 0, in = 180] (str.west);
 \end{tikzpicture}
\end{align}

Now, what is a quantum theory? First, this is a probabilistic theory, which does not predict the outcome of a given experiment, but the probabilities of different outcomes.
Second, a quantum theory does not predict just probabilities, but actually probability amplitudes.
Such amplitudes can be added, giving rise to interferences. 

Last, the \textbf{\boldmath field}\index{field} in field theory is a variable defined on some space, which can describe an arbitrary number of particles propagating in that space.
In particular, field theories can describe the creation and annihilation of particles.
For example, a ``height of water'' field can be defined on the surface of the ocean, in order to describe arbitrary numbers of water waves.
A field can in some respects be thought of as an infinite collection of elementary objects: in our example, individual waves of definite shapes.
Field theory constrains how these objects behave in relation with the underlying space, for instance by requiring interactions to be local.

So, quantum field theory is particularly well-suited to predicting the outcomes of collisions of particles, whether in the cosmos or in a particle accelerator.
Such collisions can indeed create or destroy particles, so we need a field theory, and repeating the same collision can produce different outcomes, so we need a probabilistic theory. 

\subsection{Observables}

In a model, the \textbf{\boldmath observables}\index{observable} are the quantities that can in principle be measured.
The nature of the observables is in general dictated by the theory.
For example, in the theory of general relativity, the observables are light signals and motions of objects, and do not include the space-time metric. 
Which observables are relevant may depend on the physical system.
For example, in two-dimensional conformal field theory, some observables are functions on the two-dimensional space.
Such observables are relevant in condensed-matter physics and in quantum geometry, where the space has a physical interpretation.
They are not relevant in string theory, where the two-dimensional space is the world-sheet of a string. 
In quantum field theory in general, and in conformal field theory in particular, we define the observables to be 
\begin{itemize}
 \item the spectrum and
\item the correlation functions.
\end{itemize}

In quantum theory, the \textbf{\boldmath spectrum}\index{spectrum} or space of states is a vector space whose elements represent the states of a system.
The vector space structure is what leads to interferences.
A vector $\sigma$ and the action $A(\sigma)$ of an operator $A$ on $\sigma$ may be written as 
\begin{align}
\left\{\begin{array}{l} 
 \sigma = |\sigma\rangle\ , \\ A(\sigma) = A|\sigma\rangle = |A\sigma\rangle\ .
\end{array}\right. 
\end{align}

In quantum field theory, states are supposed to live on constant-time slices of space-time. If space-time is Euclidean, we can take such slices to be spheres surrounding any point $x$. If the theory is invariant under dilations, such spheres can contract to infinitesimal sizes without loss of information. Therefore, to each state $|\sigma\rangle$ and each point $x$ we associate an object $V_\sigma(x)$, and the collection of these objects when $x$ varies is a field. 

\begin{hyp}[\textbf{\boldmath State-field correspondence}\index{state-field correspondence} or state-operator correspondence]
~\label{ax:sfc}
 There is an injective linear map from the spectrum to the space of fields,
 \begin{align}
 |\sigma \rangle  \ \mapsto \ V_\sigma(x)\ .
\end{align}
\end{hyp} 
\noindent
Thinking of the state as a type of particles, the field $V_\sigma(x)$ can be thought of as measuring the presence of such particles at point $x$.
Then, an $N$-point \textbf{\boldmath correlation function}\index{correlation function} or \textbf{\boldmath $N$-point function}\index{N-point function@$N$-point function} is the probability amplitude for the interaction of $N$ particles of types $\sigma_1,\cdots \sigma_N$, located at points $x_1,\cdots x_N$:
\begin{align}
\left\langle \prod_{i=1}^N V_{\sigma_i}(x_i)\right\rangle\in \mathbb{C}\ .
\label{pva}
\end{align}
This notation for a correlation function as the expectation value of a product of fields is only formal, as we will not define fields independently of correlation functions. The product is meant to convey the idea that an $N$-point function depends linearly on each field, so that in particular 
\begin{align}
 \frac{\partial}{\partial x} \left\langle \prod_{i=1}^N V_{\sigma_i}(x_i)\right\rangle = \left\langle \frac{\partial}{\partial x} \prod_{i=1}^N V_{\sigma_i}(x_i)\right\rangle\ ,
\end{align}
for any component $x$ of a coordinate $x_i$. 

To construct a model of quantum field theory is therefore to give principles that uniquely determine a spectrum and a set of correlation functions.
To \textbf{\boldmath solve a model}\index{solve (a model)} is to actually compute the spectrum and correlation functions. 

\section{The bootstrap approach}

\subsection{Principles \label{secprin}}

The \textbf{\boldmath bootstrap approach}\index{bootstrap approach} is a method for constructing and solving theories, based on the systematic exploitation of 
\begin{enumerate}
 \item symmetry assumptions,
 \item consistency conditions.
\end{enumerate}
For example, the symmetry assumption of general covariance is the basis of the theory of general relativity, and the further symmetry assumptions of homogeneity and isotropy of the universe are used in constructing models of cosmology.
In quantum theories, an essential consistency condition is that the sum of the probabilities of all possible events is one. 

Given a set of assumptions, there may exist any number of models that obey them:
\begin{itemize}
 \item no model at all, if the assumptions are too restrictive,
\item one model, in which case we may be able to solve it,
\item a manageable number of models, in which case we may be able to classify them, and to focus on one of them by making further assumptions,
\item a huge number of models, if the assumptions are not restrictive enough.
\end{itemize}
We will now explore in turn the symmetry assumptions and consistency conditions in quantum field theory. 

\subsection{Symmetry assumptions \label{secsa}}

If a model of quantum field theory has a symmetry algebra, then the spectrum $\mathcal{S}$ must be a representation of this algebra, and can therefore be written as
\begin{align}
 \mathcal{S} = \bigoplus_\mathcal{R} m_\mathcal{R}  \mathcal{R}\ .\
\label{somr}
\end{align}
Here we have decomposed $\mathcal{S}$ as a combination of some basic representations $\mathcal{R}$ of the symmetry algebra, and the number $m_\mathcal{R} \in {\mathbb{N}}\cup\{\infty\}$ is the \textbf{\boldmath multiplicity}\index{multiplicity (of a representation)} of the representation $\mathcal{R}$ in the spectrum $\mathcal{S}$.
(Notations: $\mathbb{N}= \{0,1,2,\dots\}, \mathbb{N}^*=\{1,2,3,\dots\}$.)
A state $\sigma$ in the spectrum can then be written as 
\begin{align}
 \sigma = (\mathcal{R},v)\ ,
\label{arv}
\end{align}
where $v$ is a state in the representation $\mathcal{R}$. 
The corresponding field can be written as 
$V_\sigma(x)= V_{(\mathcal{R},v)}(x)$. 
For example, for any spin $j\in \frac12 \mathbb{N}$ the algebra $\mathfrak{sl}_2$ has a representation of dimension $2j+1$, whose states $\sigma = (j,m)$ can be labelled by their magnetic momentums $m\in\{-j, -j+1,\cdots j\}$.

The idea is now that the dependence of a correlation function $\left\langle \prod_{i=1}^n V_{(\mathcal{R}_i,v_i)}(x_i)\right\rangle$ on $v_i$ is determined by symmetry considerations, whereas the dependence on $\mathcal{R}_i$ is constrained by consistency conditions.
This idea however comes with a number of technical assumptions.
In particular, for the dependence on $v_i$ to be completely determined by symmetry, $\mathcal{R}_i$ must be an \textbf{\boldmath indecomposable representation}\index{indecomposable representation}, i.e. a representation that cannot be written as a direct sum of smaller representations.
\begin{align}
 \mathcal{R} \ \ \text{indecomposable} \quad \iff \quad 
 \Big(\mathcal{R} = \mathcal{R}'\oplus \mathcal{R}'' \implies \mathcal{R}' \in\left\{ 0,\mathcal{R}\right\}\Big)
 \ .
\end{align}
In particular, any \textbf{\boldmath irreducible representation}\index{irreducible representation}, which by definition does not have a nontrivial subrepresentation, is indecomposable,
\begin{align}
 \mathcal{R} \ \ \text{irreducible} \quad \iff \quad 
 \Big(\mathcal{R}'\subset \mathcal{R} \implies \mathcal{R}' \in\left\{ 0,\mathcal{R}\right\}\Big)
 \ .
\end{align}
(See Exercise \ref{exoirr} for an example of an indecomposable representation that is not irreducible.)

We will therefore distinguish two types of data:
\begin{itemize}
 \item \textbf{\boldmath universal data}\index{universal data}, also called 
model-independent data.
This is the information on representations of the symmetry algebra, which determines in particular the $v_i$-dependence of correlation functions.
Universal data in conformal field theory, such as conformal blocks, will be studied in Chapter \ref{secccs}. 
\item \textbf{\boldmath model-dependent data}\index{model-dependent data}, which encode how the spectrum and correlation functions of a particular model are built from the universal data.
This includes the multiplicities $m_\mathcal{R}$, which encode how the spectrum is built from representations, and the structure constants, which encode how correlation functions are built from conformal blocks.
We will study these data 
in Chapter \ref{seccbm}.  
\end{itemize}
Which data are  universal or model-dependent depends on the choice of a symmetry algebra.
If a model's symmetry algebra $\mathfrak{A}$ is actually a subalgebra of a larger symmetry algebra $\mathfrak{A}'$, then the representations of $\mathfrak{A}$ that appear in the spectrum
must combine into representations of the larger algebra $\mathfrak{A}'$.
Which combinations can appear is universal data of $\mathfrak{A}'$, but model-dependent data from the point of view of the smaller algebra $\mathfrak{A}$. 

Since a field theory is defined on some space, we can distinguish two types of symmetries: 
\begin{itemize}
 \item space-time symmetries, which act on that space,
\item internal symmetries, which do not.
\end{itemize}
Space-time symmetries include in particular the invariances under rotations and translations.
In particle physics, internal symmetries include flavour symmetry.
The distinction between space-time and internal symmetries may depend on the choice of the space: for example, supersymmetry can be an internal or space-time symmetry depending on whether the theory is formulated on an ordinary space or on a superspace.

\subsection{Consistency conditions \label{seccc}}

\subsubsection{The axioms}

The consistency conditions on the correlation functions $\left\langle\prod_{i=1}^n V_{\sigma_i}(x_i)\right\rangle$ are derived from axioms on the fields $V_{\sigma_i}(x_i)$. These axioms should be understood as pertaining to correlation functions: we write fields outside correlation functions as a matter of notational convenience only. In particular, we do not define fields as operators acting on the spectrum, as is done in formalisms such as vertex operator algebras. 
Moreover, our axioms do not include the existence of a vacuum state, a frequently encountered axiom of quantum field theory in flat space.
In Liouville theory this axiom is not satisfied, and in minimal models we can do without it. 

\begin{hyp}[\textbf{Commutativity}\index{commutativity} or locality]
\label{ax:col}
\begin{align}
 \boxed{V_{\sigma_1}(x_1)V_{\sigma_2}(x_2) = V_{\sigma_2}(x_2) V_{\sigma_1}(x_1)}\ ,
\label{comm}
\end{align}
provided $x_1\neq x_2$. 
\end{hyp}
\noindent
We are writing this axiom for a theory on a Euclidean space: if the space was Minkowskian, fields would commute if $x_1-x_2$ was space-like, rather than non-vanishing. 
This axiom implies that correlation functions do not depend on how fields are ordered.

\begin{hyp}[Existence of an \textbf{\boldmath OPE}\index{OPE} (Operator Product Expansion)]
 \label{ax:ope}
 \begin{align}
 \boxed{V_{\sigma_1}(x_1)V_{\sigma_2}(x_2) = \sum_{\sigma\in \mathcal{S}} C_{\sigma_1,\sigma_2}^{\sigma}(x_1,x_2) V_{\sigma}(x_2)}\ ,
\label{ope}
\end{align}
where the \textbf{\boldmath OPE coefficients}\index{OPE!---coefficient} $C_{\sigma_1,\sigma_2}^{\sigma}(x_1,x_2)$ are $\mathbb{C}$-valued functions, and the sum runs over a basis of the spectrum $\mathcal{S}$. 
The sum is supposed to converge if $x_1$ and $x_2$ are sufficiently close. 
\end{hyp} 
\noindent
The commutativity axiom implies that the OPE is \textbf{associative}\index{associativity (of the OPE)} and commutative. 
In terms of OPE coefficients, the associativity of the OPE reads
\begin{align}
 \sum_{\sigma_s\in \mathcal{S}} C_{\sigma_1,\sigma_2}^{\sigma_s}(x_1,x_2) C_{\sigma_s,\sigma_3}^{\sigma_4}(x_2,x_3) = \sum_{\sigma_t\in \mathcal{S}} C_{\sigma_1,\sigma_t}^{\sigma_4}(x_1,x_3)C_{\sigma_2,\sigma_3}^{\sigma_t}(x_2,x_3)\ ,
\label{cccc}
\end{align}
for any choice of the four states $(\sigma_1,\sigma_2,\sigma_3,\sigma_4)$.
(See Exercise \ref{exoten} for a similar calculation in a technically simpler context.) This condition can be represented schematically as 
\begin{align}
\sum_{s} 
 \begin{tikzpicture}[scale = .2, baseline=(current  bounding  box.center), very thick]
  \draw (-4,3) node [above] {$1$} -- (0, -1) -- (4, 3) node [above] {$3$};
  \draw (0, 3) node [above] {$2$} -- (-2, 1) -- node [below left] {$s$} (0, -1) -- (0, -4) node [below] {$4$};
 \end{tikzpicture}
= \quad
\sum_{t} 
\begin{tikzpicture}[scale = .2, baseline=(current  bounding  box.center), very thick]
  \draw (-4,3) node [above] {$1$} -- (0, -1) -- (4, 3) node [above] {$3$};
  \draw (0, 3) node [above] {$2$} -- (2, 1) -- node [below right] {$t$} (0, -1) -- (0, -4) node [below] {$4$};
  \end{tikzpicture}
\label{sdsd}
\end{align}
where each node corresponds to an OPE coefficient. 
In terms of OPE coefficients, it is less straightforward to write commutativity than associativity of the OPE,
because we broke the symmetry $x_1\leftrightarrow x_2$ by writing the OPE in terms of $V_\sigma(x_2)$. Had we used $V_\sigma(\frac{x_1+x_2}{2})$ instead, commutativity would have reduced to the simple condition $C_{\sigma_1,\sigma_2}^{\sigma}(x_1,x_2) = C_{\sigma_2,\sigma_1}^{\sigma}(x_2,x_1)$, but associativity would have been complicated. (See Exercise \ref{exocva}.)

\subsubsection{How the axioms constrain correlation functions}

By performing multiple OPEs, an $N$-point function (with $N\geq 3$) can always be reduced to a combination of two-point functions, for instance
\begin{align}
 \left\langle \prod_{i=1}^4 V_{\sigma_i}(x_i) \right\rangle = \sum_{\sigma\in \mathcal{S}} C_{\sigma_1,\sigma_2}^{\sigma}(x_1,x_2)\sum_{\sigma'\in \mathcal{S}} C_{\sigma,\sigma_3}^{\sigma'}(x_2,x_3)\Big\langle V_{\sigma'}(x_3)V_{\sigma_4}(x_4)\Big\rangle\ .
\end{align}
We could go further and reduce the two-point function to a sum of one-point functions.
This is however unnecessary, because conformal symmetry will determine the two-point function, so that we can consider it a known quantity.
So, in order to compute correlation functions, all we need to know (in addition to the spectrum $\mathcal{S}$) is the OPE coefficient $C_{\sigma_1,\sigma_2}^{\sigma}(x_1,x_2)$.
This can be determined by solving associativity \eqref{cccc} and commutativity conditions, provided the model has enough symmetry for 
\begin{itemize}
 \item constraining the dependence of the OPE coefficient $C_{\sigma_1,\sigma_2}^{\sigma}(x_1,x_2)$ on $(x_1,x_2)$, and
\item decomposing the spectrum $\mathcal{S}$ into a reasonable number of representations.
\end{itemize}
These conditions are fulfilled in certain conformal field theories.
In particular, as we will see in Chapter \ref{secccs}, conformal symmetry is enough for determining the $x$-dependence of OPE coefficients. 

\subsubsection{Fusion rules}

Let us discuss in more detail how symmetry constrains OPEs. By definition, symmetry transformations act within representations of the symmetry algebras. Given three representations $(\mathcal{R}_1,\mathcal{R}_2,\mathcal{R}_3)$, let us write states in these representations as $\sigma=(\mathcal{R},v)$ as in Eq. \eqref{arv}: symmetry constraints on the OPE coefficients $C_{(\mathcal{R}_1,v_1),(\mathcal{R}_2,v_2)}^{(\mathcal{R}_3,v_3)}(x_1,x_2)$ are linear equations on functions of $(v_1,v_2,v_3,x_1,x_2)$. The number $N_{\mathcal{R}_1\mathcal{R}_2}^{\mathcal{R}_3}\in {\mathbb{N}} \cup \{\infty\}$ of independent solutions of these equations is the \textbf{\boldmath fusion multiplicity}\index{fusion!---multiplicity}. 
In particular, the field $V_{\sigma_3}$ can appear in a $V_{\sigma_1}V_{\sigma_2}$ OPE only if $N_{\mathcal{R}_1\mathcal{R}_2}^{\mathcal{R}_3}\neq 0$. Let us define the \textbf{\boldmath fusion product}\index{fusion!---product} or \textbf{\boldmath fusion rules}\index{fusion!---rules} for representations of the symmetry algebra as 
 \begin{align}
 \mathcal{R}_1 \times \mathcal{R}_2  = \sum_{\mathcal{R}_3} N_{\mathcal{R}_1\mathcal{R}_2}^{\mathcal{R}_3} \mathcal{R}_3 \ . 
\label{rrnr}
\end{align}
From the OPE, the fusion product inherits the properties of bilinearity, commutativity and associativity. 
In some cases, the fusion product has a constructive definition as a kind of generalization of the tensor product \cite{gab99, kr18}: then $ \mathcal{R}_1 \times \mathcal{R}_2$ is a representation, and the equality in Eq. \eqref{rrnr} is an isomorphism of representations. (We will not need such constructions.) The fusion product, in other words the structure of the category of representations, will turn out to play a more important role than the structure of the representations themselves, see Section \ref{secfr}.  

The fusion product $\mathcal{R}_1\times \mathcal{R}_2$ is said to have nontrivial multiplicities if $N_{\mathcal{R}_1\mathcal{R}_2}^{\mathcal{R}_3}\geq 2$ for some $\mathcal{R}_3$, and trivial multiplicities if $N_{\mathcal{R}_1\mathcal{R}_2}^{\mathcal{R}_3}\in \{0,1\}$.
A field is called a \textbf{\boldmath simple current}\index{simple current} if the corresponding representation $\mathcal{R}$ is such that $\mathcal{R}\times \mathcal{R}'$ is indecomposable for any indecomposable $\mathcal{R}'$, so that $\sum_{\mathcal{R}''} N_{\mathcal{R}\mathcal{R}'}^{\mathcal{R}''} =1$. 

If all fusion multiplicities are trivial, then the OPE coefficients  $C_{(\mathcal{R}_1,v_1),(\mathcal{R}_2,v_2)}^{(\mathcal{R}_3,v_3)}(x_1,x_2)$ are determined by symmetry up to overall factors that depend on the representations $\mathcal{R}_i$, but not on $v_i,x_j$. In the 
associativity equation \eqref{cccc}, the sums over the states $\sigma_s,\sigma_t$  can then be reduced to sums over the corresponding representations $\mathcal{R}_s,\mathcal{R}_t$, since everything that involves $v_s,v_t$ is determined by symmetry. 
If the symmetry algebra is large enough or the model simple enough, the number of representations in the spectrum can be small enough for the associativity equation to become tractable.
To summarize,
\begin{center}
\fbox{
\begin{minipage}{0.9\textwidth}
OPE coefficients obey linear equations from symmetry and quadratic equations \eqref{cccc} from consistency.
The numbers of OPE coefficients, consistency equations, and terms in these equations, are determined by the dimension of the spectrum and are therefore in general very large.
With the help of symmetry, these numbers can be reduced to the number of representations in the spectrum.
\end{minipage}
}
\end{center}

\subsection{So what is the Lagrangian of your model? \label{seclagr}}

In quantum field theory, the bootstrap approach is less widely used than the Lagrangian method, which is based on functional integrals over spaces of fields.
In two-dimensional conformal field theory, however, the bootstrap approach is particularly effective, while the Lagrangian method is often needlessly complicated, lacking in rigor, or downright not applicable. 
There are nevertheless cases when the Lagrangian method is useful, in particular for 
suggesting, and sometimes proving, the existence of conformal field theories. 

In the Lagrangian method, correlation functions are represented as functional integrals
\begin{align}
 \left\langle\prod_{i=1}^n V_{\sigma_i}(x_i)\right\rangle = \int D\phi\ e^{-\int dx L[\phi](x)} \ \prod_{i=1}^n \tilde{V}_{\sigma_i}[\phi](x_i)\ ,
\end{align}
where  
\begin{itemize}
 \item the integration variable is the fundamental field (or set of fields) $\phi(x)$,
\item the \textbf{\boldmath Lagrangian}\index{Lagrangian} $L[\phi](x)$ of the model is a functional of $\phi(x)$,
\item the functional $\tilde{V}_{\sigma_i}[\phi](x_i)$ of $\phi(x)$ corresponds to the field $ V_{\sigma_i}(x_i)$,
\item the integration measure $D\phi$ is usually characterized by axioms on functional integrals. 
\end{itemize}
An important advantage of the Lagrangian method is the existence of the Lagrangian itself, a relatively simple object that is not tied to any specific correlation function and encodes much information about the model.
The Lagrangian is the most common, but not the only object of this type: in certain integrable models or supersymmetric quantum field theories, much information is encoded in the geometry of a spectral curve.

In the Lagrangian method, we are thus in principle given all the correlation functions of a specific model from the start.
What we do not a priori know are the symmetry properties of the model.
Of course, we may try to choose the Lagrangian in order to ensure that certain symmetries are present.
However, the symmetries of the model also depend on the integration measure.
Axioms requiring functional integrals to behave as ordinary integrals often leave no choice for the integration measure, and 
it can happen that a transformation leaves the Lagrangian invariant, but changes the integration measure.
This transformation is then called anomalous, and is not a symmetry of the model.

Such an anomaly can also be described as a symmetry of the classical theory that is not present in the corresponding quantum theory.
This is because the Lagrangian method can be interpreted as the quantization of the classical theory that is defined by the Lagrangian.
The classical theory consists in the dynamics of the fundamental field, subject to the equations of motion associated with the Lagrangian, 
\begin{align}
 \frac{\delta }{\delta \phi(x')} \int dx L[\phi](x) = 0 \ .
\end{align}
The integration measure then contains the additional information that is necessary for building a quantum theory from the classical theory. 

This contrasts with the bootstrap approach, which does not assume a classical theory to exist.
Models constructed with the bootstrap approach may nevertheless have one or several classical limits. (See Section \ref{seclld} for the case of Liouville theory.)

\section{Conformal symmetry in two dimensions \label{secconf}}

Our subject of two-dimensional \textbf{conformal field theory}\index{conformal field theory} is defined by the existence of a space-time symmetry, namely local conformal invariance. 
We consider a two-dimensional space, and assume that it comes equipped with a metric.
The metric plays very little role in conformal field theory, because conformal invariance will allow us to fix it once and for all.
But we need the metric in order to explain what conformal invariance is.
We assume that the metric has Euclidean signature. CFT in Minkowski space is related to CFT in Euclidean space by Wick rotation, which translates the Euclidean CFT axioms into good properties of the Minkowskian CFT, including the Wightman axioms \cite{kqr21}.

\subsection{Global conformal transformations\label{secglob}}

Let us assume that the metric is the flat two-dimensional Euclidean metric, which we write in terms of complex coordinates $x=(z,\bar{z})$,
\begin{align}
 ds^2 = dz d\bar{z} \ .
\end{align}
A transformation $(z,\bar{z}) \mapsto (f(z),\overline{f(z)})$ leaves the metric invariant, provided $ds^2 = dfd\bar{f}$.  
Our metric is invariant under the familiar \textbf{\boldmath translations}\index{translation} and \textbf{\boldmath rotations}\index{rotation},
\begin{align}
f_\text{translation}(z) &= z+b \ ,
\\ 
f_\text{rotation}(z) &=a z \ , \ |a|=1\ ,
\end{align}
where $a$ and $b$ are complex constants.
Moreover, if we lift the restriction $|a|=1$ in rotations, we obtain transformations that do not leave the metric invariant, but rescale it by a factor of $|a|^2$.
For $a\in{\mathbb{R}}$, these transformations are called \textbf{\boldmath dilations}\index{dilation} or scale transformations,
\begin{align}
 f_\text{dilation}(z) &= a z\ , \ a\in {\mathbb{R}}\ .
\end{align}
While the laws of physics are (mostly) invariant under translations and rotations, they are not invariant under dilations, and the structures of matter and of the universe are strongly scale-dependent.
It is nevertheless interesting to study scale-invariant models, for at least three reasons:
\begin{enumerate}
 \item Some particular systems are invariant under scale transformations, for instance certain materials at phase transitions.
 \item By rescaling a quantum field theory or statistical model to very small or large scales in a controlled way, one often reaches a scale-invariant theory (called a fixed point of the renormalization group flow).
That new theory can be nontrivial, and helpful for understanding the original theory.
\item Moreover, many different models can lead to the same scale-invariant theory: they form a universality class. Scale-invariant theories are therefore much fewer and more general than finite-scale models.
\end{enumerate}

It may seem reasonable to study quantum field theories that are invariant under translations, rotations and dilations. 
Surprisingly, many interesting systems have much more symmetry, which makes them accessible to the powerful methods that are the subject of this text.
In the absence of physically compelling reasons for these further symmetries, we will try to provide formal justifications.

A first formal consideration is that translations, rotations and dilations are all \textbf{\boldmath conformal transformations}\index{conformal transformation}, i.e. transformations that preserve angles, and therefore that rescale the metric by a real factor.
But they are not the only conformal transformations: another one is the inversion
\begin{align}
 f_\text{inversion}(z) = \frac{1}{z}\ .
\end{align}
It may seem unwise to take this transformation as a symmetry, as it is singular at $z=0$.
This is however not a problem, if we enlarge our space with the addition of a point at $z=\infty$.
This actually amounts to compactifying the space, and working not on the complex plane but on the \textbf{\boldmath Riemann sphere}\index{Riemann!---sphere}.
This does not even prevent us from studying models on the complex plane, as such models are equivalent to models on the sphere, with one field inserted at $z=\infty$.
Combining the inversion with translations, rotations and dilations, we obtain the group of the \textbf{\boldmath global conformal transformations}\index{conformal transformation!global---} of the Riemann sphere,
\begin{align}
 \boxed{f_\text{global conformal}(z)  = \frac{az+b}{cz+d}}\ , \quad (a,b,c,d\in\mathbb{C},\ ad-bc\neq 0)\ .
\end{align}
This group is isomorphic to the group $PGL_2({\mathbb{C}})=\frac{GL_2({\mathbb{C}})}{\mathbb{C}^*}$ of the complex matrices $g$ of size $2$ modulo the relations $g\sim \lambda g\ (\forall \lambda\in\mathbb{C}^*)$, via the map
\begin{align}
 g = \left(\begin{array}{cc} a & b \\ c & d \end{array}\right) \in GL_2({\mathbb{C}}) \quad \longmapsto\quad f_g(z) = \frac{az+b}{cz+d}\ .
\label{gisl}
\end{align}
This group can alternatively be written as $PSL_2(\mathbb{C}) = \frac{SL_2(\mathbb{C})}{\mathbb{Z}_2}$.
(See Exercise \ref{exoiso}.)

\subsection{Local conformal transformations \label{secloc}}

\subsubsection{Holomorphic functions and their singularities}

Actually, any holomorphic function $h(z)$ defines a conformal transformation,
\begin{align}
 f_\text{local conformal}(z) = h(z) \quad , \quad h(z)\ \text{holomorphic}\ , 
\label{flc}
\end{align}
which transforms the metric into $dfd\bar{f} = |h'(z)|^2 dzd\bar{z}$, and is therefore angle-preserving.
One may object that the only holomorphic functions on the sphere are the functions $f_g$ \eqref{gisl} that encode the global conformal transformations.
And indeed, any further conformal transformation must have singularities, and cannot be one-to-one.
We could consider transformations that would be defined only on some subset of the Riemann sphere, where there would be no singularities.
However, in the case of infinitesimal transformations, it is possible to interpret a singularity of $h(z)$ at $z=z_0$ as meaning that a field $V_\sigma(z_0)$ transforms nontrivially.
This makes sense not only with a correlation function $\left\langle V_{\sigma}(z_0)\cdots \right\rangle $ that manifestly involves a field at $z=z_0$, 
but also with arbitrary correlation functions.
We can indeed always assume the presence at $z_0$ of a trivial field, called the identity field, which can however become nontrivial after a conformal transformation is performed.
To summarize,
\begin{center}
\fbox{
\begin{minipage}{0.9\textwidth}
Global conformal transformations only move fields around, while local conformal transformations also modify them.
\end{minipage}
}
\end{center}
So, we assume that the symmetries include the infinitely many independent \textbf{\boldmath local conformal transformations}\index{conformal transformation!local---}, in addition to the global conformal transformations.

\subsubsection{The Witt algebra}

For technical simplicity, we will work with the Lie algebra of infinitesimal conformal transformations, rather than with the Lie group of conformal transformations.
The structure of this algebra is worked out by considering transformations close to the identity, 
\begin{align}
 f_{\epsilon}(z) = z + \epsilon(z) \quad \text{with} \quad \epsilon(z) = \sum_{n\in{\mathbb{Z}}} \epsilon_n z^{n+1} \quad 
 \text{where} \quad \epsilon_n\in\mathbb{C}\ .
\label{sen}
\end{align}
To each transformation $f_\epsilon$ we associate an infinitesimal differential operator $D_\epsilon$ such that for any function $F(z)$ we have 
\begin{align}
 F(f_\epsilon(z)) = \left(1- D_\epsilon -\bar{D}_\epsilon \right)F(z) + O(\epsilon^2)\ ,
\label{ffe}
\end{align}
and we find
\begin{align}
 D_\epsilon = \sum_{n\in {\mathbb{Z}}} \epsilon_n\ell_n  \quad \text{where} \quad \boxed{\ell_n= -z^{n+1}{\frac{\partial}{\partial z}}} \ .
\label{elln}
\end{align}
The differential operators $(\ell_n)_{n\in {\mathbb{Z}}}$ generate the \textbf{\boldmath Witt algebra}\index{Witt algebra}, whose commutation relations are
\begin{align}
 \boxed{[\ell_n,\ell_m]=(n-m)\ell_{n+m}}\ .
\end{align}
Differential operators $D_\epsilon +\bar{D}_\epsilon$ that correspond to local conformal transformations are linear combinations (with real coefficients) of the operators
\begin{align}
 \ell_n + \bar\ell_n \quad , \quad i(\ell_n-\bar\ell_n) \ .
 \label{ilml}
\end{align}
Generators with $n\in\{-1,0,1\}$ correspond to global transformations, and the rest to local transformations. (See Exercise \ref{exomoz}.) 

\subsubsection{Consequences of local conformal symmetry}

The assumption of local conformal symmetry drastically restricts the dependence of the theory on the metric.
Let us restrict our attention to theories on compact Riemann surfaces, where a \textbf{\boldmath Riemann surface}\index{Riemann!---surface} is a two-dimensional orientable smooth manifold.
The topology of a compact Riemann surface is characterized by a natural number $g$ called the genus, which is the number of holes.
In the case $g=0$, the uniformization theorem states that all simply connected compact Riemann surfaces are conformally equivalent. 
This implies that the observables of a conformal field theory on any such manifold can be obtained from their values on the sphere by a change of coordinates.
In the next simplest case $g=1$, manifolds are not all conformally equivalent, but any manifold with $g=1$ is conformally equivalent to a torus $\frac{{\mathbb{C}}}{{\mathbb{Z}}+\tau{\mathbb{Z}}}$ for some value of the
complex structure modulus $\tau \in {\mathbb{C}}$.
Similarly, for $g\geq 2$, a conformal field theory depends on the metric only through $3g-3$ complex structure moduli. 

Local conformal symmetry is a natural symmetry assumption in string theory, where the world-sheet metric is an unphysical variable, and in two-dimensional gravity, where we expect background independence.
The assumption is less natural in models of condensed-matter physics, where only flat metrics are usually considered.
In any case, theories with local conformal symmetry, which we will simply call conformal field theories, should be thought of as exceptional points in the vast and unchartered space of the \textbf{\boldmath global conformal field theories}\index{conformal field theory!global---}.
We will encounter a global conformal field theory in Section \ref{seclld}, namely the light asymptotic limit of Liouville theory.
Another example is the long-range Ising model.

\subsection{The Virasoro algebra \label{secvir}}

We have studied the algebra of local conformal transformations, which acts on the geometry. 
We will now complexify and centrally extend that algebra, in order to obtain the symmetry algebra of conformal field theory, which acts on the spectrum. 
In quantum theories, symmetry algebras should be complex because spectrums are complex vector spaces. And symmetry algebras should have central extensions, because the action of a symmetry group on space, which obeys $g\cdot(g'\cdot x) = (gg')\cdot x$, needs only translate into a projective action on states, which obeys $g\cdot(g'\cdot|\sigma\rangle) = \lambda(g,g') (gg')\cdot |\sigma\rangle$ for some scalar factor $\lambda(g,g')$. 
But a projective action of a symmetry algebra is equivalent to an action of the corresponding centrally-extended algebra. 

Complexifying the algebra of local conformal transformations amounts to taking complex linear combinations of the generators \eqref{ilml}. So the complexified algebra has the complex basis $(\ell_n,\bar\ell_n)_{n\in\mathbb{Z}}$, and is the product of two commuting Witt algebras. Elements of the complexified algebras do not act on our complex plane $\mathbb{C}=\mathbb{R}^2$, where $z$ and $\bar z$ are conjugates of one another, but on the complexified complex plane ${\mathbb{C}}^2$, where $z$ and $\bar z$ are independent coordinates. 
We must however insist that the correlation functions, which are the $z$-dependent observables of the theory, do live on $\mathbb{C}$ and not on ${\mathbb{C}}^2$.
Differential operators that act on functions on ${\mathbb{C}}^2$, for example $\frac{\partial}{\partial z}$, can appear at intermediate steps of calculations, whereas correlation functions only involve functions that are well-defined on $\mathbb{C}$, for example $\sqrt{|z|}$ but not $\sqrt{z}$. 

The central extension of the Witt algebra is the 
\textbf{\boldmath Virasoro algebra}\index{Virasoro!---algebra} $\mathfrak{V}$.
Its generators are $(L_n)_{n\in {\mathbb{Z}}}$, and its commutation relations are 
\begin{align}
 \boxed{[L_n,L_m]=(n-m)L_{n+m} + \frac{c}{12} (n-1)n(n+1) \delta_{n+m,0}}\ .
\label{vir}
\end{align}
Here the \textbf{\boldmath central charge}\index{central charge} $c$ is in principle a central generator, i.e. a generator that commutes with all $L_n$. However, this generator will always be proportional to the identity when acting on the spectrum of a given CFT: we then identify it with its eigenvalue, and consider $c$ as a complex number that characterizes the CFT. 
The Virasoro algebra is the only central extension of the Witt algebra up to trivial redefinitions. (See Exercise \ref{exovir}.) Notice that the presence of the central term does not affect the generators $L_{-1},L_0,L_1$ of global conformal transformations. 

Therefore, the symmetry algebra of conformal field theory is made of two commuting copies of the Virasoro algebra. It has
\begin{itemize}
 \item \textbf{\boldmath left-moving}\index{left-moving (symmetry generators)}, chiral or holomorphic
generators $L_n$,  and
\item  \textbf{\boldmath right-moving}\index{right-moving (symmetry generators)}, anti-chiral or antiholomorphic generators $\bar{L}_n$,
\end{itemize}
 with
$[L_n,\bar{L}_m]=0$. Assuming invariance under parity, which in our two-dimensional Euclidean space is the transformation $z\to -\bar z$, 
both Virasoro algebras must have the same central charge, which is then called the central charge of the model:
\begin{align}
\renewcommand{\arraystretch}{1.3}
 \begin{tabular}{|c|c|c|c|}
 \hline
Notation &  Name & Generators &  Central\ charge
\\
\hline\hline
 $\mathfrak{V}$ & left-moving & $L_n$ & $c$
\\
\hline
$\overline{\mathfrak{V}}$ & right-moving & $\bar{L}_n$ & $c$
\\
\hline
$\mathfrak{V}\times \overline{\mathfrak{V}}$ & full & $(L_n,\bar{L}_m)$ & $(c,c)$
\\  
\hline
 \end{tabular}
\end{align}

\begin{hyp}[Symmetry algebra of two-dimensional conformal field theory]
 ~\label{ax:sa}
 We have a $\mathfrak{V}\times \overline{\mathfrak{V}}$ symmetry algebra, such that $L_n$ and $\bar{L}_n$ generate conformal transformations. 
\end{hyp} 
\noindent
This axiom is the only link between the symmetry algebra and the two-dimensional space.
Some conformal field theories have several Virasoro symmetry algebras, only one of which corresponds to conformal symmetry.
In the free boson theory, it is even possible to find Virasoro algebras with all possible values of the central charge. (See Section \ref{secaua}.)

\section{Basic consequences of conformal symmetry}

\subsection{Structure of the spectrum \label{secsots}}

\subsubsection{Decomposition into irreducible representations}

We know that the spectrum is a representation of the 
$\mathfrak{V}\times \overline{\mathfrak{V}}$ symmetry algebra.
Let us make further assumptions on its structure.

\begin{hyp}[Structure of the spectrum]
 ~\label{ax:sots}
 The spectrum decomposes into irreducible, factorizable representations of $\mathfrak{V}\times \overline{\mathfrak{V}}$. When acting on the spectrum, $L_0$ and $\bar L_0$ are diagonalizable, and $L_0+\bar L_0$ is bounded from below.
\end{hyp} 
\noindent
Let us comment on the three parts of this axiom:
\begin{itemize}
 \item 
 Factorizable representations are of the type $\mathcal{R}\otimes \bar{\mathcal{R}'}$, where $\mathcal{R}$ and $\mathcal{R}'$ are representations of $\mathfrak{V}$, and the bar in $\overline{\mathfrak{V}}$ and $\bar{\mathcal{R}'}$ distinguishes right-moving from left-moving objects. 
Then the decomposition \eqref{somr} takes the form
\begin{align}
 \boxed{\mathcal{S} = \bigoplus_{(\mathcal{R},\mathcal{R}')\in \text{Rep}(\mathfrak{V})^2} m_{\mathcal{R},\mathcal{R}'} \mathcal{R}\otimes \bar{\mathcal{R}'}}\ ,
\label{sorr}
\end{align}
where $\text{Rep}(\mathfrak{V})$ is some set of irreducible representations of the Virasoro algebra.
% NB: Many, but not all interesting spectrums are of this type: for example, the spectrum of the $GL_{1|1}$ WZW model involves indecomposable, non-factorizable representations \cite{ss05}. -- This example involves an extended symmetry algebra.
\item
The assumption that the spectrum decomposes into irreducible representations excludes indecomposable, reducible representations. This assumption implies that $L_0$ and $\bar L_0$ are diagonalizable, because the direct sum of the eigenspaces of $L_0$ is a subrepresentation.
There exist \textbf{\boldmath logarithmic conformal field theories}\index{conformal field theory!logarithmic---} where the action of $L_0$ on the spectrum is not diagonalizable. 
But we will not consider such theories (except in Exercise \ref{exolog}).
\item 
By $L_0+\bar L_0$ being bounded from below, we mean that the real parts of its eigenvalues are bounded from below, which implies that the real parts of the eigenvalues of $L_0$ are bounded from below in each $\mathcal{R}\in\text{Rep}(\mathfrak{V})$. 
This is motivated by the interpretation of the dilation generator $L_0+\bar L_0$ as the Hamiltonian, if we consider the radial coordinate $|z|$ as the Euclidean time. 
\end{itemize}

Two important special cases are
\begin{itemize}
\item \textbf{\boldmath rational models}\index{rational model}, where the spectrum involves finitely many irreducible representations,
 \item \textbf{\boldmath diagonal models}\index{diagonal!---model}, where the spectrum is of the type $\mathcal{S}=\bigoplus_{\mathcal{R}\in \text{Rep}(\mathfrak{V})} m_\mathcal{R} \mathcal{R}\otimes \bar{\mathcal{R}}$, i.e. $m_{\mathcal{R},\mathcal{R}'}\neq 0 \implies \mathcal{R}=\mathcal{R}'$ in Eq. \eqref{sorr}.
\end{itemize}

\subsubsection{Closure under fusion}

Let us now introduce an axiom about fusion of Virasoro representations. Factorizability of the symmetry algebra and of the representations indeed implies that symmetry constraints on OPEs factorize as well, which allows us to define fusion multiplicities and fusion products for representations of $\mathfrak{V}$ (rather than $\mathfrak{V}\times \overline{\mathfrak{V}}$). 

\begin{hyp}[Closure under fusion]
~\label{ax:cuf}
If the irreducible representations $\mathcal{R}_1$ and $\mathcal{R}_2$ of $\mathfrak{V}$ appear in the spectrum, then any representation $\mathcal{R}_3$ with a non-vanishing fusion multiplicity $N_{\mathcal{R}_1\mathcal{R}_2}^{\mathcal{R}_3}\neq 0$ also appears in the spectrum. 
\end{hyp} 
\noindent
In other words, we assume that if Virasoro symmetry allows a representation of $\mathfrak{V}$ to appear in an OPE, then that representation does appear. This is true in many CFTs, but there are exceptions, such as the $(E_8,A_{30})$ E-series minimal model \cite{fms97}. 
There are two similar-looking axioms that should not be adopted:
\begin{itemize}
 \item Closure under fusion of $\mathfrak{V}\times \overline{\mathfrak{V}}$ representations, instead of $\mathfrak{V}$ representations, would rule out most nontrivial models, and in particular diagonal models. 
\item Closure under tensor product, instead of fusion, would not make sense in field theory, because the definition of fusion involves the space dependence of fields. Actually, 
the tensor product of two representations of $\mathfrak{V}_c$ is a representation of $\mathfrak{V}_{2c}$ -- central charges behave additively in a tensor product. 
\end{itemize}

\subsubsection{Conformal dimensions}

Let us further discuss the roles of the operators $L_0$ and $\bar L_0$. Their eigenvalues are called \textbf{\boldmath conformal dimensions}\index{conformal dimension} or conformal weights. If 
two states $v_1$ and $v_2$ are related by the action of Virasoro generators,
\begin{align}
 v_2 = \left(\prod_i L_{n_i}\right) v_1\ ,
\label{vovt}
\end{align}
and if $v_1$ is an $L_0$-eigenstate with conformal dimension $\Delta_1$, then $v_2$ is also an eigenstate with the conformal dimension
\begin{align}
 \Delta_2 = \Delta_1 - \sum_i n_i\ ,
\label{ddsn}
\end{align}
as a consequence of the Virasoro algebra's commutation relations \eqref{vir}. Therefore, in any indecomposable representation of the Virasoro algebra, all conformal dimensions differ by integers. (See Exercise \ref{exodiffint}.)

Let us illustrate the power of local conformal symmetry by thinking of a model in terms of global conformal symmetry. 
We decompose each Virasoro representation $\mathcal{R}$ into representations of the $\mathfrak{sl}_2$ algebra of global conformal transformations with generators $(L_0,L_1,L_{-1})$,
\begin{align}
 \mathcal{R}=\bigoplus_{n\in{\mathbb{N}}} m_{\mathcal{R},n} \mathcal{D}^{\Delta+n}\ ,
\label{rbd}
\end{align}
where $m_{\mathcal{R},n}\in {\mathbb{N}}$ is a multiplicity, $\Delta$ is the lowest $L_0$-eigenvalue of the indecomposable Virasoro representation $\mathcal{R}$, and $\mathcal{D}^{\Delta+n}$ is a representation of $\mathfrak{sl}_2$ whose lowest $L_0$-eigenvalue is $\Delta+n$.
(See Exercise \ref{exodis}.)
The above decomposition is universal data of the Virasoro algebra, and 
model-dependent information from the point of view of global conformal symmetry.
This shows how constraining local conformal symmetry is, and how much more freedom there is in the spectrum of global conformal field theories.
In particular, in a global conformal field theory, there is no reason for the spectrum to contain series of
$\mathfrak{sl}_2$ representations whose lowest $L_0$-eigenvalues differ by integers. 

\subsubsection{Unitarity}

Let us now define and discuss \textbf{\boldmath unitarity}\index{unitary!---theory} of two-dimensional conformal field theories. While a theory needs not be unitary for being consistent, only unitary theories can have quantum mechanical interpretations. Unitarity indeed means that the squared norms of states are positive, and can therefore be interpreted as probabilities in quantum mechanics. On the other hand, statistical physics can give rise to non-unitary theories. (See \cite{prv18} for more details.) 

Technically, unitarity requires that 
the spectrum is a Hilbert space, i.e. has a positive definite Hermitian form. Let us first discuss the existence of a Hermitian form, positive or not. 
In conformal field theory, this Hermitian form should be compatible with the action of the Virasoro algebra.
This means that the \textbf{Hermitian conjugation}\index{Hermitian conjugation} on $\mathfrak{V}\times\overline{\mathfrak{V}}$ that is induced by the Hermitian form, is an antilinear involution $A\to A^\dagger$ such that $[A,B]^\dagger = -[A^\dagger, B^\dagger]$.
Interpreting the dilation generator $L_0+\bar L_0$ as the Hamiltonian, we assume that it is self-adjoint. It is also natural to assume that $L_0-\bar L_0$ is self-adjoint, since its eigenvalues are conformal spins, which must be half-integer and therefore real. (See Eq. \eqref{dbdz}.) So both $L_0$ and $\bar L_0$ are self-adjoint,
which implies
\begin{align}
 L^\dagger_n = L_{-n} \quad , \quad \bar L_n^\dagger = \bar L_{-n}\ ,
\label{ldn}
\end{align}
and it follows that $c\in \mathbb{R}$. (See Exercise \ref{exolnd}.) 
It also follows that the Hermitian form is compatible with the decomposition \eqref{sorr} of the spectrum as a sum of $\mathfrak{V}\times \overline{\mathfrak{V}}$ representations, in the sense that two different irreducible representations must be orthogonal to each other.

So the positivity of the Hermitian form can be examined within each representation.
A representation of the Virasoro algebra is called \textbf{\boldmath unitary}\index{unitary!---representation} if it has a positive definite Hermitian form such that 
$L^\dagger_n = L_{-n}$, and a model whose spectrum is unitary is itself called unitary. Unitarity implies that the central charge is not only real, but positive, 
\begin{align}
 c \in \mathbb{R}_{\geq 0}\ .
 \label{cpos}
\end{align}
(See Exercise \ref{exocp}.)

\subsection{Conformal bootstrap}

The application of the bootstrap approach to conformal field theories, called the \textbf{conformal bootstrap}\index{conformal bootstrap}, 
is particularly powerful in two dimensions.
This is because the algebra of conformal transformations is infinite-dimensional, 
so that the spectrum can be decomposed into a small number of large representations.
Moreover, since the symmetry algebra factorizes into two copies of the Virasoro algebra, and since the spectrum is a sum of factorizable representations,
symmetry equations factorize into $z$-dependent equations from the left-moving Virasoro algebra, and $\bar{z}$-dependent equations from the right-moving Virasoro algebra.
This leads to  \textbf{\boldmath holomorphic factorization}\index{holomorphic factorization}:
any universal quantity can be factorized into a holomorphic function of $z$, times a holomorphic function of $\bar z$. The holomorphic functions that appear are typically not entire functions, and have singularities such as poles, branch cuts and/or essential singularities, just like the functions $\frac{1}{z},\sqrt{z}, e^{-\frac{1}{z}}$. 
We will see in more detail in Section \ref{secaco} how holomorphic factorization simplifies the condition \eqref{cccc} of OPE associativity.

In two-dimensional conformal field theory, the assumption of the existence of an OPE has a natural generalization: the assumption that on any closed contour $C$, one can insert a decomposition of the identity operator,
\begin{align}
 \mathbf{1} = \sum_{\sigma\in \mathcal{S}} |\sigma \rangle \langle \sigma^*| \ ,
\label{oss}
\end{align}
where the sum runs over a basis of the spectrum, and the linear form $\langle \sigma^*|$ is defined by $\langle \sigma^*|\sigma'\rangle = \delta_{\sigma,\sigma'}$.
% NB: This dual linear form depends on the whole basis, not just on |\sigma\rangle. But to define it we do not need a scalar product.
Let us schematically explain why this implies the existence of an OPE.
Considering a contour $C$ around the positions $x_1,x_2$ of two fields $V_{\sigma_1}(x_1),V_{\sigma_2}(x_2)$, and inserting the decomposition of the identity, we obtain
\begin{align}
V_{\sigma_1}(x_1)V_{\sigma_2}(x_2)  =  \sum_{\sigma\in \mathcal{S}} \Big\langle \sigma^* \Big| V_{\sigma_1}(x_1)V_{\sigma_2}(x_2)  \Big\rangle  V_{\sigma}(x_2) \ .
\end{align}
Here we abused the state-field correspondence, by replacing the field $V_{\sigma_1}(x_1)V_{\sigma_2}(x_2)$ with the corresponding state $|V_{\sigma_1}(x_1)V_{\sigma_2}(x_2)\rangle$, and the state $|\sigma\rangle$ with the corresponding field $V_{\sigma}(x_2)$.
This can be drawn as follows:
\begin{align}
\newcommand{\myoval}[2]{\draw[dashed] (#1, #2 + 2) to [out = 0, in = 0] (#1, #2 - 2); 
			\draw (#1, #2 + 2) to [out = 180, in = 180] (#1, #2 - 2);}
\begin{tikzpicture}[scale = .45, baseline=(current  bounding  box.center)]
  \myoval{0}{0};
  \myoval{4}{0};
  \draw (0, 2) -- (6, 2);
  \draw (0, -2) -- (6, -2);
  \draw[dashed] (0, 2) -- (-2, 2);
  \draw[dashed] (0, -2) -- (-2, -2);
  \draw (6, 2) arc (90 : -90 : 2);
  \node at (2.4, 0) {$C$};
  \filldraw (7.4, 1) circle [radius = 3pt] node [left] {$V_{\sigma_1}$};
  \filldraw (7.4, -1) circle [radius = 3pt] node [left] {$V_{\sigma_2}$};
 \end{tikzpicture}
\quad
= \ \ \sum_{\sigma\in \mathcal{S}} 
\quad
\begin{tikzpicture}[scale = .45, baseline=(current  bounding  box.center)]
  \myoval{0}{0};
  \draw[dashed] (0, 2) -- (-2, 2);
  \draw[dashed] (0, -2) -- (-2, -2);
  \draw (0, 2) -- (2, 2);
  \draw (0, -2) -- (2, -2);
  \draw (2, 2) arc (90 : -90 : 2);
  \filldraw (3.7, 0) circle [radius = 3pt] node [left] {$V_{\sigma^*}$};
  \draw (8, 0) circle (2);
  \filldraw (9.4, 1) circle [radius = 3pt] node [left] {$V_{\sigma_1}$};
  \filldraw (9.4, -1) circle [radius = 3pt] node [left] {$V_{\sigma_2}$};
  \filldraw (6.3, 0) circle [radius = 3pt] node [right] {$V_{\sigma}$};
 \end{tikzpicture}
\end{align}
We thus obtain an OPE whose coefficient $\left\langle \sigma^* \left| V_{\sigma_1}(x_1)V_{\sigma_2}(x_2) \right. \right\rangle $ is essentially a three-point function.
We will see the relation between OPE coefficients and three-point functions in more detail in Section \ref{secope}. 

Inserting the decomposition of the identity \eqref{oss} on a contour $C$ amounts to cutting the space into two pieces, and in each piece replacing the resulting hole with a puncture -- an insertion of a field $V_{\sigma}(x)$ for some $x$.
If we view the space near the contour $C$ as a cylinder, this can be drawn as follows:
\begin{align}
\newcommand{\myoval}[2]{\draw[dashed] (#1, #2 + 2) to [out = 0, in = 0] (#1, #2 - 2); 
			\draw (#1, #2 + 2) to [out = 180, in = 180] (#1, #2 - 2);}
\begin{tikzpicture}[scale = .45, baseline=(current  bounding  box.center)]
  \myoval{0}{0};
  \myoval{4}{0};
  \myoval{8}{0};
  \node at (2.4, 0) {$C$};
  \draw (0, 2) -- (8, 2);
  \draw (0, -2) -- (8, -2);
  \draw[dashed] (0, 2) -- (-2, 2);
  \draw[dashed] (0, -2) -- (-2, -2);
  \draw[dashed] (8, 2) -- (10, 2);
  \draw[dashed] (8, -2) -- (10, -2);
 \end{tikzpicture}
\quad
= \ \ \sum_{\sigma\in \mathcal{S}} 
\quad
\begin{tikzpicture}[scale = .45, baseline=(current  bounding  box.center)]
  \myoval{0}{0};
  \myoval{8}{0};
  \draw[dashed] (0, 2) -- (-2, 2);
  \draw[dashed] (0, -2) -- (-2, -2);
  \draw[dashed] (8, 2) -- (10, 2);
  \draw[dashed] (8, -2) -- (10, -2);
  \draw (1.5, 2) arc (90 : -90 : 2);
  \draw (6.5, -2) arc (90 : -90: -2);
  \draw (0, 2) -- (1.5, 2);
  \draw (0, -2) -- (1.5, -2);
  \draw (6.5, 2) -- (8, 2);
  \draw (6.5, -2) -- (8, -2);
  \filldraw (4.8, 0) circle [radius = 3pt] node [right] {$V_{\sigma}$};
  \filldraw (3.2, 0) circle [radius = 3pt] node [left] {$V_{\sigma^*}$};
\end{tikzpicture}
\label{cscc}
\end{align}
In the case of a topologically trivial contour $C$, this equality amounts to using OPEs for replacing all fields within $C$ with one field $V_\sigma(x)$.
If on the other hand $C$ is a non-contractible cycle, then this equality is nontrivial even in the absence of fields.
If we studied a model on a torus, then we could obtain a nontrivial constraint on the spectrum by applying Eq. \eqref{cscc} to the partition function (the zero-point function) using two topologically inequivalent contours.
The study of constraints of this type is called the \textbf{\boldmath modular bootstrap}\index{modular bootstrap}. 

We will only consider conformal field theories on the sphere, which are consistent if they have commutative, associative OPEs. For consistency on arbitrary Riemann surfaces, we would need to impose the further condition that torus one-point functions are invariant under modular transformations \cite{ms89b}.

\subsection{Sketching the space of simple models \label{secmomo}}

We will now describe some simple models that have been solved.

\subsubsection{Diagonal models with trivial multiplicities}

We first introduce models that are diagonal and such that representations of $\mathfrak{V}\times \overline{\mathfrak{V}}$ have multiplicities no higher than one:
\begin{itemize}
 \item For any value $c\in\mathbb{C}$ of the central charge, Liouville theory
has a diagonal spectrum that is made of a continuum of representations.
Liouville theory is unitary if $c\geq 1$. 
\item A-series minimal models (MM) are diagonal, rational models that exist for central charges of the type
\begin{align}
 c_{p,q} = 1 - 6 \frac{(p-q)^2}{pq}  \quad \text{with} \quad \left\{\begin{array}{l}  p,q \ \text{coprime  integers}\ , \\ 2\leq p<q\ , \end{array}\right.  
\label{cpq}
\end{align}
in particular $c_{p,q} <1$.
There are $\frac{(p-1)(q-1)}{2}$ distinct representations in the spectrum.
The model is unitary if $q=p+1$, then $c_{p,p+1}=1-\frac{6}{p(p+1)}=0,\frac12,\frac{7}{10},\frac45, \frac67,\cdots$.
\item For any  $c\in\mathbb{C}-\{c_{p,q}\}_{p,q\in\mathbb{Z}^*}$, the generalized minimal model (GMM) has a diagonal spectrum that is made of a discrete infinity of representations. (It is not known whether such a model exists for $c=c_{p,q}$.)
\item For any $c\in \{c_{p,q}\}_{p,q\in\mathbb{N}^*}$, there is a Runkel--Watts-type theory (RWT) with the same continuous spectrum as Liouville theory, but different correlation functions. 
\end{itemize}

\subsubsection{Limits}

These models are related by a number of limits, see Section \ref{secltf}. 
We summarize these limits on the following diagram, where we place the models according to whether their spectrums are discrete or continuous, and to their central charges. 
We distinguish $c\leq 1$ from the rest of the complex $c$-plane, because Liouville theory is not analytic on $\{c\leq 1\}$, and because only $c\leq 1$ can be reached from $c\in \{c_{p,q}\}_{p,q\in \mathbb{N}^*}$. Dashed arrows are for limits that hold only for a subset of correlation functions:
\begin{align}
\begin{tikzpicture}[scale = .25, baseline=(current  bounding  box.center)]
\fill[red!50, rounded corners = 10] (-6, -10) rectangle (6, 10);
\fill[blue!15, rounded corners = 10] (7, -10) rectangle (19, 10);
\fill[blue!30, rounded corners = 10] (20, -10) rectangle (32, 10);
\node at (0, 11) {$c=c_{p,q}$};
\node at (13, 11) {$c\leq 1$};
\node at (26, 11) {$c\in \mathbb{C}$};
\node[left] at (-6.5, 7) {Discrete};
\node[left] at (-6.5, -7) {Continuous};
\draw (0, 7) node[draw, fill = white] (mm) {MM};
\draw (13, 7) node[draw, fill = white] (gmm1) {GMM};
\draw (26, 7) node[draw, fill = white] (gmm2) {GMM};
\draw (0, -7) node[draw, fill = white] (rwt) {RWT};
\draw (13, -7) node[draw, fill = white] (clo) {Liouville};
\draw (26, -7) node[draw, fill = white] (liou) {Liouville};
\draw [ultra thick, shorten <= 2mm, shorten >= 2mm, -latex, 
       out = -10, in = -170] (mm) to (gmm1);
\draw [ultra thick, shorten <= 2mm, shorten >= 2mm, -latex, dashed,
       out = 170, in = 10] (gmm1) to (mm);
\draw [ultra thick, shorten <= 2mm, shorten >= 2mm, -latex] (gmm2) to (gmm1);
\draw [ultra thick, shorten <= 2mm, shorten >= 2mm, -latex] (mm) to (clo);
\draw [ultra thick, shorten <= 2mm, shorten >= 2mm, -latex] (mm) to (rwt);
\draw [ultra thick, shorten <= 2mm, shorten >= 2mm, -latex] (rwt) to (clo);
\draw [ultra thick, shorten <= 2mm, shorten >= 2mm, -latex] (gmm1) to (clo);
\draw [ultra thick, shorten <= 2mm, shorten >= 2mm, -latex,
       out = -165, in = -15] (liou) to (rwt);
\draw [ultra thick, shorten <= 2mm, shorten >= 2mm, -latex, dashed] (liou) to (gmm2);
\end{tikzpicture}
\label{lims}
\end{align}
Let us also draw some relevant regions of the $c$-complex plane. We use light blue for $\{c\leq 1\}$, blue for the rest of the plane, and red for the discrete central charges of minimal models (with larger bars for smaller spectrums). We use darker colors for 
unitary Liouville theory and unitary minimal models:
\begin{align}
 \begin{tikzpicture}[baseline=(current  bounding  box.center), scale = .95] 
 \clip (-5,-2.5) rectangle (7,2.5);
 % Liouville theory
 \fill[blue!30] (-5, -3) rectangle (5.5, 3); 
 \fill[blue!15] (-5, -.1) rectangle (1, .1);
 \fill[blue!70] (1, -.1) rectangle (5.5, .1);
  % Minimal models
 \newcommand{\mm}[3]{\draw[ultra thick, #3] ({#1}, -{#2}) -- ({#1}, {#2});}
 \foreach \p in {2,...,14}{
 \foreach \q in {\numexpr\p+2,...,16}{
 \ifthenelse{\q<\numexpr 3*\p}{
 \mm{1-6*(\p-\q)*(\p-\q)/(\p*\q)}{2/(\p*\q)^(.7)}{red!50};
 }}};
 \foreach \p in {2,...,14}{
 \mm{1-6/(\p*(\p+1))}{2/(\p*(\p+1))^(.7)}{red!80!black};
 }
  % Landmarks 
  \draw[-latex] (-5, 0) -- (0, 0) node[above] {$0$} --   (1, 0) node[above] {$1$} 
   -- (6.5,0) node[above] {$c$};
 \end{tikzpicture}
\end{align}

\subsubsection{Other simple models}

\begin{itemize}
 \item Free bosonic theories, whose spectrums may be continuous and diagonal (non-compact case) or discrete and non-diagonal (compactified free bosons), exist for any $c\in {\mathbb{C}}$.
Their symmetry algebra is larger than the Virasoro algebra, and  
the extra symmetry leads to very simple correlation functions. (See Section \ref{secfb}.)
 \item 
(Virasoro) \textbf{minimal models}\index{minimal model} are by definition rational models, and beyond the A-series there exist two other series, called the 
D- and E-series. D- and E-series minimal models are not diagonal, and some of them involve representations of $\mathfrak{V}\times \overline{\mathfrak{V}}$  with multiplicity $2$.
The corresponding values of the central charge are still of the type $c_{p,q}$.
\end{itemize}

\subsubsection{Interpretation of the central charge}

The properties of conformal field theories, starting with their very existence, crucially depend on the value of the central charge $c$ of the underlying Virasoro algebra.
In the case of unitary theories, we observe that the size of the spectrum (as measured by the number of representations) increases with $c$, from the $c=0$ minimal model whose spectrum contains only one representation, to $c\geq 1$ Liouville theory whose spectrum contains infinitely many.
The central charge is actually additive: given two theories with spectrums $\mathcal{S}$ and $\mathcal{S}'$ and central charges $c$ and $c'$, there is a natural product theory with central charge $c+c'$ and spectrum $\mathcal{S}\otimes \mathcal{S}'$.
Moreover, as a result of the modular bootstrap, for any rational theory that is consistent on a torus, 
the behaviour of the number $N(\Delta,\bar{\Delta})$ of states with left and right conformal dimensions $\Delta,\bar{\Delta}$ is given by 
\begin{align}
\text{Cardy's formula:} \qquad \log N(\Delta,\bar{\Delta}) \underset{\Delta,\bar{\Delta}\to \infty}{\sim} \frac{2\pi}{\sqrt{6}} \Big(\sqrt{c_\text{eff} \Delta} + \sqrt{\bar c_\text{eff} \bar\Delta}\Big)\ ,
% NB: Reference: see Wikipedia
% NB: Corrections are logarithmic in \Delta, see Carlip, so writing (\Delta-c/24) instead of \Delta makes little sense.
\label{nds}
\end{align}
where we introduced the effective central charge $c_\text{eff}=c-24\Delta_0$ where $\Delta_0$ is the lowest dimension in the spectrum.
This quantitatively shows how the central charge $c$ controls the size of the spectrum.
It can be checked that Cardy's formula is compatible with the additivity of the central charge. (See Exercise \ref{exoacf}.)

\section{Exercises}

\begin{exo}[Irreducible and indecomposable representations] 
~\label{exoirr}
Consider the algebra with one generator $A$ such that $A^2 = 0$.
Consider a two-dimensional representation $\mathcal{R}$ with a basis $(v_1,v_2)$ such that $Av_1=v_2$.
Write the matrix of $A$ in the basis $(v_1,v_2)$.
Show that $\mathcal{R}$ is indecomposable, but not irreducible. 
\end{exo}

\begin{exo}[Associativity in a tensor category]
 ~\label{exoten}
Consider a set of objects $\mathcal{R}_i$ with two binary operations $\oplus$ (addition) and $\otimes$ (multiplication), such that $\otimes$ is distributive over $\oplus$, and
\begin{align}
 \mathcal{R}_i \otimes \mathcal{R}_j = \bigoplus_k c_{ij}^k \mathcal{R}_k\ ,
\end{align}
where $c_{ij}^k$ is a number.
Assuming the objects $\mathcal{R}_i$ are linearly independent, show that the associativity of the multiplication $\otimes$ amounts to a quadratic condition on the coefficients $c_{ij}^k$,
\begin{align}
 \sum_m c_{ij}^m c_{mk}^\ell = \sum_n c_{in}^\ell c_{jk}^n\ .
\end{align}
\end{exo}

\begin{exo}[Commutativity and OPE coefficients]
 ~\label{exocva}
In this exercise we study how the commutativity of the OPE constrains OPE coefficients.
\begin{enumerate}
 \item By Taylor-expanding $V_\sigma(x_1)$ around $x_2$, show that the commutativity of the OPE amounts to the equation
 \begin{align}
  \sum_{\sigma\in \mathcal{S}} C_{\sigma_1,\sigma_2}^{\sigma}(x_1,x_2) V_{\sigma}(x_2) 
  = \sum_{\sigma\in \mathcal{S}} C_{\sigma_2,\sigma_1}^{\sigma}(x_2,x_1)
  \sum_{n=0}^\infty \frac{(x_1-x_2)^n}{n!}\partial^n
  V_{\sigma}(x_2)\ .
 \end{align}
 \item Assuming that $\sigma = (\sigma_0,m)$ with $m\in\mathbb{N}$ and $V_\sigma(x) = \partial^m V_{(\sigma_0,0)}(x)$, write the commutativity of the OPE in terms of OPE coefficients. 
 \item Redefining the OPE and OPE coefficients as 
 \begin{align}
  V_{\sigma_1}(x_1)V_{\sigma_2}(x_2) = \sum_{\sigma\in \mathcal{S}} C_{\sigma_1,\sigma_2}^{\sigma}(x_1,x_2) V_{\sigma}(\tfrac{x_1+x_2}{2})\ ,
 \end{align}
 show that OPE coefficients now obey a simple commutativity condition, and a complicated associativity condition.
\end{enumerate}
\end{exo}

\begin{exo}[Group of global conformal transformations]
 ~\label{exoiso}
Show that the map \eqref{gisl} is a group morphism whose kernel is the center of $GL_2({\mathbb{C}})$, such that any global conformal transformation has a preimage in $SL_2(\mathbb{C})$.
Deduce that the group of global conformal transformations of the sphere is isomorphic to $PGL_2({\mathbb{C}})$, or equivalently $PSL_2(\mathbb{C}) = \frac{SL_2(\mathbb{C})}{\mathbb{Z}_2}$.
\end{exo}

\begin{exo}[Algebra of infinitesimal global conformal transformations]
 ~\label{exomoz}
Show that the generators \eqref{ilml} with $n\in\{-1,0,1\}$ correspond to infinitesimal global conformal transformations.
To do this, expand the map $f_g$ of Eq. \eqref{gisl} near the identity, following Eq. \eqref{sen}.
\end{exo}

\begin{exo}[Central term of the Virasoro algebra]
~\label{exovir}
 Show that the commutation relations \eqref{vir} of the Virasoro algebra define a Lie algebra, which is the only possible central extension of the Witt algebra up to trivial redefinitions.
To do this, consider commutation relations of the type
\begin{align}
 [L_n,L_m] =(n-m)L_{n+m} + f(n,m) C  \quad , \quad [C,L_n]=0\ ,
\end{align}
where $C$ is a central generator and $f(n,m)$ an arbitrary antisymmetric function.
Show that these commutation relations obey the Jacobi identities if and only if 
\begin{align}
 f(n,m) = \lambda (n^3-n)\delta_{m+n,0} + (n-m)g(n+m)\ ,
\end{align}
for some constant $\lambda$ and function $g(n)$.
Observe that the second term can be absorbed by the redefinition $L_n\mapsto L_n-g(n)C$.
\end{exo}

\begin{exo}[Spectrum of $L_0$ in an indecomposable representation]
 ~\label{exodiffint}
 Let $\mathcal{R}$ be an indecomposable representation of the Virasoro algebra where $L_0$ is diagonalizable. 
Show that the eigenvalues of $L_0$ in  $\mathcal{R}$ differ by integers.
To do this, use the decomposition $\mathcal{R}=\oplus_\Delta \mathcal{R}_\Delta$ where $\mathcal{R}_\Delta$ is the generalized eigenspace associated with the eigenvalue $\Delta$.
Given any value of $\Delta$, show that  $\oplus_{n\in{\mathbb{Z}}} \mathcal{R}_{\Delta+n}$ is a subrepresentation, and conclude. 
\end{exo}

\begin{exo}[Comparing Virasoro and $\mathfrak{sl}_2$ representations]
 ~\label{exodis}
Consider the $\mathfrak{sl}_2$ algebra with generators $(L_0,L_1,L_{-1})$, and a vector $|v\rangle$ such that $L_1|v\rangle = 0 $ and $L_0|v\rangle = \Delta |v\rangle$.
\begin{enumerate}
 \item 
Repeatedly acting with $L_{-1}$ on $|v\rangle$, show that you obtain a representation of $\mathfrak{sl}_2$, where the eigenvalues of $L_0$ belong to $\Delta+{\mathbb{N}}$, and that this representation is irreducible if $\Delta\notin -\frac12 \mathbb{N}$.
\item
If $\mathcal{R}$ is an indecomposable representation of the Virasoro algebra where $L_0$ is diagonalizable with eigenvalues in $\Delta_0+\mathbb{N}$ for $\Delta_0\notin\frac12 \mathbb{Z}$, show that $\mathcal{R}$ can be decomposed into irreducible $\mathfrak{sl}_2$ representations as in Eq. \eqref{rbd}.
\end{enumerate}
\end{exo}

\begin{exo}[Hermitian conjugates of Virasoro generators]
 ~\label{exolnd}
Assuming that there is a Hermitian form such that $L_0^\dagger = L_0$, and that $L_n^\dagger$ belongs to the Virasoro algebra, show that $L_n^\dagger = L_{-n}$. To do this, use the Virasoro commutation relations \eqref{vir}, and the identity $[A,B]^\dagger = -[A^\dagger, B^\dagger]$. Show that $L_n^\dagger = \lambda_n L_{-n}$ for some coefficients $\lambda_n$, and that there is a simple redefinition of $L_n$ such that $\lambda_n=1$. Deduce that the central charge is real.
\end{exo}

\begin{exo}[Positivity of the central charge in a unitary CFT]
 ~\label{exocp}
In a unitary CFT, consider an eigenvector $|\Delta\rangle$ of $L_0$ with the eigenvalue $\Delta$. Compute the square norm of $L_{-n}|\Delta\rangle$ for any $n\in\mathbb{N}$, and deduce that the central charge is positive \eqref{cpos}.
\end{exo}

\begin{exo}[Number of states and additivity of the central charge]
 ~\label{exoacf} 
Let us show that Cardy's formula \eqref{nds} is compatible with the additivity of the central charge.
\begin{enumerate}
 \item Construct the action of the Virasoro algebra on the spectrum $\mathcal{S}\otimes \mathcal{S}'$ of the product theory by 
\begin{align}
 L_n^\text{product} = L_n\otimes 1' + 1\otimes L_n'\ ,
\end{align}
where $L_n$ and $L_n'$ act on $\mathcal{S}$ and $\mathcal{S}'$ respectively.
Check the relation $c^\text{product} = c+c'$, and show $\Delta^\text{product} = \Delta+\Delta'$.
Compute the effective central charges $c^\text{product}_\text{eff}, \bar c^\text{product}_\text{eff}$.
\item 
Assuming that Cardy's formula holds for $\mathcal{S}$ and $\mathcal{S}'$, show that computing the large $\Delta,\bar\Delta$ behaviour of the number of states $N^\text{product}(\Delta,\bar\Delta)$ in the product theory amounts to finding the maximum of the function $f(\Delta') = \sqrt{c_\text{eff}\Delta'} +\sqrt{c'_\text{eff}(\Delta-\Delta')}$.
\item
Conclude that Cardy's formula holds for $\mathcal{S}\otimes \mathcal{S}'$.
\end{enumerate}
\end{exo}

\chapter{From representation theory to conformal blocks \label{secccs}}

This chapter is devoted to studying the linear equations that symmetry constraints impose on correlation functions.
Such equations depend on the action of the symmetry algebra on the fields -- in mathematical terms, on the representations to which the corresponding states belong. 

\section{Representations of the Virasoro algebra \label{secrep}}

\subsection{Highest-weight representations}

\subsubsection{Definition}

According to Axiom \ref{ax:sots}, the spectrum can be decomposed into irreducible Virasoro representations where $L_0$ is bounded from below. 
Let us study the structure of a representation $\mathcal{R}$ of this type. 
Let $|\Delta\rangle$ be an $L_0$-eigenstate for the lowest $L_0$-eigenvalue $\Delta$ in $\mathcal{R}$.
Since acting with $L_{n>0}$ decreases $L_0$-eigenvalues (see Eq. \eqref{ddsn}), $|\Delta\rangle$ must be a \textbf{\boldmath primary state}\index{primary state}, that is
\begin{align}
 \boxed{\left\{\begin{array}{l}  L_{n>0}|\Delta\rangle = 0\ , \\ L_0 |\Delta\rangle = \Delta |\Delta\rangle\ .\end{array}\right. }
\label{lvlv}
\end{align}
Since $\mathcal{R}$ is irreducible, $\mathcal{R}$ coincides with the subrepresentation that is generated by $|\Delta\rangle$. We denote this subrepresentation as $U(\mathfrak{V})|\Delta\rangle$, where $U(\mathfrak{V})$ is the universal enveloping algebra of the Virasoro algebra, i.e. the associative algebra that is generated by the Virasoro generators $L_n$.
Since $|\Delta\rangle$ is a primary state, it is enough to consider the algebra $U(\mathfrak{V}^+)$ of the \textbf{\boldmath creation operators}\index{creation operator} that is generated by the creation modes $\{L_n\}_{n<0}$, 
\begin{align}
 \mathcal{R} = U(\mathfrak{V})|\Delta\rangle =U(\mathfrak{V}^+)|\Delta\rangle\ .
\label{ruv}
\end{align}
(See Exercise \ref{exospan}.)
Representations of this type are called \textbf{\boldmath highest-weight representations}\index{highest-weight representation}.
A state of $\mathcal{R}$ of the type $U|\Delta\rangle$ with $U\in U(\mathfrak{V}^+)$ is called a \textbf{\boldmath descendant state}\index{descendant!---state} if it is linearly independent from $|\Delta\rangle$.

\subsubsection{Structure}

The highest-weight representation $\mathcal{R}$
comes with a natural surjective map
\begin{align}
\begin{array}{cclcl}
 \varphi_\mathcal{R} & : & U(\mathfrak{V}^+) & \rightarrow & \mathcal{R} 
\\
 &  & u & \mapsto & u|\Delta\rangle \ .
\end{array}
\label{pur}
\end{align}
Let us introduce a basis $\mathcal{L}$ of $U(\mathfrak{V}^+)$, parametrized by ordered $p$-uples $(-n_1,\cdots -n_p)$ of decreasing, strictly negative integers:
\begin{align}
\mathcal{L} =   \left\{ L_{-n_1} \cdots L_{-n_p}  \right\}_{1\leq n_1\leq n_2\leq \cdots \leq n_p} \ .
\label{lels}
\end{align}
The natural integer
\begin{align}
 N=\left|L_{-n_1} \cdots L_{-n_p}\right|=\sum_{i=1}^p n_i \ ,
\label{nsn}
\end{align}
is called the \textbf{level}\index{level (of a descendant state)} of the basis element $L_{-n_1} \cdots L_{-n_p}$, and also of the corresponding state $L_{-n_1} \cdots L_{-n_p}|\Delta\rangle$ in a highest-weight representation.
The conformal dimension of such a state is then $\Delta+N$. 

The dimension of the level-$N$ subspace of $U(\mathfrak{V}^+)$ is the number $p(N)$ of partitions of $N$.
The basis of $U(\mathfrak{V}^+)$ up to the level $N=5$ can be plotted as follows: 
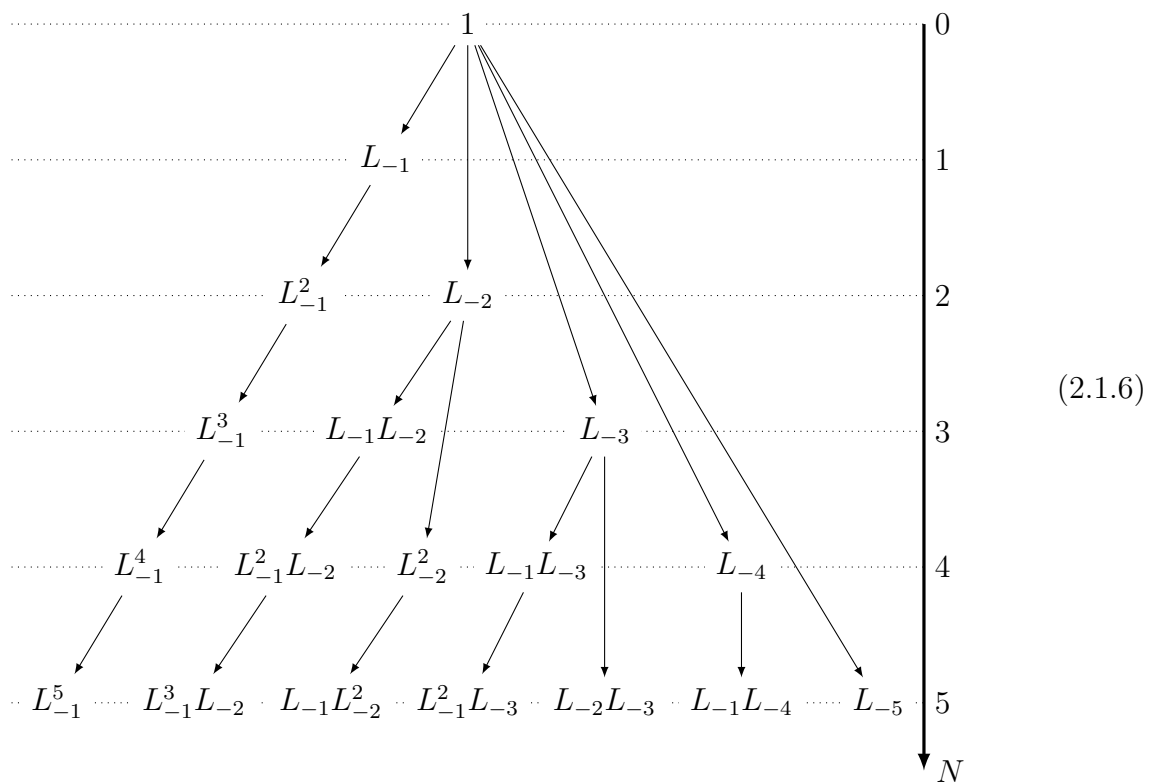
\begin{align}
 \begin{tikzpicture}[scale = .3, baseline=(current  bounding  box.center)]
  \draw[-latex, very thick] (20, 0) -- (20, -33) node [right] {$N$};
  \foreach \x in {0, ..., 5}
  {
  \draw [dotted] (-20, {-6*\x}) -- (20, {-6*\x}) node [right] {${\x}$};
  }
  \node[fill = white] at (0, 0) (0) {$1$};
  \node[fill = white] at (-3.6,-6) (1) {$L_{-1}$};
  \node[fill = white] at (-7.2, -12) (11) {$L_{-1}^2$};
  \node[fill = white] at (-10.8, -18) (111) {$L_{-1}^3$};
  \node[fill = white] at (-14.4, -24) (1111) {$L_{-1}^4$};
  \node[fill = white] at (-18, -30) (11111) {$L_{-1}^5$};
  \node[fill = white] at (0,-12) (2) {$L_{-2}$};
  \node[fill = white] at (-4,-18) (12) {$L_{-1}L_{-2}$};
  \node[fill = white] at (-8,-24) (112) {$L_{-1}^2L_{-2}$};
  \node[fill = white] at (-12,-30) (1112) {$L_{-1}^3L_{-2}$};
  \node[fill = white] at (-2,-24) (22) {$L_{-2}^2$};
  \node[fill = white] at (-6,-30) (122) {$L_{-1}L_{-2}^2$};
  \node[fill = white] at (6,-18) (3) {$L_{-3}$};
  \node[fill = white] at (3,-24) (13) {$L_{-1}L_{-3}$};
  \node[fill = white] at (0,-30) (113) {$L_{-1}^2L_{-3}$};
  \node[fill = white] at (6,-30) (23) {$L_{-2}L_{-3}$};
  \node[fill = white] at (12,-24) (4) {$L_{-4}$};
  \node[fill = white] at (12,-30) (14) {$L_{-1}L_{-4}$};
  \node[fill = white] at (18,-30) (5) {$L_{-5}$};
  \draw[-latex] (0) -- (1);
  \draw[-latex] (1) -- (11);
  \draw[-latex] (11) -- (111);
  \draw[-latex] (111) -- (1111);
  \draw[-latex] (1111) -- (11111);
  \draw[-latex] (0) -- (2);
  \draw[-latex] (0) -- (3);
  \draw[-latex] (0) -- (4);
  \draw[-latex] (0) -- (5);
  \draw[-latex] (2) -- (12);
  \draw[-latex] (12) -- (112);
  \draw[-latex] (112) -- (1112);
  \draw[-latex] (2) -- (22);
  \draw[-latex] (22) -- (122);
  \draw[-latex] (3) -- (13);
  \draw[-latex] (13) -- (113);
  \draw[-latex] (4) -- (14);
  \draw[-latex] (3) -- (23);
 \end{tikzpicture}
\end{align}
In this diagram, each arrow stands for the action of a Virasoro generator $L_{n<0}$ from the left.
Acting with generators that are not depicted on the diagram would produce linear combinations of our basis states, for example $L_{-2}L_{-1} = L_{-1}L_{-2} - L_{-3}$.

Our basis \eqref{lels}, while convenient for enumerating the states, is not distinguished by any particularly useful property.
Other bases can be used, starting with the ``reverse-ordered'' basis $\{ L_{-n_1} \cdots L_{-n_p} \}_{n_1\geq n_2\geq \cdots n_p\geq 1} $.
It can be tempting to use the fact that the two operators $L_{-1},L_{-2}$ algebraically generate $\{L_{n}\}_{n<0}$, and to consider $\{\prod_{i=1}^p L_{-n_i}\}_{n_i\in\{1,2\}}$.
While this set does span $U(\mathfrak{V}^+)$, it however does not provide a basis. (See Exercise \ref{exoot}.)

\subsection{Verma modules and degenerate representations \label{secvm}}

\subsubsection{Definition}

We define the \textbf{\boldmath Verma module}\index{Verma module} $\mathcal{V}_\Delta$ with conformal dimension $\Delta$ as the highest-weight representation that contains a primary state $|\Delta\rangle$ with conformal dimension $\Delta$, and that is linearly isomorphic to $U(\mathfrak{V}^+)$ via the map $\varphi_{\mathcal{V}_\Delta}$ of Eq. \eqref{pur}.
In other words, $\mathcal{V}_\Delta$ is the representation whose basis is given by the states  $\left\{ L_{-n_1} \cdots L_{-n_p}|\Delta\rangle\right\}_{1\leq n_1\leq n_2\leq \cdots n_p}$.
Therefore, $\mathcal{V}_\Delta$ is the largest possible highest-weight representation with lowest conformal dimension $\Delta$. 

Any  highest-weight representation $\mathcal{R}$ that is not a Verma module is called a \textbf{\boldmath degenerate representation}\index{degenerate representation}.
If $\Delta$ is the conformal dimension of the highest-weight state of $\mathcal{R}$, we have a natural surjective morphism of representations from $\mathcal{V}_\Delta$ to $\mathcal{R}$,
\begin{align}
\varphi_\mathcal{R} \varphi_{\mathcal{V}_\Delta}^{-1}\ : \  \mathcal{V}_\Delta\ \rightarrow\ \mathcal{R} \ .
\end{align}
Therefore, $\mathcal{R}$ is a quotient of $\mathcal{V}_\Delta$ by some subrepresentation $\mathcal{R}'$,
\begin{align}
 \mathcal{R} = \frac{\mathcal{V}_\Delta}{\mathcal{R}'}\ .
\label{rvrp}
\end{align}
In other words, a degenerate representation is associated with a nontrivial subrepresentation of a Verma module.
Now, in any nontrivial subrepresentation of a highest-weight representation, the $L_0$ eigenvalues are bounded from below, and there is therefore a primary state $|\chi\rangle$.
This primary state $|\chi\rangle$ is also a descendant -- it is then called a \textbf{\boldmath singular vector}\index{singular vector} or \textbf{\boldmath null vector}\index{null vector} of the highest-weight representation.
While a highest-weight representation is by construction always indecomposable, it is irreducible if and only if it has no singular vectors.

Coming back to our Verma module $\mathcal{V}_\Delta$, and assuming that it has a singular vector $|\chi\rangle$, we can define the degenerate representation $\mathcal{R} = \frac{\mathcal{V}_\Delta}{U(\mathfrak{V}^+)|\chi\rangle}$, which is a quotient of two Verma modules.
If a Verma module has several singular vectors, then it also has several corresponding subrepresentations, and therefore a number of possible quotients, depending on which subrepresentation is chosen as the denominator representation $\mathcal{R}'$ in Eq. \eqref{rvrp}.

\subsubsection{Looking for singular vectors at low levels}

We start at the level $N=1$: is $|\chi\rangle=L_{-1}|\Delta\rangle$ a singular vector? The states $L_n|\chi\rangle$ with $n\geq 2$ automatically vanish because they have negative levels, and we are left with computing
\begin{align}
 L_1|\chi\rangle = L_1 L_{-1}|\Delta\rangle = [L_1,L_{-1}]|\Delta\rangle = 2L_0 |\Delta\rangle = 2\Delta|\Delta\rangle\ .
\end{align}
So the Verma module $\mathcal{V}_\Delta$ has a singular vector at the level $N=1$ if and only if $\Delta = 0$, and in particular $\mathcal{V}_0$ is reducible.
Let us now look for singular vectors at the level $N=2$.
The level-$2$ descendants of a primary state $|\Delta\rangle$ are of the type
\begin{align}
 |\chi\rangle = \left(a_{1,1} L_{-1}^2 + a_2 L_{-2}\right) |\Delta\rangle\ ,
\end{align}
where $a_{1,1}$ and $a_2$ are complex coefficients.
We compute 
\begin{align}
 L_1|\chi\rangle &= \left((4\Delta+2)a_{1,1} + 3a_2\right) L_{-1}|\Delta\rangle\ ,
\\
L_2 |\chi \rangle &= \left(6\Delta a_{1,1}+(4\Delta+\tfrac12 c) a_2\right)|\Delta\rangle\ .
\end{align}
Then $L_1|\chi\rangle=L_2 |\chi \rangle=0$ is a system of two linear equations for the two unknowns $(a_{1,1},a_2)$, whose determinant is 
\begin{align}
 D_2(\Delta) = 4(2\Delta+1)^2 +(c-13)(2\Delta+1) +9\ . 
\label{dud}
\end{align}
Singular vectors at the level $N=2$ exist if and only if
\begin{align}
D_2(\Delta)=0 \quad \iff \quad \Delta = \frac{5-c\pm \sqrt{(c-25)(c-1)}}{16}\ .
\label{dcscc}
\end{align}
As our last explicit example, we look for a singular vector at the level $N=3$,
\begin{align}
 |\chi\rangle = \left(a_{1,1,1} L_{-1}^3 + a_{1,2}L_{-1}L_{-2} + a_3 L_{-3}\right) |\Delta\rangle\ .
\end{align}
The nontrivial relations that $|\chi\rangle$ must obey in order to be a singular vector are $L_1|\chi\rangle =L_2|\chi\rangle= L_3 |\chi\rangle=0$.
However, since $L_3 = [L_2,L_1]$, the relation $L_3|\chi\rangle=0$ is actually redundant, and we need only compute
\begin{align}
 L_1|\chi\rangle &= \Big((6\Delta+6)a_{1,1,1}+3a_{1,2}\Big)L_{-1}^2|\Delta\rangle +\Big((2\Delta+4)a_{1,2}+4a_3\Big)L_{-2}|\Delta\rangle\ ,
\\
L_2|\chi\rangle &= \Big((18\Delta+6)a_{1,1,1}+(4\Delta+\tfrac12c +9)a_{1,2}+5a_3\Big)L_{-1}|\Delta\rangle\ .
\end{align}
This leads to a system of three linear equations for the three unknowns $(a_{1,1,1},a_{1,2},a_3)$, whose determinant is
\begin{align}
 D_3(\Delta) = 12\Big(3(\Delta+1)^2+(c-13)(\Delta+1)+12\Big)\ .
\end{align}
Singular vectors at the level $N=3$ exist if and only if 
\begin{align}
 D_3(\Delta) = 0 \quad \iff \quad \Delta = \frac{7-c\pm\sqrt{(c-25)(c-1)}}{6}\ .
\end{align}

\subsubsection{Alternative notations for $c$ and $\Delta$}

Let us simplify these formulas by introducing alternative notations for $c$ and $\Delta$.
We introduce two new notations for the central charge $c$: the \textbf{\boldmath background charge}\index{background charge} $Q$ and the \textbf{\boldmath coupling constant}\index{coupling constant} $b$,
\begin{align}
\boxed{ c= 1+6Q^2 }\quad , \quad \boxed{Q = b+\frac{1}{b}}\ ,
\label{cqb}
\end{align}
or equivalently
\begin{align}
 b = \sqrt{\frac{c-1}{24}} + \sqrt{\frac{c-25}{24}}\ .
\end{align}
Let us indicate the correspondences between certain values of $c,Q$ and $b$. We pay particular attention to the values $c=1$ and $c=25$, which are the critical points of the map $c(b)$, and play an important role in the properties of Virasoro representations:
\begin{align}
\renewcommand{\arraystretch}{1.7}
 \begin{array}{|l|c||c|c|c|c|c|c|c|}
  \hline
  \text{central\ charge} & c & \mathbb{C} & \leq 1 & 1-6\frac{(p-q)^2}{pq} & 1 & [1,25] & 25 & \geq 25 
\\
\hline
\text{background\ charge} & Q & \mathbb{C} & i{\mathbb{R}} & i\frac{p-q}{\sqrt{pq}} & 0 & [0,2] & 2 & \geq 2 
\\
\hline
\text{coupling\ constant} &
b& \mathbb{C}^* & i{\mathbb{R}} & i\sqrt{\frac{p}{q}} & i & e^{i\mathbb{R}} & 1 & {\mathbb{R}^*}
\\
\hline
 \end{array}
\label{cqbval}
\end{align}
where for each generic value of $c$ we choose among two corresponding values $\pm Q$ for $Q$ and four corresponding values $\pm b^{\pm 1}$ for $b$.
The conformal dimension $\Delta$ can be written in terms of a 
new parameter $P$ called the \textbf{\boldmath momentum}\index{momentum}, such that 
\begin{align}
 \boxed{\Delta(P) = \frac{Q^2}{4} - P^2}\ ,
\label{daq}
\end{align}
which is defined up to the \textbf{\boldmath reflection}\index{reflection}
\begin{align}
 P \mapsto -P\ .
\label{arqa}
\end{align}
We will sometimes write $\mathcal{V}_P=\mathcal{V}_{-P}$ for the Verma module $\mathcal{V}_{\Delta(P)}$ with conformal dimension $\Delta(P)$. In the literature, it is often the quantity $\alpha = \frac{Q}{2}+P$ or $\alpha=\frac{Q}{2}-P$ that is called the momentum, but most formulas are simpler in terms of $P$.

\subsubsection{Singular vectors at all levels}

In terms of the coupling constant $b$, the singular vectors at the levels $N=1,2,3$, and the corresponding conformal dimensions $\Delta$ are 
\begin{align}
\renewcommand{\arraystretch}{1.3}
\begin{array}{|c|c|c|c|}
\hline 
N & \langle r,s\rangle & \Delta_{\langle r,s\rangle} & L_{\langle r,s\rangle} 
\\
\hline\hline
1 & \langle 1,1\rangle & 0  & L_{-1}
\\
\hline
\multirow{2}{*}{2} & 
\langle 2,1\rangle & -\frac12 -\frac{3}{4} b^2  & \frac{1}{b^2}L_{-1}^2 + L_{-2}
\\
\cline{2-4}
& \langle 1,2\rangle & -\frac12 - \frac{3}{4b^2}  & b^2L_{-1}^2 + L_{-2} 
\\
\hline
\multirow{2}{*}{3} &
\langle 3,1 \rangle &  -1 -2 b^2  & \frac{1}{4b^2}L_{-1}^3 + L_{-1}L_{-2}+(b^2-\frac12)L_{-3}
\\
\cline{2-4}
& \langle 1,3 \rangle &  -1 - \frac{2}{b^2}  & \frac14 b^2L_{-1}^3 + L_{-1}L_{-2} + (\frac{1}{b^2}-\frac12)L_{-3}
\\
\hline
\end{array}
\label{lot}
\end{align}
Singular vectors at levels $N\geq 4$ can similarly be computed. (See Exercises \ref{exolf} and \ref{exohl}.) The general result is \cite{fms97}: for any factorization $N=rs$ of $N$ into two positive integers, there is a number $\Delta_{\langle r,s \rangle}$ such that the Verma module $\mathcal{V}_{\Delta_{\langle r,s \rangle}}$ has a singular vector at the level $N$, which we denote as
\begin{align}
 |\chi_{\langle r,s \rangle}\rangle = L_{\langle r,s \rangle} |\Delta_{\langle r,s \rangle}\rangle\ .
\label{lrs}
\end{align}
The general formula for $\Delta_{\langle r,s \rangle}$ is
\begin{align}
 \Delta_{\langle r,s \rangle} = \frac14\left(Q^2-(rb+sb^{-1})^2\right)\ .
\label{drs}
\end{align}
This corresponds to the momemtums $\pm P_{\langle r,s \rangle}$, where
\begin{align}
 \boxed{P_{\langle r,s \rangle} = \frac12\left(rb + sb^{-1}\right) }\ . 
\label{ars}
\end{align}
(For a simple derivation using fusion rules, see Section \ref{secfr}.)
Conversely, any singular vector that is not itself a descendant of another singular vector is of the type $|\chi_{\langle r,s \rangle}\rangle$.
However, singular vectors that are descendants of other singular vectors are not necessarily of this type. (See 
Exercise \ref{exosv}.) We define the \textbf{\boldmath maximally degenerate representation}\index{degenerate representation!maximally---} $\mathcal{R}_{\langle r,s \rangle}$ as the quotient of $\mathcal{V}_{\Delta_{\langle r,s \rangle}}$ by the subrepresentation generated by all the singular vectors. 
Then $\mathcal{R}_{\langle r,s \rangle}$ is irreducible.
For generic values of $c$, $|\chi_{\langle r,s \rangle}\rangle$ is the only singular vector of $\mathcal{V}_{\Delta_{\langle r,s \rangle}}$, and 
\begin{align}
 \mathcal{R}_{\langle r,s\rangle} =\frac{\mathcal{V}_{\Delta_{\langle r,s \rangle}}}{U(\mathfrak{V}^+) |\chi_{\langle r,s \rangle}\rangle }
= \frac{\mathcal{V}_{\Delta_{\langle r,s \rangle}}}{\mathcal{V}_{\Delta_{\langle -r,s \rangle}} }\ ,
\end{align}
where $\Delta_{\langle -r,s \rangle}$ is actually the conformal dimension of $|\chi_{\langle r,s \rangle}\rangle$, due to the identity
\begin{align}
 \Delta_{\langle r,s \rangle} + rs = \Delta_{\langle -r,s \rangle}\ .
\label{dmr}
\end{align}
For particular values of $c$ such that $\mathcal{V}_{\Delta_{\langle r,s \rangle}}$ has more than one singular vector, the structure of $ \mathcal{R}_{\langle r,s\rangle}$ is more complicated, see Exercises \ref{exosv} and \ref{exochar}.

\subsection{Unitarity}\label{secuni}

\subsubsection{Unitarity and singular vectors}

Let us discuss which highest-weight representations of the Virasoro algebra are unitary, starting with Verma modules. Let us assume there is a Hermitian form (also called a scalar product) such that $L_n^\dagger = L_{-n}$ Eq. \eqref{ldn}.
This conjugation rule allows us to deduce scalar products of descendant states, from the square norm of the primary state $|\Delta\rangle$ of a Verma module $\mathcal{V}_\Delta$, which we normalize to $1$ and write  $\langle \Delta|\Delta\rangle =1$. 
For example,
\begin{align}
 \langle L_{-1} \Delta|L_{-1}\Delta\rangle = \langle \Delta |L_{-1}^\dagger L_{-1}|\Delta\rangle =\langle \Delta |L_1 L_{-1}|\Delta\rangle = 2\langle \Delta|L_0|\Delta\rangle = 2\Delta\ . 
\label{levo} 
\end{align}
Since $L_0$ is self-conjugate, different $L_0$-eigenspaces are orthogonal, and $\mathcal V_\Delta$ is unitary if and only if the scalar product is positive definite on each eigenspace. 
Given a basis $\{v_i\}$ of the level-$N$ eigenspace $\mathcal{R}_N$, the scalar product on $\mathcal{R}_N$ is characterized by the Gram matrix $M^{(N)}$ defined by $M^{(N)}_{ij}=\langle v_i|v_j\rangle$, and in particular the scalar product is positive definite if and only if all the eigenvalues of $M^{(N)}$ are strictly positive. This implies in particular $\det M^{(N)}>0$.

It is easy to see that $\det M^{(N)}$ must be a polynomial function of $\Delta$ and $c$.
In order to determine its sign, we should study its zeros. These zeros are determined by the property:
\begin{center}
 \begin{minipage}{0.9\textwidth}
 $\det M^{(N)}=0$ if and only if there is a singular vector at a level $N'\leq N$.  
 \end{minipage}
\end{center}
(See Exercise \ref{exodmn}.) 
We therefore know the zeros $\Delta_{\langle r,s \rangle}$ of $\det M^{(N)}$ as a function of $\Delta$ Eq. \eqref{drs}, and we obtain the \textbf{\boldmath Kac determinant}\index{Kac!---determinant} formula
\begin{align}
 \boxed{\det M^{(N)} \propto \prod_{\begin{smallmatrix} r,s\geq 1 \\ rs \leq N \end{smallmatrix}} (\Delta-\Delta_{\langle r,s \rangle})^{p(N-rs)}} \ ,
\end{align}
where the constant of proportionality is a $\Delta,c$-independent positive number, and $p(N')$ is the number of partitions of $N'$, in particular $p(0)=1$.
The multiplicity of the zero $\Delta_{\langle r,s \rangle}$ is $p(N-rs)$ because this is the number of linearly independent descendants at the level $N$ of a singular vector at the level $rs$.  

\subsubsection{Unitarity of Verma modules and degenerate representations}

According to Eq. \eqref{levo}, the positivity of $M^{(1)}$ implies $\Delta >0$. 
The unitarity of $\mathcal{V}_\Delta$ will now depend on the value of the central charge $c$, which must be real 
as we already saw in Section \ref{secsots}:
\begin{enumerate}
 \item 
If $c > 1$, then $\Delta_{\langle r,s \rangle}$ \eqref{drs} cannot be a real, strictly positive number. 
Therefore, eigenvalues of $\det M^{(N)}$ do not change signs on the half-line $c > 1$.
In the limit $c \to \infty$, the matrix $M^{(N)}$ is positive definite. (See Exercise \ref{exoun}.) 
Therefore, $\mathcal{V}_\Delta$ is unitary. 
\item 
If $c=1$, then the Verma module $\mathcal{V}_\Delta$ is unitary if $\Delta > 0$ by continuity from the $c>1$ case, unless $\Delta=\Delta_{\langle r,s \rangle}$ for some $\langle r,s \rangle$, that is unless $\Delta =\frac14 n^2$ with $n\in {\mathbb{N}}$.
\item 
If $c<1$, then the Verma module $\mathcal{V}_\Delta$ is never unitary. 
This is because for a given $\Delta>0$, the values of $c$ such that $\Delta = \Delta_{\langle r,s \rangle}$ for some $\langle r,s \rangle$ accumulate at $c=1^-$. By tracking sign changes, one can show that $\det M^{(N)}\leq 0$ for some $N$ \cite{fms97}. 
\end{enumerate}

We still have to consider the unitarity of degenerate highest-weight representations of the type $\mathcal{R}=\frac{\mathcal{V}_\Delta}{\mathcal{R}'}$, where the subrepresentation $\mathcal{R}'$ contains at least one singular vector. 
For $\mathcal{R}$ to be unitary, the subrepresentation $\mathcal{R}'$ must include all negative-norm states, that is all eigenstates of Gram matrices with negative eigenvalues.

If $c > 1$, the only degenererate representation with a positive conformal dimension is $\mathcal{R}_{\langle 1,1 \rangle}$, and actually $\Delta_{\langle 1,1 \rangle}=0$. This representation is unitary, because all the zero-norm states in $\mathcal{V}_0 = \underset{\Delta \to 0^+}{\lim} \mathcal{V}_\Delta$ are descendants of the level one singular vector. If $c=1$, all degenerate representations have positive conformal dimensions, and again they are unitary.

If now $c<1$, it turns out that in order to have a unitary quotient, a Verma module must have not only one but actually two independent singular vectors \cite{fms97}. 
Since the existence of a singular vector in $\mathcal{V}_\Delta$ implies a relation between $\Delta$ and the central charge $c$, the existence of two singular vectors implies a constraint on $c$. 
As we will see in Section \ref{secmmf}, the resulting values of $c$ correspond to the minimal models, whose central charges are given in Eq. \eqref{cpq}. 
However, the existence of two singular vectors is a necessary, but not yet a sufficient condition for a Verma module to have a unitary quotient.
The final result is that for $c<1$ unitary quotients exist only if the central charge $c$ takes values that correspond to the unitary minimal models,
\begin{align}
 c_{p,p+1} = 1-\frac{6}{p(p+1)}  \quad \text{with} \quad 2\leq p\ ,
\label{cpp}
\end{align}
For these values of $c$,
the conformal dimensions of Verma modules that have unitary quotients are $\Delta=\Delta_{\langle r,s \rangle}=\Delta_{\langle p-r, p+1-s \rangle}$ where
\begin{align}
   1\leq s\leq r\leq p-1 \ .
\label{srp}
\end{align}
These unitary quotients are the maximally degenerate representations $\mathcal{R}_{\langle r,s \rangle}$, which take the form
\begin{align}
 \mathcal{R}_{\langle r,s \rangle} = \frac{\mathcal{V}_{\Delta_{\langle r,s \rangle}}}{U(\mathfrak{V}^+)|\chi_{\langle r,s \rangle}\rangle + U(\mathfrak{V}^+)|\chi_{\langle p-r,p+1-s \rangle}\rangle}\ ,
\label{rrs}
\end{align}
where the sum of the two subrepresentations is not a direct sum, because their intersection is nonzero. One way to prove that such representations are unitary is to identify them with representations of quotients of affine Lie algebras, in the context of the description of minimal models as quotients of WZW models \cite{fms97}.

These results are summarized in the following table:
\begin{align}
\renewcommand{\arraystretch}{1.3}
 \begin{tabular}{|c|c|c|c|}
  \hline
central charge & $c<1$ & $c=1$ & $c>1$
\\
\hline
$\mathcal{V}_\Delta$ unitary? & no & $\Delta>0$ and $\Delta \neq \frac14 n^2$ & $\Delta>0$
\\
\hline
$\mathcal{R}_{\langle r,s\rangle}$ unitary? & see Eqs. \eqref{cpp} and \eqref{srp} &  yes & $\langle r,s \rangle = \langle 1,1\rangle$
\\
\hline
 \end{tabular}
\end{align}

\section{Fields and correlation functions \label{secfcf}}

Using the state-field correspondence, the action of the Virasoro algebra on states gives rise to a natural action of the Virasoro algebra on fields at any given point $z$. We denote the action of a Virasoro generator as $L_n^{(z)}$ or $L_n$, so that 
\begin{align}
 L_nV_\sigma(z) = L_n^{(z)}V_\sigma(z) = V_{L_n\sigma}(z)\ .
\label{lnzv}
\end{align}
(By definition, $ L_n^{(z_2)} V_{\sigma_1}(z_1)V_{\sigma_2}(z_2) = V_{\sigma_1}(z_1) L_nV_{\sigma_2}(z_2)$.)
Let a \textbf{\boldmath primary field}\index{primary field} be a field that corresponds to a primary state.
A primary field
$V_\Delta(z)$ of conformal dimension $\Delta$ obeys
\begin{align}
\boxed{
 \left\{\begin{array}{l}  L_{n>0} V_\Delta(z) = 0 \ , 
\\
L_0 V_\Delta(z) = \Delta V_\Delta(z) \ .
\end{array}\right. 
}
\label{ldld}
\end{align}
Fields such as  $\left(\prod_{i=1}^p L_{-n_i}\right) V_\Delta(z)$ that correspond to descendants of a primary state are then called  \textbf{descendant fields}\index{descendant!---field}.

Having two Virasoro symmetry algebras $\mathfrak{V}$ and $\overline{\mathfrak{V}}$, we also introduce left and right primary fields $V_{\Delta,\bar\Delta}(z)$, which are primary with respect to both algebras. Then $\Delta$ and $\bar\Delta$ are respectively called the left and right conformal dimensions of $V_{\Delta,\bar\Delta}(z)$. A \textbf{diagonal primary field}\index{diagonal!---primary field} $V_{\Delta,\Delta}(z)$ is a primary field whose left and right dimensions coincide.

In order to solve a model whose spectrum is known, we should compute the correlation functions of primary and descendant fields.
An $N$-point function of primary fields $\left\langle \prod_{i=1}^N V_{\Delta_i,\bar\Delta_i}(z_i) \right\rangle$ is a function of their positions and conformal dimensions.
$N$-point functions of descendant fields are of the type $\left\langle \prod_{i=1}^N \left(\left(\prod_{j=1}^{p_i} L_{-n_{i,j}}\prod_{\bar j=1}^{\bar p_i} \bar L_{-\bar n_{i,\bar j}}\right)V_{\Delta_i,\bar\Delta_i}(z_i)\right) \right\rangle$, with $p_i,\bar p_i\geq 0$ and $n_{i,j},\bar n_{i,\bar j}\geq 1$. 

\begin{hyp}[Single-valuedness of correlation functions]
 ~\label{ax:svcf}
 Correlation functions are single-valued functions, and in particular have trivial monodromies when the fields move around one another.
\end{hyp}
\noindent
This axiom is important enough that it deserves an explicit statement, although strictly speaking it follows from our definition of correlation functions.
Relaxing this axiom can be fruitful, in particular by allowing the existence of parafermionic fields.

\subsection{The energy-momentum tensor \label{secem}}

In order to derive equations that constrain the correlation functions, we will now explain how the action of conformal symmetry is encoded in a field, the energy-momentum tensor.

\subsubsection{Definition}

\begin{hyp}[Dependence of fields on the position]
 ~\label{ax:dvz}
 For any field $V_\sigma(z)$,
 \begin{align}
 \boxed{{\frac{\partial}{\partial z} V_\sigma(z)} = L_{-1} V_\sigma (z) }  \quad \text{and} \quad {\frac{\partial}{\partial \bar z} V_\sigma(z)} = \bar L_{-1} V_\sigma (z)\ ,
\label{lvpv}
\end{align}
consistently with the interpretation of $L_{-1},\bar L_{-1}$ as generators of translations. 
\end{hyp}
\noindent
The action of the left-moving Virasoro algebra will therefore give rise to holomorphic derivatives $\frac{\partial}{\partial z}$, and the solutions of the corresponding differential equations will be holomorphic functions of $z$. The analogous results for the right-moving Virasoro algebra are obtained by $z\to \bar z$ and $\frac{\partial}{\partial z} \to \frac{\partial}{\partial \bar z}$. 

From Axiom \ref{ax:dvz} we can immediately deduce how $L_n^{(z)}$ depends on $z$. Applying the axiom to the two fields $V_\sigma(z)$ and $L_n^{(z)}V_\sigma(z)$, we find $\frac{\partial}{\partial \bar z} L_n^{(z)}=0$ and 
\begin{align}
 \frac{\partial}{\partial z} L_n^{(z)} = [L_{-1},L_n^{(z)}] = -(n+1)L_{n-1}^{(z)}\ .
 \label{pll}
\end{align}
(See Exercise \ref{exowitt} for the corresponding Witt algebra identity.)
This shows that the Virasoro generators at different points $(L_n^{(z_1)})_{n\in \mathbb{Z}}$ and $(L_n^{(z_2)})_{n\in \mathbb{Z}}$ are linearly related, and should be understood of two bases of the same space of symmetry generators. Let us now introduce  a generating function $T(y)$ for $(L_n^{(z)})_{n\in\mathbb{Z}}$, such that the equations that determine the $z$-dependence of $L_n^{(z)}$ become $\frac{\partial}{\partial z} T(y) = \frac{\partial}{\partial \bar z} T(y) = 0$. The appropriate generating function is the \textbf{\boldmath energy-momentum tensor}\index{energy-momentum tensor}
\begin{align}
 \boxed{T(y) = \sum_{n\in{\mathbb{Z}}} \frac{L_n^{(z)}}{(y-z)^{n+2}}}\ .
\label{tsl}
\end{align}
(The series should converge if $y$ is close enough to $z$.) Calling $T(y)$ the energy-momentum tensor is a stretch of terminology which is specific to two-dimensional CFT: when specialized to a two-dimensional space with complex coordinates, the general definition of the energy-momentum tensor actually yields the matrix 
$\left(\begin{smallmatrix} T_{yy} = T(y) & T_{y\bar y} \\ T_{\bar y y} & T_{\bar y \bar y} = \bar T(y) \end{smallmatrix}\right)$, where in particular $\bar T(y)= \sum_{n\in{\mathbb{Z}}} \frac{\bar L_n^{(z)}}{(\bar y-\bar z)^{n+2}}$. 
% (We refrain from using the notation $\bar T(\bar y)$ for a locally antiholomorphic field and reserving $T(y)$ for a holomorphic field, while fields that are neither holomorphic nor antiholomorphic would have to be written as $V_\sigma(z,\bar z)$.)

\subsubsection{Analytic properties}

The energy-momentum tensor encodes the action of the Virasoro algebra at all points $z$. 
From $T(y)$, we can indeed recover the Virasoro generators at any point,
\begin{align}
\boxed{ L_n^{(z)} = \frac{1}{2\pi i}\oint_{z} dy (y-z)^{n+1} T(y)} \ .
\label{lit}
\end{align}
So, for any field $V_\sigma(z)$, we have
\begin{align}
 T(y)V_\sigma(z) = \sum_{n\in{\mathbb{Z}}} \frac{L_n V_\sigma(z)}{(y-z)^{n+2}}\ .
\label{tv}
\end{align}
This is actually a special case of the general OPE \eqref{ope}, where the sum over the spectrum of the theory reduces to a combination of fields of the type of $L_n V_\sigma(z)$, as follows from the definition of $T(y)$ as a combination of symmetry generators.
This OPE simplifies if $V_\sigma(z)$ is a primary field. Using Eqs. \eqref{ldld} and \eqref{lvpv}, we have in this case
\begin{align}
 \boxed{T(y) V_\Delta(z) = \frac{\Delta V_\Delta(z)}{(y-z)^2} + \frac{\partial V_\Delta(z)}{y-z} + O(1)}\ ,
\label{tvp}
\end{align}
where $O(1)$ is regular in the limit $y\to z$.
In many calculations, $T(y)$ is determined by its poles and residues, and the regular term does not contribute.
This is because by definition $T(y)$ is holomorphic at any point $y\in\mathbb{C}$ where no field is present.
Since our field theory lives on the Riemann sphere, $T(y)$ should be holomorphic at $y=\infty$ as well. Let us specify what this means. (See Exercise \ref{exoti} for a justification.)

\begin{hyp}[Holomorphy of the energy-momentum tensor at infinity]
 \label{ax:hti}
 \begin{align}
 \boxed{T(y) \underset{y\to \infty}{=} O\left(\frac{1}{y^4}\right)}\ ,
\label{tyi}
\end{align}
\end{hyp} 
\noindent

\subsubsection{Interpretation}

The commutation relations \eqref{vir} of the Virasoro algebra are equivalent to the following OPE of the field $T(y)$ with itself,
\begin{align}
 \boxed{T(y)T(z) = \frac{\frac{c}{2}}{(y-z)^4} + \frac{2T(z)}{(y-z)^2} + \frac{\partial T(z)}{y-z} + O(1)}\ .
\label{tt}
\end{align}
The proof of this equivalence is sketched in Exercise \ref{exott}. 
The term $\frac{c}{2}$ should be understood as $\frac{c}{2}I$ where $I$ is the
central generator of the Virasoro algebra.
We can consider $I$ as a field called the 
\textbf{\boldmath identity field}\index{identity field}, whose presence does not affect correlation functions, and which is in particular $z$-independent,
\begin{align}
 \left\langle I \cdots \right\rangle = \left\langle \cdots \right\rangle \ .
\label{ivac}
\end{align}
The energy-momentum tensor itself can be seen as a descendant of the identity field. (See Exercise \ref{exoit}.)

The energy-momentum tensor is a \textbf{\boldmath symmetry field}\index{symmetry field}: a field that encodes the action of symmetry transformations, but does not necessarily correspond to a state in the spectrum. 
That field nevertheless obeys the same axioms of commutativity and existence of an OPE, as the fields $V_\sigma(z)$.
More generally, let a \textbf{Virasoro field}\index{Virasoro!---field} be a holomorphic field $T(y)$ that obeys the OPE \eqref{tt}. 
From such a field, we can deduce a Virasoro symmetry algebra via Eq. \eqref{lit}, so $T(y)$ is a symmetry field. However, that Virasoro symmetry may be unrelated to conformal symmetry, in which case it does not obey Axiom \ref{ax:dvz}. 
Most results in Section \ref{secccs} will hold for Virasoro fields and non-conformal Virasoro symmetries, including: Ward identities, nontrivial factors of conformal blocks, and fusion rules. We will however work under Axiom \ref{ax:dvz}, as this simplifies some calculations.
% NB: The right place for dealing with Virasoro fields would be a part on W algebras.

Finally, notice that all our formulas are consistent with dimensional analysis, if we adopt the following dimensions:
\begin{align}
 \boxed{[z]=-1,\quad [T]=2, \quad [L_n]=-n, \quad [V_\Delta]=\Delta}\ .
\label{zaz}
\end{align}
Here we anticipate that the dimension of $V_\Delta$ is given by the eigenvalue of $L_0$ when acting on that primary field.

\subsection{Ward identities \label{secswi}}

We now derive linear equations for correlation functions of primary and/or descendant fields $V_{\sigma_i}(z_i)$, called the Virasoro \textbf{\boldmath Ward identities}\index{Ward identity} or conformal Ward identities.
These identities follow from the properties of the energy-momentum tensor $T(z)$, which translate into the following properties of the correlation function $\left\langle T(z)\prod_{i=1}^N V_{\sigma_i}(z_i) \right\rangle$ as a function of $z$:
\begin{itemize}
 \item it is holomorphic on ${\mathbb{C}}-\{z_1,\cdots z_N\}$,
\item its behaviour at $z=z_i$ is controlled by the OPE \eqref{tv},
\item its behaviour at $z=\infty$ is controlled by Axiom \ref{ax:hti}.
\end{itemize}
For any meromorphic function $\epsilon(z)$, with no poles outside $\{z_1,\cdots z_N\}$, we therefore have
\begin{align}
 \oint_\infty dz\ \epsilon(z) \left\langle T(z)\prod_{i=1}^N V_{\sigma_i}(z_i) \right\rangle = 0 \quad \text{provided} \quad \epsilon(z)\underset{z\to\infty}{=} O(z^2)\ ,
\label{oiz}
\end{align}
where the contour of integration encloses all the points $z_1,\cdots z_N$.
We will distinguish two types of symmetry equations:
\begin{itemize}
 \item \textbf{\boldmath global Ward identities}\index{Ward identity!global---}, which are obtained if $\epsilon(z)$ is holomorphic, thus a polynomial of degree two, 
\item \textbf{\boldmath local Ward identities}\index{Ward identity!local---}, which are obtained if $\epsilon(z) \underset{z\to\infty}{=} O(\frac{1}{z})$, so that $\epsilon(z)$ must have poles.
\end{itemize}
All symmetry equations can be obtained as linear combinations of such local and global Ward identities.

\subsubsection{Local Ward identities}

A spanning set of local Ward identities can be obtained by taking $\epsilon(z) = \frac{1}{(z-z_i)^{n-1}}$ with $n\geq 2$ and $i=1, \cdots , N$ in Eq. \eqref{oiz}. 
Since the integrand has singularities only at $z_1,\cdots z_N$, we have $\oint_\infty=\sum_{i=1}^N \oint_{z_i}$.
Using the 
$TV_\sigma$ OPE \eqref{tv}, we obtain
\begin{align}
 \left\langle \left( L_{-n}^{(z_i)} + (-1)^{n+1}\sum_{j\neq i}\sum_{p=-1}^\infty \frac{\binom{p+n-1}{p+1}}{(z_i-z_j)^{n+p}} L_p^{(z_j)}\right)  \prod_{j=1}^N V_{\sigma_j}(z_j) \right\rangle = 0\ ,
\label{lwi}
\end{align}
where $\binom{p+n-1}{p+1}$ is a binomial coefficient.
The sum over $p$ has finitely many nonzero terms, because $L_p^{(z_j)}V_{\sigma_j}(z_j)=0$ if $p$ exceeds the level $N_j$ of the descendant state $\sigma_j$. 
Using Eq. \eqref{lvpv}, and the fact that $\sigma_j$ is an $L_0$-eigenstate, the terms with $p=-1,0$ involve differential operators, whereas the terms with $p\geq 1$ involve annihilation operators. 
Therefore, the local Ward identity amounts to writing $\left\langle  L_{-n}^{(z_i)} \prod_{j=1}^N V_{\sigma_j}(z_j) \right\rangle$, whose total level is $n+\sum_j N_j$, in terms of correlation functions whose total levels do not exceed $\sum_jN_j$. By induction on the total level,
\begin{center}
\fbox{
\begin{minipage}{0.9\textwidth}
the repeated use of local Ward identities yields an expression for any $N$-point function of descendant fields, as a differential operator of the corresponding $N$-point function of primary fields.
\end{minipage}
}
\end{center}
This result is a consequence of the OPE $T(y)V_\Delta(z)=(\text{differential operator})V_\Delta(z) +O(1) $ \eqref{tvp}. It no longer holds if we enlarge the Virasoro algebra into a W-algebra, or for non-conformal Virasoro symmetry i.e. in the absence of Axiom \ref{ax:dvz}.

If the fields with indices $j\neq i$ are primary, the local Ward identity simplifies to
\begin{align}
 \boxed{\left\langle L_{-n}^{(z_i)}V_{\sigma_i}(z_i)\prod_{j\neq i} V_{\Delta_j}(z_j) \right\rangle =
\sum_{j\neq i} \left(-\frac{1}{z_{ji}^{n-1}} {\frac{\partial}{\partial z_j}}  + \frac{n-1}{z_{ji}^n} \Delta_j\right)
\left\langle V_{\sigma_i}(z_i)\prod_{j\neq i} V_{\Delta_j}(z_j) \right\rangle} \ ,
\label{lmn}
\end{align}
where we assume $n\geq 1$, and we use the notation $z_{ji}=z_j-z_i$. (See Exercise \ref{exodma}.) Actually, inserting $T(z)$ itself in an $N$-point function of primary fields also amounts to acting with a differential operator. This operator is uniquely determined by its poles and residues, which are given by the $TV_\Delta$ OPE \eqref{tvp}, and we find
\begin{align}
 \boxed{\left\langle T(z) \prod_{i=1}^N V_{\Delta_i}(z_i)\right\rangle = \sum_{i=1}^N \left(\frac{\Delta_i}{(z-z_i)^2} + \frac{1}{z-z_i}{\frac{\partial}{\partial z_i}}\right)\left\langle  \prod_{i=1}^N V_{\Delta_i}(z_i)\right\rangle }\ .
\label{dtz}
\end{align}

\subsubsection{Global Ward identities}

Taking $\epsilon(z)\in \{1,z,z^2\}$ in Eq. \eqref{oiz}, we find 
\begin{align}
\left\langle \sum_{i=1}^N L_{-1}^{(z_i)} \prod_{i=1}^N V_{\sigma_i}(z_i) \right\rangle &= 0 \ ,
\label{slz}
\\
\left\langle \sum_{i=1}^N \left( L_0^{(z_i)} + z_i L_{-1}^{(z_i)}\right) \prod_{i=1}^N V_{\sigma_i}(z_i) \right\rangle & = 0 \ ,
\label{sllz}
\\
\left\langle \sum_{i=1}^N \left( L_1^{(z_i)} + 2z_i L_0^{(z_i)} + z_i^2 L_{-1}^{(z_i)}\right) \prod_{i=1}^N V_{\sigma_i}(z_i) \right\rangle & = 0\ .
\label{slllz}
\end{align}

\subsection{Global conformal symmetry}\label{secgcs}

\subsubsection{Infinitesimal global conformal transformations}

We will now study the consequences of the global Ward identities, under the assumption that the fields $V_{\sigma_i}(z_i)$ are primary fields. Since however $L_{n\geq 2}$ do not appear in the global Ward identities, the results will also be valid if the fields $V_{\sigma_i}(z_i)$ are quasi-primary fields, where we define a \textbf{\boldmath quasi-primary field}\index{quasi-primary field} or $\mathfrak{sl}_2$-primary field with conformal dimension $\Delta$ as a field $V_\Delta(z)$ such that 
\begin{align}
 \left\{\begin{array}{l}  L_1 V_\Delta(z) = 0 \ , 
\\
L_0 V_\Delta(z) = \Delta V_\Delta(z) \ .
\end{array}\right. 
\label{lolz}
\end{align}
For example, the energy-momentum tensor $T(z)$ is a quasi-primary field with conformal dimension two, but not a primary field. (See Exercise \eqref{exoit}.) 

Using Axiom \ref{ax:dvz}, the global Ward identities for an $N$-point function $\left\langle \prod_{i=1}^N V_{\Delta_i}(z_i)\right\rangle$ of quasi-primary fields reduce to a system of three differential equations, 
\begin{align}
\forall a \in \{0,+,-\}\quad , \quad 
 \left(\sum_{i=1}^N D_{z_i}^{-\Delta_i}(t^a)\right) \left\langle \prod_{i=1}^N V_{\Delta_i}(z_i)\right\rangle & = 0\ ,
\label{spz}
\end{align}
where we define the differential operators
\begin{align}
\renewcommand{\arraystretch}{1.3}
 \left\{ \begin{array}{rl} D_x^{j}(t^-) & = -{\frac{\partial}{\partial x}}\ ,
\\
D_x^{j}(t^0) &  = x{\frac{\partial}{\partial x}} -j\ ,
\\
D_x^{j}(t^+) & = x^2{\frac{\partial}{\partial x}} - 2j x \ .
\end{array}\right. 
\label{ddz}
\end{align}
% NB: Our convention for the spin is such that finite-dimensional representations have spins 0, 1/2, 1, ... Similarly, our convention for the level is such that the SU_2 WZW models has positive integer level.
These operators form a representation of the Lie algebra \textbf{\boldmath $\mathfrak{sl}_2$} \index{sl2@$\mathfrak{sl}_2$ (Lie algebra)} with the generators $(t^0,t^+,t^-)$ and commutation relations 
\begin{align}
 [t^0,t^\pm ] =\pm t^\pm \quad , \quad [t^+,t^-]=2t^0\ .
\label{ttpm}
\end{align}
The algebra $\mathfrak{sl}_2$ also has a representation in terms of traceless matrices of size two,
\begin{align}
\renewcommand{\arraystretch}{1.3}
\left\{ \begin{array}{rl}
 M(t^+)& = \left(\begin{smallmatrix} 0 & 1 \\ 0& 0 \end{smallmatrix} \right) \ , 
\\
 M(t^0) &= \frac12\left(\begin{smallmatrix} 1 & 0 \\ 0 & -1 \end{smallmatrix}\right)\ ,
\\
M(t^-) & = \left(\begin{smallmatrix} 0 & 0 \\ 1 & 0 \end{smallmatrix} \right) \ .
\end{array} \right.
\label{mta}
\end{align}
Notice that the differential terms of $D^j_x(t^a)$ correspond to the generators $(\ell_{-1},\ell_0,\ell_1)$ \eqref{elln} of global conformal transformations (up to signs).
This shows that global Ward identities encode the covariance of correlation functions under infinitesimal global conformal transformations.

\subsubsection{Finite global conformal transformations}

The global Ward identities for 
finite global conformal transformations \eqref{gisl} are found by exponentiating the $\mathfrak{sl}_2$ matrices $M(t^a)$ into elements of the Lie group $SL_2({\mathbb{C}})$,
\begin{align}
\left\langle \prod_{i=1}^N V_{\Delta_i,\bar\Delta_i}(z_i)\right\rangle  = \left\langle \prod_{i=1}^N T_g V_{\Delta_i,\bar\Delta_i}(z_i) \right\rangle \ , 
\label{vtv}
\end{align}
where we define the image $T_gV_{\Delta,\bar\Delta}(z)$ of a quasi-primary field by
\begin{align}
 \boxed{T_g V_{\Delta,\bar\Delta}(z) = (cz+d)^{-2\Delta} (\bar c\bar z+\bar d)^{-2\bar\Delta}V_{\Delta,\bar\Delta}\left(\frac{az+b}{cz+d}\right)} \quad \text{with} \quad g = \left(\begin{smallmatrix} a & b \\ c & d \end{smallmatrix}\right) \in SL_2({\mathbb{C}})\ .
\label{tgv}
\end{align}
(See Exercise \ref{exoqp} for an interpretation in terms of representations of $\mathfrak{sl}_2$.) Writing this equation for a left and right quasi-primary field $V_{\Delta,\bar\Delta}(z)$ instead of a left quasi-primary field $V_\Delta(z)$ allows us to specify the locally antiholomorphic factor $(\bar c\bar z+\bar d)^{-2\bar\Delta}$, instead of leaving it undetermined.

We can now determine the behaviour of a quasi-primary field $V_{\Delta,\bar\Delta}(z)$ at $z=\infty$, as this is the same as the behaviour of $T_{\left(\begin{smallmatrix} 0 & 1 \\ -1 & 0 \end{smallmatrix}\right)}V_{\Delta,\bar\Delta}(z)= (-z)^{-2\Delta}(-\bar z)^{-2\bar\Delta}V_{\Delta,\bar\Delta}(-\frac{1}{z})$.
Since $V_{\Delta,\bar\Delta}(-\frac{1}{z})$ is smooth at $z=\infty$, we obtain
\begin{align}
 \boxed{V_{\Delta,\bar\Delta}(z) \underset{z\to \infty}{=} O\left(z^{-2\Delta}\bar z^{-2\bar\Delta}\right)}\ .
\label{vdz}
\end{align}
This is consistent with the assumed behaviour \eqref{tyi} of the energy-momentum tensor, which is a quasi-primary field of dimension $\Delta=2$.
Moreover, this suggests that we may define a field at $z=\infty$ by 
\begin{align}
 V_{\Delta,\bar\Delta}(\infty) = \underset{z\to \infty}{\lim} z^{2\Delta}\bar z^{2\bar\Delta} V_{\Delta,\bar\Delta}(z)\ . 
\end{align}
The behaviour \eqref{vdz} of $V_{\Delta,\bar\Delta}(z)$ at $z=\infty$ holds provided no other field is present at $z=\infty$.
To determine the behaviour of a correlation function $\left\langle V_{\Delta,\bar\Delta}(z) V_\sigma(\infty)\cdots \right\rangle$ at $z=\infty$, we would need to know the OPE $V_{\Delta,\bar\Delta}(z) V_{\sigma}(\infty)$.

One may then wonder whether local Ward identities are infinitesimal expressions of the covariance of correlation functions under local conformal transformations \eqref{flc}.
For any holomorphic function $h(z)$, we can define the image of a primary field by
\begin{align}
 T_h V_{\Delta,\bar\Delta}(z) = h'(z)^{\Delta}\overline{h'(z)}^{\bar \Delta}V_{\Delta,\bar\Delta}(h(z))\ ,
\label{thv}
\end{align}
which generalizes the image of a quasi-primary field under a global conformal transformation \eqref{tgv}.
However, this makes sense only in domains where $h'(z)\notin \{0, \infty\}$. On such domains, the analog of Eq. \eqref{vtv} holds. We did not encounter such restrictions in the case of infinitesimal local conformal transformations: for $h(z)$ close to the identity, we have $h'(z)\neq 0$, and a singularity $z_0$ where $h'(z_0)=\infty$ gave rise to contributions of $L_{n\leq -2}$ descendants of the field at $z_0$.

\subsubsection{Solving global Ward identities}

Let us discuss the solutions of the three global Ward identities \eqref{spz} for an $N$-point function of primary fields
$
\left\langle \prod_{i=1}^N V_{\Delta_i}(z_i) \right\rangle\ .
$
Viewing the derivatives $\frac{\partial}{\partial z_i}\left\langle \prod_{i=1}^N V_{\Delta_i}(z_i) \right\rangle$ as $N$ unknowns, the properties of the system depend a lot on the value of $N$:
\begin{itemize}
 \item $\boxed{N=0}$\ : The global Ward identities are trivial ($0=0$).
The zero-point function, which may be called the sphere partition function, is actually a number rather than a function, and carries no information beyond the value of the central charge \cite{car01}.
% NB: Citing Cardy's review on percolation, rather than the ancient, paywalled article by Cardy and Peschel.

\item $\boxed{N=1}$\ :  The global Ward identities amount to $\frac{\partial}{\partial z_1}\left\langle V_{\Delta_1}(z_1)\right\rangle  =\Delta_1 \left\langle V_{\Delta_1}(z_1)\right\rangle =0$.
So the one-point function is constant, and this constant vanishes unless our primary field is degenerate with a level one null vector.

\item $\boxed{N=2}$\ : We still have more Ward identities than unknowns, and we can obtain a condition on $\Delta_i$ by eliminating the derivatives from the Ward identities.
The elegant way to do this is to use $\epsilon(z)=(z-z_1)(z-z_2)$ in Eq. \eqref{oiz}, which leads to 
\begin{align}
 (z_1-z_2)(\Delta_1-\Delta_2)\Big\langle V_{\Delta_1}(z_1)V_{\Delta_2}(z_2)\Big\rangle =0\ .
\end{align}
Assuming $\left\langle V_{\Delta_1}(z_1)V_{\Delta_2}(z_2)\right\rangle\neq 0$, we must have 
\begin{align}
 \Delta_1 = \Delta_2 \ . 
\label{ded}
\end{align}
The remaining Ward identities then amount to 
\begin{align}
 \frac{\partial}{\partial z_1}\Big\langle V_{\Delta_1}(z_1)V_{\Delta_2}(z_2)\Big\rangle = - \frac{\partial}{\partial z_2}\Big\langle V_{\Delta_1}(z_1)V_{\Delta_2}(z_2)\Big\rangle = -\frac{2\Delta_1}{z_1-z_2} \Big\langle V_{\Delta_1}(z_1)V_{\Delta_2}(z_2)\Big\rangle\ .
\end{align}
Using the notation $z_{12}=z_1-z_2$, the solution is
\begin{align}
 \Big\langle V_{\Delta_1}(z_1)V_{\Delta_2}(z_2)\Big\rangle \propto z_{12}^{-2\Delta_1}\ ,
 \label{fzz}
\end{align}
where the proportionality factor is an arbitrary antiholomorphic function. We determine this function by assuming that our fields are not only left primaries but also right primaries, and find
\begin{align}
 \boxed{ \Big\langle V_{\Delta_1,\bar{\Delta}_1}(z_1) V_{\Delta_2,\bar{\Delta}_2}(z_2)\Big\rangle = B_{1}\delta_{\Delta_1,\Delta_2}\delta_{\bar{\Delta}_1,\bar{\Delta}_2} z_{12}^{-2\Delta_1}\bar z_{12}^{-2\bar \Delta_1} }\ .
 \label{eq:2pt}
\end{align}
Here the symbol $\delta_{\Delta_1,\Delta_2}$ is a Kronecker delta or a Dirac delta function, depending on whether the conformal dimensions take discrete or continuous values. We have assumed that the fields have no multiplicities, i.e. two different primary fields cannot have the same dimensions. And we have introduced the \textbf{two-point structure constant}\index{structure constant!two-point---} $B_{1}$, which depends on $(\Delta_1,\bar\Delta_1)$, or equivalently on $(\Delta_2,\bar\Delta_2)$. We could perform a field renormalization such that $B_i=1$, but this would be incompatible with the analyticity of the three-point structure constant in Liouville theory, as we will see in Section \ref{seccbe}.

\item $\boxed{N=3}$\ : The three global Ward identities determine the dependence of three-point functions on the three coordinates $z_i$, without constraining the conformal dimensions $\Delta_i$. Using $\epsilon(z)= (z-z_2)(z-z_3)$ in Eq. \eqref{oiz}, we indeed obtain
\begin{align}
  \left(  \frac{\partial}{\partial z_1} + \frac{\Delta_1+\Delta_2-\Delta_3}{z_1-z_2} +\frac{\Delta_1-\Delta_2+\Delta_3}{z_1-z_3} \right) \Big\langle V_{\Delta_1}(z_1)V_{\Delta_2}(z_2)V_{\Delta_3}(z_3)\Big\rangle= 0\ .
\end{align}
A solution of this equation, and of the analogous equations for the dependences on $z_2$ and $z_3$, is 
\begin{align}
 \boxed{\mathcal{F}^{(3)}(\Delta_1,\Delta_2,\Delta_3|z_1,z_2,z_3) = z_{12}^{\Delta_3-\Delta_1-\Delta_2} z_{23}^{\Delta_1-\Delta_2-\Delta_3} z_{31}^{\Delta_2-\Delta_3-\Delta_1}}\ .
\label{fzzz}
\end{align}
(See Exercise \ref{exolog} for the generalization to logarithmic conformal field theory.)
A three-point function of left and right primary fields is therefore of the type
\begin{align}
 \boxed{ \left\langle \prod_{i=1}^3 V_{\Delta_i,\bar{\Delta}_i}(z_i) \right\rangle = C_{123}
 \left|\mathcal{F}^{(3)}(\Delta_1,\Delta_2,\Delta_3|z_1,z_2,z_3)\right|^2 }\ ,
\label{cff}
\end{align}
where the $z_i$-independent factor
$C_{123}$ is called a \textbf{\boldmath three-point structure constant}\index{structure constant!three-point---}, and the modulus square notation means 
\begin{align}
|f(\Delta,z)|^2 = f(\Delta,z)f(\bar{\Delta},\bar{z})\ .
\label{eq:msn}
\end{align}

\item $\boxed{N\geq 4}$\ : The three differential equations are not enough for controlling the dependence on the $N$ variables $z_i$, and their general solution is 
\begin{align}
 \left\langle\prod_{i=1}^N V_{\Delta_i,\bar\Delta_i}(z_i)\right\rangle = \left|\prod_{i<j} z_{ij}^{\delta_{ij}}\right|^2 F(x_1,x_2,\cdots x_{N-3}) \ ,
\label{xfxn}
\end{align}
where the exponents $\delta_{ij}$ are numbers such that 
\begin{align}
 \sum_{j< i} \delta_{ji} +\sum_{i<j}\delta_{ij} = -2\Delta_i\ ,
 \label{sdd}
\end{align}
and $F(x_1,x_2,\cdots x_{N-3})$ is an arbitrary function of the \textbf{\boldmath cross-ratios}\index{cross-ratio}
\begin{align}
 x_i = \frac{(z_i-z_{N-2})(z_{N-1}-z_N)}{(z_i-z_{N-1})(z_{N-2}-z_N)}\ ,
\end{align}
which are invariant under the global conformal transformations \eqref{gisl}.

Then $F(x_1,x_2,\cdots x_{N-3})$ is related to an $N$-point function with $(z_{N-2},z_{N-1},z_N)=(0,\infty,1)$, as we now illustrate in the case $N=4$.

\item $\boxed{N=4}$\ : In this case there is only one cross-ratio 
\begin{align}
 x=\frac{z_{12}z_{34}}{z_{13}z_{24}}\ ,
\label{xe}
\end{align}
and the general solution is 
\begin{align}
 \left\langle\prod_{i=1}^4 V_{\Delta_i,\bar\Delta_i}(z_i)\right\rangle = \Big| 
 z_{13}^{-2\Delta_1}z_{23}^{\Delta_1-\Delta_2-\Delta_3+\Delta_4}
 z_{34}^{\Delta_1+\Delta_2-\Delta_3-\Delta_4} z_{24}^{-\Delta_1-\Delta_2+\Delta_3-\Delta_4}
 \Big|^2 F(x)\ ,
\label{zgg}
\end{align}
for an arbitrary function $F(x)$.
The exponents in the prefactor are a particular solution of Eq. \eqref{sdd}, chosen so that 
\begin{align}
 F(x) =\Big\langle V_{\Delta_1,\bar\Delta_1}(x)V_{\Delta_2,\bar\Delta_2}(0)V_{\Delta_3,\bar\Delta_3}(\infty) V_{\Delta_4,\bar\Delta_4}(1)\Big\rangle \ .
\label{fx}
\end{align}
Choosing another solution of the $4$ equations for $6$ unknowns $(\delta_{ij})_{i<j}$ would amount to replacing the function $F(x)$ with $x^\lambda (1-x)^\mu F(x)$ for some numbers $\lambda,\mu$. For the behaviour of $F(x)$ under field permutations, see Exercise \ref{exoperm}.
\end{itemize}

\subsubsection{Conformal spins and single-valuedness}

Let us define the \textbf{conformal spin}\index{spin (conformal)} of the field $V_{\Delta,\bar\Delta}(z,\bar z)$ to be the number $S=\Delta-\bar\Delta$. This number determines how that field behaves under rotations, 
\begin{align}
 T_{\left(\begin{smallmatrix} e^{i\theta} & 0 \\ 0 & e^{-i\theta} \end{smallmatrix}\right)} V_{\Delta,\bar{\Delta}}(z) = e^{2i\theta(\Delta-\bar\Delta)} V_{\Delta,\bar{\Delta}}(e^{2i\theta} z)\ .
\end{align}
A primary field is diagonal if and only if its conformal spin is zero.

In order for the three-point function \eqref{cff} to be single-valued as required by Axiom \ref{ax:svcf}, we need $S_1\pm S_2\pm S_3 \in\mathbb{Z}$, equivalently
\begin{align}
 \boxed{S_i \in \frac12{\mathbb{Z}} \qquad \text{and} \qquad S_1+S_2+S_3\in\mathbb{Z}}\ .
\label{dbdz}
\end{align}
Primary fields with spins $S\in\frac12+\mathbb{Z}$ are called fermionic. Allowing fermionic fields would require us to modify our commutativity Axiom \ref{ax:col}, as fermions anticommute. Considering indeed a two-point function \eqref{eq:2pt} of a fermionic field with itself, we have $\left< V(z_1)V(z_2)\right> = -\left<V(z_2)V(z_1)\right>$, which suggests $\{V(z_1),V(z_2)\}=0$. For an exposition of fermionic CFTs, including fermionic extensions of minimal models, see \cite{rw20}.
It is actually possible to relax the single-valuedness axiom, and have fields with spins $S\notin \frac12\mathbb{Z}$, such as parafermionic fields. But we will only consider fields with integer spins $S\in\mathbb{Z}$.

The conformal spin then controls how three-point structure constants behave under permutations. For $\sigma$ a permutation of $(1,2,3)$, we indeed have 
\begin{align}
 \left|\mathcal{F}^{(3)}(\Delta_{\sigma(1)},\Delta_{\sigma(2)},\Delta_{\sigma(3)}|z_{\sigma(1)},z_{\sigma(2)},z_{\sigma(3)})\right|^2 = \operatorname{sign}(\sigma)^{S_1+S_2+S_3} \left|\mathcal{F}^{(3)}(\Delta_1,\Delta_2,\Delta_3|z_1,z_2,z_3)\right|^2 .
\end{align}
Since the three-point function is invariant under permutations of the fields, the structure constant must therefore behave as 
\begin{align}
 \boxed{ C_{\sigma(1)\sigma(2)\sigma(3)} = \operatorname{sign}(\sigma)^{S_1+S_2+S_3} C_{123} }\ .
 \label{css}
\end{align}
For example, if fields $1$ and $2$ are identical, then $C_{123}=C_{213}$, and the three-point function can only be nonzero if the third field has even spin $S_3\in 2\mathbb{Z}$.

\subsection{Operator product expansions \label{secope}}

Let us study how local conformal symmetry constrains OPEs. We start with a generic OPE of the type $V_{\sigma_1}(z_1)V_{\sigma_2}(z_2) = \sum_{\sigma_3} C_{\sigma_1,\sigma_2}^{\sigma_3}(z_1,z_2) V_{\sigma_3}(z_2)$. Let us insert $\oint_C dz (z-z_2)^{n+1} T(z)$ on both sides of the OPE, where the contour $C$ encircles both $z_1$ and $z_2$. Assuming $n\geq -1$, and using the $TV$ OPE \eqref{tv}, this yields the \textbf{OPE Ward identity}\index{Ward identity!---for OPEs}
\begin{align}
 \left(L_n^{(z_2)}+\sum_{m=-1}^{n}\binom{m+1}{n+1} z_{12}^{n-m}L_{m}^{(z_1)}\right)V_{\sigma_1}(z_1)V_{\sigma_2}(z_2) = \sum_{\sigma_3} C_{\sigma_1,\sigma_2}^{\sigma_3}(z_1,z_2) L_n V_{\sigma_3}(z_2)\ .
\end{align}

\subsubsection{Global Ward identities $n\in\{-1,0\}$}

Using Axiom \ref{ax:dvz}, the $n=-1$ identity becomes 
 \begin{align}
  \left(\frac{\partial}{\partial z_1} +\frac{\partial}{\partial z_2}\right) V_{\sigma_1}(z_1)V_{\sigma_2}(z_2) = \sum_{\sigma_3} C_{\sigma_1,\sigma_2}^{\sigma_3}(z_1,z_2) \frac{\partial}{\partial z_2} V_{\sigma_3}(z_2)\ .
 \end{align}
Using the OPE again on the left-hand side, this implies $\left(\frac{\partial}{\partial z_1} +\frac{\partial}{\partial z_2}\right) C_{\sigma_1,\sigma_2}^{\sigma_3}(z_1,z_2)=0$ -- translation invariance of the OPE coefficients. 
Assuming the fields $V_{\sigma_i}$ are  $L_0,\bar L_0$-eigenvectors with the dimensions $(\Delta_{\sigma_i}, \bar\Delta_{\sigma_i})$, the $n=0$ identity becomes 
\begin{align}
 \left(z_{12}\frac{\partial}{\partial z_1} + \Delta_{\sigma_1}+\Delta_{\sigma_2}\right) V_{\sigma_1}(z_1)V_{\sigma_2}(z_2) = \sum_{\sigma_3} C_{\sigma_1,\sigma_2}^{\sigma_3}(z_1,z_2) \Delta_{\sigma_3} V_{\sigma_3}(z_2)\ .
\end{align}
Using the OPE again on the left-hand side, this
determines how OPE coefficients behave under dilations. 
Therefore, the two global Ward identities determine the dependence of OPE coefficients on $z_1,z_2$,
\begin{align}
 C_{\sigma_1,\sigma_2}^{\sigma_3}(z_1,z_2) = C_{\sigma_1,\sigma_2}^{\sigma_3} 
 \left| z_{12}^{\Delta_{{\sigma_3}}-\Delta_{\sigma_1}-\Delta_{\sigma_2}} \right|^2 
 \ ,
 \label{eq:coz}
\end{align}
in agreement with dimensional analysis Eq. \eqref{zaz}.

\subsubsection{Local Ward identities $n\geq 1$}

For simplicity, let us specialize to an OPE of two primary fields $V_{\Delta_1}(z_1)V_{\Delta_2}(z_2)$. 
The left-hand side of the OPE Ward identity reduces to the terms with $L_{-1}^{(z_1)}$ and $L_0^{(z_1)}$, and therefore to $\left(z_{12}^{n+1}\frac{\partial}{\partial z_1} + (n+1)\Delta_1z_{12}^n\right) V_{\Delta_1}(z_1)V_{\Delta_2}(z_2)$. 
Let us generically write the OPE as $V_{\Delta_1}(z_1)V_{\Delta_2}(z_2) = z_{12}^{-\Delta_1-\Delta_2}\mathcal{O}(z_1,z_2)$, where $\mathcal{O}(z_1,z_2)$ is a linear combination of fields at $z_2$, with coefficients that depend on $z_{12}$. In this notation, the $n=0$ global Ward identity is 
\begin{align}
 \left(z_{12}\frac{\partial}{\partial z_1} - L_0^{(z_2)}\right) \mathcal{O}(z_1,z_2) = 0 \ .
\end{align}
Using this identity for eliminating $z_1$-derivatives in the $n\geq 1$ identity, we obtain
\begin{align}
 \Bigg\{ L_n^{(z_2)} - z_{12}^n \left( n\Delta_1-\Delta_2+L_0^{(z_2)}\right)\Bigg\} \mathcal{O}(z_1,z_2) = 0 \ .
 \label{eq:llo}
\end{align}
Let us specialize to the case where $\mathcal{O}(z_1,z_2)$ is a combination of primary fields and their descendants, and focus on the contribution of one primary field with dimension $\Delta_3$:
\begin{align}
 \mathcal{O}_{\Delta_3}(z_1,z_2) = z_{12}^{\Delta_3}\sum_{L\in\mathcal{L}} z_{12}^{|L|} C^{L|\Delta_3\rangle}_{12} LV_{\Delta_3}(z_2)\ , 
\end{align}
where $\mathcal{L}$ is a basis of the space of creation operators $U(\mathfrak{V}^+)$.
In this case, for any level $N\geq n$, we extract the coefficient of $z_{12}^{\Delta_3+N}$ in Eq. \eqref{eq:llo}, and we obtain the local Ward identity
\begin{align}
 \sum_{|L|=N-n} C^{L|\Delta_3\rangle}_{12}(\Delta_3+N-n+n\Delta_1-\Delta_2)L V_{\Delta_3}(z_2)
 = 
 \sum_{|L|=N} C^{L|\Delta_3\rangle}_{12}L_nL V_{\Delta_3}(z_2)\ .
 \label{eq:lwo}
\end{align}
This system of linear equations determines the coefficients $C^{L|\Delta_3\rangle}_{12}$ of descendant fields in terms of the coefficient $C^{|\Delta_3\rangle}_{12}=C^3_{12}$ of the primary field,
\begin{align}
 C^{L|\Delta_3\rangle}_{12} 
 = 
 C^{3}_{12} f^{\Delta_3,L}_{\Delta_1,\Delta_2}\ ,
\end{align}
where  $f^{\Delta_3,L}_{\Delta_1,\Delta_2}$ are universal coefficients such that $f^{\Delta_3,1}_{\Delta_1,\Delta_2} =1$.
Let us show this for $|L|\leq 2$. The cases $(N,n)=(1,1),(2,2),(2,1)$ of \eqref{eq:lwo} respectively yield 
\begin{align}
 \Delta_3 + \Delta_1-\Delta_2 & = 2\Delta_3 f^{L_{-1}} \ ,
 \label{flfo}
 \\
 \Delta_3+ 2\Delta_1-\Delta_2 & = 6\Delta_3 f^{L_{-1}^2} + (4\Delta_3+\tfrac{c}{2})f^{L_{-2}}\ ,
 \\
 (\Delta_3+1+\Delta_1-\Delta_2) f^{L_{-1}} & = 2(2\Delta_3+1) f^{L_{-1}^2} + 3 f^{L_{-2}}\ ,
 \label{flff}
\end{align}
where we temporarily use the notation $f^L=f^{\Delta_3,L}_{\Delta_1,\Delta_2}$.
The first equation determines $f^{L_{-1}}$, unless $\Delta_3=0$. The next two equations determine
$f^{L_{-1}^2}$ and $f^{L_{-2}}$, unless $\Delta_3 \in \{\Delta_{\langle 1,2 \rangle}, \Delta_{\langle 2,1 \rangle}\}$.
More generally, given a value of the level $N\geq 2$, the equations with $n=1,2$ determine $\{f^L\}_{|L|=N}$, unless $\Delta_3\in\{\Delta_{\langle r,s \rangle}\}_{rs=N}$. (See Exercise \ref{exohf}.) If $\Delta_3\in\{\Delta_{\langle r,s \rangle}\}_{r,s\in\mathbb{N}^*}$, then for generic values of $\Delta_1,\Delta_2$ there is no solution for $\{f^L\}_{L\in\mathcal{L}}$, which means that $V_{\Delta_3}$ cannot appear in the OPE $V_{\Delta_1}V_{\Delta_2}$.

\subsubsection{Structure of the OPE}

Let us write the OPE of two left and right primary fields, while neglecting the contributions of descendant fields:
\begin{align}
 V_{\Delta_1,\bar\Delta_1}(z_1) V_{\Delta_2,\bar\Delta_2}(z_2) 
 = 
 \sum_{\Delta_3,\bar\Delta_3} C^{3}_{12}
 \left| z_{12}^{\Delta_3-\Delta_1-\Delta_2}\right|^2 \Big( V_{\Delta_3,\bar\Delta_3}(z_2) + O(z_{12})\Big)\ .
\end{align}
Inserting this in a three-point function, and using the two-point function \eqref{eq:2pt}, this leads to
\begin{align}
 \left<  \prod_{i=1}^3 V_{\Delta_i,\bar\Delta_i}(z_i) \right> 
 = 
 B_3C^3_{12} \left| z_{12}^{\Delta_3-\Delta_1-\Delta_2} \Big( z_{23}^{-2\Delta_3} + O(z_{12}) \Big) \right|^2\ .
\end{align}
Comparing this with the expression \eqref{cff} for the three-point function, we find the expression of  $C^3_{12}$ in terms of the two- and three-point structure constants, 
\begin{align}
 C^3_{12}= (-1)^{S_1-S_2+S_3}\frac{C_{123}}{B_3}\ .
 \label{cftt}
\end{align}
Here we wrote the sign prefactor under the single-valuedness assumption \eqref{dbdz}. Under our further assumption that the spins $S_i$ are integer, the prefactor can be written as $(-1)^{\sum S_i}$.  
Let us now write the complete expression of the OPE of two primary fields:
\begin{align}
 \boxed{ V_{\Delta_1,\bar\Delta_1}(z_1) V_{\Delta_2,\bar\Delta_2}(z_2) 
 = 
 \sum_{\Delta_3,\bar\Delta_3} (-1)^{\sum S_i}\frac{C_{123}}{B_3} \left| z_{12}^{\Delta_3-\Delta_1-\Delta_2}\sum_{L\in\mathcal{L}} z_{12}^Lf^{\Delta_3,L}_{\Delta_1,\Delta_2} L \right|^2 V_{\Delta_3,\bar\Delta_3}(z_2)
 }\ ,
 \label{vvs}
\end{align}
where  the modulus square notation \eqref{eq:msn} means a product of left-moving $z,\Delta,L$ and right-moving $\bar z, \bar \Delta, \bar L$ quantities.

Since OPE Ward identities uniquely determine the contributions of the descendants of a given primary field, fusion multiplicities are trivial in fusion products of highest-weight representations of the Virasoro algebra. By definition, the multiplicity of the Verma module $\mathcal{V}_{\Delta_3}$ in the fusion product $\mathcal{V}_{\Delta_1}\times \mathcal{V}_{\Delta_2}$ indeed coincides with the dimension of the space of solutions of the OPE Ward identities, and that dimension is one or zero.

The case $z_2=\infty$ can be obtained from the case $z_2\in\mathbb{C}$ by a global conformal transformation, and we find 
\begin{align}
 V_{\Delta_1,\bar\Delta_1}(z_1) V_{\Delta_2,\bar\Delta_2}(\infty) =  \sum_{\Delta_3,\bar\Delta_3}(-1)^{\sum S_i}\frac{C_{123}}{B_3}\left| z_1^{\Delta_2-\Delta_1-\Delta_3}\right|^2 \Big( V_{\Delta_3,\bar\Delta_3}(\infty) + O\left(z_1^{-1}\right)\Big) \ .
 \label{iope}
\end{align}
In the special case where there is no field at infinity, so that $\Delta_2=0$ and $\Delta_1=\Delta_3$, this is consistent with the behaviour \eqref{vdz} of $V_{\Delta_1,\bar\Delta_1}(z_1)$ near $z_1=\infty$.

\section{Degenerate fields}\label{sec:degf}

A primary field that has a vanishing descendant is called a \textbf{\boldmath degenerate field}\index{degenerate field}.
In other words, a degenerate field corresponds to the highest-weight state of a degenerate representation.
The degenerate field that corresponds to the highest-weight state of the degenerate representation $\mathcal{R}_{\langle r,s \rangle}$ will be denoted $V_{\langle r,s\rangle}(z)$, and obeys the equation 
\begin{align}
 L_{\langle r,s \rangle} V_{\langle r,s \rangle}(z) = 0 \ , 
\label{lrsv}
\end{align}
where $L_{\langle r,s \rangle}$ is defined in Eq. \eqref{lrs}.
This field differs from the field $V_{\Delta_{\langle r,s \rangle}}(z)$, which corresponds to the highest-weight state of the Verma module $\mathcal{V}_{\Delta_{\langle r,s \rangle}}$, and obeys no such equation.

Inserting Eq. \eqref{lrsv} into correlation functions, we obtain linear equations for correlation functions that involve the degenerate field $V_{\langle r,s\rangle}$. Such linear equations are called \textbf{\boldmath null vector equations}\index{null vector!---equation}. 

\subsection{Fusion rules}\label{secfr}

Let us focus on null vector equations for three-point functions. 
We will show that the null vector equation $\left< L_{\langle r,s \rangle} V_{\langle r,s \rangle} V_{\Delta_2}V_{\Delta_3}\right>=0$ leads to
a constraint on $\Delta_2$ and $\Delta_3$. Using the relation between OPEs and three-point functions, this can be interpreted as the fusion rule of the degenerate representation $\mathcal{R}_{\langle r,s \rangle}$.

\subsubsection{Three-point functions of descendant fields}

Conformal symmetry allows us to write three-point functions of descendant fields as three-point functions of primary fields, times the universal factors
\begin{align}
g^{L}_{\Delta_1,\Delta_2,\Delta_3} = 
 \frac{ \left< L V_{\Delta_1}(0)V_{\Delta_2}(\infty)V_{\Delta_3}(1)\right> }{  \left<  V_{\Delta_1}(0)V_{\Delta_2}(\infty)V_{\Delta_3}(1)\right>}\ .
 \label{glvv}
\end{align}
In the case of $L_{-1}^n$ descendants, this only involves taking derivatives of three-point conformal blocks, 
\begin{align}
 g^{L_{-1}^n}_{\Delta_1,\Delta_2,\Delta_3} & =\left. \frac{\frac{\partial^n}{\partial z_1^n} \mathcal{F}^{(3)}(\Delta_i|z_i)}{\mathcal{F}^{(3)}(\Delta_i|z_i)}\right|_{(z_1,z_2,z_3)=(0,\infty,1)} =(\Delta_1-\Delta_2+\Delta_3)_n \ ,
 \label{gln}
\end{align}
where we use the notation
\begin{align}
 (x)_n = \frac{\Gamma(x+n)}{\Gamma(x)} = \prod_{i=0}^{n-1}(x+i)\ .
\label{xn}
\end{align}
For more general descendants, we use local Ward identities \eqref{lmn}. For example,
\begin{align}
 g^{L_{-2}}_{\Delta_1,\Delta_2,\Delta_3}
 = \Delta_1-\Delta_2+2\Delta_3\ .
 \label{glt}
\end{align}
Let us rewrite the null vector equations for three-point functions as 
\begin{align}
 \left< V_{\langle r,s \rangle} V_{\Delta_2}V_{\Delta_3}\right> \neq 0 \quad \implies \quad 
 g^{L_{\langle r,s\rangle}}_{\Delta_{\langle r,s\rangle},\Delta_2,\Delta_3} = 0\ ,
\end{align}
and proceed to solve this equation in the three simplest cases. 

\subsubsection{Basic degenerate representations $\mathcal{R}_{\langle 1,1 \rangle}, \mathcal{R}_{\langle 2,1 \rangle}, \mathcal{R}_{\langle 1,2 \rangle}$}

We begin with the case $\langle r,s \rangle = \langle 1,1 \rangle$.
Using Eq. \eqref{lot} for $\Delta_{\langle 1,1 \rangle}$ and $L_{\langle 1,1 \rangle}$, we obtain
$g^{L_{-1}}_{0,\Delta_2,\Delta_3} = 0$.
Using Eq. \eqref{gln} for $g^{L_{-1}}$, this implies 
$
 \Delta_2=\Delta_3
$,
which is equivalent to the fusion rule
\begin{align}
 \boxed{\mathcal{R}_{\langle 1,1\rangle} \times \mathcal{V}_\Delta = \mathcal{V}_\Delta} \ .
 \label{roof}
\end{align}
In fact, $V_{\langle 1,1\rangle}$ is an identity field, see Exercise \ref{exoid}.

Next, let us consider the case $\langle r,s \rangle = \langle 2,1 \rangle$.
Using Eq. \eqref{lot}, we get
$g^{L_{-2} +\frac{1}{b^2} L_{-1}^2 }_{-\frac12 -\frac34 b^2,\Delta_2,\Delta_3} = 0$.
Using Eqs. \eqref{gln} and \eqref{glt} for $g^{L_{-1}^2}$ and $g^{L_{-2}}$, this implies
\begin{align}
 2(\Delta_2-\Delta_3)^2 + b^2 (\Delta_2+\Delta_3) - \frac12 - b^2 -\frac38 b^4= 0\ .
\end{align}
In terms of the momentums $P_2,P_3$ \eqref{daq} that correspond to the conformal dimensions $\Delta_2,\Delta_3$, this equation becomes
\begin{align}
\prod_{\pm,\pm} \left(P_2\pm P_3 \pm \frac{b}{2}\right) = 0 \ .
\end{align}
Up to reflections of $P_2$ or $P_3$, we must therefore have
$
 P_3 = P_2 \pm \frac{b}{2}
$.
We would similarly find $P_3 = P_2 \pm \frac{1}{2b}$ if we considered the case $\langle r,s\rangle = \langle 1,2\rangle$. 
The resulting fusion rules are 
\begin{align}
\boxed{ \mathcal{R}_{\langle 2,1 \rangle}\times \mathcal{V}_P = \sum_\pm \mathcal{V}_{P \pm \frac{b}{2}} }\quad , \quad 
\boxed{ \mathcal{R}_{\langle 1,2 \rangle}\times \mathcal{V}_P = \sum_\pm \mathcal{V}_{P \pm \frac{1}{2b}} }\ .
\label{rot}
\end{align}
These fusion rules could be derived by directly analyzing OPEs, instead of three-point functions. (See Exercise \ref{exooit}.) 

\subsubsection{Higher degenerate representations $\{\mathcal{R}_{\langle r,s\rangle}\}_{rs>2}$}

The higher degenerate representations' existence, momentums and fusion rules are in principle dictated by their null vectors. We know how to write these null vectors explicitly if $r=1$ or $s=1$, but not in general \cite{fms97}.
Instead, we will deduce all we need from the basic degenerate representations and their fusion rules, thanks to the associativity of the fusion product. 
This amounts to working with the category of representations of the Virasoro algebra, rather than with the structure of the representations. 

The basic idea is to build $\mathcal{R}_{\langle 3, 1\rangle}$ as the combination $\mathcal{R}_{\langle 2,1 \rangle} \times \mathcal{R}_{\langle 2,1 \rangle} - \mathcal{R}_{\langle 1,1 \rangle}$.
From the product $\mathcal{R}_{\langle 2,1 \rangle}\times \mathcal{V}_P$, we first deduce 
\begin{align}
 \left(\mathcal{R}_{\langle 2,1 \rangle}\times \mathcal{R}_{\langle 2,1 \rangle}\right)\times\mathcal{V}_P = \mathcal{R}_{\langle 2,1 \rangle}\times \left(\mathcal{R}_{\langle 2,1 \rangle}\times\mathcal{V}_P\right) = 
 \mathcal{V}_{P-b} + 2\cdot \mathcal{V}_P + \mathcal{V}_{P+b}\ .
\end{align}
Since we obtain a sum of finitely many Verma modules, $\mathcal{R}_{\langle 2,1 \rangle}\times \mathcal{R}_{\langle 2,1 \rangle}$ must be a sum of finitely many degenerate representations. But we know that $\mathcal{R}_{\langle 2,1 \rangle}\times \mathcal{R}_{\langle 2,1 \rangle}$ is a sum of two representations with the momentums $\{P_{\langle 2,1 \rangle} \pm \frac{b}{2}\} = \{P_{\langle 1,1\rangle}, P_{\langle 3,1\rangle}\}$, where $P_{\langle r,s\rangle}$ is defined in Eq. \eqref{ars}. 
We deduce that these two representations are degenerate. In particular, if we did not know it already, we could deduce that there exists a degenerate representation $\mathcal{R}_{\langle 3, 1\rangle}=\mathcal{R}_{\langle 2,1 \rangle}\times \mathcal{R}_{\langle 2,1 \rangle} -\mathcal{R}_{\langle 1,1\rangle}$ with the momentum $P_{\langle 3,1\rangle}$ and the fusion rule 
\begin{align}
 \mathcal{R}_{\langle 3, 1\rangle}\times \mathcal{V}_P = \left(\mathcal{R}_{\langle 2,1 \rangle}\times \mathcal{R}_{\langle 2,1 \rangle}\right)\times\mathcal{V}_P - \mathcal{R}_{\langle 1,1\rangle}\times \mathcal{V}_P =  \mathcal{V}_{P-b} + \mathcal{V}_P + \mathcal{V}_{P+b}\ .
\end{align}
Iterating this procedure, let us construct $\{\mathcal{R}_{\langle r,s\rangle}\}_{rs>2}$  and their fusion rules by recursion on $r,s\in\mathbb{N}^*$. 
We first find the fusion rule 
\begin{align}
\mathcal{R}_{\langle 2,1\rangle}\times \mathcal{R}_{\langle r,s\rangle} = \mathcal{R}_{\langle r-1,s\rangle} + \mathcal{R}_{\langle r+1,s\rangle }\ , \quad (r\geq 1)\ ,
\label{rtod}
\end{align}
which allows us to deduce $\mathcal{R}_{\langle r+1,s\rangle}$ from $\mathcal{R}_{\langle 2,1\rangle}, \mathcal{R}_{\langle r-1,s\rangle}$ and $\mathcal{R}_{\langle r,s\rangle}$. Similarly, the recursion on the second index $s$ relies on the fusion rule 
\begin{align}
 \mathcal{R}_{\langle 1,2\rangle}\times \mathcal{R}_{\langle r,s\rangle} = \mathcal{R}_{\langle r,s-1\rangle} + \mathcal{R}_{\langle r,s+1\rangle }\ , \quad (s\geq 1)\ .
\end{align}
Using the associativity of the fusion products $\mathcal{R}_{\langle 2,1\rangle}\times \mathcal{R}_{\langle r,s\rangle}\times \mathcal{V}_P$ and $\mathcal{R}_{\langle 1,2\rangle}\times \mathcal{R}_{\langle r,s\rangle}\times \mathcal{V}_P$, the recursion yields
\begin{align}
 \boxed{\mathcal{R}_{\langle r,s \rangle}\times \mathcal{V}_P = \sum_{i=-\frac{r-1}{2}}^{\frac{r-1}{2}} \sum_{j=-\frac{s-1}{2}}^{\frac{s-1}{2}} \mathcal{V}_{P + ib+jb^{-1}}}
 \ ,
\label{rtv}
\end{align}
where the summation index $i$ takes half-integer values if $r$ is even, and integer values if $r$ is odd. 
This fusion product is therefore a sum of 
$rs$ terms.

\subsubsection{Fusion product of two degenerate representations}

In order to determine $\mathcal{R}_{\langle r_1,s_1 \rangle} \times \mathcal{R}_{\langle r_2,s_2 \rangle}$, we could use a recursion on $r_1,s_1$ using the associativity of $\mathcal{R}_{\langle 2,1\rangle}\times \mathcal{R}_{\langle r_1,s_1 \rangle} \times \mathcal{R}_{\langle r_2,s_2 \rangle}$ and $\mathcal{R}_{\langle 1,2\rangle}\times\mathcal{R}_{\langle r_1,s_1 \rangle} \times \mathcal{R}_{\langle r_2,s_2 \rangle}$. 
We will however employ a more direct method. 
Using the fusion rule \eqref{rtv}, we find 
\begin{align}
 \mathcal{R}_{\langle r_1,s_1 \rangle} \times \mathcal{V}_{P_{\langle r_2,s_2\rangle}} = \sum_{r_3\overset{2}{=}r_2-r_1+1}^{r_1+r_2-1}\ \sum_{s_3\overset{2}{=}s_2-s_1+1}^{s_1+s_2-1} \mathcal{V}_{P_{\langle r_3,s_3 \rangle}}\ ,
\\
\mathcal{V}_{P_{\langle r_1,s_1 \rangle}} \times \mathcal{R}_{\langle r_2,s_2\rangle} = \sum_{r_3\overset{2}{=}r_1-r_2+1}^{r_1+r_2-1}\ \sum_{s_3\overset{2}{=}s_1-s_2+1}^{s_1+s_2-1} \mathcal{V}_{P_{\langle r_3,s_3 \rangle}}\ ,
\end{align}
where a superscript $\overset{2}{=}$ indicates that the corresponding sum runs by increments of $2$.
These two fusion products encode the vanishing of the null vectors of $\mathcal{R}_{\langle r_1,s_1 \rangle}$ and $\mathcal{R}_{\langle r_2,s_2\rangle}$ respectively.
By definition of fusion rules, the product $\mathcal{R}_{\langle r_1,s_1 \rangle} \times \mathcal{R}_{\langle r_2,s_2 \rangle}$ is constrained by the vanishing of the null vectors of both degenerate representations. 
Therefore, we have 
\begin{align}
 \mathcal{R}_{\langle r_1,s_1 \rangle} \times \mathcal{R}_{\langle r_2,s_2 \rangle}\ \overset{\text{formal}}{=}\  \left(\mathcal{R}_{\langle r_1,s_1 \rangle} \times \mathcal{V}_{P_{\langle r_2,s_2\rangle}} \right) \bigcap \left(\mathcal{V}_{P_{\langle r_1,s_1 \rangle}} \times \mathcal{R}_{\langle r_2,s_2\rangle}\right)\ .
\end{align}
This relation is formal in the sense that it determines the momentums of the representations that appear in $\mathcal{R}_{\langle r_1,s_1 \rangle} \times \mathcal{R}_{\langle r_2,s_2 \rangle}$, but does not say whether these representations are Verma modules or degenerate representations. 
However, we do know that these representations must all be degenerate, because $\mathcal{R}_{\langle r_1,s_1 \rangle} \times \mathcal{R}_{\langle r_2,s_2 \rangle}\times \mathcal{V}_P$ is a finite sum of Verma modules by associativity.
Therefore,
\begin{align}
 \boxed{\mathcal{R}_{\langle r_1,s_1 \rangle} \times \mathcal{R}_{\langle r_2,s_2 \rangle} = \sum_{r_3\overset{2}{=}|r_1-r_2|+1}^{r_1+r_2-1}\ \sum_{s_3\overset{2}{=}|s_1-s_2|+1}^{s_1+s_2-1} \mathcal{R}_{\langle r_3,s_3 \rangle}}\ .
\label{rrsr}
\end{align}
So the fusion product $\mathcal{R}_{\langle r_1,s_1 \rangle} \times \mathcal{R}_{\langle r_2,s_2 \rangle}$ is a sum of $\min(r_1,r_2)\times \min(s_1,s_2)$ degenerate representations.
This can be rewritten as 
\begin{align}
\mathcal{R}_{\langle r_1,s_1 \rangle} \times \mathcal{R}_{\langle r_2,s_2 \rangle} = \sum_{r_3,s_3=1}^\infty f_{r_1,r_2,r_3} f_{s_1,s_2,s_3} \mathcal{R}_{\langle r_3,s_3 \rangle}\ ,
 \label{rrrsss}
\end{align}
with the coefficients
\begin{align}
f_{r_1,r_2,r_3} = \left\{\begin{array}{l}  1 \quad \text{if} \quad 
 \forall i\neq j\neq k \ , \quad r_i+r_j-r_k \in 1 + 2{\mathbb{N}}\ .
\\ 0 \quad \text{otherwise} \ .\end{array}\right.  
\label{frrr}
\end{align}

\subsection{BPZ differential equations \label{secbpz}}

Using the representation \eqref{lmn} of creation operators as differential operators, null vector equations for correlation functions of primary fields become differential equations, called \textbf{\boldmath BPZ equations}\index{BPZ equation} after Belavin, Polyakov and Zamolodchikov. 
For example, in the cases of the degenerate fields $V_{\langle 1,1 \rangle}(x)$ and $V_{\langle 2,1 \rangle}(x)$, Eq. \eqref{lrsv} gives rise to the following BPZ equations: 
\begin{align}
 {\frac{\partial}{\partial x}} \left\langle V_{\langle 1,1 \rangle}(x) \prod_{i=1}^{N-1} V_{\Delta_i}(z_i) \right\rangle = 0 \ ,
\label{pvoo} 
\end{align}
\begin{align}
\boxed{\left( \frac{\partial^2}{\partial x^2}  +b^2 \sum_{j=1}^{N-1} \left[\frac{1}{x-z_j}{\frac{\partial}{\partial z_j}}+ \frac{\Delta_j}{(x-z_j)^2} \right]\right)\left\langle V_{\langle 2,1 \rangle}(x) \prod_{i=1}^{N-1} V_{\Delta_i}(z_i) \right\rangle = 0} \ .
\label{pvot}
\end{align}
More generally, a degenerate field $V_{\langle r,s \rangle}(x)$ with a vanishing descendant at level $rs$ leads to a BPZ equation of order $rs$. 

In the case of a three-point function ($N=3$), the dependence on $z_i$ is already completely determined by global conformal symmetry, see Eq. \eqref{fzzz}.
The BPZ equation therefore leads to constraints on the conformal dimensions $\Delta_1$ and $\Delta_2$ -- that is, to the fusion rules. (See Exercise \ref{exotob}.) If $N\geq 4$, taking the global conformal symmetry into account as in Eq. \eqref{xfxn}, it is possible to rewrite a BPZ equation as a differential equation for a function of $N-3$ variables.
In particular, a BPZ equation for a four-point function boils down to an ordinary differential equation -- a differential equation for a function of a single variable.  

\subsubsection{Case of a four-point function with the field $V_{\langle 2,1 \rangle}$}

Let us derive the ordinary differential equation for $\left\langle V_{\langle 2,1 \rangle}(x)\prod_{i=1}^3 V_{\Delta_i}(z_i)\right\rangle$.
Inserting the identity $\oint_\infty dz\ \epsilon(z) T(z) =0$ with $\epsilon(z) = \frac{(z-z_1)(z-z_2)(z-z_3)}{z-x}$, and using Eq. \eqref{lrsv}, we obtain a differential equation with derivatives with respect to $x$ alone, 
\begin{multline}
  \Bigg\{ \prod_{i=1}^3(x-z_i)\left(-\frac{1}{b^2}\frac{\partial^2}{\partial x^2} +\sum_{i=1}^3 \frac{1}{x-z_i} {\frac{\partial}{\partial x}} \right) + (3x-z_1-z_2-z_3)\Delta_{\langle 2,1 \rangle} 
  \\
 +\frac{z_{12}z_{13}}{z_1-x}\Delta_1 + \frac{z_{21}z_{23}}{z_2-x}\Delta_2+\frac{z_{31}z_{32}}{z_3-x}\Delta_3\Bigg\} 
\left\langle V_{\langle 2,1 \rangle}(x)\prod_{i=1}^3 V_{\Delta_i}(z_i)\right\rangle  = 0\ .
\label{uode}
\end{multline}
Setting $(z_1,z_2,z_3)=(0,\infty,1)$, this amounts to the equation 
\begin{align}
  \left\{ \frac{x(1-x)}{b^2}\frac{\partial^2}{\partial x^2} + (2x-1){\frac{\partial}{\partial x}} +\Delta_{\langle 2,1 \rangle} +\frac{\Delta_1}{x}-\Delta_2 + \frac{\Delta_3}{1-x}\right\} \mathcal{F}(x)=0\ ,
\label{sode}
\end{align}
for 
\begin{align}
 \mathcal{F}(x) = \left\langle V_{\langle 2,1 \rangle}(x)V_{\Delta_1}(0)V_{\Delta_2}(\infty)V_{\Delta_3}(1)\right\rangle\ .
\label{fxv}
 \end{align}
(The $r$-th order differential equation for $\left\langle V_{\langle r,1 \rangle}(x)V_{\Delta_1}(0)V_{\Delta_2}(\infty)V_{\Delta_3}(1)\right\rangle $ can be found in \cite{flno09}.)

\subsubsection{Qualitative properties of the second-order BPZ equation}

Since the equation \eqref{sode} has coefficients that diverge at $x=0,1,\infty$, its solutions can be non-analytic at these points. 
More precisely, these points are regular singular points.
A \textbf{regular singular point}\index{regular singular point} $x_0$ is characterized by the existence of a basis of solutions of the type  
\begin{align}
 \mathcal{F}(x)=(x-x_0)^\lambda\left(1+\sum_{n=0}^\infty c_n (x-x_0)^n\right)\ ,
\label{zxl}
\end{align}
where $\lambda$ is called a \textbf{\boldmath characteristic exponent}\index{characteristic exponent}.
The characteristic exponents of the equation \eqref{sode} at the regular singular point $x=0$ are obtained by inserting the ansatz $\mathcal{F}(x) = x^\lambda(1+O(x))$, which leads to 
\begin{align}
 \lambda^2 - bQ\lambda + b^2 \Delta_1 = 0 \ .
\end{align}
The roots of this equation have simple expressions in terms of the momentum $P_1$,
\begin{align}
 \lambda_\pm = b\left(\frac{Q}{2}\mp P_1\right) \ .
\label{lpm}
\end{align}
In conformal field theory, coincidences of fields lead to 
singularities of correlation functions, with characteristic exponents that are constrained by the corresponding OPEs.
In the case of our four-point function \eqref{fxv} at $x=0$, the fusion rule \eqref{rot} leads to an OPE of the type
\begin{align}
 V_{\langle 2,1 \rangle}(x)V_{P}(0) = \sum_\pm C_\pm(P) x^{\Delta\left(P\pm \frac{b}{2}\right) -\Delta(P) - \Delta_{\langle 2,1 \rangle}} \left(V_{P\pm\frac{b}{2}}(0) + O(x)\right)\ ,
\end{align}
where $C_\pm(P)$ are some constants, and the powers of $x$ are dictated by Eq. \eqref{eq:coz}. 
Computing these powers, we recover the characteristic exponents
\begin{align}
 \lambda_\pm = \Delta\left(P\pm \frac{b}{2}\right) -\Delta(P) - \Delta_{\langle 2,1 \rangle} \ ,
\end{align}
so that each characteristic exponent corresponds to a primary field in the OPE.
(See also Exercise \ref{exotbf}.) The corresponding solutions are called $s$-channel \textbf{degenerate conformal blocks}\index{conformal block!degenerate---}, denoted $\mathcal{F}^{(s)}_\pm(x)$, and depicted as
\begin{align}
 \mathcal{F}^{(s)}_\pm(x)  =  
 \begin{tikzpicture}[baseline=(current  bounding  box.center), very thick, scale = .4]
\draw (-1,2) node [left] {$P_1$} -- (0,0) -- node [above] {$P_1\pm \frac{b}{2}$} (4,0) -- (5,2) node [right] {$P_2$};
\draw[dashed] (-1,-2) node [left] {$\langle 2,1 \rangle$} -- (0,0);
\draw (4,0) -- (5,-2) node [right] {$P_3$};
\end{tikzpicture}
\ =\ 
\begin{tikzpicture}[baseline=(current  bounding  box.center), very thick, scale = .8]
\draw[dashed] (0, -2) -- node [below left] {$\langle 2,1 \rangle$} (-2, 0); 
\draw (-2, 0) -- node [above left] {$P_1$} (0, 2) -- node [above right] {$P_2$} (2, 0) -- node [below right] {$P_3$} (0, -2) -- node [left] {$P_1\pm \frac{b}{2}\! \! $} (0, 2); 
\end{tikzpicture}
\  \ ,
\label{gpic}
\end{align}
where the degenerate field is represented as a dashed line.
These degenerate conformal blocks are special cases of the more general conformal blocks of Section \ref{secaco}.

\subsubsection{Alternative derivation from fusion rules}

It turns out that the second-order BPZ equation is the only meromorphic second-order differential equation with three regular singular points at $x=0,1,\infty$, and characteristic exponents of the type \eqref{lpm}. So this equation can be derived from the fusion rules, without knowing the level two null vectors. (See Exercise \ref{exoefr}.) The mathematical interpretation is that the second-order BPZ equation corresponds to a rigid Fuchsian system. Higher-order BPZ equations do not correspond to rigid Fuchsian systems: for a discussion of the theory of rigid Fuchsian systems as applied to differential equations from CFT, see \cite{bhs17}.

\subsection{Hypergeometric conformal blocks \label{sechcb}}

\subsubsection{Hypergeometric equation, hypergeometric function}

 Let us define the \textbf{\boldmath hypergeometric equation}\index{hypergeometric!---equation} with parameters $A,B,C$, 
\begin{align}
 \left\{ x(1-x) \frac{\partial^2}{\partial x^2} +\left[C-(A+B+1)x\right]{\frac{\partial}{\partial x}} - AB \right\} \mathcal{G}(x)= 0\ .
\label{dzp}
\end{align}
Using a change of parameters and unknown function,
\begin{align}
\mathcal{G}(x) = x^{j_1} (x-1)^{j_3} \mathcal{H}(x)
 \ , \quad \renewcommand{\arraystretch}{1.3}\left\{\begin{array}{l}  j_1 = -\frac{C}{2} \ , \\ j_2=\frac{B-A-1}{2} \ , \\  j_3= \frac{C-A-B-1}{2}\ , \end{array}\right.    
\quad \left\{\begin{array}{l}  A = -1-j_1-j_2-j_3\ , \\ B = -j_1+j_2-j_3\ , \\ C = -2j_1\ , \end{array}\right. 
\end{align}
we obtain the twisted hypergeometric equation,
\begin{align}
 \left\{ x(1-x) \frac{\partial^2}{\partial x^2}  -\frac{j_1(j_1+1)}{x}+ j_2(j_2+1)+\frac{j_3(j_3+1)}{x-1}\right\} \mathcal{H}(x) = 0\ .
\label{hj}
\end{align}
The characteristic exponents of the hypergeometric equation at the regular singular point $x=0$ are $\lambda\in \{0,1-C\}$, and the solution $\mathcal{G}(x)$ that corresponds to $\lambda=0$
 is the \textbf{\boldmath hypergeometric function}\index{hypergeometric!---function}
\begin{align}
 F(A,B,C,x) = \sum_{n=0}^\infty \frac{(A)_n(B)_n}{n!(C)_n}x^n\ ,
\label{fsn}
\end{align}
where $(A)_n = \frac{\Gamma(A+n)}{\Gamma(A)}$.
The solution with the characteristic exponent $\lambda=1-C$ is $x^{1-C}F(A-C+1,B-C+1,2-C,x)$.
The hypergeometric function obeys the identities 
\begin{align}
 F(A,B,C,x) = F(B,A,C,x) = (1-x)^{C-A-B} F(C-A,C-B,C,x)\ .
\label{fff}
\end{align}

\subsubsection{Solving a hypergeometric BPZ equation}

We have the following equivalences of differential equations:
\begin{align}
  \text{BPZ equation \eqref{sode} for\ \ } & \mathcal{F}(x)
\\ 
\iff\ \text{hypergeometric equation for\ \ }&  \mathcal{G}(x)= x^{-b(\frac{Q}{2}+P_1)}(1-x)^{-b(\frac{Q}{2}+P_3)}\mathcal{F}(x) 
\label{heg}
\\
\iff\  \text{twisted\ hypergeometric equation for\ \ }& \mathcal{H}(x)=(x(1-x))^{-\frac{b^2}{2}}\mathcal{F}(x) \ ,
\label{fgs}
\end{align}
provided the parameters are related as
\begin{align}
\renewcommand{\arraystretch}{1.3}
\left\{\begin{array}{l}   A = \frac12 + b(P_1+P_2+P_3) \ , \\
      B = \frac12 + b(P_1-P_2+P_3) \ , \\
      C = 1 + 2bP_1 \ ,
\end{array}\right. 
\label{abc}
\end{align}
or equivalently $P_i = -b^{-1}(j_i +\frac12)$.
(See Exercise \ref{exohge}.) 
Therefore, the solutions of the equation \eqref{sode} that correspond to the characteristic exponents $\lambda_\pm$ \eqref{lpm} at $x=0$ are 
\begin{align}
\renewcommand{\arraystretch}{1.3}
\left\{\begin{array}{l}  \mathcal{F}^{(s)}_-(x) = x^{b(\frac{Q}{2}+P_1)} (1-x)^{b(\frac{Q}{2}+P_3)} F(A,B,C,x)\ ,
\\ \mathcal{F}^{(s)}_+(x)   = x^{b(\frac{Q}{2}-P_1)} (1-x)^{b(\frac{Q}{2}+P_3)} F(1-C+A,1-C+B,2-C,x) \ ,
\end{array}\right. 
\label{gpm}
\end{align}
where the superscript in $\mathcal{F}^{(s)}_\pm(x)$ stands for the $s$-channel basis of solutions.
These formulas are consistent with the invariance of the original differential equation \eqref{sode} under reflections of $P_i$: reflection of $P_1$ exchanges the two solutions, and reflections of $P_2$ and $P_3$ leave each solution invariant, as follows from Eq. \eqref{fff}.
The corresponding solutions of the twisted equation \eqref{hj} are 
\begin{align}
\renewcommand{\arraystretch}{1.3}
\left\{\begin{array}{l}  \mathcal{H}^{(s)}_-(x)  = x^{-j_1}(1-x)^{-j_3} F(-j_1-j_2-j_3-1,-j_1+j_2-j_3,-2j_1,x)\ ,
\\
 \mathcal{H}^{(s)}_+(x)  = x^{j_1+1}(1-x)^{-j_3} F(j_1-j_2-j_3,j_1+j_2-j_3+1,2j_1+2,x)\ .
\end{array}\right. 
\label{fpm}
\end{align}

\subsubsection{Fusing matrix}

Let us define the $t$-channel basis of solutions, whose elements are of the type \eqref{zxl} near the regular singular point $x_0=1$.
Such solutions are obtained from the $s$-channel solutions by the exchanges $\left\{\begin{smallmatrix} 1\leftrightarrow 3\\ x\leftrightarrow 1-x \end{smallmatrix}\right.$, and we find 
\begin{align}
\renewcommand{\arraystretch}{1.3}
 \left\{\begin{array}{l}  \mathcal{F}^{(t)}_-(x) = x^{b(\frac{Q}{2}+P_1)} (1-x)^{b(\frac{Q}{2}+P_3)} F(A,B,A+B-C+1,1-x) \ ,
\\ \mathcal{F}^{(t)}_+(x) =x^{b(\frac{Q}{2}+P_1)} (1-x)^{b(\frac{Q}{2}-P_3)} F(C-A,C-B,C-A-B+1,1-x)\ ,
\end{array}\right. 
\label{tpm}
\end{align}
whose diagrammatic representations are 
\begin{align}
 \mathcal{F}^{(t)}_\pm(x)  =  
 \begin{tikzpicture}[baseline=(current  bounding  box.center), very thick, scale = .4]
 \draw (-2,3) node [left] {$P_1$} -- (0,2) -- node [left] {$P_3\pm \frac{b}{2}$} (0,-2);
 \draw[dashed] (0, -2) -- (-2, -3) node [left] {$\langle 2,1 \rangle$};
\draw (0,2) -- (2,3) node [right] {$P_2$};
\draw (0,-2) -- (2, -3) node [right] {$P_3$};
\end{tikzpicture}
\ =\
\begin{tikzpicture}[baseline=(current  bounding  box.center), very thick, scale = .7]
\draw (-2, 0) -- node [above left] {$P_1$} (0, 2) -- node [above right] {$P_2$} (2, 0) -- node [below right] {$P_3$} (0, -2);
\draw[dashed] (0, -2) -- node [below left] {$\langle 2,1 \rangle$} (-2, 0);
\draw (-2, 0) -- node [below] {$P_3\pm \frac{b}{2}$} (2, 0); 
\end{tikzpicture}
\ \ .
\label{tpic}
\end{align}
In order to find the relation between our two bases of solutions, we need the formula
\begin{multline}
 F(A,B,C,x) = \frac{\Gamma(C)\Gamma(C-A-B)}{\Gamma(C-A)\Gamma(C-B)} F(A,B,A+B-C+1,1-x) 
\\
 + \frac{\Gamma(C)\Gamma(A+B-C)}{\Gamma(A)\Gamma(B)} (1-x)^{C-A-B}F(C-A,C-B,C-A-B+1,1-x)\ .
\end{multline}
Together with Eq. \eqref{fff}, this implies 
\begin{align}
 \mathcal{F}^{(s)}_{\epsilon_1}(x) = \sum_{\epsilon_3=\pm} F_{\epsilon_1,\epsilon_3} \mathcal{F}^{(t)}_{\epsilon_3}(x)\ ,
\label{gfg}
\end{align}
where we introduce the \textbf{degenerate fusing matrix}\index{fusing matrix!degenerate---} 
\begin{align}
F_{\epsilon_1,\epsilon_3} = \frac{\Gamma(1-2b\epsilon_1P_1)\Gamma(2b\epsilon_3P_3)}{\prod_\pm \Gamma(\frac12+b(-\epsilon_1P_1\pm P_2+\epsilon_3P_3))}\ ,
\label{fmd}
\end{align}
whose determinant is 
\begin{align}
 \det F = -\frac{P_1}{P_3}\ .
\label{detf}
\end{align}

\subsubsection{Single-valued solutions}

Let us consider a four-point function that involves a diagonal degenerate field $V_{\langle 2,1 \rangle}$, and therefore obeys not only the BPZ equation, but also the complex conjugate equation. Our four-point function must therefore be a combination of hypergeometric conformal blocks:
\begin{multline}
 \left\langle V_{\langle 2,1 \rangle}(x)V_{\Delta_1,\bar\Delta_1}(0)V_{\Delta_2,\bar\Delta_2}(\infty)V_{\Delta_3,\bar\Delta_3}(1)\right\rangle 
 \\
 = \sum_{\epsilon_1,\bar{\epsilon}_1=\pm} c^{(s)}_{\epsilon_1,\bar{\epsilon}_1} \mathcal{F}_{\epsilon_1}^{(s)}(x) \bar{\mathcal{F}}_{\bar{\epsilon}_1}^{(s)}(\bar{x}) = \sum_{\epsilon_3,\bar{\epsilon}_3=\pm} c^{(t)}_{\epsilon_3,\bar{\epsilon}_3} \mathcal{F}_{\epsilon_3}^{(t)}(x) \bar{\mathcal{F}}_{\bar{\epsilon}_3}^{(t)}(\bar{x})\ ,
\end{multline}
where $c^{(s)}_{\epsilon_1,\bar{\epsilon}_1}$ and $c^{(t)}_{\epsilon_3,\bar{\epsilon}_3}$ are $x$-independent coefficients.
These two sets of coefficients are related by the change of basis \eqref{gfg},
\begin{align}
\forall \epsilon_3,\bar{\epsilon}_3 \in \{+,-\}\ , \quad 
 \sum_{\epsilon_1,\bar{\epsilon}_1=\pm} c^{(s)}_{\epsilon_1,\bar{\epsilon}_1} F_{\epsilon_1,\epsilon_3} \bar{F}_{\bar{\epsilon}_1,\bar{\epsilon}_3} = c^{(t)}_{\epsilon_3,\bar{\epsilon}_3}\ .
 \label{febe}
\end{align}
We now assume that our three fields $V_{\Delta_i,\bar\Delta_i}$ are diagonal, so that $\bar{\mathcal{F}} = \mathcal{F}$ and $\bar F = F$. (For the non-diagonal case, see \cite{mr17}.)
For generic dimensions $\Delta_i$, single-valuedness near $x=0$ implies $c^{(s)}_{+,-}=c^{(s)}_{-,+}=0$, and single-valuedness near $x=1$ similarly implies $c^{(t)}_{+,-}=c^{(t)}_{-,+}=0$.
Thus we can write
\begin{align}
 \left\langle V_{\langle 2,1 \rangle}(x)V_{\Delta_1}(0)V_{\Delta_2}(\infty)V_{\Delta_3}(1)\right\rangle = \sum_{\epsilon_1=\pm} c^{(s)}_{\epsilon_1} \mathcal{F}_{\epsilon_1}^{(s)}(x) \mathcal{F}_{\epsilon_1}^{(s)}(\bar{x}) = \sum_{{\epsilon_3}=\pm} c^{(t)}_{\epsilon_3} \mathcal{F}^{(t)}_{\epsilon_3}(x) \mathcal{F}^{(t)}_{\epsilon_3}(\bar{x})\ ,
\label{zsc}
\end{align}
where we slightly simplified the notation by writing $c^{(s)}_\epsilon = c^{(s)}_{\epsilon,\epsilon}$. 
In particular, using Eq. \eqref{febe} in the case $(\epsilon_3,\bar{\epsilon}_3)=(+, -)$, we find 
\begin{align}
 \boxed{\frac{c^{(s)}_+}{c^{(s)}_-}  = -\frac{F_{-+}F_{--}}{F_{++}F_{+-}}
 = \gamma(2bP_1)\gamma(1+2bP_1)\prod_{\pm,\pm}\gamma\left(\tfrac12+b(-P_1\pm P_2\pm P_3)\right)}\ ,
\label{spsm}
\end{align}
where we introduce the function
\begin{align}
 \gamma(x) = \frac{\Gamma(x)}{\Gamma(1-x)}\ .
\label{gx}
\end{align}
This relation determines our four-point function up to an $x$-independent factor.

\section{Conformal blocks and crossing symmetry \label{secaco}}

\subsection{Definition and basic properties of conformal blocks}

\subsubsection{Definition}

\textbf{\boldmath Conformal blocks}\index{conformal block} are to correlation functions what spherical harmonics are to atomic orbitals: special functions that encode the symmetries of the system.
Correlation functions are therefore combinations of conformal blocks, and model-dependent quantities: for example, a three-point function of primary fields \eqref{cff} is a combination of 
the three-point block $\mathcal{F}^{(3)}(\Delta_1,\Delta_2,\Delta_3|z_1,z_2,z_3)$, and a structure constant.
We have also encountered degenerate four-point blocks that obey differential equations in Section \ref{sec:degf}.
And on the torus, there exist nontrivial zero-point blocks, which coincide with the characters of representations. 

Conformal blocks associated with a correlation function obey the Ward identities for that correlation function. 
Since the Ward identities are holomorphically factorized, the blocks can be factorized into left-moving and right-moving conformal blocks. Left-moving conformal blocks are solutions of the left-moving Ward identities, and are locally holomorphic functions of the fields' positions. From now on, by conformal block we will mean left-moving conformal block. 

Conformal blocks are sometimes defined as arbitrary solutions of the Ward identities. We will be more specific, and define them as the elements of particular bases of solutions. The bases in question are defined by using OPEs for computing correlation functions. In particular, using the OPE of $V_{\sigma_1}(z_1)V_{\sigma_2}(z_2)$ and Ward identities, any four-point function of primary or descendant fields can be decomposed as 
\begin{align}
 \left\langle \prod_{i=1}^4 V_{\sigma_i}(z_i)\right\rangle 
 = \sum_{\Delta_s,\bar{\Delta}_s} \frac{C_{12s} C_{s34}}{B_s} \left| \mathcal{F}^{(s)}_{\Delta_s}(\sigma_i|z_i)\right|^2\ ,
\label{fsd}
\end{align}
where all quantities that are not two- and three-point structure constants have been combined into the \textbf{\boldmath $s$-channel}\index{s-channel@$s$-channel} four-point conformal block $\mathcal{F}^{(s)}_{\Delta_s}(\sigma_i|z_i)$.
The appearance of the factorized quantity $\left| \mathcal{F}^{(s)}_{\Delta_s}(\sigma_i|z_i)\right|^2 $ (which involves the modulus square notation \eqref{eq:msn}) is a manifestation of holomophic factorization, and the right-moving conformal block $\mathcal{F}^{(s)}_{\bar\Delta_s}(\bar\sigma_i|\bar z_i)$ is given by the same function as the left-moving conformal block, applied to right-moving quantities. 

More precisely, deducing the conformal block decomposition \eqref{fsd} from the OPE $V_{\sigma_1}(z_1)V_{\sigma_2}(z_2)$ involves splitting the sum in the OPE into two sums:
\begin{itemize}
 \item the sum over conformal dimensions $\Delta_s,\bar{\Delta}_s$, which stand for representations of the Virasoro algebra,
 \item a sum over all states in a given irreducible representation, which defines the conformal block. Since the relative contributions of these states are determined by Ward identities, the conformal block is a universal quantity.
\end{itemize}
Both sums have to converge whenever the OPE $V_{\sigma_1}(z_1)V_{\sigma_2}(z_2)$ converges.
However, we will shortly see that conformal blocks can be defined for all positions $z_i$ by analytic continuation. Moreover, the sum over conformal dimensions actually converges for all positions $z_i$ in minimal models (because it is a finite sum) and in Liouville theory (see Section \ref{seceul}).

Using the OPE of $V_{\sigma_1}(z_1)V_{\sigma_4}(z_4)$, we would obtain \textbf{\boldmath $t$-channel}\index{t-channel@$t$-channel} four-point conformal block $\mathcal{F}^{(t)}_{\Delta_t}(\sigma_i|z_i)$. And with the OPE of $V_{\sigma_1}(z_1)V_{\sigma_3}(z_3)$, we would obtain \textbf{\boldmath $u$-channel}\index{u-channel@$u$-channel} four-point conformal block $\mathcal{F}^{(u)}_{\Delta_u}(\sigma_i|z_i)$. The three inequivalent OPEs give rise to three different decompositions of the same four-point function, and therefore to three bases of conformal blocks. The equality of the different decompositions leads to constraints on the spectrum and correlation functions, called \textbf{\boldmath crossing symmetry}\index{crossing symmetry}.

\subsubsection{Dependence on field positions}

For simplicity, we consider the conformal blocks that are associated to a four-point function of primary fields $\left<\prod_{i=1}^4V_{\Delta_i}(z_i)\right>$. 
Thanks to global conformal symmetry, we restrict our attention to the case $(z_1,z_2,z_3,z_4)=(x,0,\infty, 1)$. Our four-point blocks read
\begin{align}
 \mathcal{F}^{(s)}_{\Delta_s}(\Delta_i|x)=\mathcal{F}^{(s)}_{\Delta_s}(\Delta_i|x,0,\infty,1)\ ,
\end{align}
from which $\mathcal{F}^{(s)}_{\Delta_s}(\Delta_i|z_i)$ can be recovered  using Eqs. \eqref{zgg} and \eqref{fx}.
Once $s$-channel blocks are known, $t$-channel blocks can be obtained by a permutation of the arguments, 
\begin{align}
 \mathcal{F}^{(t)}_{\Delta_t}(\Delta_1,\Delta_2,\Delta_3,\Delta_4|z_1,z_2,z_3,z_4) = \mathcal{F}^{(s)}_{\Delta_t}(\Delta_1,\Delta_4,\Delta_3,\Delta_2|z_1,z_4,z_3,z_2)\ . 
\label{gtgs}
\end{align}
Taking global conformal symmetry into account, this becomes
\begin{align}
 \mathcal{F}^{(t)}_{\Delta_t}(\Delta_1,\Delta_2,\Delta_3,\Delta_4|x) = \mathcal{F}^{(s)}_{\Delta_t}(\Delta_1,\Delta_4,\Delta_3,\Delta_2|1-x)\ .
\end{align}

Let us summarize the analytic properties of 
the $s$-channel conformal block $\mathcal{F}^{(s)}_{\Delta_s}(\Delta_i|x)$ as a function of $x$, before giving supporting arguments:
\begin{enumerate}
 \item The $s$-channel conformal block has a regular singular point at $x=0$, where it behaves as 
 \begin{align}
 \mathcal{F}^{(s)}_{\Delta_s}(\Delta_i|x) \underset{x\to 0}{=} x^{\Delta_s-\Delta_1-\Delta_2}\Big( 1+ O(x)\Big)\ .
\end{align}
\item The factor $\big( 1+ O(x)\big)$ is a power series in $x$, whose radius of convergence is $1$.
\item The conformal block has an analytic continuation to $\mathbb{C}-\{0,1\}$. 
\item For generic values of the parameters $c,\Delta_s,\Delta_i$, the block has nontrivial monodromy around the singularities at $x=0,1,\infty$, and the singularities at $x=1,\infty$ are essential.
\end{enumerate}
Similarly, $t$- and $u$-channel conformal blocks have singularities at $x=0,1,\infty$, including 
regular singularities at $x=1$ and $x=\infty$ respectively.

The behaviour of the $s$-channel conformal block near $x=0$ follows from its definition from an OPE. The presence of a field at $x=1$ suggests that the radius of convergence of the resulting series is $1$, but it is not known how to prove this, or how to prove the existence of an analytic continuation to $\mathbb{C}-\{0,1\}$. These properties are however manifest in the case of the hypergeometric conformal blocks of Section \ref{sechcb}. 

Virasoro four-point blocks only have singularities at $x=0,1,\infty$. Four-point blocks of larger symmetry algebras can have additional singularities. The KZ-BPZ relation \eqref{dyy} shows that conformal blocks of the affine Lie algebra $\hat{\mathfrak{sl}}_2$ have additional singularities that depend on isospin variables. Certain families of conformal blocks of the $W_3$ algebra have an additional singularity at $x=-1$ \cite{fr11}. The existence of additional singularities might well be a generic feature of conformal blocks of larger symmetry algebras. Vanishing null vectors however tend to remove the additional singularities, while making the blocks more tractable.

\subsubsection{Dependence on conformal dimensions}

A conformal block in general has a pole when the $s$-channel dimension takes a degenerate value $\Delta_s=\Delta_{\langle r,s\rangle}$, because the OPE has a pole there. (See Section \ref{secope}.) 
This means that a field with this conformal dimension cannot appear in the decomposition of our four-point function, unless the residue of the conformal block's pole vanishes. The residue turns out to vanish if the fusion rules allow $\mathcal{R}_{\langle r,s \rangle}\times \mathcal{V}_{\Delta_1}\to \mathcal{V}_{\Delta_2}$ or $\mathcal{R}_{\langle r,s \rangle}\times \mathcal{V}_{\Delta_3}\to \mathcal{V}_{\Delta_4}$. 
In (generalized) minimal models, $s$-channel dimensions take degenerate values, and the vanishing of residues allows conformal blocks to be finite nonetheless.

Now what happens if our four-point function involves degenerate fields? 
The conformal block $\mathcal{F}^{(s)}_{\Delta_s}(\Delta_i|z_i)$ is a smooth function of $\Delta_i$, and has a finite value for $\Delta_1 = \Delta_{\langle r,s\rangle}$. 
If $\Delta_2$ and $\Delta_s$ happen to be related as dictated by the fusion rule of the degenerate field $V_{\langle r,s\rangle}$, then this finite value obeys a differential equation of order $rs$, as required by the degenerate field. (See Section \ref{secbpz}.) 
In other words, a conformal block involving a degenerate representation coincides with the conformal block involving the corresponding Verma module.
The degenerate representation only manifests itself by constraining which conformal blocks can contribute (via its fusion rules).

\subsection{Computing four-point conformal blocks}

\subsubsection{Pedestrian computation}

Inserting the OPE \eqref{vvs} in our four-point function, we obtain
\begin{multline}
 \Big< V_{\Delta_1,\bar\Delta_1}(x) V_{\Delta_2,\bar\Delta_2}(0)V_{\Delta_3,\bar\Delta_3}(\infty)V_{\Delta_4,\bar\Delta_4}(1)\Big>
 \\
 = 
 \sum_{\Delta_s,\bar\Delta_s} \frac{C_{12s}}{B_s} 
 \left| x^{\Delta_s-\Delta_1-\Delta_2}\sum_{L\in\mathcal{L}} x^Lf^{\Delta_s,L}_{\Delta_1,\Delta_2} \right|^2
 \Big< L\bar L V_{\Delta_s,\bar\Delta_s}(0) V_{\Delta_3,\bar\Delta_3}(\infty) V_{\Delta_4,\bar\Delta_4}(1) \Big>\ .
 \label{4ope}
\end{multline}
Using the universal factors $g^L_{\Delta_s,\Delta_3,\Delta_4}$ \eqref{glvv}, this can be written as a conformal block decomposition \eqref{fsd}, where the blocks are 
\begin{align}
 \mathcal{F}^{(s)}_{\Delta_s}(\Delta_i|x) = x^{\Delta_s-\Delta_1-\Delta_2}\sum_{L\in\mathcal{L}} f_{\Delta_1,\Delta_2}^{\Delta_s,L} g^{L}_{\Delta_s,\Delta_3,\Delta_4}x^{|L|}\ .
\label{gsd}
\end{align}
Explicitly, using Eqs. \eqref{flfo}-\eqref{flff} for $f_{\Delta_1,\Delta_2}^{\Delta_s,L}$ and Eqs. \eqref{gln}-\eqref{glt} for $g^L_{\Delta_s,\Delta_3,\Delta_4}$, we find 
\begin{multline}
 \mathcal{F}^{(s)}_{\Delta_s}(\Delta_i|x) 
= x^{\Delta_s - \Delta_1 - \Delta_2}\Bigg\{ 1 
+ \frac{(\Delta_s+\Delta_1-\Delta_2)(\Delta_s+\Delta_4-\Delta_3)}{2\Delta_s} x  
\\
+ \frac{1}{D_2(\Delta_s)}
\begin{bmatrix} (\Delta_s+\Delta_1-\Delta_2)_2 \\ \Delta_s+2\Delta_1-\Delta_2 \end{bmatrix}^T
\begin{bmatrix} 2+\frac{c}{4\Delta_s} & -3 \\ -3 & 4\Delta_s+2 \end{bmatrix}
\begin{bmatrix} (\Delta_s+\Delta_4-\Delta_3)_2 \\ \Delta_s+2\Delta_4-\Delta_3 \end{bmatrix}
 x^2 + O(x^3)\Bigg\}\ ,
 \label{eq:fsexp}
\end{multline}
where we use the level-$2$ determinant $D_2(\Delta_s)=4(2\Delta_s+1)^2 +(c-13)(2\Delta_s+1) +9$ from Eq. \eqref{dud}.

\subsubsection{Zamolodchikov's recursion}

For practical computations of conformal blocks, the formula \eqref{gsd} is not very efficient: the number of terms grows quickly with the level $|L_s|$, and the cofficients $f_{\Delta_1,\Delta_2}^{\Delta_s,L_s}$ are not known explicitly.
Fortunately, there is also
\textbf{Zamolodchikov's recursion}\index{Zamolodchikov's recursion} (or recursive representation) \cite{zz90, ccy17}, which converges faster, and for all $x\in\mathbb{C}-\{1\}$, not just $|x|<1$.
Instead of a power series in the cross-ratio $x$, this representation is a power series in the \textbf{nome}\index{nome}
\begin{align}
 q(x) = \exp -\pi \frac{F(\frac12,\frac12,1,1-x)}{F(\frac12,\frac12,1,x)}  \ , 
\end{align}
where $F(\frac12,\frac12,1,x)$ is a special case of the hypergeometric function \eqref{fsn}. 
We have 
\begin{align}
 q(0)=0 \ \ , \ \ q(\tfrac12) = e^{-\pi} \ \ , \ \ q(1)=1 \ \ , \ \ q(x)\underset{x\to 0}{=} \frac{x}{16}+\frac{x^2}{32} +O(x^3)\ .
\end{align}
Zamolodchikov's recursive representation reads 
\begin{align}
 \mathcal{F}^{(s)}_{\Delta_s}(\Delta_i|x) 
=  (16q)^{\Delta_s -\frac{Q^2}{4}} x^{\frac{Q^2}{4}-\Delta_1-\Delta_2} (1-x)^{\frac{Q^2}{4}-\Delta_1-\Delta_4} \theta_3(q)^{3Q^2-4(\Delta_1+\Delta_2+\Delta_3+\Delta_4)} H_{\Delta_s}(\Delta_i|q)\ .
\end{align}
Here $\theta_3(q)$ is the Jacobi theta function
\begin{align}
 \theta_3(q) = \sum_{n\in{\mathbb{Z}}} q^{n^2}\ ,
\end{align}
 and the nontrivial factor is the function
\begin{align}
 H_{\Delta_s}(\Delta_i|q) = \sum_{k=0}^\infty \prod_{j=1}^k \sum_{m_j,n_j=1}^\infty \frac{(16q)^{m_jn_j}R_{m_j,n_j}}{\Delta_{\langle m_{j-1},-n_{j-1} \rangle}-\Delta_{\langle m_j,n_j\rangle}}\ ,
\label{hdq}
\end{align}
where by convention $\Delta_{\langle m_0,-n_0\rangle} = \Delta_s$, and we introduce the coefficients
\begin{align}
 R_{m,n} = \frac{2P_{\langle 0,0\rangle} P_{\langle m,n\rangle}}{\prod_{r=1-m}^m \prod_{s=1-n}^n 2P_{\langle r,s\rangle}}
\prod_{r\overset{2}{=}1-m}^{m-1} \prod_{s\overset{2}{=}1-n}^{n-1} \prod_\pm (P_2\pm P_1 + P_{\langle r,s\rangle}) (P_3\pm P_4 +P_{\langle r,s\rangle})\ .
\end{align}
(We used the notation $\overset{2}{=}$ from Section \ref{secfr} for indices that run by increments of $2$.
Moreover, $P$ is the momentum, and $P_{\langle r,s \rangle}$ is given by Eq. \eqref{ars}.) In Zamolodchikov's  recursion, conformal blocks are manifestly invariant under reflections of all momentums.

Zamolodchikov's recursion owes its name to the original definition of the function $H_{\Delta_s}(\Delta_i|q)$ \eqref{hdq} by a recursive formula,
\begin{align}
 H_{\Delta_s}(\Delta_i|q) = 1 + \sum_{m,n=1}^\infty \frac{(16q)^{mn}}{\Delta_s-\Delta_{\langle m,n\rangle}} R_{m,n} H_{\Delta_{\langle m,-n\rangle}}(\Delta_i|q)\ .
 \label{hrec}
\end{align}
This formula makes it manifest that conformal blocks have poles for $\Delta_s \in \{\Delta_{\langle r,s\rangle}\}_{r,s\in {\mathbb{N}}^*}$. However, the recursive representation also has unphysical singularities at 
values of the central charge of the type $c= c_{p, q}$ \eqref{cpq}: singularities of some terms that cancel when the terms are added. This makes it difficult to use the recursive representation for such values of the central charge, and in particular for minimal models \cite{rib18}.

\subsubsection{Other representations}

The recursive representation that we described can be understood as a large $\Delta_s$ expansion, and it makes $\Delta_s$-poles manifest. Al. Zamolodchikov has derived another recursive representation that is a large $c$ expansion, and makes $c$-poles manifest. That $c$-recursive representation is a power series in the cross-ratio $x$, rather than the nome $q$. For both recursive representations in the case of $N$-point conformal blocks (and not just $4$-point blocks), see \cite{ccy17}.

Moreover, conformal blocks have a combinatorial representation that follows from the AGT relation \cite{aflt10}. 
This representation is a power series in $x$, and therefore converges more slowly than the $\Delta_s$-recursive representation, and only for $|x|<1$.
Moreover, invariance under reflections is not manifest, and individual terms have spurious poles in $\Delta_s$.
However, an advantage of the combinatorial representation is that it has an interpretation as a sum over descendants in a particular basis.

\subsection{Crossing symmetry and the fusing matrix}\label{seccsfm}

\subsubsection{The crossing symmetry equation}

Let us analyze the crossing symmetry equation that relates the $s$- and $t$-channel decompositions of a four-point function. We introduce diagrammatic notations for the conformal blocks, 
\begin{align}
 \mathcal{F}^{(s)}_{\Delta_s}  =  
\begin{tikzpicture}[baseline=(current  bounding  box.center), very thick, scale = .3]
\draw (-1,2) node [left] {$2$} -- (0,0) -- node [above] {$s$} (4,0) -- (5,2) node [right] {$3$};
\draw (-1,-2) node [left] {$1$} -- (0,0);
\draw (4,0) -- (5,-2) node [right] {$4$};
\end{tikzpicture}
=
\begin{tikzpicture}[baseline=(current  bounding  box.center), very thick, scale = .4]
\draw (0, -2) -- node [below left] {$1$} (-2, 0) -- node [above left] {$2$} (0, 2) -- node [above right] {$3$} (2, 0) -- node [below right] {$4$} (0, -2) -- node [left] {$s$} (0, 2); 
\end{tikzpicture}
\quad\ ,\quad\
 \mathcal{F}^{(t)}_{\Delta_t}  =  
\begin{tikzpicture}[baseline=(current  bounding  box.center), very thick, scale = .3]
 \draw (-2,3) node [left] {$2$} -- (0,2) -- node [left] {$t$} (0,-2) -- (-2, -3) node [left] {$1$};
\draw (0,2) -- (2,3) node [right] {$3$};
\draw (0,-2) -- (2, -3) node [right] {$4$};
\end{tikzpicture}
=
\begin{tikzpicture}[baseline=(current  bounding  box.center), very thick, scale = .4]
\draw (-2, 0) -- node [above left] {$2$} (0, 2) -- node [above right] {$3$} (2, 0) -- node [below right] {$4$} (0, -2) -- node [below left] {$1$} (-2, 0) -- node [below] {$t$} (2, 0); 
\end{tikzpicture}
\ \ ,
\label{ftfs}
\end{align}
and write crossing symmetry as
\begin{align}
 \sum_{\Delta_s,\bar{\Delta}_s} \frac{C_{12s} C_{s34}}{B_s} \left| 
 \begin{tikzpicture}[baseline=(current  bounding  box.center), very thick, scale = .3]
\draw (-1,2) node [left] {$2$} -- (0,0) -- node [above] {$s$} (4,0) -- (5,2) node [right] {$3$};
\draw (-1,-2) node [left] {$1$} -- (0,0);
\draw (4,0) -- (5,-2) node [right] {$4$};
\end{tikzpicture} 
\right|^2 = \sum_{\Delta_t,\bar{\Delta}_t} \frac{C_{23t}C_{t41}}{B_t} \left|
\begin{tikzpicture}[baseline=(current  bounding  box.center), very thick, scale = .3]
 \draw (-2,3) node [left] {$2$} -- (0,2) -- node [left] {$t$} (0,-2) -- (-2, -3) node [left] {$1$};
\draw (0,2) -- (2,3) node [right] {$3$};
\draw (0,-2) -- (2, -3) node [right] {$4$};
\end{tikzpicture}
\right|^2\ .
\label{csd}
\end{align}
Considering the conformal blocks as known quantities, this is an equation for the spectrum and the two- and three-point structure constants, which is quadratic in the three-point structure constant.
This equation is equivalent to the associativity of the OPE. In order to show that the theory is consistent, it suffices to moreover check the commutativity of the OPE, equivalently the behaviour \eqref{css} of the three-point structure constant under permutations. 
% NB: in particular this implies equality with the u-channel.

The crossing symmetry equation is invariant under two types  of transformations: 
\begin{itemize}
\item \textbf{field renormalizations}\index{renormalization!field---} $V_i(z_i)\to \lambda_i V_i(z_i)$, where $\lambda_i$ depends on the field's dimensions $(\Delta_i,\bar\Delta_i)$ but not on its position $z_i$,
 \item \textbf{structure constant renormalizations}\index{renormalization!structure constant---} $C_{ijk}\to \kappa C_{ijk}$. 
\end{itemize}
Taken together, these transformations act on the structure constants as 
\begin{align}
 B_i \to \lambda_i^2 B_i \quad , \quad C_{ijk} \to \kappa \lambda_i\lambda_j\lambda_k C_{ijk}\ .
\end{align}
Field renormalizations can be reduced to sign ambiguities by setting $B_i=1$. 
And 
whenever there exists an identity field $I$, structure constant renormalizations can be eliminated by the requirement $C_{III}=1$, and more generally $C_{Iii}=1$.

The crossing symmetry equation \eqref{csd}, while it may in principle determine the three-point structure constant modulo renormalizations, is in practice often intractable, because
\begin{enumerate}
 \item it can involve sums over large sets of possible conformal dimensions in the $s$- and $t$-channels, 
 \item the conformal blocks are rather complicated. 
\end{enumerate}
So, while this equation is useful for testing the consistency of proposals for the spectrum and three-point structure constant, it is often useless for deriving specific proposals in the first place. However, it is possible to restrict the $s$- and $t$-channels conformal dimensions to finite sets, and to simplify the conformal blocks, by taking one of the four fields to be degenerate. (See Section \ref{sec:degf}.) If the other three fields remain generic, this still leads to constraints on the generic three-point structure constant. We will use such constraints for solving Liouville theory and minimal models.

\subsubsection{The fusing matrix}

By definition, $s$-channel and $t$-channel four-point blocks are two bases of the same space of solutions of the four-point Ward identities, whose elements are respectively parametrized by $\Delta_s$ and $\Delta_t$.
So there must exist a linear, invertible relation between the two bases, 
\begin{align}
 \mathcal{F}^{(s)}_{\Delta_s}(\sigma_i|z_i) = \int_{i\mathbb{R}}dP_t\ F_{\Delta_s,\Delta_t}\begin{bmatrix} \Delta_2 & \Delta_3 \\ \Delta_1 & \Delta_4 \end{bmatrix} \mathcal{F}^{(t)}_{\Delta_t}(\sigma_i|z_i)\ ,
 \label{eq:fusrel}
\end{align}
whose $z_i$-independent kernel $F_{\Delta_s,\Delta_t}\begin{bmatrix} \Delta_2 & \Delta_3 \\ \Delta_1 & \Delta_4 \end{bmatrix}$ is called the \textbf{\boldmath fusing matrix}\index{fusing matrix} of the Virasoro algebra. (Alternative names: fusion kernel, fusion matrix.)
The parameters of the fusing matrix are in principle representations of the Virasoro algebra: the conformal dimensions that we wrote stand for the corresponding Verma modules. The measure of integration is $dP_t$, and actually the expression \eqref{eq:fm} for the fusing matrix involves momentums rather than conformal dimensions; we nevertheless use conformal dimensions in our notation for the fusing matrix, as it is by definition invariant under reflections of momentums.
A diagrammatic notation for the fusing matrix is 
\begin{align}
 F_{\Delta_s,\Delta_t}\begin{bmatrix} \Delta_2 & \Delta_3 \\ \Delta_1 & \Delta_4 \end{bmatrix} = \ \ 
\begin{tikzpicture}[baseline=(current  bounding  box.center), very thick, scale = .65]
\draw (0, -2) -- node [below left] {$1$} (-2, 0) -- node [above left] {$2$} (0, 2) -- node [above right] {$3$} (2, 0) -- node [below right] {$4$} (0, -2) -- (0, 2);
\draw (-2, 0) -- (-.2, 0); \draw (.2, 0) -- (2, 0);
\node at (-.25, .6) {$s$};
\node at (-.6, -.35) {$t$};
\end{tikzpicture}
\ \ .
\end{align}
The fusing matrix can be written in terms of the Barnes double Gamma function $\Gamma_b$ as \cite{tv12} 
\begin{multline}
 F_{\Delta_s,\Delta_t}\begin{bmatrix} \Delta_2 & \Delta_3 \\ \Delta_1 & \Delta_4 \end{bmatrix} = \left(\prod_{\pm}\frac{\Gamma_b(Q\pm 2P_s)}{\Gamma_b(\pm 2P_t)}\right) \frac{\Delta_+(P_1,P_4,P_t)\Delta_+(P_2,P_3,P_t)}{\Delta_-(P_1,P_2,P_s)\Delta_-(P_3,P_4,P_s)}
 \\
 \times \int_{\frac{Q}{4}+i\mathbb{R}}du \ S_b(u-P_{12s})S_b(u-P_{s34})S_b(u-P_{23t})S_b(u-P_{t14}) \hspace{3cm}
 \\
 \times S_b(\tfrac{Q}{2}-u+P_{1234}) S_b(\tfrac{Q}{2}-u+P_{st13}) S_b(\tfrac{Q}{2}-u+P_{st24})S_b(\tfrac{Q}{2}-u) \ .
 \label{eq:fm}
\end{multline}
In this formula, we use momentums $P_i$ and sums of momentums such as $P_{12s}=P_1+P_2+P_s$. We also introduced combinations of double Gamma functions: the double Sine function $S_b(x) = \frac{\Gamma_b(x)}{\Gamma_b(Q-x)}$, and the products
\begin{align}
 \Delta_\epsilon(P_1,P_2,P_3) =\prod_{\substack{\epsilon_1,\epsilon_2,\epsilon_3=\pm \\ \epsilon_1\epsilon_2\epsilon_3=\epsilon}} \Gamma_b\left(\tfrac{Q}{2}+\textstyle{\sum}_i\epsilon_iP_i\right)\ .
\end{align}
The analytic properties of the fusing matrix, and in particular the poles of the integrand, are such that the relation \eqref{eq:fusrel} between the two bases reduces to the relation \eqref{gfg} between hypergeometric conformal blocks, in the limit $P_1\to P_{\langle 2,1\rangle},\ P_s\to P_2 \pm \frac{b}{2}$.

All possible changes of bases for $N$-point conformal blocks on the sphere can be expressed in terms of two matrices: the fusing matrix, and the diagonal matrix that describes how $s$-channel blocks behave under a permutation of the first two fields,
\begin{align}
 \mathcal{F}^{(s)}_{\Delta_s}(\Delta_1,\Delta_2,\Delta_3,\Delta_4|z_1,z_2,z_3,z_4) = e^{i\pi(\Delta_s-\Delta_1-\Delta_2)} \mathcal{F}^{(s)}_{\Delta_s}(\Delta_2,\Delta_1,\Delta_3,\Delta_4|z_2,z_1,z_3,z_4)\ .
 \label{eq:bot}
\end{align}
(See Exercise \ref{exobot} for a proof and discussion of this relation.)
Some changes of bases can be written in several possible ways, giving rise to consistency constraints on the fusing matrix. All consistency constraints actually boil down to finitely many relations \cite{ms89b}, the least trivial being 
the \textbf{pentagon relation}\index{pentagon relation}, 
\begin{multline}
 \int_{i\mathbb{R}} dP_{23}\ 
 F_{\Delta_{12},\Delta_{23}}\begin{bmatrix} \Delta_2 & \Delta_3 \\ \Delta_1 & \Delta_{45} \end{bmatrix}
 F_{\Delta_{45},\Delta_{51}}\begin{bmatrix} \Delta_{23} & \Delta_4 \\ \Delta_1 & \Delta_5 \end{bmatrix}
 F_{\Delta_{23},\Delta_{34}}\begin{bmatrix} \Delta_3 & \Delta_4 \\ \Delta_2 & \Delta_{51} \end{bmatrix}
 \\
 = 
 F_{\Delta_{45},\Delta_{34}}\begin{bmatrix} \Delta_3 & \Delta_4 \\ \Delta_{12} & \Delta_5 \end{bmatrix}
 F_{\Delta_{12},\Delta_{51}}\begin{bmatrix} \Delta_2 & \Delta_{34} \\ \Delta_1 & \Delta_5 \end{bmatrix}
 \ .
\end{multline}
The pentagon relation follows from the existence of two ways of writing the linear relation between the two bases of five-point blocks that are depicted as filled pentagons in the following picture. One way uses two fusing matrices, the other way uses three fusing matrices:
\begin{align}
 \begin{tikzpicture}[baseline=(current  bounding  box.center), scale = .8]
  \foreach\i in {0, 1, 2, 3, 4}{
 \begin{scope}[shift = +(36+72*\i:3.5)]
   \ifthenelse{\equal{\i}{2}\OR\equal{\i}{4}}{  
   \fill[color = blue!20] (-36:1) -- (36:1) -- (108:1) -- (180:1) -- (-108:1) -- cycle;
   }
   \draw (-36:1) -- (36:1); % Useless line to avoid a bug.
   \draw (-108+144*\i:1) -- (36+144*\i:1) -- (180+144*\i:1);
   \foreach\j in {1, 2, 3, 4, 5}{
   \begin{scope}[rotate = 72*(\j-1)]
   \draw (-36:1) -- (36:1);
   \node at (1.1, 0) {$\j$};
   \end{scope}
   }
  \end{scope}
  }
  \foreach\i in {0, 1, 2}{
  \draw [-latex, ultra thick, red] (10+72*\i:3) -- (-10+72*\i:3);
  }
 \foreach\i in {3, 4}{
  \draw [-latex, ultra thick, red] (-10+72*\i:3) -- (10+72*\i:3);
  }
 \end{tikzpicture}
\end{align}

\subsubsection{Crossing symmetry in terms of the fusing matrix}

It is possible to eliminate the conformal blocks from the crossing symmetry equation \eqref{csd}, using the fusing relation and the linear independence of blocks within a given basis,
\begin{align}
 \sum_{\Delta_s,\bar{\Delta}_s} \frac{C_{12s} C_{s34}}{B_s}
F_{\Delta_s,\Delta_t}\begin{bmatrix} \Delta_2 & \Delta_3 \\ \Delta_1 & \Delta_4 \end{bmatrix}
F_{\bar{\Delta}_s,\bar{\Delta}_t}\begin{bmatrix} \bar{\Delta}_2 & \bar{\Delta}_3 \\ \bar{\Delta}_1 & \bar{\Delta}_4 \end{bmatrix}
=  \frac{C_{23t}C_{t41}}{B_t}\ .
\end{align}
This can be further simplified if the theory is diagonal.
In this case, we have $\Delta_i=\bar{\Delta}_i$ for $i=1,2,3,4$, and we should insert factors $\delta_{\Delta_s,\bar\Delta_s}$ (left-hand side) and $\delta_{\Delta_t,\bar\Delta_t}$ (right-hand side) in the above formula.
Using the inverse fusing matrix, we obtain 
\begin{align}
 \frac{C_{12s} C_{s34}}{B_s}
F_{\Delta_s,\Delta_t}\begin{bmatrix} \Delta_2 & \Delta_3 \\ \Delta_1 & \Delta_4 \end{bmatrix}
= \frac{C_{23t}C_{t41}}{B_t} 
F^{-1}_{\Delta_t,\Delta_s}\begin{bmatrix} \Delta_2 & \Delta_3 \\ \Delta_1 & \Delta_4 \end{bmatrix}
\ .
\label{ffm}
\end{align}
(Replacing one of the four Verma modules $\mathcal{V}_{\Delta_i}$ with the degenerate representation $\mathcal{R}_{\langle 2,1\rangle}$, this would reduce to Eq. \eqref{spsm}.)
This formulation of crossing symmetry can be useful, because the fusing matrix is a simpler object than the conformal blocks. Moreover, this formulation implies that in a diagonal theory, the structure constants are uniquely determined by crossing symmetry. (See Exercise \ref{exoudt}.) 
And we will now show that the solution of this equation can be written in terms of the fusing matrix. 
To do this, let us consider a pentagon relation where the Verma module $\mathcal{V}_{\Delta_{45}}$ is replaced with the identity representation $\mathcal{R}_{\langle 1,1\rangle}$. From the fusion rule \eqref{roof}, we deduce the triviality of the fusing matrix element
\begin{align}
 F_{\Delta_s,\Delta_t}\begin{bmatrix} \Delta_2 & \Delta_3 \\ \Delta_1 & \langle 1,1\rangle \end{bmatrix} = \delta_{\Delta_s,\Delta_3}\delta_{\Delta_t,\Delta_1}\ .
\end{align}
The pentagon relation therefore reduces to 
\begin{align}
 F_{\langle 1,1\rangle, \Delta_{4}} \begin{bmatrix} \Delta_s & \Delta_3\\ \Delta_s & \Delta_3\end{bmatrix} 
 F_{\Delta_s,\Delta_t}\begin{bmatrix} \Delta_2 & \Delta_3 \\ \Delta_1 & \Delta_4 \end{bmatrix} = 
 F_{\langle 1,1\rangle, \Delta_{t}} \begin{bmatrix} \Delta_2 & \Delta_3\\ \Delta_2 & \Delta_3\end{bmatrix}
 F_{\Delta_2,\Delta_4}\begin{bmatrix} \Delta_1 & \Delta_t \\ \Delta_s & \Delta_3 \end{bmatrix}\ .
\end{align}
This equation governs the behaviour of the fusing matrix under a permutation of its arguments.
Moreover, by definition, the fusing matrix is invariant under simpler permutations that leave $\Delta_s, \Delta_t$ untouched,
\begin{align}
 F_{\Delta_s,\Delta_t}\begin{bmatrix} \Delta_2 & \Delta_3 \\ \Delta_1 & \Delta_4 \end{bmatrix} = 
 F_{\Delta_s,\Delta_t}\begin{bmatrix}\Delta_1 & \Delta_4 \\  \Delta_2 & \Delta_3  \end{bmatrix} =
 F_{\Delta_s,\Delta_t}\begin{bmatrix} \Delta_3 & \Delta_2 \\ \Delta_4 & \Delta_1 \end{bmatrix}  \ .
 \end{align}
 And the inverse fusing matrix can be rewritten in terms of the fusing matrix,
 \begin{align}
 F^{-1}_{\Delta_t,\Delta_s}\begin{bmatrix} \Delta_2 & \Delta_3 \\ \Delta_1 & \Delta_4 \end{bmatrix} = 
 F_{\Delta_s,\Delta_t}\begin{bmatrix} \Delta_3 & \Delta_4 \\ \Delta_2 & \Delta_1 \end{bmatrix} \ .
\end{align}
Combining the last three equations, we obtain
\begin{align}
 & F_{\langle 1,1\rangle, \Delta_{4}} \begin{bmatrix} \Delta_s & \Delta_3\\ \Delta_s & \Delta_3\end{bmatrix}  
 F_{\langle 1,1\rangle, \Delta_{s}} \begin{bmatrix} \Delta_2 & \Delta_1\\ \Delta_2 & \Delta_1\end{bmatrix}  
 F_{\Delta_s,\Delta_t}\begin{bmatrix} \Delta_2 & \Delta_3 \\ \Delta_1 & \Delta_4 \end{bmatrix}
 \nonumber
 \\
=\ & 
F_{\langle 1,1\rangle, \Delta_{t}} \begin{bmatrix} \Delta_2 & \Delta_3\\ \Delta_2 & \Delta_3\end{bmatrix}
F_{\langle 1,1\rangle, \Delta_{4}} \begin{bmatrix} \Delta_t & \Delta_1\\ \Delta_t & \Delta_1\end{bmatrix}
F^{-1}_{\Delta_t,\Delta_s}\begin{bmatrix} \Delta_2 & \Delta_3 \\ \Delta_1 & \Delta_4 \end{bmatrix}
\ .
\end{align}
This shows that a solution of the crossing symmetry equation \eqref{ffm} can be built as
\begin{align}
 C_{123} = \mu_{\Delta_1} \nu_{\Delta_2}\nu_{\Delta_3}F_{\langle 1,1\rangle, \Delta_{1}} \begin{bmatrix} \Delta_2 & \Delta_3\\ \Delta_2 & \Delta_3\end{bmatrix}\ ,
 \label{cmnnf}
\end{align}
provided there exist functions $\mu,\nu$ of the conformal dimension such that 
the right-hand side is symmetric under permutations of $\Delta_1,\Delta_2,\Delta_3$. 
This right-hand side can actually be computed as a limit of the general expression \eqref{eq:fm} of the fusing matrix,
\begin{align}
 F_{\langle 1,1\rangle, \Delta_{t}} \begin{bmatrix} \Delta_1 & \Delta_4\\ \Delta_1 & \Delta_4\end{bmatrix} 
 & = 
 \lim_{P_2\to P_1}\lim_{P_s\to -\frac{Q}{2}} \lim_{P_3\to P_4} 
 F_{\Delta_s,\Delta_t}\begin{bmatrix} \Delta_2 & \Delta_3 \\ \Delta_1 & \Delta_4 \end{bmatrix} \ ,
 \label{eq:flim}
 \\
  &\propto \frac{1}{\prod_{\pm,\pm,\pm} \Gamma_b(\frac{Q}{2} \pm P_t\pm P_1\pm P_4)}\ ,
  \label{eq:fdozz}
\end{align}
where we neglect factors that can be absorbed into field renormalizations. (See Exercise \ref{exoflfg}.) This  coincides with the DOZZ formula \eqref{caaa} for the three-point function of Liouville theory, given the relation 
\begin{align}
 \Upsilon_b(x) = \frac{1}{\Gamma_b(x)\Gamma_b(Q-x)}\ ,
 \label{eq:upga}
\end{align}
between the Upsilon function that appears in the DOZZ formula, and the double Gamma function.

This derivation of the DOZZ formula is rather formal: proving the equations that we used, starting with the existence of the linear  relation \eqref{eq:fusrel} between $s$- and $t$-channel conformal blocks, requires a good control over the analytic properties of conformal blocks \cite{tv12}, which is beyond the scope of this text. Moreover, we have deduced the formula from the much more complicated expression for the fusing matrix, which we had to admit. In Chapter \ref{seccbm} we will solve Liouville theory in a more constructive and elementary way, which can be generalized to more complicated CFTs. We will also discuss under which assumptions Liouville theory is unique, starting with the assumption that there is at most one primary field with a given conformal dimension -- an assumption that we implicitly made when writing the crossing symmetry equation \eqref{csd} as a sum over conformal dimensions.

Rather than an appealing way of solving Liouville theory, the present discussion is a demonstration of its fundamental nature, as its three-point structure constant is a special case of the fusing matrix. Actually, the full fusing matrix coincides (up to simple factors) with Liouville theory's boundary three-point structure constant \cite{pt01}, and is sometimes called the Liouville fusing matrix.

\section{Exercises}

\begin{exo}[Spanning set for a highest-weight representation]
 ~\label{exospan}
For $|\Delta\rangle$ a primary state, prove the equality $U(\mathfrak{V})|\Delta\rangle =U(\mathfrak{V}^+)|\Delta\rangle$ in Eq. \eqref{ruv}.
To do this, prove that any state of the type $\prod_{i=1}^p L_{n_i}|\Delta\rangle$ with $n_1,\cdots n_p\in {\mathbb{Z}}$ belongs to $U(\mathfrak{V}^+)|\Delta\rangle$.
The proof can be done by hand in the cases $p=0,1,2$, and then by induction on $p$.
\end{exo}

\begin{exo}[Alternative spanning set for a Verma module]
 ~\label{exoot}
In $U(\mathfrak{V}^+)$, write the states belonging to  $\{\prod_{i=1}^p L_{-n_i}\}_{n_i\in\{1,2\}}$ up to the level $N=5$.
For which levels do we obtain a basis? Explain the observed results by studying whether $L_{-1}$ and $L_{-2}$ are algebraically independent.
In particular, show that the Virasoro commutation relations \eqref{vir} imply a relation of the type
\begin{align}
 [L_{-1},[L_{-1},[L_{-1},L_{-2}]]] \propto [L_{-2},[L_{-1},L_{-2}]]\ .
 \label{llll}
\end{align}
\end{exo}

\begin{exo}[Singular vectors at the level $N=4$]
~\label{exolf}
 Compute the singular vectors at the level $N=4$, and write the results in a table analogous to \eqref{lot}.
In particular, show that the five coefficients of the singular vectors obey five linear equations, and that the determinant of the system is a polynomial of degree three in the conformal dimension $\Delta$. 
If $\Delta_{\langle 2,2 \rangle} = \Delta_{\langle 4, 1\rangle}$, do the corresponding singular vectors coincide?
% NB: Answer: yes.
\end{exo}

\begin{exo}[Singular vectors at higher levels]
~\label{exohl}
Let $|\chi\rangle$ be
a null vector in the $p(N)$-dimensional level-$N$ subspace of a Verma module.
\begin{enumerate}
 \item How many equations do the constraints $L_1|\chi\rangle = L_2|\chi\rangle = 0$ provide? Count the excess equations using the pentagonal number theorem, and explain their existence using Eq. \eqref{llll}.
 \item
 Write a computer program for computing singular vectors at arbitrary levels, and check that the number of Verma modules that have a level $N$ singular vector coincides with the number of factorizations of $N$ into two positive integers.
\end{enumerate}
\end{exo}

\begin{exo}[Singular vectors that are not of the type $|\chi_{\langle r,s \rangle}\rangle$]
 ~\label{exosv}
Let us consider the Virasoro algebra with the coupling constant $b^2=-\frac{q}{p}$ where $p,q$ are strictly positive, coprime integers.
\begin{enumerate}
 \item 
Prove the identities 
\begin{align}
 \Delta_{\langle r,s \rangle}=\Delta_{\langle r+p,s+q \rangle}=\Delta_{\langle p-r,q-s \rangle}\ .
\end{align}
Under suitable assumptions on $r$ and $s$, show that $\mathcal{V}_{\Delta_{\langle r,s \rangle}}$ has two singular vectors $|\chi_{\langle r,s \rangle}\rangle$ and $|\chi_{\langle p-r,q-s \rangle}\rangle$.
\item
Show that each one of the two states  $|\chi_{\langle r,s \rangle}\rangle$ and $|\chi_{\langle p-r,q-s \rangle}\rangle$ has a descendant that is itself a singular vector at the level $pq+qr-ps$ in $\mathcal{V}_{\Delta_{\langle r,s \rangle}}$.
Assuming that these two singular vectors are in fact identical, enumerate all the singular vectors of $\mathcal{V}_{\Delta_{\langle r,s \rangle}}$.
\item
Which ones of these singular vectors are of the type $|\chi_{\langle r',s' \rangle}\rangle$? Show that the singular vector at the level $pq+qr-ps$ is not of this type in $\mathcal{V}_{\Delta_{\langle r,s \rangle}}$, although it is of this type when considered as a singular vector of the Verma modules generated by $|\chi_{\langle r,s \rangle}\rangle$ and $|\chi_{\langle p-r,q-s \rangle}\rangle$.
\end{enumerate}
\end{exo}

\begin{exo}[Characters of Virasoro representations]
 ~\label{exochar}
For a representation $\mathcal{R}$ of the Virasoro algebra, let us define the character 
\begin{align}
 \operatorname{ch}_\mathcal{R}(y) = \operatorname{Tr}_\mathcal{R} y^{L_0-\frac{c}{24}}\ .
\end{align}
\begin{enumerate}
 \item 
Show that the character of a Verma module is 
\begin{align}
 \operatorname{ch}_{\mathcal{V}_P}(y) = \frac{y^{-P^2}}{\eta(y)}\ ,
\end{align}
where $\eta(y) = y^{\frac{1}{24}}\prod_{n=1}^\infty(1-y^n)$ is the Dedekind eta function, and $P$ is the momentum. 
\item
Deduce that for generic values of the central charge, the character of a maximally degenerate representation is 
\begin{align}
 \operatorname{ch}_{\mathcal{R}_{\langle r,s\rangle}} = \frac{y^{-P^2_{\langle r,s\rangle}} - y^{-P^2_{\langle -r,s\rangle}}}{\eta(y)}\ .
\end{align}
\item
Let us assume $b^2=-\frac{q}{p}$ where $p,q$ are strictly positive integers, and $1\leq r\leq p-1$ and $1\leq s\leq q-1$. Using the results of Exercise \ref{exosv}, show that 
\begin{align}
 \operatorname{ch}_{\mathcal{R}_{\langle r,s\rangle}} = 
 \sum_{k\in\mathbb{Z}} \frac{y^{-P^2_{\langle r, s+2qk\rangle}} - y^{-P^2_{\langle r, -s+2qk\rangle}}}{\eta(y)}\ .
\end{align}
\end{enumerate}
\end{exo}

\begin{exo}[Singular vectors and Gram matrices]
 ~\label{exodmn}
 In a Verma module with a Hermitian form, let $M^{(N)}$ be the level $N$ Gram matrix. Show that the following statements are equivalent:
 \begin{itemize}
  \item $\det M^{(N)}=0$,
  \item there is a level $N$ state that is orthogonal to all states,
  \item there is a nontrivial subrepresentation that has a nonzero state at the level $N$,
  \item there is a singular vector at a level $N'\leq N$.
 \end{itemize}
(To construct the nontrivial subrepresentation, consider the space of states that are orthogonal to all states in the Verma module.)

These equivalences hold in Verma modules of the Virasoro algebra, but not necessarily of larger symmetry algebras. In general, $\det M^{(N)}$ can vanish not only due to singular vectors, but also to subsingular vectors: states that become singular vectors only in quotients of the Verma module \cite{dv95}. Subsingular vectors can exist when a zero-mode generator of the symmetry algebra has a non-diagonalizable action in a Verma module \cite{rad13}.
\end{exo}

\begin{exo}[Unitarity of Virasoro representations]
~\label{exoun}
Show that the Gram matrix $M^{(N)}$ of the level-$N$ subspace of the Verma module $\mathcal{V}_\Delta$ is positive definite in the limit $c\to \infty$, provided  $\Delta >0$. 
To do this, show that in this limit the diagonal elements of $M^{(N)}$ coincide with its eigenvalues, provided $M^{(N)}$ is written in a basis of creation operators of the type $(\prod_i L_{-n_i}) L_{-1}^p$ with $n_i\geq 2$.
In other words, the Virasoro algebra \eqref{vir} effectively reduces to a sum of commuting finite-dimensional subalgebras, 
\begin{align}
 \underset{c\to \infty}{\lim} \mathfrak{V}  
=  \operatorname{Span}(L_{-1},L_0,L_1) \oplus \bigoplus_{n=2}^\infty \operatorname{Span} (L_{-n}, L_n)\ .
\end{align}
\end{exo}

\begin{exo}[Dependence of the Witt algebra on the position]
 ~\label{exowitt}
For any reference point $z_0\in\mathbb{C}$, we define the generators
\begin{align}
 \ell_n^{(z_0)} = -(z-z_0)^{n+1}{\frac{\partial}{\partial z}}\ .
\end{align}
\begin{enumerate}
 \item Show that $\ell_n^{(z_0)}$ obey the commutation relations of the Witt algebra. To which value of $z_0$ do the original generators \eqref{elln} correspond?
Compute $\frac{\partial}{\partial z_0} \ell_n^{(z_0)}$, and compare with Eq. \eqref{pll}.
\item For any function $f:\mathbb{C}\to \mathbb{C}$, we define the field $V_f(z_0)$ as the function $z\mapsto f(z-z_0)$. Compute $\frac{\partial}{\partial z_0} V_f(z_0)$, and compare with Eq. \eqref{lvpv}.
\end{enumerate}
\end{exo}

\begin{exo}[Behaviour of the energy-momentum tensor at infinity]
 ~\label{exoti}
 If $z$ has dimension $-1$, what is the dimension of $L_{-1}$ according to Eq. \eqref{lvpv}? Then what is the dimension of $T(y)$? Deduce that the differential $T(y)dy^2$ is dimensionless, and should be holomorphic at infinity. Taking $\frac{1}{y}$ to be the natural coordinate at infinity, compare $T(y)dy^2$ with the holomorphic differential $\left(d(\frac{1}{y})\right)^2$, and deduce Eq. \eqref{tyi}.
\end{exo}

\begin{exo}[Virasoro algebra and OPE] 
~\label{exott}
Show that the $T(y)T(z)$ OPE \eqref{tt}, the commutativity axiom $T(y)T(z) = T(z)T(y)$, and the expansion \eqref{tsl} of $T(y)$ into modes $L_n^{(z_0)}$, imply that such modes obey the Virasoro commutation relations \eqref{vir} for any choice of $z_0$.
To do this, write 
\begin{align}
 [L_n^{(z_0)},L_m^{(z_0)}] = -\frac{1}{4\pi^2} \left(\oint_{z_0} dy \oint_{z_0} dz - \oint_{z_0} dz \oint_{z_0} dy\right) (y-z_0)^{n+1}(z-z_0)^{m+1} T(y)T(z)\ ,
\end{align}
and use contour manipulations to show that 
\begin{align}
 \oint_{z_0} dy \oint_{z_0} dz - \oint_{z_0} dz \oint_{z_0} dy = \oint_{z_0} dy \oint_y dz\ .
\end{align}
Explain why regular terms in the $T(y)T(z)$ OPE do not contribute to the result.
\end{exo}

\begin{exo}[From the identity field to the energy-momentum tensor]
 ~\label{exoit}
By comparing it with the general $T(y)V_\sigma(z)$ OPE \eqref{tv}, show that the $T(y)T(z)$ OPE \eqref{tt} encodes the equations
\begin{align}
 L_{-1}T(z) &= \partial T(z) \ ,
\label{lmt}
\\
L_0 T(z) &= 2 T(z)\ ,
\\
L_1 T(z) &= 0 \ ,
\\
L_2 T(z) &= \frac{c}{2} I\ ,
\\
L_{n\geq 3} T(z) &= 0 \ .
\label{lgt}
\end{align}
With the help of Eq. \eqref{pll}, show that these equations follow from 
\begin{align}
 \partial I & = 0\ ,
\label{piz}
\\
 L_{n\geq -1} I& = 0\ ,
\\
L_{-2}^{(z)}I & = T(z)\ ,
\label{let}
\end{align}
\end{exo}

\begin{exo}[Creation operators as differential operators]
 ~\label{exodma}
 Check that the representation \eqref{lmn} of creation operators $L_{-n}^{(z_i)}$ (with $n\geq 1$) as differential operators in $z_1,\cdots , z_N$, is consistent with the commutation relations of the Virasoro algebra. 
\end{exo}

\begin{exo}[Quasi-primary fields and representations of $\mathfrak{sl}_2$]
 ~\label{exoqp}
By the state-field correspondence, a quasi-primary field $V_\Delta(z)$ corresponds to a representation of the algebra of global conformal transformations $\mathfrak{sl}_2^{(z)} = \operatorname{Span}(L_{-1}^{(z)},L_0^{(z)},L_1^{(z)})$.
\begin{enumerate}
 \item 
Using the definition \eqref{lolz}, identify this representation as a highest-weight representation.
\item
How is this compatible with Eq. \eqref{tgv}, which suggests that $V_\Delta(z)$ transforms in a generic representation of $SL_2({\mathbb{C}})$? To solve the apparent contradiction, consider how the algebra $\mathfrak{sl}_2^{(z)}$ depends on the choice of $z$, and how $\mathfrak{sl}_2^{(z')}$ with $z'\neq z$ acts on representations of $\mathfrak{sl}_2^{(z)}$.
\end{enumerate}
\end{exo}

\begin{exo}[Logarithmic conformal field theory]
 ~\label{exolog}
Consider a finite-dimensional vector space $E$ with a linear action of $L_0$, and the representation $\mathcal{R}=U(\mathfrak{V}^+)E$ of the Virasoro algebra obtained by assuming $L_{n>0}E=0$.
\begin{enumerate}
 \item 
Assume that $\mathcal{R}$ is indecomposable but reducible: what does this mean for the action of $L_0$ on $E$, and the corresponding matrix? 
\item
Let $V_v(z)$ be the field that corresponds to a vector $v\in E$: write the global Ward identities for correlation functions of such fields, and prove 
\begin{align}
 {\frac{\partial}{\partial z_i}} \left\langle z_{12}^{L_0^{(1)}+L_0^{(2)}-L_0^{(3)}} z_{23}^{L_0^{(2)}+L_0^{(3)}-L_0^{(1)}} z_{13}^{L_0^{(1)}+L_0^{(3)}-L_0^{(2)}} \prod_{i=1}^3 V_{v_i}(z_i)\right\rangle = 0\ .
\end{align}
\item
Compare the three-point function $\left\langle  \prod_{i=1}^3 V_{v_i}(z_i)\right\rangle$ with the three-point function  of primary fields \eqref{fzzz}.
In the simplest case where $E$ is two-dimensional, show that the three-point function involves not only powers of $z_{ij}$, but also logarithms.
\item
Cite an example of a reducible Verma module, and conclude that 
a conformal field theory can involve indecomposable, reducible representations without being logarithmic. 
\end{enumerate}
\end{exo}

\begin{exo}[Behaviour of four-point functions under field permutations]
 ~\label{exoperm}
Due to the commutativity of fields \eqref{comm}, the four-point function of primary fields \eqref{zgg} is invariant under field permutations. Deduce the behaviour of the reduced four-point function \eqref{fx} under permutations of the dimensions $(\Delta_i,\bar \Delta_i)$. Using the notation 
\begin{align}
 F_{ijkl}(x) =\Big\langle V_{\Delta_i,\bar\Delta_i}(x)V_{\Delta_j,\bar\Delta_j}(0)V_{\Delta_k,\bar\Delta_k}(\infty) V_{\Delta_l,\bar\Delta_l}(1)\Big\rangle \ ,
\end{align}
with in particular $F(x) = F_{1234}(x)$, you should find, in the case of transpositions, 
\begin{align}
 F(x) 
 &= \left|(x-1)^{-\Delta_1-\Delta_2+\Delta_3-\Delta_4}\right|^2 F_{2134}\left(\tfrac{x}{x-1}\right)
 =\left|(x-1)^{-2\Delta_1}\right|^2 F_{1243}\left(\tfrac{x}{x-1}\right)\ ,
 \label{ftotf}
 \\
 & = \left|x^{-2\Delta_1}\right|^2 F_{1324}\left(\tfrac{1}{x}\right) 
  = \left|x^{-\Delta_1-\Delta_2+\Delta_3-\Delta_4}\right|^2 F_{4231}\left(\tfrac{1}{x}\right) 
  \ ,
  \\
&  = \left|x^{-\Delta_1-\Delta_2+\Delta_3+\Delta_4}(x-1)^{-\Delta_1+\Delta_2+\Delta_3-\Delta_4}\right|^2 F_{3214}(1-x)
  = F_{1432}(1-x)\ .
\end{align}
\end{exo}

\begin{exo}[Computing OPE coefficients]
 ~\label{exohf}
The OPE coefficients $f^{\Delta,L}_{\Delta_1,\Delta_2}$ at the level $|L|= 2$ are determined by the linear equations \eqref{eq:lwo}.
\begin{enumerate}
 \item Compute these coefficients,
and compare your results with the available literature.
\item
Write the equations for the OPE coefficients at the level $|L|=3$. 
\item
Show that $f^{\Delta,L}_{\Delta_1,\Delta_2}$ is uniquely determined for generic values of $\Delta$, 
by counting the equations and unknowns in Eq. \eqref{eq:lwo}, as in Exercise \ref{exohl}.
\item
Discuss how the presence of a null vector affects the equations for $f^{\Delta,L}_{\Delta_1,\Delta_2}$, starting with the case $\Delta=0$. 
\end{enumerate}
\end{exo}

\begin{exo}[$V_{\langle 1,1\rangle}$ is an identity field]
~\label{exoid}
Using $\frac{\partial}{\partial z_1} V_{\langle 1,1\rangle}(z_1)=0$, show that the OPE of $V_{\langle 1,1\rangle}$ with another primary field is of the form 
\begin{align}
 V_{\langle 1,1\rangle}(z_1)V_\Delta(z_2) = C_\Delta V_\Delta(z_2)\ ,
\end{align}
where the subleading terms vanish. Using associativity, show that the constant $C_\Delta$ actually does not depend on $\Delta$. Deduce that, up to a factor $C=C_\Delta$, the field $V_{\langle 1,1\rangle}$ is an identity field.
\end{exo}

\begin{exo}[Fusion rules from OPEs]
 ~\label{exooit}
Rederive the fusion rule \eqref{rot} by analyzing the corresponding OPE.
If $C$ is a contour around both $z_1$ and $z_2$, insert $\oint_C dz\, T(z)$ and $\oint_C dz \frac{1}{z-z_2}T(z)$ on both sides of that OPE, and compute the leading terms of the OPEs $V_{\Delta_1}(z_1) LV_{\langle 2,1 \rangle}(z_2)$ with $L\in\{L_{-1},L_{-1}^2,L_{-2}\}$, before using $L_{\langle 2,1 \rangle} V_{\langle 2,1 \rangle} (z_2) =0$. 
\end{exo}

\begin{exo}[Third-order BPZ equation]
 ~\label{exotob}
 Write the BPZ equation for an $N$-point function involving a degenerate field $V_{\langle 1,3 \rangle}(x)$.
In the case $N=3$, rederive the relevant fusion rule.
\end{exo}

\begin{exo}[Third-order BPZ equation for a four-point function]
 ~\label{exotbf}
 Write the third-order BPZ ordinary differential equation for the four-point function $\left\langle V_{\langle 3,1 \rangle}(x)V_{\Delta_1}(0)V_{\Delta_2}(\infty)V_{\Delta_3}(1)\right\rangle $.
Check that the characteristic exponents at $x=0,1,\infty$ are consistent with the fusion rules. 
\end{exo}

\begin{exo}[BPZ equations from fusion rules]
 ~\label{exoefr}
Let us assume that we do not know the structures of the Virasoro algebra or its representations, but only the fusion rules. We are interested in the four-point function $\mathcal{F}(x) = \left\langle V_{\langle 2,1 \rangle}(x)V_{\Delta_1}(0)V_{\Delta_2}(\infty)V_{\Delta_3}(1)\right\rangle$, where the degenerate field $V_{\langle 2,1 \rangle}$ is defined by the fusion rule \eqref{rot}. Since this fusion rule has two terms, we conjecture that $\mathcal{F}(x)$ obeys a second-order differential equation of the type 
\begin{align}
 \left\{\frac{\partial^2}{\partial x^2} + a(x) \frac{\partial}{\partial x} + b(x) \right\} \mathcal{F}(x) = 0 \ .
\end{align}
\begin{enumerate}
\item From the OPEs and fusion rules of $V_{\langle 2,1\rangle}$, deduce that $\mathcal{F}(x)$ has regular singular points at $x=0,1,\infty$, and compute its characteristic exponents. (Use Eq. \eqref{iope} for the OPE $V_{\langle 2,1 \rangle}(x)V_{\Delta_2}(\infty)$.)
\item Using the ansatz $\mathcal{F}(x)\underset{x\to 0}{=}x^\lambda(1+O(x))$, show that $a(x)$ has a simple pole at $x=0$, and that $b(x)$ has a double pole. Using the ansatz $\mathcal{F}(x)\underset{x\to \infty}{=}x^\lambda(1+O(x^{-1}))$, show that near $x=\infty$ we have $a(x)=O(x^{-1})$ and $b(x)= O(x^{-2})$. Assuming that $a(x)$ and $b(x)$ are meromorphic functions with no poles outside $\{0,1,\infty\}$, write these functions in terms of five coefficients, and deduce that the six characteristic exponents obey the relation
\begin{align}
 \sum_\pm \left(\lambda^{(0)}_\pm + \lambda^{(1)}_\pm -\lambda^{(\infty)}_\pm\right) = 1\ .
\end{align}
\item Show that this relation is satisfied by the characteristic exponents that follow from the fusion rules, and compute the functions $a(x)$ and $b(x)$. Compare the resulting differential equation with the BPZ equation \eqref{sode}.
 \item Show that the third-order BPZ equation is not completely determined by its characteristic exponents. In particular, show that adding a term proportional to $\frac{1}{x^2(x-1)^2}$ to the differential operator $\frac{\partial^3}{\partial x^3} + \cdots$ does not affect the characteristic exponents.
\end{enumerate}

\end{exo}

\begin{exo}[Hypergeometric form of the second-order BPZ equation for a four-point function]
 ~\label{exohge}
 Check that the change of unknown function $\mathcal{F}(x)=x^{\lambda}(1-x)^{\mu}\mathcal{G}(x)$ in the second-order BPZ equation \eqref{sode} leads to a hypergeometric equation \eqref{dzp} for $\mathcal{G}(x)$, provided $\lambda$ and $\mu$ are suitably chosen.
What are the four possible choices for the pair $(\lambda,\mu)$?
Check that Eq. \eqref{heg} corresponds to one of these choices.
\end{exo}

\begin{exo}[Behaviour of conformal blocks under a permutation]
 ~\label{exobot}
Let us prove and discuss the relation \eqref{eq:bot} that describes how an $s$-channel conformal block behaves under the permutation of the first two fields.
\begin{enumerate}
 \item Deduce the relation \eqref{eq:bot} from the existence and uniqueness of the decomposition a four-point function into conformal blocks, given a choice of basis of conformal blocks.
 \item 
 With the help of Eq. \eqref{ftotf}, show that this relation is equivalent to 
 \begin{multline}
  \mathcal{F}^{(s)}_{\Delta_s}(\Delta_1,\Delta_2,\Delta_3,\Delta_4|x) 
  \\
  = e^{i\pi(\Delta_s-\Delta_1-\Delta_2)} (1-x)^{-\Delta_1-\Delta_2+\Delta_3-\Delta_4} \mathcal{F}^{(s)}_{\Delta_s}\left(\Delta_2,\Delta_1,\Delta_3,\Delta_4\middle|\tfrac{x}{x-1}\right)\ .
 \end{multline}
 \item Check that this relation is compatible with the expansion \eqref{eq:fsexp}.
 \item Show that this relation is compatible with Zamolodchikov's recursion. To do this, you may use the identities 
 \begin{align}
  q\left(\tfrac{x}{x-1}\right)=-q(x) \quad , \quad \theta_3(-q) = (x-1)^{\frac14}\theta_3(q)\ ,
 \end{align}
 and derive the identity
 \begin{align}
  \left. R_{m,n}\right|_{P_1\leftrightarrow P_2} = (-1)^{mn} R_{m,n}\ .
 \end{align}
\end{enumerate}

\end{exo}

\begin{exo}[Uniqueness of the three-point structure constant in diagonal theories]
 ~\label{exoudt}
 Let us show that the crossing symmetry equation \eqref{ffm} uniquely determines the two- and three-point structure constant modulo field and structure constant renormalizations.
 \begin{enumerate}
  \item 
  We set the two-point structure constant to one by a field renormalization, and assume that we have two solutions $C^\pm_{123}$ for the three-point structure constant. 
  Show that the ratio $\rho_{123}=\frac{C^+_{123}}{C^-_{123}}$ obeys $\rho_{12s}\rho_{s34} = \rho_{23t}\rho_{t41}$.
  \item Show that $\rho_{12s}^2 = \rho_{12t}^2$, and deduce that $\rho_{123}\in\{-1,1\}$ modulo a structure constant renormalization.
  \item Show that $\frac{\rho_{12s}}{\rho_{12s'}} = \frac{\rho_{34s}}{\rho_{34s'}} $, and deduce that our ratio factorizes as $\rho_{123} = \lambda_1\lambda_2\lambda_3$. Conclude.
 \end{enumerate} 
\end{exo}

\begin{exo}[A special case of the fusing matrix of Verma modules]
 ~\label{exoflfg}
 Let us prove that the limit \eqref{eq:flim} of the fusing matrix exists, and is given by Eq. \eqref{eq:fdozz}. 
 \begin{enumerate}
 \item Study how $s$-channel conformal blocks behave in our limit, and deduce the relation \eqref{eq:flim} between the fusing matrix of Verma modules, and the fusing matrix that involves the degenerate representation $\mathcal{R}_{\langle 1,1\rangle}$. 
  \item Accepting that the double Gamma function $\Gamma_b(x)$ has simple poles for $x\in -b\mathbb{N}-b^{-1}\mathbb{N}$, study the behaviour of the poles of the integrand in the expression \eqref{ffm} of the fusing matrix. Deduce that in our limit, the integral reduces to its residue at $u=P_{t14}$.
  \item Compute the limit of the integral, and of the prefactors. Check that the result is Eq. \eqref{eq:fdozz} up to field renormalizations, and conclude that we found a solution of the crossing symmetry equation.
 \end{enumerate}
\end{exo}

% NB: Ensuring the coming chapter begins on an odd-numbered page, in case we want to print the document in two halves.

\cleardoublepage

\chapter{Liouville theory and minimal models \label{seccbm}}

In this chapter we introduce and solve Liouville theory and minimal models, which are the simplest nontrivial two-dimensional conformal field theories, and appear in many applications. In order to complete the picture of Section \ref{secmomo}, we also discuss the less well-known generalized minimal models and Runkel--Watts-type theories. 

\section{Liouville theory \label{secliou}}

\subsection{Definition and spectrum \label{secspe}}

We want to define \textbf{Liouville theory}\index{Liouville!---theory} as the simplest possible nontrivial theory with a continuous spectrum. More specifically, we assume
\begin{enumerate}
 \item that Liouville theory is a family of conformal field theories, parametrized by the central charge $c\in\mathbb{C}$;
\item that each theory has a continuous spectrum, where representations of the symmetry algebra $\mathfrak{V}\times\overline{\mathfrak{V}}$ have multiplicities zero or one;
\item and that correlation functions are meromorphic functions of the coupling constant and of the fields' momentums.
\end{enumerate}
Let us first show that these assumptions determine the spectrum of Liouville theory.

\subsubsection{Spectrum}

Since the spectrum is continuous, it must involve Verma modules -- the other type of highest-weight representations, degenerate representations, form a discrete set. 
According to Axiom \ref{ax:sots}, real parts of conformal dimensions are bounded from below, let us guess their lower bound. Remember that fusion rules of degenerate fields are analytic if expressed in terms of momentums rather than conformal dimensions: this is why we use momentums in our meromorphicity assumption. The relation \eqref{daq} between momentums and dimensions has a unique critical point $P=0$ such that $\Delta'(P)=0$: this provides a natural lower bound, and leads to the following values for the momentums and dimensions:
\begin{align}
 P \in i{\mathbb{R}} \quad \iff \quad \Delta \in \frac{c-1}{24}+\mathbb{R}_+\ .
\label{aqd}
\end{align}
Other guesses for the spectrum might also seem plausible: in particular, for $c>1$, unitarity would allow $\Delta> 0$ rather than $\Delta \geq \frac{c-1}{24}$. 
The decisive argument in favour of our spectrum will be the consistency of the resulting CFT, i.e. 
crossing symmetry of the resulting four-point functions, see Section \ref{seceul}.

The representation $\mathcal{V}_\Delta\otimes \overline{\mathcal{V}}_{\bar{\Delta}}$ can appear in the spectrum only if both 
$\Delta$ and $\bar{\Delta}$ obey the condition \eqref{aqd}. 
Moreover, single-valuedness of correlation functions requires $\Delta-\bar{\Delta}\in {\frac12\mathbb{Z}} $, see Eq. \eqref{dbdz}.
By continuity, $\Delta-\bar{\Delta}$ must be a constant, and this constant must be zero for both $\Delta$ and $\bar{\Delta}$ to span the whole half-line $\left[\frac{c-1}{24},\infty\right[$.
Therefore, the spectrum must be diagonal,
\begin{align}
 \boxed{ \mathcal{S}= \frac12\int_{i{\mathbb{R}}} dP\ \mathcal{V}_P \otimes \overline{\mathcal{V}}_P =\int_{i{\mathbb{R}_+}} dP\ \mathcal{V}_P \otimes \overline{\mathcal{V}}_P = \int_{\frac{c-1}{24}}^\infty d\Delta\ \mathcal{V}_\Delta\otimes \overline{\mathcal{V}}_\Delta} \ ,
\label{sad}
\end{align}
where the  factor $\frac12$ and the restriction of the integration domain to the half-line $P\in i\mathbb{R}_+$ are two ways of
eliminating the redundancy that comes from the reflection relation $\mathcal{V}_P=\mathcal{V}_{-P}$. This spectrum is unitary if $c>1$.

A subtlety occurs if our spectrum involves representations with dimensions $\Delta_{\langle r, s\rangle}$, which happens if and only if $c\leq 1$ i.e. $Q\in i\mathbb{R}$. 
We would then have singularities in the decomposition \eqref{fsd} of four-point functions into conformal blocks, as the conformal block $\mathcal{F}_{\Delta_s}^{(s)}(\sigma_i|z_i)$ has a pole at $\Delta_s = \Delta_{\langle r, s\rangle}$. 
The solution is simply to replace the line $P\in i\mathbb{R}$ with the slightly shifted line 
\begin{align}
 (c\leq 1) \qquad P \in i\mathbb{R} + \epsilon\ ,
\end{align}
and the four-point function will turn out to be independent of the value of the regularizing parameter $\epsilon \in \mathbb{R}^*$. 
This subtlety with the spectrum is a first hint that something special happens if $c\leq 1$ -- we will find subtleties with the three-point structure constants too.

\subsubsection{Fields and correlation functions}

We work with momentums on the half-line $P\in i\mathbb{R}_+$, so that $\Delta(P_1)=\Delta(P_2)\iff P_1=P_2$. 
Then the two-point function takes the form
\begin{align}
 \Big\langle V_{P_1}(z_1) V_{P_2}(z_2)\Big\rangle = B_{P_1}\delta(P_1-P_2) |z_{12}|^{-4\Delta(P_1)}\ , 
\label{vvc}
\end{align}
where $B_P$ is the two-point structure constant. 
According to Eq. \eqref{cff}, the three-point function takes the form
\begin{align}
\left\langle \prod_{i=1}^3 V_{P_i}(z_i)\right\rangle = C_{P_1,P_2,P_3}\ |z_{12}|^{2(\Delta_3-\Delta_1-\Delta_2)} |z_{23}|^{2(\Delta_1-\Delta_2-\Delta_3)} |z_{31}|^{2(\Delta_3-\Delta_2-\Delta_1)}\ ,
\label{vvv}
\end{align}
where $\Delta_i = \Delta(P_i)$ is given in Eq. \eqref{daq}, and $C_{P_1,P_2,P_3}$ is the three-point structure constant. 
Omitting the dependence on $z_i$, and neglecting the descendant fields, we schematically rewrite the two- and three-point functions, and the OPE, as follows: 
\begin{align}
 &\boxed{\left\langle V_{P_1}V_{P_2} \right\rangle = B_{P_1}\delta(P_1-P_2) }\ ,
\label{vvss}
\\
 &\boxed{ \left\langle V_{P_1}V_{P_2}V_{P_3} \right\rangle = C_{P_1,P_2,P_3} }\ ,
\label{vvvs}
\\
 &\boxed{V_{P_1}V_{P_2} \sim \int_{i{\mathbb{R}_+}} dP \frac{C_{P_1,P_2,P}}{B_P} V_P}\ .
\label{vvi}
\end{align}

\subsubsection{Degenerate fields}

Although degenerate representations do not appear in the spectrum of Liouville theory, assuming  that they exist is crucial for analytically solving the theory using the conformal bootstrap method. This is because the crossing symmetry equations become simpler in the presence of degenerate fields, as we explained in Section \ref{seccsfm}.

\begin{hyp}[Existence of degenerate fields]
 ~\label{ax:edf}
There exist diagonal degenerate fields $V_{\langle r,s \rangle}$ for $r,s\in\mathbb{N}^*$, and correlation functions thereof.
\end{hyp}
\noindent
Equivalently, we could assume that the two degenerate fields $V_{\langle 2,1\rangle}$ and $V_{\langle 1,2\rangle}$ exist, and construct the remaining degenerate fields by repeatedly performing OPEs of these two degenerate fields.

Axiom \ref{ax:edf} means that there exist correlation functions that involve degenerate fields, and that obey our axioms on fields and correlation functions. However, Axiom \ref{ax:ope} on the existence of an OPE involves a sum over the whole spectrum, which would be inconsistent with the degenerate fields' fusion rules. 
Let us work out the natural extension of that axiom to degenerate fields, and in particular determine which fields appear in the OPE $V_{\langle r,s \rangle} V_P$. First and foremost, these fields must obey the fusion rules
\eqref{rtv}. The fusion rules apply independently to left- and right-moving representations, and  a priori they allow non-diagonal fields, for example a field with left momentum $P-\frac{b}{2}$ and right momentum $P+\frac{b}{2}$ could appear in the OPE $V_{\langle 2,1\rangle} V_P$. However, for generic value of $P$, such non-diagonal fields have spins that are not half-integer, and therefore violate the single-valuedness condition on spins Eq. \eqref{dbdz}.
Therefore, our OPE must involve only diagonal fields, and must be of the type
\begin{align}
 \boxed{V_{\langle r,s \rangle} V_P \sim \sum_{i=-\frac{r-1}{2}}^{\frac{r-1}{2}} \sum_{j=-\frac{s-1}{2}}^{\frac{s-1}{2}}  C_{i,j}^{\langle r,s \rangle}(P) V_{P + ib+jb^{-1}}}\ .
\label{vrsv}
\end{align}
For generic values of $b$, we have $P\in i\mathbb{R} \centernot\implies P+ib+jb^{-1}\in i\mathbb{R}$, and the field $ V_{P + ib+jb^{-1}}$ may not belong to the spectrum. However, we now resort to our assumption that correlation functions are meromorphic functions of momentums, and intepret $ V_{P + ib+jb^{-1}}$ as an analytic continuation of the fields with imaginary momentums.
In particular, the OPEs of the degenerate fields $V_{\langle 2,1\rangle}$ and $V_{\langle 1,2\rangle}$ can be written as 
\begin{align}
 &\boxed{V_{\langle 2,1 \rangle} V_P \sim C_+(P) V_{P+\frac{b}{2}} + C_-(P) V_{P-\frac{b}{2}}}\ ,
\label{vot}
\\
& \boxed{V_{\langle 1,2 \rangle} V_P \sim \tilde{C}_+(P) V_{P+\frac{1}{2b}} + \tilde{C}_-(P) V_{P-\frac{1}{2b}}}\ .
 \label{vto}
\end{align}
The coefficients $C_+(P)$ and $C_-(P)$ are not independent, and can be deduced from one another, see Exercise \ref{exocpcm}.

\subsection{Degenerate crossing symmetry equations \label{seccbe}}

Let us now investigate the consequences of the crossing symmetry of four-point functions. For simplicity, we focus on crossing symmetry of four-point functions that involve one degenerate field $V_{\langle 2,1\rangle}$ or $V_{\langle 1, 2\rangle}$. This will nevertheless turn out to uniquely determine the three-point structure constant $C_{P_1,P_2,P_3}$. 

In order to write the degenerate crossing symmetry equations, we need to constrain the OPE coefficients of the degenerate fields $V_{\langle 2,1\rangle}$ and $V_{\langle 1, 2\rangle}$. 
To do this, we will first consider four-point functions that involve two degenerate fields. 

\subsubsection{What we learn from $
 \left\langle V_{\langle 2,1 \rangle}(x) V_P(0) V_{P}(\infty) V_{\langle 2,1 \rangle}(1) \right\rangle$}
 
Using the degenerate OPE \eqref{vot} and the two-point function \eqref{vvss}, we find that this four-point function has an $s$-channel decomposition of the type \eqref{zsc}, with the following conformal blocks and structure constants:
\begin{align}
 \begin{tikzpicture}[baseline=(current  bounding  box.center), very thick, scale = .6]
\draw (-1,2) node [left] {$P$} -- (0,0) -- node [above] {$P-\frac{b}{2}$} (4,0) -- (5,2) node [right] {$P$};
\draw (-1,-2) node [left] {$\langle 2,1\rangle$} -- (0,0);
\draw (4,0) -- (5,-2) node [right] {$\langle 2,1\rangle$};
\node at (1, -4) {$c_{-}^{(s)} = \colorboxed{red}{C_-(P)}\, \colorboxed{red}{B_{P-\tfrac{b}{2}}}\, \colorboxed{red}{C_-(P)} $};
\draw[dashed, ->, red] (-.7,-3.2) to [out=90, in=-70] (.2, -.3);
\draw[dashed, ->, red] (4.6,-3.2) to [out=90, in=-110] (3.8, -.3);
\draw[dashed, ->, red] (1.9,-3.2) to [out=90, in=-110] (2, -.3);
\end{tikzpicture} 
\qquad\quad
\begin{tikzpicture}[baseline=(current  bounding  box.center), very thick, scale = .6]
\draw (-1,2) node [left] {$P$} -- (0,0) -- node [above] {$P+\frac{b}{2}$} (4,0) -- (5,2) node [right] {$P$};
\draw (-1,-2) node [left] {$\langle 2,1\rangle$} -- (0,0);
\draw (4,0) -- (5,-2) node [right] {$\langle 2,1\rangle$};
\node at (1, -4) {$c_{+}^{(s)} = \colorboxed{red}{C_+(P)}\, \colorboxed{red}{B_{P+\tfrac{b}{2}}}\, \colorboxed{red}{C_+(P)} $};
\draw[dashed, ->, red] (-.7,-3.2) to [out=90, in=-70] (.2, -.3);
\draw[dashed, ->, red] (4.6,-3.2) to [out=90, in=-110] (3.8, -.3);
\draw[dashed, ->, red] (1.9,-3.2) to [out=90, in=-110] (2, -.3);
\end{tikzpicture} 
\label{eq:ddgg}
\end{align} 
Crossing symmetry and single-valuedness of the four-point function imply that the two structure constants $c_\pm^{(s)}$ obey Eq. \eqref{spsm}. In our case, this equation reduces to 
\begin{align}
 \frac{C_+(P)^2B_{P+\tfrac{b}{2}}}{C_-(P)^2 B_{P-\tfrac{b}{2}}}
 =  \frac{\gamma(2bP)}{\gamma(-2bP)}
 \frac{\gamma(-b^2-2bP)}{\gamma(-b^2+2bP)} \ .
 \label{eq:shiftd}
\end{align}
This is all that we will need to know about the degenerate OPE coefficients $C_\pm(P)$. A dual equation can be derived for $\tilde{C}_\pm (P)$, using a  four-point function of the type  $
 \left\langle V_{\langle 1,2 \rangle} V_P V_{P} V_{\langle 1,2 \rangle} \right\rangle$ instead of $
 \left\langle V_{\langle 2,1 \rangle} V_P V_{P} V_{\langle 2,1 \rangle} \right\rangle$. 
 That dual equation is obtained by performing 
 the substitutions $C_\pm \to \tilde{C}_\pm$ and $b\to \frac{1}{b}$,
 \begin{align}
 \frac{\tilde{C}_+(P)^2B_{P+\tfrac{1}{2b}}}{\tilde{C}_-(P)^2 B_{P-\tfrac{1}{2b}}}
 =  \frac{\gamma(2b^{-1}P)}{\gamma(-2b^{-1}P)}
 \frac{\gamma(-b^{-2}-2b^{-1}P)}{\gamma(-b^{-2}+2b^{-1}P)} \ .
\end{align}
Additional constraints on degenerate OPE coefficients can be derived using a four-point function of the type $
 \left\langle V_{\langle 2,1\rangle} V_P V_{\langle 1, 2\rangle} V_{P+\frac{Q}{2}}\right\rangle$, see
Exercise \ref{exonorm}. But we will not need these additional constraints.

\subsubsection{What we learn from $\left\langle V_{\langle 2,1 \rangle}(x)V_{P_1}(0)V_{P_2}(\infty)V_{P_3}(1)\right\rangle$}

Using the degenerate OPE \eqref{vot} and the three-point function \eqref{vvvs}, we find that this four-point function has an $s$-channel decomposition of the type \eqref{zsc}, with the following conformal blocks and structure constants:
\begin{align}
 \begin{tikzpicture}[baseline=(current  bounding  box.center), very thick, scale = .6]
\draw (-1,2) node [left] {$P_1$} -- (0,0) -- node [above] {$P_1-\frac{b}{2}$} (4,0) -- (5,2) node [right] {$P_2$};
\draw (-1,-2) node [left] {$\langle 2,1\rangle$} -- (0,0);
\draw (4,0) -- (5,-2) node [right] {$P_3$};
\node at (1.5, -4) {$c_{-}^{(s)} = \colorboxed{red}{C_-(P_1)}\, \colorboxed{red}{C_{P_1-\frac{b}{2},P_2,P_3}}  $};
\draw[dashed, ->, red] (.7,-3.2) to [out=90, in=-70] (.2, -.3);
\draw[dashed, ->, red] (3.5,-3.2) to [out=90, in=-110] (3.8, -.3);
\end{tikzpicture} 
\qquad\qquad
\begin{tikzpicture}[baseline=(current  bounding  box.center), very thick, scale = .6]
\draw (-1,2) node [left] {$P_1$} -- (0,0) -- node [above] {$P_1+\frac{b}{2}$} (4,0) -- (5,2) node [right] {$P_2$};
\draw (-1,-2) node [left] {$\langle 2,1\rangle$} -- (0,0);
\draw (4,0) -- (5,-2) node [right] {$P_3$};
\node at (1.5, -4) {$c_{+}^{(s)} = \colorboxed{red}{C_+(P_1)}\, \colorboxed{red}{C_{P_1+\frac{b}{2},P_2,P_3}}  $};
\draw[dashed, ->, red] (.7,-3.2) to [out=90, in=-70] (.2, -.3);
\draw[dashed, ->, red] (3.5,-3.2) to [out=90, in=-110] (3.8, -.3);
\end{tikzpicture} 
\label{cs}
\end{align}
Crossing symmetry and single-valuedness of the four-point function imply that the two structure constants $c_\pm^{(s)}$ obey Eq. \eqref{spsm}. In our case, this equation reads
\begin{align}
 \frac{C_+(P_1) C_{P_1+\frac{b}{2},P_2,P_3}}{C_-(P_1) C_{P_1-\frac{b}{2},P_2,P_3} } 
 =\gamma(2bP_1) \gamma(1+2bP_1)\prod_{\pm,\pm} \gamma(\tfrac12 -bP_1 \pm bP_2 \pm bP_3)\ .
 \label{eq:shift}
\end{align}
This equation determines how the three-point structure constant behaves under the shifts $P_1\to P_1+b$. 
Using a four-point function of the type  $
 \left\langle V_{\langle 1,2 \rangle} V_{P_1}V_{P_2}V_{P_3}\right\rangle$ instead of $
 \left\langle V_{\langle 2,1 \rangle} V_{P_1}V_{P_2}V_{P_3} \right\rangle$, we would obtain a dual equation for the shift $P_1\to P_1+\frac{1}{b}$. That dual equation is obtained from the above equation by the substitutions $C_\pm \to \tilde{C}_\pm$ and $b\to \frac{1}{b}$.

\subsubsection{Normalization-independent quantities} 

Crossing symmetry equations are invariant under field and structure constant renormalizations, and can therefore only determine renormalization-invariant combinations of structure constants.
We could focus on combinations of two- and three-point structure constants that are invariant under field renormalizations. However, the combination $\frac{C_{P_1,P_2,P_3}}{\sqrt{B_{P_1}B_{P_2}B_{P_3}}}$ involves a square root and is therefore not meromorphic, and the combination $\frac{C_{P_1,P_2,P_3}^2}{B_{P_1}B_{P_2}B_{P_3}}$ loses track of the sign of the three-point structure constant. 
Rather, we will decompose the three-point structure constant as 
\begin{align}
 C_{P_1,P_2,P_3} = N_{P_1}N_{P_2}N_{P_3}C'_{P_1,P_2,P_3}\ ,
 \label{cnnnc}
\end{align}
where the normalization factor $N_{P}$ transforms as $N_{P}\to \lambda_PN_{P}$ under a field renormalization $V_P\to \lambda_PV_P$, while  $C'_{P_1,P_2,P_3}$ is invariant. In other words, $C'_{P_1,P_2,P_3}$ is the three-point structure constant in a reference normalization which we can in principle choose arbitrarily. We now choose a normalization such that $C'_{P_1,P_2,P_3}$ is as simple as possible, by assuming that the shift equation \eqref{eq:shift} splits into the two equations 
\begin{align}
\frac{C'_{P_1+\frac{b}{2},P_2,P_3}}{C'_{P_1-\frac{b}{2},P_2,P_3} } 
 &=b^{8bP_1}\prod_{\pm,\pm} \gamma(\tfrac12 -bP_1 \pm bP_2 \pm bP_3)\ ,
 \label{cps}
 \\
\frac{C_+(P_1) N_{P_1+\frac{b}{2}}}{C_-(P_1) N_{P_1-\frac{b}{2}} } 
 &=b^{-8bP_1}\gamma(2bP_1) \gamma(1+2bP_1) \ .
 \label{cns}
\end{align}
Combining the second equation with the shift equation \eqref{eq:shiftd}, we obtain a shift equation for the combination $N^2B^{-1}$,
\begin{align}
 \frac{\left(N^2B^{-1}\right)_{P+\frac{b}{2}}}{\left(N^2B^{-1}\right)_{P-\frac{b}{2}}} = b^{-16bP} \frac{\gamma(2bP)}{\gamma(-2bP)} \frac{\gamma(-b^2+2bP)}{\gamma(-b^2-2bP)}\ .
 \label{nbs}
\end{align}
When splitting the shift equation, we have separated the factors that only depend on $P_1$ from the rest, and added the prefactors $b^{\pm 8bP_1}$. These prefactors are here to ensure that the resulting shift equations are compatible with the dual equations, in other words that the shifts by $b$ and $\frac{1}{b}$ commute. Next we will elaborate on this point, and solve the shift equations.

\subsection{The three-point structure constant \label{sectpf}}

For $b^2\in \mathbb{R}$, the shift equation \eqref{cps} and its dual under $b\to \frac{1}{b}$ 
determine the renormalized three-point structure constant up to a momentum-independent factor. 
This is because any smooth function with two incommensurable periods must be constant, and we assumed the three-point structure constant to be a smooth function of the momentums and of $b$.
So we will distinguish three cases: two cases with $b^2\in\mathbb{R}$, and the case $b^2\notin\mathbb{R}$. In each case, we plot $b$ and $\frac{1}{b}$ as vectors in the complex plane:
\begin{equation}
 \begin{tikzpicture}[baseline=(current  bounding  box.center), scale = .7]
\draw (0, 2) node[left]{$i$} -- (0, 1) node[below left] {$0$} -- (1, 1) node[below] {$1$};
\draw [thick, latex-latex,red] (4,3) -- (4,1) node[fill, circle, minimum size = 1mm, inner sep = 0]{} -- (4,-.3);
\draw [thick, latex-latex, red] (8,3) -- (7,1) node[fill, circle, minimum size = 1mm, inner sep = 0]{}-- (7.6,-.2);
\draw [thick, latex-latex, red] (12,1) -- (10,1) node[fill, circle, minimum size = 1mm, inner sep = 0]{} -- (11.3,1) ;
\node at (4, -1.5){$\begin{array}{c} b\in i\mathbb{R} \\ c\leq 1 \end{array}$};
\node at (7.5, -1.5){$\begin{array}{c} b^2\notin \mathbb{R} \\ c\in\mathbb{C} \end{array}$};
\node at (11, -1.5){$\begin{array}{c} b\in \mathbb{R} \\ c\geq 25 \end{array}$};
 \end{tikzpicture}
\end{equation}

\subsubsection{The special function that we need}

In all cases, in order to solve Eq. \eqref{cps} together with the dual equation, we need a function $\Upsilon_b(x)$ such that 
\begin{align}
 \frac{\Upsilon_b(x+b)}{\Upsilon_b(x)} \sim \gamma(bx)\quad \text{and} \quad \frac{\Upsilon_b(x+\frac{1}{b})}{\Upsilon_b(x)} \sim \gamma(\tfrac{x}{b})\ ,
\end{align}
where the $\sim$ sign indicates that some simple factors may be missing.
Adding simple factors is actually necessary for ensuring the compatibility of these equations, so that the following two expressions for $ \frac{\Upsilon_b(x+b+\frac{1}{b})}{\Upsilon_b(x)}$ coincide:
\begin{align}
  \frac{\Upsilon_b(x+b+\frac{1}{b})}{\Upsilon_b(x+b)} \frac{\Upsilon_b(x+b)}{\Upsilon_b(x)} = \frac{\Upsilon_b(x+\frac{1}{b}+b)}{\Upsilon_b(x+\frac{1}{b})} \frac{\Upsilon_b(x+\frac{1}{b})}{\Upsilon_b(x)} \ .
\end{align}
The missing simple factors can be determined with the help of the identity
\begin{align}
 \gamma(x+1) = -x^2 \gamma(x) \ .
\end{align}
If $b>0$, the compatible equations that result from inserting the missing factors are 
\begin{align}
  \boxed{\frac{\Upsilon_b(x+b)}{\Upsilon_b(x)} = b^{1-2bx} \gamma(bx)}\quad \text{and} \quad \boxed{\frac{\Upsilon_b(x+\frac{1}{b})}{\Upsilon_b(x)} = b^{\frac{2x}{b}-1} \gamma(\tfrac{x}{b})}\ ,
\label{upup}
\end{align}
These equations define a unique (up to a constant factor) \textbf{Upsilon function}\index{Upsilon function} $\Upsilon_b(x)$, which turns out to be defined for any $b$ such that $\Re b > 0$ by analytic continuation. 
This function can be constructed from the Barnes double Gamma function using Eq. \eqref{eq:upga}.
On the other hand, if $ib>0$, the compatible equations are 
\begin{align}
 \frac{\hat{\Upsilon}_b(x+b)}{\hat{\Upsilon}_b(x)} = (ib)^{1-2bx} \gamma(bx)\quad \text{and} \quad \frac{\hat{\Upsilon}_b(x+\frac{1}{b})}{\hat{\Upsilon}_b(x)} = (ib)^{\frac{2x}{b}-1} \gamma(\tfrac{x}{b})\ .
\end{align}
The solution $\hat\Upsilon_b(x)$ of these equations can actually be constructed from the function $\Upsilon_b(x)$,
\begin{align}
 \boxed{ \hat{\Upsilon}_b(x) = \frac{1}{\Upsilon_{ib}(-ix+ib)} }\ ,
\label{tub}
\end{align}
which exists for $\Im b < 0$. 
For any $c\notin ]-\infty, 1] \cup [25,\infty[$, there is a choice of $b$ such that $\Re b>0$ and $\Im b<0$. 
Both functions $\Upsilon_b(x)$ and $\hat\Upsilon_b(x)$ exist and obey shift equations that are essentially the same, so their ratio is essentially an elliptic function \cite{zam05}. 

\subsubsection{Properties of the Upsilon function}

The equations \eqref{upup} that determine how $\Upsilon_b(x)$ behaves under shifts $x\to x+b$ and $x\to x+\frac{1}{b}$, also determine how $\Upsilon_b(Q-x)$ behaves under the same shifts.
The resulting shift equations for $\Upsilon_b(Q-x)$ turn out to be identical to the shift equations for $\Upsilon_b(x)$, which suggests 
\begin{align}
 \boxed{\Upsilon_b(x) = \Upsilon_b(Q-x)}\ .
\label{upq}
\end{align}
Given that $\gamma(x)$ has simple poles for $x\in -{\mathbb{N}}$ and zeros for $x\in 1+{\mathbb{N}}$, the shift equations \eqref{upup} constrain the poles and zeros of  $\Upsilon_b(x)$.
Both $\frac{\Upsilon_b(x+b)}{\Upsilon_b(x)}$ and $\frac{\Upsilon_b(x+\frac{1}{b})}{\Upsilon_b(x)}$ have a pole at $x=0$, which suggests that $\Upsilon_b(x)$ has a simple zero at $x=0$.
Then the shift equations imply that 
\begin{align}
 \boxed{\Upsilon_b(x)\ \ \text{has simple zeros for}\ \ x\in  \left(-b{\mathbb{N}} -\tfrac{1}{b}{\mathbb{N}} \right) \cup \left( Q+b{\mathbb{N}} + \tfrac{1}{b}{\mathbb{N}}\right)} \ .
\label{xbn}
\end{align}
These zeros lie in two infinite cones with tips at $x=0$ and $x=Q$, which we represent as follows:
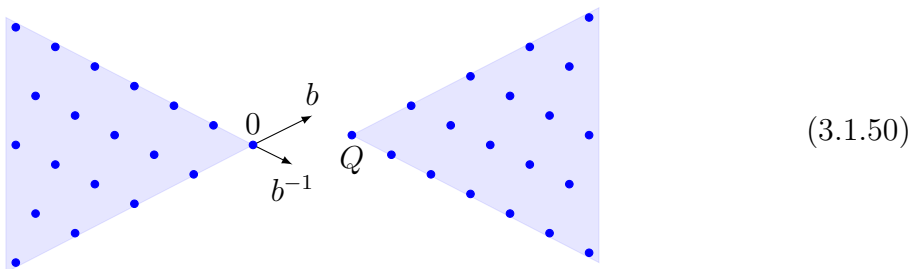
\begin{align}
\begin{tikzpicture}[scale = 1.3, baseline=(current  bounding  box.center)]
\node[above] at (0, 0) {$0$};
\node[below] at (1, .1) {$Q$};
\draw [latex-latex] (.6, .3) node[above] {$b$} -- (0, 0) -- (.4, -.2) node[below] {$b^{-1}$};
\begin{scope}[rotate = 90]
\filldraw[blue, opacity = .1] (0,0) -- (-1.3, 2.5) -- (1.3, 2.5) -- cycle;
\clip (-1.4, -.2) -- (1.4, -.2) -- (1.4, 2.5) -- (-1.4, 2.5) -- cycle;
\foreach \x in {0, 1,...,4}{
  \foreach \y in {0, 1,...,6}{
    \node[draw,circle,inner sep=1pt,fill,blue] at (-.3*\x +.2*\y, .6*\x +.4*\y) {};
  }}
  \end{scope}
\begin{scope}[shift = {(1, .1)}, rotate = -90]
\filldraw[blue, opacity = .1] (0,0) -- (-1.3, 2.5) -- (1.3, 2.5) -- cycle;
\clip (-1.4, -.2) -- (1.4, -.2) -- (1.4, 2.5) -- (-1.4, 2.5) -- cycle;
\foreach \x in {0, 1,...,4}{
  \foreach \y in {0, 1,...,6}{
    \node[draw,circle,inner sep=1pt,fill,blue] at (-.3*\x +.2*\y, .6*\x +.4*\y) {};
  }}  
\end{scope}  
 \end{tikzpicture}
\label{lines}
\end{align}
It turns out that these zeros account for all the poles and zeros of $\frac{\Upsilon_b(x+b)}{\Upsilon_b(x)}$ and $\frac{\Upsilon_b(x+\frac{1}{b})}{\Upsilon_b(x)}$ in the equations \eqref{upup}, so that $\Upsilon_b(x)$ has no poles and is analytic on ${\mathbb{C}}$. 

There is an explicit expression for the Upsilon function in the strip $0<\Re x<\Re Q$, 
\begin{align}
 \log\Upsilon_b(x) = \int_0^\infty \frac{dt}{t} \left[\left(\tfrac{Q}{2}-x\right)^2 e^{-2t} -\frac{\sinh^2\left(\left(\frac{Q}{2}-x\right)\!t\right)}{\sinh (bt)\sinh\left(\frac{t}{b}\right)}\right]\ .
\label{lup}
\end{align}
If $x$ is outside the strip, the value of $\Upsilon_b(x)$ can be found by combining the above integral expression with the shift equations \eqref{upup}.
An alternative formula for the Upsilon function, which makes the zeros manifest, is 
\begin{align}
 \Upsilon_b(x) = \lambda_b^{(\frac{Q}{2}-x)^2}\prod_{m,n=0}^\infty f\left(\frac{\frac{Q}{2}-x}{\frac{Q}{2}+mb+nb^{-1}}\right) \quad \text{with} \quad f(x)=(1-x^2)e^{x^2}\ ,
\end{align}
where $\lambda_b$ is an irrelevant $b$-dependent constant. (The dependence of structure constants on $\lambda_b$ can be absorbed into a field renormalization.)
This product formula is valid for all values of $x\in\mathbb{C}$, but it does not converge very fast.

\subsubsection{Expression of the three-point structure constant}

Using the functions $\Upsilon_b(x)$ and $\hat\Upsilon_b(x)$, we can write solutions $C'$ and $\hat C'$ of the shift equation \eqref{cps} (and its dual) for the three-point structure constant, 
\begin{align}
 \boxed{C'_{P_1,P_2,P_3} \ \underset{c\notin ]-\infty, 1]}{=}\ \prod_{\pm,\pm} \Upsilon_b\left(\tfrac{Q}{2}+P_1\pm P_2 \pm P_3\right)^{-1} }\ ,
 \label{cp}
 \\
 \boxed{\hat C'_{P_1,P_2,P_3} \ \underset{c\notin [25,\infty[}{=}\ \prod_{\pm,\pm} \hat\Upsilon_b\left(\tfrac{Q}{2}+P_1\pm P_2 \pm P_3\right)^{-1}}\ .
 \label{hcp}
\end{align}
These solutions are 
\begin{itemize}
 \item invariant under $b\to \frac{1}{b}$, due to the invariance of $\Upsilon_b(x)$ itself,
 \item invariant under reflection of each momentum $P_i\to -P_i$, due to Eq. \eqref{upq},
 \item and symmetric under permutations of the momentums.
\end{itemize}
Similarly, we can write solutions of the shift equation \eqref{nbs} (and its dual) for the normalization-independent combination $N^2B^{-1}$,
\begin{align}
 \boxed{\left(N^2B^{-1}\right)_P =\prod_\pm \Upsilon_b(\pm 2P)} \quad , \quad \boxed{\left(\hat N^2\hat B^{-1}\right)_P = \prod_\pm \hat\Upsilon_b(\pm 2P) } \ .
\end{align}
This is again invariant under $b\to \frac{1}{b}$, and under the reflection $P\to -P$.

We have determined the normalization-independant combinations of structure constants, and this is enough for solving Liouville theory in the conformal bootstrap approach. However, in order to ease the comparison with other approaches, it can be convenient to consider specific field normalizations. 
In the normalization such that $B_P=1$, the three-point structure constant must be real if the theory is unitary \cite{rib14b}: and indeed, we find
\begin{align}
\renewcommand{\arraystretch}{1.3}
 \left\{\begin{array}{l} c > 1\ , \\ P,P_i\in i\mathbb{R}\ , \end{array}\right. \quad \implies \quad \left\{\begin{array}{l} C'_{P_1,P_2,P_3}\in\mathbb{R}\ , \\ \left(N^2B^{-1}\right)_P\in \mathbb{R}_+\ , \end{array}\right.
\end{align}
which implies $C_{P_1,P_2,P_3}\in \mathbb{R}$. However, in this normalization, $N_P$ has square root branch cuts, which violates our assumption that correlation functions are meromorphic. Then let us  consider the normalization such that 
\begin{align}
 N_P = \mu^{P}\Upsilon_b(2P)\ ,
 \label{nop}
\end{align}
where $\mu$ is an arbitrary parameter called the \textbf{\boldmath cosmological constant}\index{cosmological constant} -- a name that comes from the interpretation of Liouville theory in terms of two-dimensional gravity.
The resulting expression for the three-point structure constant is
\begin{align}
 C_{P_1,P_2,P_3} =  \frac{\mu^{-\frac{Q}{2}}\Upsilon_b'(0)\prod_{i=1}^3 \mu^{P_i}\Upsilon_b(2P_i) }{\prod_{\pm,\pm} \Upsilon_b\left(\tfrac{Q}{2}+P_1\pm P_2 \pm P_3\right)} \ .
\label{caaa}
\end{align}
This is called the \textbf{\boldmath DOZZ formula}\index{DOZZ formula} for Dorn, Otto, A.
Zamolodchikov and Al. Zamolodchikov. 
The DOZZ formula was originally derived using the Lagrangian formulation of Liouville theory \cite{zz95}. 
That formulation led to normalizations that agree with ours, 
up to a redefinition of the cosmological constant, and a reflection of the momentums. 

The DOZZ formula is valid for $c\notin ]-\infty, 1]$. The analogous choice of normalizations for $c\notin [25,\infty[$ leads to the three-point structure constant 
\begin{align}
 \hat C_{P_1,P_2,P_3} =  \frac{\mu^{-\frac{Q}{2}}\hat\Upsilon_b(0)\prod_{i=1}^3 \mu^{P_i}\hat\Upsilon_b(2P_i) }{\prod_{\pm,\pm} \hat\Upsilon_b\left(\tfrac{Q}{2}+P_1\pm P_2 \pm P_3\right)} \ .
\label{hc}
\end{align}
The normalizations that we have chosen are such that
\begin{align}
 \underset{P_1+P_2+P_3=\frac{Q}{2}}{\operatorname{Res}} C_{P_1,P_2,P_3} = 1 \qquad , \qquad \hat C_{P_1,P_2,P_3}\underset{P_1+P_2+P_3=\frac{Q}{2}}{=} 1\ .
\label{chco}
 \end{align}
% Corresponds to \sum\alpha = Q if \alpha = Q/2 - P

\subsubsection{The reflection relation}

While the representations in Liouville theory's spectrum can be parametrized by momentums $P\in i\mathbb{R}_+$, the analytic properties of correlation functions allow us to consider fields with arbitrary complex momentums. 
In particular, for a given momentum $P$, it makes sense to consider the reflected momentum $-P$, which corresponds to the same representation of the Virasoro algebra. 
Our definition of Liouville theory includes the assumption that multiplicities are trivial: this implies that the fields $V_P(z)$ and $V_{-P}(z)$ correspond to the same primary state, and must be the same up to a factor $R_P$ called the \textbf{\boldmath reflection coefficient}\index{reflection!---coefficient}. In other words, we must have the \textbf{\boldmath reflection relation}\index{reflection!---relation}
\begin{align}
V_P(z)= R_P V_{-P}(z)\ .
\label{vrv}
\end{align}
The reflection coefficient must obey $R_PR_{-P}=1$, in particular $R_0\in \{-1,1\}$. Allowing arbitrary complex momentums, and taking the reflection relation into account, the two-point function \eqref{vvss} becomes 
\begin{align}
 \left<V_{P_1}V_{P_2}\right> = B_{P_1}\Big[ \delta(P_1-P_2) + R_{P_2}\delta(P_1+P_2) \Big]\ ,
\end{align}
Under reflection, our structure constants and normalization factor behave as 
\begin{align}
 B_P=R_P^2B_{-P} \ , \quad C_{P_1,P_2,P_3} = R_{P_1} C_{-P_1,P_2,P_3} \ , \quad N_P=R_PN_{-P}\ .
\end{align}
Normalization-independent quantities are invariant under reflection.

Let us now focus on the particular normalization \eqref{nop}, such that the three-point function is given by the DOZZ formula. This normalization obeys $N_PN_{-P} = \left(N^2B^{-1}\right)_P$, which implies $R_P=B_P$ and 
\begin{align}
 \left<V_{P_1}V_{P_2}\right> = R_{P_1}\delta(P_1-P_2) + \delta(P_1+P_2)\ .
 \label{vvrdd}
\end{align}
The expressions of the reflection coefficient for $c\notin ]-\infty, 1]$ and $c\notin [25,\infty[$ are respectively
\begin{align}
 R_P = \mu^{2P}\frac{\Upsilon_b(2P)}{\Upsilon_b(-2P)} \quad , \quad \hat R_P = \mu^{2P}\frac{\hat{\Upsilon}_b(2P)}{\hat{\Upsilon}_b(-2P)}\ .
 \end{align}
We have  $\hat\Upsilon_b(0)\neq 0$ so that $\hat R_0=1$. On the other hand, we have $\Upsilon_b(0)=0$, so that $R_0=-1$. This implies $V_{P=0}(z)=0$, which is consistent with $C_{0,P_2,P_3}=0$.
Writing $\Upsilon_b(-2P)=\Upsilon_b(2P+Q)$, and using the shift equations \eqref{upup}, we can rewrite the reflection coefficient in terms of Gamma functions, and we find 
\begin{align}
 R_P &= -\left[b^{2b-\frac{2}{b}}\mu\right]^{2P} \frac{\Gamma(-2bP)\Gamma(-2b^{-1}P)}{\Gamma(2bP)\Gamma(2b^{-1}P)} \ ,
 \label{ram}
\\
 \hat R_P &= \left[(ib)^{2b-\frac{2}{b}}\mu\right]^{2P} \frac{\Gamma(-2bP)\Gamma(-2b^{-1}P)}{\Gamma(2bP)\Gamma(2b^{-1}P)} \ .
\end{align}

\subsection{Existence and uniqueness}\label{seceul}

\subsubsection{Existence}

We derived the three-point structure constants $C$ and $\hat C$ by solving degenerate crossing symmetry equations. 
For Liouville theory to exist, we also need generic four-point functions $\left< \prod_{i=1}^4 V_{P_i}\right>$ to obey the crossing symmetry equation \eqref{csd}. In the case of Liouville theory, this boils down to 
\begin{align}
 \int_{i\mathbb{R}_+} dP_s \frac{C_{12s}C_{s34}}{B_s} \left|\mathcal{F}_{P_s}^{(s)} \right|^2 = \int_{i\mathbb{R}_+} dP_t \frac{C_{23t}C_{t41}}{B_t} \left|\mathcal{F}_{P_t}^{(t)} \right|^2\ ,
\end{align}
where $\mathcal{F}_{P_s}^{(s)}$ and $\mathcal{F}_{P_t}^{(t)}$ are $s$- and $t$-channel conformal blocks.
To begin with, the convergence of the integrals over $s$- and $t$-channel momentums is easily deduced from the expressions for the structure constants and conformal blocks, see Exercise \ref{exocvg}.
Then, the equality between the $s$- and the $t$-channel decompositions can be explored numerically \cite{rs15}. 
It turns out that for each value of the central charge $c\in\mathbb{C}$, either $C$ or $\hat C$ leads to crossing symmetric four-point functions, and is therefore the three-point structure constant of Liouville theory:
\begin{align}
\renewcommand{\arraystretch}{1.3}
 \begin{tabular}{|l||c|c|}
  \hline
  central charge &  $c\in ]-\infty, 1]$ & $c\notin ]-\infty, 1]$
  \\
  \hline
  structure constant & $\hat C$  & $C$ 
  \\
  \hline
 \end{tabular}
 \label{ccr}
\end{align}
In addition, there is a proposed proof \cite{tes03b} of crossing symmetry for $c\notin ]-\infty, 1]$.
Let us also mention that Liouville theory is consistent on the torus \cite{hjs09, rs15}. 

So $C$ is a valid structure constant wherever it is defined, whereas $\hat C$ is valid only for $c\in ]-\infty, 1]$ although it is defined for $c\notin [25,\infty[$.
To understand this, let us study the analytic properties of the corresponding four-point functions,
\begin{align}
\renewcommand{\arraystretch}{2}
 \left< \prod_{i=1}^4 V_{P_i}\right> = 
 \left\{\begin{array}{ll} 
         \frac12 \int_{i\mathbb{R}} dP_s \frac{C_{12s}C_{s34}}{B_s} \left|\mathcal{F}_{P_s}^{(s)} \right|^2 \quad & \text{if } c\notin ]-\infty, 1]\ ,
         \\
         \frac12 \int_{i\mathbb{R}+\epsilon} dP_s \frac{\hat C_{12s}\hat C_{s34}}{\hat B_s} \left|\mathcal{F}_{P_s}^{(s)} \right|^2 \quad & \text{if } c\in ]-\infty, 1]\ .
        \end{array}
 \right.
 \label{vfcch}
\end{align}
Although the structure constants $C$ and $\hat C$ have rather different analytic properties (see Exercise \ref{exo4a}), what matters is the behaviour of the poles of the $s$-channel conformal blocks $\mathcal{F}_{P_s}^{(s)}$. 
These poles correspond to degenerate values of the momentum $P_s$, and are therefore found in two cones with tips at $P_s= \pm\frac{Q}{2}$. 
Consider the relative positions of these cones (blue), and of the integration line for $P_s$ (red), depending on the value of the central charge $c$:
\begin{align}
 \newcommand{\polewedge}[3]{
\begin{scope}[#1]
\node[blue, draw,circle,inner sep=1pt,fill] at (0, 0) {};
%\node[#3] at (0,0) {#2};
\filldraw[opacity = .1, blue] (0,0) -- (4, -4) -- (4, 4) -- cycle;
\end{scope}
}
\begin{array}{ccc}
\begin{tikzpicture}[scale = .4, baseline=(current  bounding  box.center)]
  \draw[-latex] (-3,0) -- (0, 0) -- (4,0) node [above] {$P_s$};
  \draw[ultra thick, blue, opacity = .3] (0, -4.5) -- (0, 4.5);
  \draw (0, -4.5) -- (0, 4.5);
  \draw[ultra thick, red] (.5, -4.5) -- (.5, 4.5);
\node[above left] at (.2,-.2) {$0$};
\node[blue, draw,circle,inner sep=1pt,fill] at (0, 2) {};
\node[blue, draw,circle,inner sep=1pt,fill] at (0, -2) {};
%\node[above left] at (0,1.8) {$\frac{Q}{2}$};
%\node[left] at (0,-2) {$-\frac{Q}{2}$};
 \end{tikzpicture}
 & 
 \begin{tikzpicture}[scale = .4, baseline=(current  bounding  box.center)]
  \draw[-latex] (-4,0)  --  (6,0) node [above] {$P_s$};
  \draw[ultra thick, red] (1, -4.5)  -- (1, 4.5);
  \node[above left] at (1.2,-.2) {$0$};
  \polewedge{rotate = 180, shift = {(0,.4)}}{$-\frac{Q}{2}$}{below};
  \polewedge{shift = {(2, .4)}}{$\frac{Q}{2}$}{above};
 \end{tikzpicture}
 &
 \begin{tikzpicture}[scale = .4, baseline=(current  bounding  box.center)]
 \draw[ultra thick, blue, opacity = .3] (0,0) -- (-4,0);
 \draw[ultra thick, blue, opacity = .3] (2,0) -- (6,0);
  \draw[-latex] (-4,0) -- (6,0) node [above] {$P_s$};
  \draw[ultra thick, red] (1, -4.5) -- (1, 4.5);
  \node[blue, draw,circle,inner sep=1pt,fill] at (0, 0) {};
\node[above left] at (1.2,-.2) {$0$};
\node[blue, draw,circle,inner sep=1pt,fill] at (2, 0) {};
 \end{tikzpicture}
 \vspace{3mm}
 \\
 c\in ]-\infty, 1] & c\notin ]-\infty, 1] \cup [25,\infty[ & c\in [25,\infty[
\end{array}
\end{align}
For any $c\notin ]-\infty, 1]$, the poles of the conformal blocks stay safely away from the integration line, and actually the poles of the structures constants behave similarly. 
So the four-point function is analytic as a function of $c$ in that domain. 
In particular, crossing symmetry for $c\in [25,\infty[$ implies crossing symmetry for $c\notin ]-\infty, 1]$ by analyticity of both $s$- and $t$-channel expressions for the four-point function. 
In contrast, when
$c$ approaches the half-line $]-\infty, 1]$, a whole cone of poles crosses the integration line before collapsing into the imaginary axis. 
So we cannot continue the four-point function from $c\in ]-\infty, 1]$ to a larger region (or vice-versa).

\subsubsection{Uniqueness}

Let us discuss the uniqueness of Liouville theory, as defined by our assumptions in Section \ref{secspe}. 
We saw that three-point structures constants are uniquely determined by shift equations if $c\in ]-\infty, 1] \cup [25,\infty[$, so that Liouville theory is unique for these values of $c$. 
For the other values of $c$, we have $b^2\notin\mathbb{R}$, and we can find alternative solutions of the equations \eqref{upup} for $\Upsilon_b(x)$, by multiplying $\Upsilon_b(x)$ with elliptic functions -- meromorphic functions $\theta(x)$ such that $\theta(x+b)=\theta(x+\frac{1}{b}) = \theta(x)$.
This leads to alternative solutions of the shift equations for structure constants.
But these alternative solutions are excluded by the assumption that correlation functions are meromorphic functions of $b$, as this assumption determines correlation functions for $c\notin ]-\infty, 1]$ from their values for $c\in [25,\infty[$.

This uniqueness statement relies on the assumption that degenerate fields exist. We can drop this assumption and have a stronger uniqueness statement, if we remember that structure constants of diagonal theories are uniquely determined by crossing symmetry, see Eq. \eqref{cmnnf}. It would be interesting to probe the limits of uniqueness, i.e. to understand whether uniqueness survives when we weaken the assumptions on the spectrum and analyticity of correlation functions. We will now mention two cases in which uniqueness is known to break down:
\begin{itemize}
 \item We can relax the assumption that fields have multiplicities zero or one, and allow finite multiplicities. A simple theory where all fields have the same multiplicity $N\in\mathbb{N}^*$ can be built by taking $N$ non-interacting copies of Liouville theory.
 \item We can relax the assumption that correlation functions are meromorphic functions of the coupling constant and of the fields' momentums. Then we find that for $b^2=-\frac{q}{p}\in \mathbb{Q}_{<0}$, there exists a solution of crossing symmetry that differs from the Liouville three-point structure constant $\hat{C}_{P_1,P_2,P_3}$ by a non-analytic factor with values in $\{0,1\}$ \cite{rs15},
 \begin{align}
  \sigma_{P_1,P_2,P_3} = \left\{\begin{array}{cl} 1 & \text{if} \ \ \prod_{\pm,\pm} \sin\pi\left(\frac12(p-q)+\sqrt{pq}(P_1\pm P_2\pm P_3)\right) < 0\ , 
  \\ 0 & \text{else}\ . \end{array} \right.
  \label{sigma}
 \end{align}
This structure constant makes sense for real momentums $P_i$.
We will call the corresponding theories \textbf{Runkel--Watts-type theories}\index{Runkel--Watts-type theory}, for reasons that will become apparent in Section \ref{secltf}. See Exercise \ref{exonurw} for more details on how uniqueness is circumvented in this case. 
\end{itemize}

\subsection{Analytic continuation of the OPE}\label{secacl}

We will now sketch how OPEs behave when momentums are analytically continued beyond the imaginary axis, in particular when they reach the values that correspond to degenerate fields.

\subsubsection{Beyond the imaginary axis}

Using the invariance of normalization-independant quantities under the reflection $P\to -P$, we write the Liouville OPE \eqref{vvi} as an integral over the whole imaginary axis,
\begin{align}
 V_{P_1}V_{P_2} \sim \frac12\int_{i{\mathbb{R}}} dP \frac{C_{P_1,P_2,P}}{B_P} V_P\ .
 \label{vvir}
\end{align}
As a function of $P$, the OPE coefficient $\frac{C_{P_1,P_2,P}}{B_P}$ can have poles whose positions depend on $P_1,P_2$. 
And these poles control how the OPE behaves when we analytically continue $P_1,P_2$ beyond the imaginary axis. In the case $c\in ]-\infty, 1]$, the poles of the OPE coefficient $\frac{\hat C_{P_1,P_2,P}}{\hat B_P}$ do not depend on $P_1,P_2$, and our OPE remains valid as written for any $P_1,P_2\in\mathbb{C}$ such that $\hat N_{P_1}$ and $\hat N_{P_2}$ are finite.

Let us focus on the case $c\notin ]-\infty,1]$. Then the three-point structure constant has $P_1,P_2$-dependent simple poles due to the normalization-independent factor $C'_{P_1,P_2,P}$ \eqref{cp}. 
Let us plot these poles in the complex $P$-plane, assuming $P_1,P_2\in i\mathbb{R}$. The poles lie on cones that we draw in the fashion of Eq. \eqref{lines}. We assume $\Re Q>0$, and we also draw the line  of integration $P\in i{\mathbb{R}}$, together with the two lines $P\in \pm\frac{Q}{2}+i\mathbb{R}$:
\begin{align}
 \newcommand{\polewedge}[3]{
\begin{scope}[#1]
\node[blue, draw,circle,inner sep=1pt,fill] at (0, 0) {};
\node[#3] at (0,0) {#2};
\filldraw[opacity = .1, blue] (0,0) -- (5, -2.5) -- (5, 2.5) -- cycle;
\end{scope}
}
 \begin{tikzpicture}[baseline=(current  bounding  box.center)]
  \draw[-latex] (-5,0) --  (.5,0) node [below left] {$0$} -- (6,0) node [below] {$P$};
  \clip (-5, -3.5) -- (-5, 3.5) -- (6, 3.5) -- (6, -3.5) -- cycle;
  \draw (-1, -4.5) -- (-1, 4.5);
  \draw (2, -4.5) -- (2, 4.5);
  \draw[ultra thick, red] (.5, -4.5) -- (.5, 4.5);
  \polewedge{shift = {(-1, .7)}, rotate = 180}{$-\frac{Q}{2}+P_1-P_2$}{left};
  \polewedge{shift = {(-1, -.7)}, rotate = 180}{$-\frac{Q}{2}-P_1+P_2$}{left};
  \polewedge{shift = {(-1, 1.8)}, rotate = 180}{$-\frac{Q}{2}+P_1+P_2$}{left};
  \polewedge{shift = {(-1, -1.8)}, rotate = 180}{$-\frac{Q}{2}-P_1-P_2$}{left};
  \polewedge{shift = {(2, .8)}}{$\frac{Q}{2}+P_1-P_2$}{right};
  \polewedge{shift = {(2, -.6)}}{$\frac{Q}{2}-P_1+P_2$}{right};
  \polewedge{shift = {(2, 1.9)}}{$\frac{Q}{2}+P_1+P_2$}{right};
  \polewedge{shift = {(2, -1.7)}}{$\frac{Q}{2}-P_1-P_2$}{right};
 \end{tikzpicture}
\end{align}
If we continue $P_1,P_2$ such that $|\Re(P_1+P_2)|>\frac12\Re Q$ or $|\Re(P_1-P_2)|>\frac12\Re Q$, then some poles cross the line of integration. By reflection symmetry, it is enough to consider the poles that come from the left. The finite set of the poles that come from the left and end up on the right of the imaginary axis is
\begin{align}
 \mathcal{P}_{P_1,P_2} = \left\{\Re P >0\right\} \cap \bigcup_{\pm,\pm} \left(-\frac{Q}{2}\pm P_1\pm P_2 -b\mathbb{N}-b^{-1}\mathbb{N}\right)\ .
\end{align}
Then the OPE becomes
\begin{align}
 V_{P_1}V_{P_2} \sim \frac12\int_{i{\mathbb{R}}} dP \frac{C_{P_1,P_2,P}}{B_P} V_P + \sum_{P_k\in \mathcal{P}_{P_1,P_2}} \frac{\underset{P=P_k}{\operatorname{Res}} C_{P_1,P_2,P}}{B_{P_k}} V_{P_k}\ ,
 \label{acope}
\end{align}
where the discrete terms have lost their factor $\frac12$ due to the contributions of the poles from the right. (See Exercise \ref{exoaur} for examples of the resulting fusion rules.)

Our analytically continued OPE is valid for generic values of $P_1,P_2$, such that $C_{P_1,P_2,P}$ only has simple poles.
We will now study what happens in a special case where some poles coincide.

\subsubsection{Degenerate momentums}

Let us study how the Liouville OPE behaves when $P_2 \to P_{\langle r,s\rangle}$, where the degenerate momentum $P_{\langle r,s \rangle}$ is given in Eq. \eqref{ars}.
We still assume $c\notin ]-\infty,1]$, and we moreover choose the field normalization \eqref{nop} such that the three-point structure constant is given by the DOZZ formula. 
With this field normalization, we have $N_{P_{\langle r,s\rangle}}=0$, and therefore $\lim_{P_2 \to P_{\langle r,s\rangle}} C_{P_1,P_2,P}=0$.
This seems to imply that the limit of our OPE is zero. But actually our OPE has a finite limit, in other words the OPE coefficient does not vanish when considered as a distribution. 

To see this, let us follow how the poles behave when $P_2 \to P_{\langle r,s\rangle}$. In this limit, some poles from the left coincide with some poles from the right. In particular, $rs$ poles in $-\frac{Q}{2}+P_1+P_2-b\mathbb{N}-b^{-1}\mathbb{N}$ coincide with as many poles in $\frac{Q}{2}+P_1-P_2+b\mathbb{N}+b^{-1}\mathbb{N}$. Let us draw these coinciding poles in the case ${\langle r,s\rangle} = {\langle 3,4\rangle}$. For simplicity, we also deform the integration line so that it is crossed by all the coinciding poles from the left, and only those poles:
\begin{align}
\newcommand{\polewedge}[4]{
\begin{scope}[#1]
\node[blue, draw,circle,inner sep=1pt,fill] at (0, 0) {};
\node[#3] at (0,0) {#2};
\filldraw[opacity = .1, blue] (0,0) -- (8, -4) -- (8, 4) -- cycle;
\end{scope}
}
 \begin{tikzpicture}[baseline=(current  bounding  box.center)]
  \draw[-latex] (-5,0) -- (0.5, 0) node[above left] {$0$} -- (6,0) node [above] {$P$};
  \draw (0.5, -4) -- (0.5, 4);
 \clip (-5, -4) -- (-5, 4) -- (6, 4) -- (6, -4) -- cycle;
  \polewedge{shift = {(1.7, 2)}, rotate = 180}{}{below right};
  \node[right] at (1.2, 1.6) {$-\frac{Q}{2}+P_1+P_2$};
  \polewedge{shift = {(-1.7, -2)}, rotate = 180}{$-\frac{Q}{2}-P_1-P_2$}{left};
  \polewedge{shift = {(-1.7, 1.9)}, rotate = 180}{$-\frac{Q}{2}+P_1-P_2$}{left};
  \polewedge{shift = {(1.7, -1.9)}, rotate = 180}{}{below right};
  \node[right] at (1.2, -2.3) {$-\frac{Q}{2}-P_1+P_2$};
  \polewedge{shift = {(2.7, 2.1)}}{$\frac{Q}{2}+P_1+P_2$}{right};
  \polewedge{shift = {(-.7, -1.9)}}{$\frac{Q}{2}-P_1-P_2$}{above left};
  \polewedge{shift = {(-.7, 2)}}{$\frac{Q}{2}+P_1-P_2$}{above left};
  \polewedge{shift = {(2.7, -1.8)}}{$\frac{Q}{2}-P_1+P_2$}{right};
  \foreach \x in {0, ..., 2}{
  \foreach \y in {0, ..., 3}{
  \node[red, draw, circle,inner sep = 1.5pt, fill] at (-.7 + .6*\x + .4*\y, 2 + .3*\x -.2*\y) {};
  \node[red, draw, circle,inner sep = 1.5pt, fill] at (-.7 + .6*\x + .4*\y, -1.9 + .3*\x -.2*\y) {};
  }}
  \draw[ultra thick, red, rounded corners = 6] (.5, -4.5) -- (.5, -2.7) -- (-1.1, -1.9) -- (.5, -1.1) -- (.5, 1.2) -- (-1.1, 2) -- (.5, 2.8) -- (.5, 4.5);
 \end{tikzpicture}
\end{align}
The integral along the deformed line goes to zero as $P_2\to P_{\langle r,s \rangle}$. On the other hand, the residues of the coinciding poles have finite nonzero limits, because the vanishing factor $N_{P_2}$ is compensated by an infinite factor due to the coincidence of poles. We obtain 
\begin{align}
\underset{P_2\to P_{\langle r,s \rangle}}{\lim} V_{P_1}V_{P_2} \sim \sum_{i=-\frac{r-1}{2}}^{\frac{r-1}{2}} \sum_{j=-\frac{s-1}{2}}^{\frac{s-1}{2}}
 \left(\underset{P_2\to P_{\langle r,s \rangle}}{\lim}\ \underset{P=P_1+P_2-P_{\langle r,s \rangle}+ib+jb^{-1}}{\operatorname{ Res}}\ \frac{C_{P_1,P_2,P}}{B_P}\right) V_{P_1 + ib+jb^{-1}}\ .
\label{crs}
\end{align}
This OPE is of the type of the degenerate OPE \eqref{vrsv}, so that $V_{P_2}$ becomes a degenerate field, whose fusion rule \eqref{rtv} we have now rederived from the Liouville OPE.
Moreover, we have obtained a formula for the degenerate OPE coefficient in terms of the two- and three-point structure constants. Let us insert the explicit expressions of these structure constants in this formula. The calculation is slightly simplified by using the identity $\frac{C_{P_1,P_2,P}}{B_P} = C_{P_1,P_2,-P}$ which holds in our normalization, and we find
\begin{multline}
 C_{i,j}^{\langle r,s \rangle}(P) = \mu^{-\frac{Q}{2}+P_{\langle r,s\rangle}-ib-jb^{-1}} 
 \frac{\Upsilon_b'(0)\Upsilon_b'(2P_{\langle r,s \rangle})}{\prod_{\pm}\Upsilon'_b(\frac{Q}{2}\pm P_{\langle r,s\rangle} +ib+jb^{-1})} 
 \\ \times 
 \frac{\Upsilon_b(2P)\Upsilon_b(Q+2P+2ib+2jb^{-1})}{\prod_\pm \Upsilon_b(\frac{Q}{2}+2P\pm P_{\langle r,s\rangle} +ib+jb^{-1})}
 \label{cijrs}
 \ ,
\end{multline}
where taking limits and residues produces derivatives of the Upsilon function at its zeros.
We could expand the ratios of Upsilon functions into products of $\gamma$ functions using Eq. \eqref{upup}.
In the special case $i=\frac{r-1}{2}, j=\frac{s-1}{2}$, we have $ib+jb^{-1} = P_{\langle r,s\rangle}-\frac{Q}{2} $, so that 
\begin{align}
 C_{\frac{r-1}{2},\frac{s-1}{2}}^{\langle r,s \rangle}(P) = 1 \ ,
 \label{cco}
\end{align}
which is a special case of Eq. \eqref{chco}.
In particular, in the cases of the degenerate fields $V_{\langle 2,1\rangle}$ and $V_{\langle 1,2\rangle}$, we have $C_+(P)=\tilde{C}_+(P)=1$. This could have been derived more directly from crossing symmetry of four-point functions with two degenerate fields. 
(See Exercise \ref{exodoc}.) 

In the case $c\in]-\infty, 1]$, the OPE coefficient has no poles that could lead to discrete terms in the limit $P_2\to P_{\langle r,s\rangle}$, and the normalization factor $\hat N_{P_2} = \mu^{P_2}\hat\Upsilon_b(2P_2)$ does not vanish in that limit. As a result, the OPE \eqref{vvir} remains valid as written, and $\lim V_{P_2} = V_{P_{\langle r,s\rangle}}$ is not degenerate. Moreover, the non-vanishing null vector $L_{\langle r,s\rangle}\bar{L}_{\langle r,s\rangle} V_{P_{\langle r,s\rangle}}$ is a diagonal primary field of dimension $\Delta_{\langle r,-s\rangle}$, and must therefore be proportional to $V_{P_{\langle r,-s\rangle}}$. Coming back to the case $c\notin ]-\infty,1]$, there is an analogous relation called a higher equation of motion, which involves $\left.\frac{\partial}{\partial P} V_P\right|_{P=P_{\langle r,s\rangle}}$, and whose coefficient of proportionality is known \cite{zam03}. To summarize:
\begin{align}
\renewcommand{\arraystretch}{1.3}
 \begin{tabular}{|r||c|c|}
  \hline
  central charge &  $c\in ]-\infty, 1]$ & $c\notin ]-\infty, 1]$
  \\
  \hline
  $\underset{P\to P_{\langle r,s \rangle}}{\lim } V_P = $ & $ V_{P_{\langle r,s \rangle}}$  & $V_{\langle r,s \rangle}$ 
  \\
  \hline
  $V_{P_{\langle r,-s\rangle}} \propto$ 
  & $L_{\langle r,s\rangle}\bar{L}_{\langle r,s\rangle} V_{P_{\langle r,s\rangle}}$ 
  &  $L_{\langle r,s\rangle}\bar{L}_{\langle r,s\rangle} \left.\frac{\partial}{\partial P} V_P\right|_{P=P_{\langle r,s\rangle}}$
  \\
  \hline
 \end{tabular}
 \label{cicni}
\end{align}

\section{Minimal models \label{secvmm}}

Let us look for conformal field theories such that
\begin{itemize}
\item the theory is rational, i.e.
the spectrum is made of finitely many representations;
\item each representation of the algebra $\mathfrak{V}\times \overline{\mathfrak{V}}$ has multiplicity zero or one;
\item the theory is diagonal.
\end{itemize}
Diagonality is a simplifying assumption, which excludes the non-diagonal D- and E-series minimal models \cite{fms97}.
Having multiplicities zero or one is another simplifying assumption, but it is not well-known what we would find by relaxing it. 
Rationality is the crucial axiom here -- and at first sight it seems difficult for rational theories to exist.
The OPE of two fields $V_1V_2\sim \sum_{s \in \mathcal{S}_{12}} C_{12s} V_s$ typically involves infinitely many primary fields $V_s$, and can only be finite in two cases:
\begin{itemize}
 \item if the set $\mathcal{S}_{12}$ of allowed representations is finite as a consequence of fusion rules,
 \item or failing that, if $C_{12s}$ vanishes for all but a finite number of values of $s$. 
\end{itemize}
The second alternative amounts to imposing infinitely many constraints on the three-point structure constant, on top of crossing symmetry. 
This cannot plausibly lead to nonzero solutions, except in special cases. 
We therefore accept that all fields that can appear do appear, in other words that the spectrum is closed under fusion, as required by Axiom \ref{ax:cuf}.
We then have to ensure that the set $\mathcal{S}_{12}$ is finite.
This requires us to allow
only degenerate representations in the spectrum, i.e. representations whose fusion products \eqref{rrsr} are finite sums. 

However, having finite fusion products is still not enough for a having a finite spectrum. 
Actually, for generic central charges, no nontrivial finite set of degenerate representations of the Virasoro algebra is closed under fusion.
Starting with the degenerate representation $\mathcal{R}_{\langle 1,2 \rangle}$ for instance, the multiple fusion products $\mathcal{R}_{\langle 1,2 \rangle}\times \mathcal{R}_{\langle 1,2 \rangle}\times \cdots \times \mathcal{R}_{\langle 1,2 \rangle}$ span the infinite set of representations $\{\mathcal{R}_{\langle 1,s \rangle}\}_{s\in{\mathbb{N}}^*}$.
Our task would be easier if we would weaken our assumptions and replace the rationality of the theory with the discreteness of the spectrum, allowing the number of representations to be countable instead of finite.
We would obtain generalized minimal models, whose spectrums contain all degenerate representations, and which exist for generic values of the central charge. 

What if we insist on the original assumption of rationality? We will see that there are finite sets 
of degenerate representations that close under fusion, from which we will build the A-series minimal models.
Such sets are made of
\textbf{\boldmath doubly degenerate representations}\index{degenerate representation!doubly---}, i.e. maximally degenerate representations of the type $\mathcal{R}_{\langle r,s \rangle}=\mathcal{R}_{\langle r',s' \rangle}$ with $\langle r,s \rangle\neq \langle r',s' \rangle$. 
Since 
\begin{align}
\renewcommand{\arraystretch}{1.3}
 \left.\begin{array}{r} 
 \Delta_{\langle r,s \rangle}=\Delta_{\langle r',s' \rangle} \\ 
        \langle r,s \rangle\neq \langle r',s' \rangle
       \end{array} \right\} \quad \implies \quad b^2 \in\mathbb{Q}\ ,
 \label{ddibq}
\end{align}
doubly degenerate representations exist only for certain discrete values of the central charge. 

\subsection{Generalized minimal models \label{secgmm}}

For any complex value of the central charge such that $b^2\notin \mathbb{Q}$, we define a \textbf{\boldmath generalized minimal model}\index{minimal model!generalized---} by its spectrum,
\begin{align}
 \boxed{\mathcal{S} = \bigoplus_{r,s=1}^\infty \mathcal{R}_{\langle r,s \rangle}\otimes \bar{\mathcal{R}}_{\langle r,s \rangle}}\ .
\end{align}
The representations $\mathcal{R}_{\langle r,s \rangle}$ are all distinct due to Eq. \eqref{ddibq}, and it is not known whether  generalized minimal models exist for rational $b^2$. 

We could obtain other consistent models by using subsets of the degenerate representations, provided these subsets were closed under fusion, for instance $\mathcal{S}=\mathcal{\mathcal{R}}_{\langle 1,1 \rangle}\otimes \bar{\mathcal{R}}_{\langle 1,1 \rangle}$, $\mathcal{S}=\bigoplus_{r=1}^\infty \mathcal{R}_{\langle r,1 \rangle}\otimes \bar{\mathcal{R}}_{\langle r,1 \rangle}$ or $\mathcal{S}=\bigoplus_{r,s=1}^{\infty} \mathcal{R}_{\langle 2r+1,2s+1 \rangle}\otimes \bar{\mathcal{R}}_{\langle 2r+1,2s+1 \rangle}$.
Solving the generalized minimal model will also solve all these submodels. 

\subsubsection{Correlation functions}

Due to Eq. \eqref{ddibq}, different representations have different conformal dimensions, and the two-point function must be of the type 
\begin{align}
 \left\langle V_{\langle r_1,s_1 \rangle} V_{\langle r_2,s_2 \rangle} \right\rangle = B_{\langle r_1,s_1 \rangle} \delta_{r_1,r_2} \delta_{s_1,s_2}\ ,
\label{vvdd}
\end{align}
where $B_{\langle r_1,s_1\rangle}$ is the two-point structure constant.

When writing the three-point function, we take care to explicitly include a factor $f_{r_1,r_2,r_3} f_{s_1,s_2,s_3}$ \eqref{frrr}, which enforces the degenerate fusion rules:
\begin{multline}
 \left\langle \prod_{i=1}^3 V_{\langle r_i,s_i \rangle}(z_i) \right\rangle = C_{\langle r_1,s_1\rangle ,\langle r_2,s_2\rangle ,\langle r_3,s_3 \rangle}  
\\ \times 
f_{r_1,r_2,r_3} f_{s_1,s_2,s_3} |z_{12}|^{2(\Delta_3-\Delta_1-\Delta_2)} |z_{23}|^{2(\Delta_1-\Delta_2-\Delta_3)} |z_{31}|^{2(\Delta_2-\Delta_3-\Delta_1)}\ .
\end{multline}
This involves the model-dependent three-point structure constant $C_{\langle r_1,s_1\rangle ,\langle r_2,s_2\rangle ,\langle r_3,s_3 \rangle}$, times universal factors. We should resist the temptation to include the factor $f_{r_1,r_2,r_3} f_{s_1,s_2,s_3}$ in the definition of the three-point structure constant: this would cause conceptual confusion by combining model-dependent with universal data, and would prevent us from directly deriving the three-point structure constant from Liouville theory.

\subsubsection{Relation with Liouville theory and uniqueness}

The three-point structure constants of generalized minimal models and of Liouville theory obey the same crossing symmetry equations. We can therefore deduce our three-point structure constants from Liouville theory, by sending momentums to degenerate values, 
\begin{align}
 C_{\langle r_1,s_1\rangle ,\langle r_2,s_2\rangle ,\langle r_3,s_3 \rangle} 
 &= \hat{C}_{P_{\langle r_1,s_1 \rangle}, P_{\langle r_2,s_2 \rangle} ,P_{\langle r_3,s_3 \rangle} } \ ,
 \label{chc}
 \\
& = \underset{P_1\to P_{\langle r_1,s_1 \rangle}}{\lim}\ 
\underset{P_2\to P_{\langle r_2,s_2 \rangle}}{\lim}\ 
\underset{\sum_i P_i = \sum_i P_{\langle r_i,s_i \rangle} }{\operatorname{ Res}} C_{P_1,P_2,P_3}\ .
\label{clc}
\end{align}
This is certainly a solution of the crossing symmetry equations in generalized minimal models, 
and it must be unique according to the arguments of Exercise \ref{exoudt}, which apply to all diagonal theories. 
(See Exercise \ref{exofix} for a more explicit proof in this case.) 

If the fusion rules are violated, our three-point structure constant may or may not vanish. For example, $C_{\langle 1,1\rangle , \langle 1,1 \rangle,\langle 3,3 \rangle} \neq 0$.
Therefore, while the three-point structure constant can be deduced from Liouville theory, the three-point function cannot, as the factor 
$f_{r_1,r_2,r_3} f_{s_1,s_2,s_3}$ that enforces the fusion rules has no known derivation from Liouville theory. 
This makes it difficult to deduce crossing symmetry in generalized minimal models from crossing symmetry in Liouville theory. 

Instead, crossing symmetry in generalized minimal models can be checked numerically \cite{rs15}, and this shows that generalized minimal models are consistent on the sphere. 
On the other hand, nothing guarantees that generalized minimal models are consistent on the torus. Actually, torus correlation functions that involve only the identity field $V_{\langle 1,1\rangle}$ reduce to the partition function, which is infinite. And there is no reason for the other torus correlation functions to be better behaved.

\subsubsection{Two- and three-point structure constants}

Let us deduce the three-point structure constant from the DOZZ formula \eqref{caaa} using Eq. \eqref{clc}:
\begin{align}
 C_{\langle r_1,s_1\rangle ,\langle r_2,s_2\rangle ,\langle r_3,s_3 \rangle} =   \frac{\mu^{-\frac{Q}{2}}\Upsilon'_b(0)\prod_{i=1}^3 \mu^{P_{\langle r_i,s_i \rangle}}\Upsilon'_b(2P_{\langle r_i,s_i \rangle}) }{\prod_{\pm,\pm} \Upsilon'_b\left(\tfrac{Q}{2}+P_{\langle r_1,s_1 \rangle}\pm P_{\langle r_2,s_2 \rangle} \pm P_{\langle r_3,s_3 \rangle}\right)} \ .
\label{crisi}
\end{align}
This is the Liouville structure constant $C_{P_1,P_2,P_3}$, where all Upsilon factors are replaced with their derivatives at some of their zeros.
Knowing the zeros \eqref{xbn} of the Upsilon function, we deduce that 
all Upsilon factors have zeros whenever
the momentums are degenerate and the degenerate fusion rules \eqref{rrrsss} are obeyed, in particular
\begin{align}
 \forall \pm, \pm\ , \quad \Upsilon_b\left(\tfrac{Q}{2}+P_{\langle r_1,s_1 \rangle}\pm P_{\langle r_2,s_2 \rangle} \pm P_{\langle r_3,s_3 \rangle}\right) = 0\ .
\end{align}
However, the Upsilon factors in the denominator can also have zeros when fusion rules are violated, and these additional zeros are responsible for the expression \eqref{clc} being nonzero in such cases, for example $C_{\langle 1,1\rangle , \langle 1,1 \rangle,\langle 3,3 \rangle} \neq 0$.

The three-point structure constant can be rewritten in terms of Gamma functions, using the following consequence of Eq. \eqref{upup},
\begin{align}
 \frac{\Upsilon_b'(-rb-sb^{-1})}{\Upsilon_b'(0)} = (-1)^{rs} b^{s-r} b^{s(s+1)b^{-2}-r(r+1)b^2} P_r(b^2)P_s(b^{-2}) Q_{r,s}(b)\ ,
\end{align}
where we defined
\begin{align}
 P_r(x) = \prod_{i=1}^r \gamma(1+ix) \qquad \text{and} \qquad Q_{r,s}(b) = \prod_{i=1}^r \prod_{j=1}^s (ib+jb^{-1})^2\ .
\end{align}
Assuming that the degenerate fusion rules \eqref{rrrsss} are obeyed, we find 
\begin{align}
 C_{\langle r_1,s_1\rangle ,\langle r_2,s_2\rangle ,\langle r_3,s_3 \rangle} =  \frac{-(-b^2)^{r_0-s_0}\left[b^{2b-\frac{2}{b}}\mu\right]^{br_0 + \frac{s_0}{b}}}{P_{r_0}(b^2)P_{s_0}(b^{-2})Q_{r_0,s_0}(b)} \prod_{i=1}^3 \frac{P_{r_i-1}(b^2)P_{s_i-1}(b^{-2}) Q_{r_i-1,s_i-1}(b)}{P_{r_0-r_i}(b^2)P_{s_0-s_i}(b^{-2}) Q_{r_0-r_i,s_0-s_i}(b)} \ ,
\label{cpqb}
\end{align}
where we introduce the integers
\begin{align}
 r_0 = \frac{r_1+r_2+r_3-1}{2} \quad , \quad s_0=\frac{s_1+s_2+s_3-1}{2}\ .
\end{align}
If we choose the cosmological constant as $\mu = b^{\frac{2}{b}-2b}$, we obtain a three-point function that is manifestly meromorphic as a function of $b^2\in {\mathbb{C}}^*$.
This contrasts with the three-point function of Liouville theory, which is
meromorphic only for $b\notin i\mathbb{R}$. 

The generalized minimal models' two-point structure constant can similarly be deduced from the Liouville two-point structure constant, equivalently from the Liouville reflection coefficient \eqref{ram}, by 
\begin{align}
B_{\langle r,s\rangle} = B_{P_{\langle r,s\rangle}} = R_{P_{\langle r,s\rangle}}\ .
\end{align}
Rather than using the relation with Liouville theory, the generalized minimal models' structure constants can of course be rederived from first principles. (See Exercise \ref{exosssc} for examples.)

\subsection{Kac table and fusion rules \label{secmmf}}

After our digression on generalized minimal models with their generic central charges, let us study the doubly degenerate representations that are used for building minimal models. We will first determine the central charges for which doubly degenerate representations exist. 

\subsubsection{Kac table}

Let us consider the doubly degenerate representation 
\begin{align}
 \mathcal{R}=\mathcal{R}_{\langle r,s \rangle}=\mathcal{R}_{\langle r',s' \rangle}\ .
 \label{rerer}
\end{align}
The equality of conformal dimension $\Delta_{\langle r,s \rangle}=\Delta_{\langle r',s' \rangle}$ implies that 
the momentums are equal or opposite, 
$P_{\langle r,s \rangle}=\pm P_{\langle r',s' \rangle}$. Using Eq. \eqref{ars} for $P_{\langle r,s\rangle}$, this implies that $b^2$ is a rational number, 
\begin{align} 
b^2=-\frac{q}{p} \in \mathbb{Q} \quad \Rightarrow \quad c\in ]-\infty, 1] \cup [25,\infty[\ .
\end{align}
Then, in addition to the relation $\Delta_{\langle r,s \rangle}=\Delta_{\langle -r,-s \rangle}$ which holds for any central charge, we have the relation
\begin{align}
 \forall r,s\in \mathbb{C}\ , \quad \Delta_{\langle r,s \rangle}=\Delta_{\langle p-r,q-s \rangle}\ ,
 \label{ddr}
\end{align}
In order to recover the known minimal models with their central charges $c_{p,q}$ \eqref{cpq}, it is actually enough to restrict to integers $p,q$ that are coprime and obey $p,q\geq 2$.
In this case, there exists at least one doubly degenerate representation $\mathcal{R}=\mathcal{R}_{\langle r,s \rangle}=\mathcal{R}_{\langle r',s' \rangle}$ such that 
\begin{align}
 p = r+r' \quad , \quad q = s+s'\ .
 \label{prrp}
\end{align}
Let us investigate the fusion product of $\mathcal{R}$ with  the Verma module $\mathcal{V}_P$. 
We will see that this fusion product is zero for almost all values of $P$. Whenever $\mathcal{R}\times \mathcal{V}_P=0$, we consider that $\mathcal{R}$ cannot coexist with a representation of momentum $P$, as correlation functions that involve both representations would vanish. 

According to Eq. \eqref{rtv}, the representation $\mathcal{V}_{P'} $ can appear in the fusion product $\mathcal{R}\times \mathcal{V}_P$ only if there are half-integers $(i,j,i',j')\in I_r\times I_s\times I_{r'}\times I_{s'}$ with $I_r=\left\{i\in \frac{r-1}{2}+\mathbb{Z}\middle| |i|\leq \frac{r-1}{2}\right\}$, such that  
\begin{align}
 \Delta(P') = \Delta\left(P+ ib+jb^{-1}\right) = \Delta\left(P+i'b+j'b^{-1}\right) \ .
 \label{dppdp}
\end{align}
This implies that the two momentums $P+ib+jb^{-1}$ and $ P+i'b+j'b^{-1}$ are either equal or opposite. If they were equal, we would have $q(2i-2i')=p(2j-2j')$, where $2i-2i'$ and $2j-2j'$ are integers. Since $p$ and $q$ are coprime, we would have $p|2i-2i'$, which would imply $i=i'$ because $|2i-2i'|\leq r+r'-2=p-2$. We would deduce that $r-r'$ is even, and therefore that $p$ is even. Similarly, we would find that $q$ is even, which would contradict our assumption that $p$ and $q$ are coprime.

So our two momentums must be opposite. Defining the integers $r'' = \frac{p}{2}-i-i'$ and $s''=\frac{q}{2}-j-j'$, this implies
\begin{align}
 \mathcal{R}\times \mathcal{V}_P \neq 0 \iff    P=P_{\langle r'',s'' \rangle}\quad \text{with}\quad  \left\{\begin{array}{l}  r'' \in [1,p-1]\ , \\ s'' \in [1,q-1]\ . \end{array}\right. 
\label{rpsq}
\end{align}
At a given central charge, this set of values of $(r'', s'')$ does not depend on the choice of $\mathcal{R}$, and is called the \textbf{\boldmath Kac table}\index{Kac!---table}.
And due to the identity \eqref{ddr},
the corresponding degenerate representations $\mathcal{R}_{\langle r'', s'' \rangle}$ are actually doubly degenerate: the set of these representations is also called the Kac table.

\subsubsection{Fusion rules}

The fusion rules \eqref{rrsr} of degenerate representations are valid at arbitrary central charges, and do not depend on the central charge. In order to deduce the fusion rules of doubly degenerate representations in the Kac table, we only need to take the identification \eqref{ddr} into account, with the result
\begin{align}
  \boxed{\mathcal{R}_{\langle r_1,s_1 \rangle} \times \mathcal{R}_{\langle r_2,s_2 \rangle} = \sum_{r_3\overset{2}{=}|r_1-r_2|+1}^{\min(r_1+r_2,2p-r_1-r_2)-1}\ \sum_{s_3\overset{2}{=}|s_1-s_2|+1}^{\min(s_1+s_2,2q-s_1-s_2)-1} \mathcal{R}_{\langle r_3,s_3 \rangle}}\ .
\label{rrmm}
\end{align}
(See Exercise \ref{exofus}.) In particular, the Kac table is closed under fusion.
Equivalently, 
the condition for three doubly degenerate representations $\mathcal{R}_{\langle r_i,s_i \rangle}$ to be intertwined by fusion is 
\begin{align}
 \left\{\begin{array}{l}  2p-r_1-r_2-r_3\in 1+2{\mathbb{N}}\ , \\
 r_i+r_j-r_k \in 1 + 2{\mathbb{N}}\ , \\
2q-s_1-s_2-s_3\in 1+2{\mathbb{N}}\ , \\
 s_i+s_j-s_k \in 1 + 2{\mathbb{N}}\ , \end{array}\right.  
\quad \text{or} \quad
 \left\{\begin{array}{l}  r_1+r_2+r_3-p\in 1+2{\mathbb{N}}\ , \\
 p+r_k-r_i-r_j\in 1+2{\mathbb{N}} \ , \\
s_1+s_2+s_3-q\in 1+2{\mathbb{N}}\ , \\
 q+s_k-s_i-s_j\in 1+2{\mathbb{N}} \ ,\end{array}\right. 
 \label{mmfr}
\end{align}
where we assume $i\neq j\neq k$.

\subsection{A-series minimal models \label{secamm}}

\subsubsection{Spectrum}

The diagonal combination of all representation in the Kac table is the spectrum of an \textbf{A-series minimal model}\index{minimal model!A-series---},
\begin{align}
 \boxed{ \mathcal{S}_{p,q} = \frac12 \bigoplus_{r=1}^{p-1} \bigoplus_{s=1}^{q-1} \mathcal{R}_{\langle r,s \rangle}\otimes \bar{\mathcal{R}}_{\langle r,s \rangle} } \ ,
\label{smin}
\end{align}
where the factor $\frac12$ is here to avoid counting each representation $\mathcal{R}_{\langle r,s \rangle}=\mathcal{R}_{\langle p-r,q-s \rangle}$ twice.
As in the case of generalized minimal models, we could define submodels whose spectrums would be based on smaller sets of representations, provided these smaller sets were closed under fusion.
However, the minimal models have the distinction of being consistent not only on the sphere, but also on all other Riemann surfaces \cite{fms97}.

It is easy to see that minimal models with $|p-q|>1$, and generalized minimal models, cannot be unitary, due to the presence of fields with negative conformal dimensions in their spectrums. (See Exercise \ref{exoneg}.)
On the other hand, minimal models with $|p-q|=1$ are unitary, see Eq. \eqref{srp}.

\subsubsection{Correlation functions}

In minimal models, we adopt the field normalization such that the two-point structure constant is one. We could not do this in Liouville theory or in generalized minimal models, as this would have made correlation functions non-meromorphic as functions of $b$. But this issue is moot in minimal models, as $b$ takes discrete values. So we call $\check{V}_{\langle r,s\rangle}$ the Kac table fields in this normalization, and the two-point function is 
\begin{align}
 \boxed{\left\langle \check{V}_{\langle r_1,s_1 \rangle} \check{V}_{\langle r_2,s_2 \rangle} \right\rangle = \delta_{r_1,p-r_2}\delta_{s_1,q-s_2} + \delta_{r_1,r_2} \delta_{s_1,s_2}} \ .
\end{align}
Compared with the two-point function of a generalized minimal model \eqref{vvdd}, we have an extra term due to the identity \eqref{ddr}. This identity actually accounts for all coincidences of conformal dimensions in the Kac table. In our normalization, two fields with the same dimension are identical, 
\begin{align}
 \check{V}_{\langle r,s \rangle} = \check{V}_{\langle p-r,q-s \rangle}\ .
  \label{vvr}
\end{align}
Let us deduce the three-point structure constant from the generalized minimal models' three-point structure constant. To do this, we use the relation between our field normalization, and the normalization we had in generalized minimal modelds,
\begin{align}
 \check{V}_{\langle r,s \rangle} = \frac{1}{\sqrt{R_{P_{\langle r,s \rangle}}}} V_{\langle r,s \rangle}\ , 
\end{align}
where $R_P$ is the Liouville reflection coefficient \eqref{ram}. We moreover renormalize the three-point structure constant with a factor $\sqrt{R_{P_{\langle 1,1 \rangle}}}$, in order to ensure that $V_{\langle 1,1\rangle}$ remains an identity field. This leads to the three-point structure constant 
\begin{align}
 \check{C}_{\langle r_1,s_1\rangle ,\langle r_2,s_2\rangle ,\langle r_3,s_3 \rangle} = \sqrt{\frac{R_{P_{\langle 1,1 \rangle}}}{\prod_{i=1}^3 R_{P_{\langle r_i,s_i \rangle}}}}  C_{\langle r_1,s_1\rangle ,\langle r_2,s_2\rangle ,\langle r_3,s_3 \rangle}\ ,
\label{tcc}
\end{align}
where the three-point structure constant of generalized minimal models $C_{\langle r_1,s_1\rangle ,\langle r_2,s_2\rangle ,\langle r_3,s_3 \rangle}$ was given in Eq. \eqref{cpqb}. 
Then $\check{C}_{\langle r_1,s_1\rangle ,\langle r_2,s_2\rangle ,\langle r_3,s_3 \rangle}$ does not depend on the cosmological constant, and we have 
\begin{align}
\boxed{ \check{C}_{\langle 1,1\rangle , \langle r,s \rangle,\langle r,s \rangle}=1}\ .
\end{align}

% NB: 3pt function essentially agrees with Dotsenko-Fateev result, as transcribed in hep-th/9112029 Eq. (15). 

\subsubsection{Trivial model $(p,q)=(3,2) \implies c = 0$}

This unitary minimal model has only one field -- the identity $I=\check{V}_{\langle 1,1 \rangle}=\check{V}_{\langle 2,1 \rangle}$.
The fusion rule $I\times I = I$ and three-point structure constant $\check{C}_{III} = 1$ are trivial.

\subsubsection{Yang--Lee singularity $(p,q)=(5,2) \implies c=-\tfrac{22}{5}$}

This minimal model has two independent fields, whose names and conformal dimensions $\Delta_{\langle r,s\rangle}$ are 
\begin{align}
 \begin{tikzpicture}[scale = .5, baseline=(current  bounding  box.center)]
  \draw[thick] (-1,-1) -- (-1, 1) -- (7, 1) -- (7, -1) -- cycle;
  \foreach \x in {1, ..., 4}{
  \node at ({2*(\x-1)}, 0) {$\langle \x,1\rangle $};
  }
 \end{tikzpicture}
 \ \ 
 \begin{tikzpicture}[scale = .5, baseline=(current  bounding  box.center)]
  \draw[thick] (-1,-1) -- (-1, 1) -- (7, 1) -- (7, -1) -- cycle;
  \node at (0, 0) {$I$};
  \node at (2, 0) {$\phi$};
  \node at (4, 0) {$\phi$};
  \node at (6, 0) {$I$};
  \end{tikzpicture}
  \ \ 
  \begin{tikzpicture}[scale = .5, baseline=(current  bounding  box.center)]
  \draw[thick] (-1,-1) -- (-1, 1) -- (7, 1) -- (7, -1) -- cycle;
  \node at (0, 0) {$0$};
  \node at (2, 0) {$-\frac15$};
  \node at (4, 0) {$-\frac15$};
  \node at (6, 0) {$0$};
  \end{tikzpicture}
\end{align}
Since the model is diagonal, we can use the symbols $I$ and $\phi$ not only for fields, but also for the corresponding representations of the Virasoro algebra.
The fusion rules are 
\begin{align}
 \begin{array}{l}
  I\times I = I \ ,
\\ I\times \phi= \phi\ ,
\\ \phi \times \phi = I + \phi\ .
 \end{array}
\end{align}
The only three-point structure constant not involving the identity field is 
\begin{align}
 \check{C}_{\phi\phi\phi} = i\sqrt{3\frac{\Gamma(-\frac35)}{\Gamma(-\frac15)^3} \frac{\Gamma(\frac15)^3}{\Gamma(\frac35)}}\ .
\label{cppp}
\end{align}
(See Exercise \ref{exocppp}.)

\subsubsection{\textbf{\boldmath Ising model}\index{Ising model} $(p,q)=(4,3) \implies c=\tfrac12$}

This unitary minimal model has three independent fields, 
\begin{align}
 \begin{tikzpicture}[scale = .5, baseline=(current  bounding  box.center)]
  \draw[thick] (-1,-1) -- (-1, 3) -- (5, 3) -- (5, -1) -- cycle;
  \foreach \x in {1, ..., 3}{
  \foreach \y in {1, 2}{
  \node at ({2*(\x-1)}, {2*(\y-1)}) {$\langle \x,\y\rangle $};
  }}
 \end{tikzpicture}
 \ \ 
 \begin{tikzpicture}[scale = .5, baseline=(current  bounding  box.center)]
  \draw[thick] (-1,-1) -- (-1, 3) -- (5, 3) -- (5, -1) -- cycle;
  \node at (0, 0) {$I$};
  \node at (2, 0) {$\sigma$};
  \node at (4, 0) {$\epsilon$};
  \node at (0, 2) {$\epsilon$};
  \node at (2, 2) {$\sigma$};
  \node at (4, 2) {$I$};
 \end{tikzpicture}
 \ \ 
 \begin{tikzpicture}[scale = .5, baseline=(current  bounding  box.center)]
  \draw[thick] (-1,-1) -- (-1, 3) -- (5, 3) -- (5, -1) -- cycle;
  \node at (0, 0) {$0$};
  \node at (2, 0) {$\frac{1}{16}$};
  \node at (4, 0) {$\frac12$};
  \node at (0, 2) {$\frac12$};
  \node at (2, 2) {$\frac{1}{16}$};
  \node at (4, 2) {$0$};
 \end{tikzpicture}
\end{align}
The fusion rules are 
\begin{align}
\begin{array}{l}
 I\times I = I \ ,
\\ I\times \epsilon = \epsilon\ ,
\\ I\times \sigma = \sigma\ ,
\end{array}
\hspace{2cm}
\begin{array}{l}
 \epsilon\times \epsilon = I\ ,
\\ \epsilon\times \sigma = \sigma\ ,
\\ \sigma \times \sigma = I + \epsilon\ .
\end{array}
\end{align}
We notice that there are two simple currents: the identity field $I$, and the field $\epsilon$.
Similarly, any A-series minimal model with $p,q>2$ has two simple currents $\check{V}_{\langle 1,1 \rangle}$ and $\check{V}_{\langle 1,q-1 \rangle}=\check{V}_{\langle p-1,1 \rangle}$. 

The only three-point structure constant not involving the identity field is 
\begin{align}
 \check{C}_{\sigma\sigma\epsilon} =\frac12\ .
\label{csse}
\end{align}
(See Exercise \ref{exocsse}.)
% NB: agrees with the ZZ book

Finally, let us comment on the interpretation of this minimal model in terms of statistical physics.
The Ising model of statistical physics describes two-state spins on a lattice.
Our field $\sigma$ describes these spins, while the field $\epsilon$ describes the energy.
Flipping spins is a symmetry of the model, and this symmetry manifests itself as the invariance of the fusion rules under $\sigma\to -\sigma$. The statistical model actually gives rise to infinitely many primary fields which do not belong to the minimal model.

\subsection{Limits to and from Liouville theory}\label{secltf}

In Liouville theory, generalized minimal models and A-series minimal models, the existence of the degenerate fields $V_{\langle 2,1\rangle}$ and $V_{\langle 1,2\rangle}$ ultimately determines the three-point structure constants. These CFTs are distinguished by their spectrums: diagonal in all cases, and respectively continuous, discrete infinite, and finite. These spectrums are related to one another by certain limits: for example, the Kac table of the $(p, q)$ minimal model becomes infinite as $p,q\to \infty$, and degenerate fields can be recovered from Liouville fields by taking limits of their momentums. (See Section \ref{secacl}.) This suggests that our CFTs themselves are related to one another by these limits. 

\subsubsection{From minimal models to Liouville theory}

Minimal models exist for central charges $c_{p,q}$ \eqref{cpq} that are dense in the half-line $]-\infty, 1]$, let us consider the limit $c_{p, q}\to c\in ]-\infty, 1]$ with $p,q\to\infty$.
We consider a doubly degenerate representation $\mathcal{R}_{\langle r,s\rangle}=\mathcal{R}_{\langle p-r,q-s\rangle}$, and keep the indices $r,s$ fixed. 
Then our representation always has a vanishing null vector at level $rs$. However, the level $(p-r)(q-s)$ of the second null vector goes to infinity, and the limit of our doubly degenerate representation is a degenerate representation, which we still denote as $\mathcal{R}_{\langle r,s\rangle}$:
 \begin{align}
  \lim_{c_{p,q}\to c} \mathcal{R}_{\langle r,s\rangle} = \mathcal{R}_{\langle r,s\rangle} \ .
 \end{align}
The fusion rules \eqref{rrmm} of doubly degenerate representations then reduce to the fusion rules of degenerate representations. Therefore, the limit of minimal models is the generalized minimal model with central charge $c$. 

For any central charge $c\in ]-\infty,1]$ such that $b^2\notin \mathbb{Q}$, the degenerate momentums $P_{\langle r,s\rangle}$ are dense in $i\mathbb{R}$. 
We can send $r,s$ to infinity so that $P_{\langle r,s\rangle}$ tends to a given imaginary value $P$. Then the null vector of $\mathcal{R}_{\langle r,s\rangle}$ goes to infinite level, and our degenerate representation becomes a Verma module,
\begin{align}
 \lim_{P_{\langle r,s\rangle} \to P} \mathcal{R}_{\langle r,s\rangle} 
 \ \underset{b^2\in \mathbb{R}_{<0}-\mathbb{Q}}{=} \
 \mathcal{V}_{P}\ .
\end{align}
Applying this treatment to the momentums in a generalized minimal model, we must recover Liouville theory.

The situation is more subtle if we take the limit $c_{p,q}\to c$ where  $c$ is again of the type $c_{p_0,q_0}$ with $p_0,q_0\in \mathbb{N}^*$. In this case, minimal models have several inequivalent nontrivial limits where the momentums are sent to continuous values. In some of these limits, we recover Liouville theory. However, in the case $c=1$, assuming $q=p+1$ while keeping $r-s$ and $\frac{r}{q}$ fixed 
does not lead to Liouville theory. Rather, we obtain another CFT with the same spectrum as Liouville theory and a non-analytic three-point structure constant, called Runkel--Watts theory \cite{rw01}. There is a Runkel--Watts-type theory for any central charges of the type $c_{p_0,q_0}$ \cite {rs15}, whose three-point structure constant includes the non-analytic factor \eqref{sigma}.
We expect that this theory can again be reached as a particular limit of minimal models.

\subsubsection{From Liouville theory to minimal models}

Starting from Liouville theory, we first want to recover generalized minimal models by making the fields degenerate, and then recover minimal models by sending the central charge to rational values. 

According to Eq. \eqref{cicni}, we can make the Liouville primary fields degenerate by sending their momentums to degenerate values, provided $c\notin ]-\infty, 1]$. However, as pointed out in Section \ref{secgmm}, we do not recover all the fusion rules of degenerate fields. This implies that after taking the limit, the decomposition of a four-point function into conformal blocks can involve terms that would not occur in generalized minimal models. The existence and nature of these terms depends on the four-point function, and on the order in which we make the fields degenerate. We conclude that from Liouville theory, we recover only certain four-point functions of generalized minimal models. 

The problem is much the same when we start from generalized minimal models, and send the central charge to $c_{p, q}$. In order to recover the corresponding minimal model, we need the degenerate fields to become doubly degenerate. At the level of correlation functions, this means that we need the doubly degenerate fusion rules to emerge. However, this occurs only in some cases, and we recover only certain four-point functions of the minimal model \cite{rib18}. 

Finally, what happens if we directly send $c\to c_{p, q}$ in Liouville theory, without first making the fields degenerate? Of course, nothing much happens if $c\in ]-\infty, 1]$ to begin with: we just recover Liouville theory at $c=c_{p, q}$. If however we start with $c\notin ]-\infty, 1]$, the limit $c\to c_0\in ]-\infty, 1]$ is not expected to exist, as the three-point structure constant has an essential singularity on $]-\infty, 1]$. However, it turns out that the limit of Liouville theory for $c\to 1$ is Runkel--Watts theory \cite{sch03}, and more generally the limit for $c\to c_{p,q}$ with $p,q\in\mathbb{N}^*$ is a Runkel--Watts-type theory \cite{mce07}.

\subsubsection{Summary of results}

Let us schematically summarize the limits we have been discussing, see also the diagram \eqref{lims}. We denote (generalized) minimal models as (G)MM, and Runkel--Watts-type theories as RWT. We begin with limits of minimal models:
\begin{align}
 \lim_{\substack{p,q\to \infty \\ c_{p,q}\to c \\ r_i,s_i\text{ fixed}}} \text{MM}_{p,q} &= \text{GMM}_{c}\ ,
\\
 \lim_{P_{\langle r_i,s_i\rangle} \to P_i} \text{GMM}_{c\neq c_{p,q}\in]-\infty,1[} &= \text{Liouville}_{c} \ , 
 \\
 \lim_{\substack{p,q\to \infty \\ c_{p,q}\to c_{p_0,q_0} \\ P_{\langle r_i,s_i\rangle} \to P_i}} \text{MM}_{p,q}
 & \supset \left\{\text{Liouville}_{c_{p_0,q_0}}, \text{RWT}_{p_0,q_0}\right\}\ ,
 \\
 \lim_{\substack{p,q\to \infty \\ c_{p,q}\to c }} \text{RWT}_{p,q} & = \text{Liouville}_{c}\ .
\end{align}
 We then summarize limits where the momentums or the central charge go from generic to discrete values. 
\begin{align}
 \lim_{P_i\to P_{\langle r_i,s_i\rangle}} \text{Liouville}_{c\in ]-\infty,1]} &= \text{Liouville}_{c}\ ,
 \\
 \lim_{c\to c_{p,q}} \text{Liouville}_{c\in ]-\infty,1]} & = \text{Liouville}_{c_{p,q}}\ ,
 \\
 \lim_{c\to c_{p,q}} \text{Liouville}_{c\notin ]-\infty,1]} & = \text{RWT}_{p,q}\ ,
\end{align}
The $7$ limits that we wrote until now are relatively straighforward consequences of the analytic properties of structure constants and conformal blocks, together with the convergence of conformal block decompositions of correlation functions. The main subtlety is the emergence of the three-point structure constants of Runkel--Watts-type theories, but this subtlety does not affect conformal blocks. On the other hand, the following two limits involve an interplay between singularities of structure constants and conformal blocks. These limits hold only for subsets of four-point functions, and these subsets are not fully known:
\begin{align}
 \lim_{P_i\to P_{\langle r_i,s_i\rangle}} \text{Liouville}_{c\notin ]-\infty,1]} &\sim \text{GMM}_c\ ,
 \\
 \lim_{c\to c_{p,q}} \text{GMM}_c & \sim \text{MM}_{p, q}\ .
\end{align}

\section{Exercises}

\begin{exo}[Degenerate OPE coefficients and the two-point structure constant]
 ~\label{exocpcm}
 Write the OPE coefficients $C_\pm(P)$ \eqref{vot} in terms of two- and three-point structure constants. In particular, show that 
 \begin{align}
  C_+(P)B_{P+\tfrac{b}{2}} = C_{\langle 2,1\rangle, P, P+\frac{b}{2}}\ . 
 \end{align}
 Using the invariance of the three-point structure constant $C_{\langle 2,1\rangle, P,P'}$ under permutations of its last two arguments, show that $C_+(P)$ and $C_-(P)$ are related as
 \begin{align}
  C_+(P)B_{P+\tfrac{b}{2}} = C_-(P+\tfrac{b}{2}) B_{P}\ .
 \end{align}
\end{exo}

\begin{exo}[Constraints on degenerate OPE coefficients from crossing symmetry]
 ~\label{exonorm}
Consider the four-point function $
 \left\langle V_{\langle 2,1\rangle}(x) V_P(0) V_{\langle 1, 2\rangle}(\infty) V_{P+\frac{Q}{2}}(1)\right\rangle$.
 \begin{enumerate}
  \item Show that the $s$- and $t$-channel decompositions of this four-point function each involve only one 
conformal block,
\begin{align}
\mathcal{F}^{(s)}_+ = 
\begin{tikzpicture}[baseline=(current  bounding  box.center), very thick, scale = .7]
\draw[dashed] (2, 0) -- node [above right] {\!$\langle 1,2 \rangle$} (0, 2);
\draw[dashed] (-2, 0)
-- node [below left] {$\langle 2,1 \rangle$\!} (0, -2); 
\draw (-2, 0) -- node [above left] {$P$\!} (0, 2) 
-- node [left] {$P+\frac{b}{2}\!  $} (0, -2)
-- node [below right] {\!$P+\frac{Q}{2}$} (2, 0);
\end{tikzpicture}
\quad , \quad 
\mathcal{F}^{(t)}_- = 
\begin{tikzpicture}[baseline=(current  bounding  box.center), very thick, scale = .7]
\draw[dashed] (2, 0) -- node [above right] {\!$\langle 1,2 \rangle$} (0, 2);
\draw[dashed] (-2, 0)
-- node [below left] {$\langle 2,1 \rangle$\!} (0, -2); 
\draw (0, 2) -- node [above left] {$P$\!} (-2, 0)
-- node [above] {$P+\frac{1}{2b}  $} (2, 0)
-- node [below right] {\!$P+\frac{Q}{2}$} (0, -2);
\end{tikzpicture}
\ . 
\end{align}
\item Compute these hypergeometric blocks using Eqs. \eqref{gpm} and \eqref{tpm}, and show in particular that $\mathcal{F}^{(s)}_+ = \mathcal{F}^{(t)}_-$.

\item Show that the coefficients of the two decompositions are expressed in terms of the OPE coefficients of degenerate fields as 
\begin{align}
 c_+^{(s)} &= C_+(P)\tilde{C}_+(P+\tfrac{b}{2})B_{P+\tfrac{Q}{2}} \ ,
 \\
 c_-^{(t)} &= \tilde{C}_+(P)C_+(P+\tfrac{1}{2b})B_{P+\tfrac{Q}{2}}\ .
\end{align}
From $c_+^{(s)}=c_-^{(t)}$, deduce a relation between $C_+$ and $\tilde{C}_+$.
\item Study how $C_+(P)$ and $\tilde{C}_+(P)$ behave under a field renormalization, and show that they renormalize to one provided
\begin{align}
 C_+(P) = \frac{\lambda_{P+\frac{b}{2}}}{\lambda_P} \quad \text{and} \quad 
\tilde{C}_+(P) = \frac{\lambda_{P+\frac{1}{2b}}}{\lambda_P} \ .
\end{align}
Using the relation between $C_+$ and $\tilde{C}_+$, show that these two equation for $\lambda_P$ are compatible, and deduce that there exists a field normalization such that 
\begin{align}
 C_+(P) = \tilde{C}_+(P) = 1\ .
 \label{cmpo}
\end{align}
 \end{enumerate}
\end{exo}

\begin{exo}[Convergence of the conformal block decomposition in Liouville theory]
 ~\label{exocvg}
 Following \cite{rs15}, let us show that the integral over $s$-channel momentums converges in the decomposition \eqref{vfcch} of the four-point function.
 \begin{enumerate}
  \item Using its integral representation \eqref{lup}, show that the Upsilon function has the asymptotic behaviour
  \begin{align}
   \log \Upsilon_b(\tfrac{Q}{2}+P) \underset{P\to i\infty}{=} P^2\log |P| -\frac32 P^2 + o(P^2)\ ,
  \end{align}
  and deduce the behaviour of the combination of structure constants $\frac{C_{12s}C_{s34}}{B_s}$ as $P_s\to i\infty$.
  \item Using Zamolodchikov's recursion, show that $s$-channel conformal blocks have the asymptotic behaviour
  \begin{align}
   \log\mathcal{F}_{\Delta_s}^{(s)}(\Delta_i|x) \underset{\Delta_s\to \infty}{=} \Delta_s\log(16q) + O(1)\ .
  \end{align}
  \item Deduce that the integral over $s$-channel momentums converges for $|q|<1$.
 \end{enumerate}
\end{exo}

\begin{exo}[Behaviour of Liouville four-point functions at coinciding points]
~\label{exo4a} 
We want to determine how the Liouville four-point function $\left<\prod_{i=1}^4 V_{P_i}(z_i)\right>$ behaves in the limit $z_1\to z_2$, using the $s$-channel decomposition \eqref{vfcch}.
\begin{enumerate}
 \item Show that for $z_1\to z_2$, the integral over $s$-channel momentums localizes near $P_s=0$. Show that the three-point structure constants $C$ and $\hat C$ behave differently near this value of the momentum.
 \item Show that 
 \begin{align}
  \left<\prod_{i=1}^4 V_{P_i}(z_i)\right> \underset{z_1\to z_2}{\sim} \left\{
  \begin{array}{ll} 
  |z_{12}|^{\frac{Q^2}{2} -2\Delta_{P_1}-2\Delta_{P_2}} \left|\log|z_{12}|\right|^{-\frac32} &\quad \text{if \ } c\notin ]-\infty,1]\ ,
  \\
  |z_{12}|^{\frac{Q^2}{2} -2\Delta_{P_1}-2\Delta_{P_2}} \left|\log|z_{12}|\right|^{-\frac12} &\quad \text{if \ } c\in ]-\infty,1]\ .
  \end{array}\right.
 \end{align}
How does the first subleading correction behave?
\item Compare with the behaviour of a four-point function in a rational theory. 
For an interpretation of these different behaviours in random energy models, see \cite{clrs16}.
\end{enumerate}
\end{exo}

\begin{exo}[How Runkel--Watts-type theories circumvent the uniqueness of diagonal CFTs]
 ~\label{exonurw}
The aim of this exercise is to understand how Liouville theory and Runkel--Watts-type theories can have different three-point structure constants, circumventing in particular the relation \eqref{cmnnf} between the three-point structure constant and the fusing matrix. We focus on the case $c=1$ for technical simplicity. 
\begin{enumerate}
 \item In the case $c=1$, the conformal blocks and fusing matrix have considerably simpler expressions than in the general case. Study the analyticity properties of the fusing matrix from the article \cite{ilt13}, and deduce that Eq. \eqref{cmnnf} cannot hold in Runkel--Watts theory. 
 \item Review the derivation of Eq. \eqref{cmnnf} in the cases of Liouville theory with $c=1$, and of Runkel--Watts theory. Paying particular attention to integration contours, poles of conformal blocks, and analyticity of structure constants, deduce that the derivation works in Liouville theory and fails in Runkel--Watts theory.
\end{enumerate}
\end{exo}

\begin{exo}[Fusion rules of all unitary representations if $c>1$]
 ~\label{exoaur}
 Let us investigate the fusion rules that follow from the analytically continued Liouville OPE \eqref{acope}.
 \begin{enumerate}
  \item Assuming $b>0$ so that $c\geq 25$, write the fusion rules of the Verma modules with momentums $\Pi\in ]0,\frac{Q}{2}[$, i.e. of the unitary representation that do not belong to the spectrum \eqref{sad}.  Check that you find
\begin{align}
 \mathcal{V}_\Pi \times \mathcal{V}_P  
&= \int_{i\mathbb{R}_+} dP'\ \mathcal{V}_{P'}\ ,
\\
 \mathcal{V}_{\Pi_1} \times \mathcal{V}_{\Pi_2} 
&= \sum_{\Pi'\in \left]0,\frac{Q}{2}\right[\cap \left(\Pi_1+\Pi_2 -\frac{Q}{2}- b\mathbb{N}-b^{-1}\mathbb{N}\right)} \mathcal{V}_{\Pi'}  
+ \int_{i\mathbb{R}_+} dP'\ \mathcal{V}_{P'}\ .
\end{align}
\item Given a unitary Verma module $\mathcal{V}_\Delta$ of conformal dimension $\Delta>0$, show that $\mathcal{V}_\Delta \times \mathcal{V}_\Delta$ belongs to the Liouville spectrum if $\Delta \geq \frac{c-1}{32}$, and has discrete terms if $\Delta \in ]0,\frac{c-1}{32}[$.
\item If $1<c<25$, show that there is at most one discrete term, and determine its momentum. Deduce that the set of unitary representations is closed under fusion.
 \end{enumerate}
\end{exo}

\begin{exo}[Degenerate OPE coefficients in the DOZZ normalization]
 ~\label{exodoc}
 Let us compute the degenerate OPE coefficients $C_\pm(P)$, assuming that fields are normalized as in Eq. \eqref{nop}.
\begin{enumerate}
 \item Further studying single-valued solutions of the hypergeometric functions, show that in addition to Eq. \eqref{spsm}, we have 
 \begin{align}
  \frac{c_+^{(s)}}{c_-^{(t)}} = -\frac{1}{\det F}\frac{F_{-+}}{F_{+-}} = \frac{\gamma(2bP_1)}{\gamma(-2bP_3)}\prod_\pm \gamma\left(\tfrac12-b(P_1\pm P_2+P_3)\right)\ .
 \end{align}
\item Specializing to a four-point function with two degenerate fields,
show that in addition to Eq. \eqref{eq:ddgg} for $c_+^{(s)}$, we have the equation
$
 c_-^{(t)} \propto B_{P}
$ for the coefficient of the $t$-channel field $V_{\langle 1,1\rangle}$. 
(Here and in later equations, 
we neglect $P$-independent factors.) Deduce that 
\begin{align}
 C_+(P)^2 \propto \gamma(2bP)\gamma(-2bP-b^2) \frac{B_P}{B_{P+\frac{b}{2}}}\ .
\end{align}
\item Using Eq. \eqref{ram} for $B_P=R_P$, deduce $C_+(P)\propto 1$, and show that the DOZZ normalization is such that Eq. \eqref{cmpo} holds. 
\item
Study how the degenerate OPE \eqref{vot} behaves under reflections, and show that $C_-(P) = R_PR_{-P+\frac{b}{2}}C_+(-P)$. Using again Eq. \eqref{ram} for $R_P$, deduce
\begin{align}
 C_-(P) = \mu^b b^{2bQ} \gamma(-2bP)\gamma(2bP-b^2)\ .
\end{align}
\end{enumerate}
 \end{exo}

 \begin{exo}[Uniqueness of generalized minimal models]
 ~\label{exofix}
Let us prove the uniqueness of the three-point structure constants \eqref{chc} of generalized minimal models.
We start by considering special cases of crossing symmetry equations: the shift equations that we used for determining the Liouville structure constant.
 \begin{enumerate}
  \item Show that the shift equations determine how the three-point structure constants $C_{\langle r_1,s_1\rangle ,\langle r_2,s_2\rangle ,\langle r_3,s_3 \rangle}$ change when the indices $r_i,s_i$ are shifted by two units.
  \item Deduce that together with permutation symmetry, the shift equations determine all three-point structure constants in terms of the five fundamental quantities 
\begin{align}
 C_{\langle 1,1\rangle,\langle 1,1\rangle, \langle 1,1\rangle}
 \ , \ 
 C_{\langle 1,1\rangle,\langle 1,2\rangle, \langle 1,2\rangle}
 \ , \
 C_{\langle 1,1\rangle,\langle 2,1\rangle, \langle 2,1\rangle}
 \ , \
 C_{\langle 1,1\rangle,\langle 2,2\rangle, \langle 2,2\rangle}
 \ , \
 C_{\langle 1,2\rangle,\langle 2,1\rangle, \langle 2,2\rangle}\ .
\end{align}
Determine the first four quantities by assuming that $V_{\langle 1,1\rangle}$ is the identity field, and that the two-point structure constant is one.
\item Using a judiciously chosen four-point function, show that $C^2_{\langle 1,2\rangle,\langle 2,1\rangle, \langle 2,2\rangle}$ is determined in terms of the other structure constants. 
\item Show that the remaining sign ambiguity in $C_{\langle 1,2\rangle,\langle 2,1\rangle, \langle 2,2\rangle}$ can be absorbed in a field renormalization, and conclude. 
 \end{enumerate}
 \end{exo}

\begin{exo}[Computation of some simple structure constants]
 ~\label{exosssc}
 In this exercise we compute structure constants of the type $\check{C}_P = \check{C}_{\langle 3,1\rangle, P,P}$, in the field normalization where two-point functions are one. 
\begin{enumerate}
\item Write the $s$-channel decomposition of the four-point function $\left< V_{\langle 2,1\rangle} V_{\langle 2,1\rangle} V_PV_P\right>$, and deduce $\check{C}_P \check{C}_{\langle 2,1\rangle}$ from Eq. \eqref{spsm}.
\item Using the identities $\gamma(x+1)=-x^2\gamma(x)$ and $\gamma(2x)= 2^{4x-1}\gamma(x)\gamma(x+\tfrac12)$, check that
\begin{align}
 \check{C}_P \check{C}_{\langle 2,1\rangle} &= 2^{8b^2+4} \prod_\pm \frac{\gamma(-b^2\pm 2bP)}{\gamma(-b^2\pm \frac12)} \ ,
 \\
 \check{C}_{\langle 3,1\rangle} &= 2^{-2-4b^2}(1+2b^2)\frac{\gamma(-\tfrac12-2b^2)}{\gamma(\tfrac12-b^2)}\sqrt{-\frac{\gamma(-b^2)}{\gamma(-1-3b^2)}}\ .
 \label{cto}
\end{align}
 \item As an independent verification, deduce $\check{C}_P$ from the formula \eqref{cijrs}, and compare.
\end{enumerate}
\end{exo}

\begin{exo}[Fusion rules of doubly degenerate representations]
 ~\label{exofus}
Deduce the fusion rules \eqref{rrmm} of doubly degenerate representations from the degenerate fusion rule \eqref{rrsr}. 
Hint: since $p$ or $q$ is odd, knowing the parity of $r, r', s, s'$ is enough for ruling out one of the two coincidences $\mathcal{R}_{\langle r,s\rangle} = \mathcal{R}_{\langle r',s'\rangle}$ and $\mathcal{R}_{\langle r,s\rangle} = \mathcal{R}_{\langle p-r',q-s'\rangle}$.
\end{exo}

\begin{exo}[Unitarity of (generalized) minimal models]
 ~\label{exoneg}
From Section \ref{secuni} on unitarity, assume we only remember that highest-weight representations with conformal dimensions $\Delta <0$ cannot be unitary. 
\begin{enumerate}
 \item 
Assuming $b^2 < -1$, show that the condition for the conformal dimension $\Delta_{\langle r,s \rangle}$ \eqref{drs} of a degenerate representation to be negative is 
\begin{align}
 \Delta_{\langle r,s \rangle} < 0 \quad \iff \quad \frac{r-1}{s-1} < -\frac{1}{b^2} < \frac{r+1}{s+1} \ .
\end{align}
\item
If $b^2$ is irrational, show that this occurs for infinitely many choices of $\langle r,s \rangle$, by applying the Dirichlet approximation theorem to $-\frac{1}{b^2}$. 
Deduce that the corresponding generalized minimal model is not unitary. 
\item
In the minimal model case $b^2 = -\frac{q}{p}$, show that 
\begin{align}
 \bigcup_{r=1}^{p-1}\bigcup_{s=1}^{q-1} \left] \frac{r-1}{s-1} , \frac{r+1}{s+1} \right[  = \left]0, \frac{p}{p+1}\right[ \ .
\end{align}
(Pay special attention to the case $s=r+1$.) Deduce that if $|q-p|>1$, then the corresponding minimal model is not unitary.
\end{enumerate}
\end{exo}

\begin{exo}[Structure constant of the $(5,2)$ minimal model]
 ~\label{exocppp}
 Let us compute the nontrivial three-point structure constant $\check{C}_{\phi\phi\phi}$ of the $(5,2)$ minimal model.
 \begin{enumerate}
  \item Using Eq. \eqref{tcc}, write
\begin{align}
 \check{C}_{\phi\phi\phi} = \sqrt{\frac{R_{P_{\langle 1,1 \rangle}}}{R_{P_{\langle 2,1 \rangle}}^3}} C_{\langle 2,1\rangle , \langle 2,1 \rangle,\langle 3,1 \rangle}\ .
\end{align}
\item Notice that $C_{\langle 2,1\rangle , \langle 2,1 \rangle,\langle 3,1 \rangle} = C_+(P_{\langle 2,1 \rangle})=1$ due to Eqs. \eqref{vot} and \eqref{cco}.
Choosing an appropriate value for $\mu$, compute
\begin{align}
 R_{P_{\langle 1,1 \rangle}} = -\frac{\Gamma(-\frac35)\Gamma(\frac32)}{\Gamma(\frac35)\Gamma(-\frac32)} \ , \quad R_{P_{\langle 2,1 \rangle}} = -\frac{\Gamma(-\frac15)\Gamma(\frac12)}{\Gamma(\frac15)\Gamma(-\frac12)}\ , 
\end{align}
and deduce the explicit formula \eqref{cppp} for $\check{C}_{\phi\phi\phi}$.
\item Rederive the same result using Eq. \eqref{cto}.
 \end{enumerate}

\end{exo}

\begin{exo}[Structure constant of the $(4,3)$ minimal model]
 ~\label{exocsse}
 Let us compute the nontrivial three-point structure constant $\check{C}_{\sigma\sigma\epsilon} $ of the $(4,3)$ minimal model.
 \begin{enumerate}
  \item Show that 
\begin{align}
 \check{C}_{\sigma\sigma\epsilon} = \frac{1}{R_{P_{\langle 2,1 \rangle}}}\sqrt{\frac{R_{P_{\langle 1,1 \rangle}}}{R_{P_{\langle 1,2 \rangle}}}} C_{\langle 2,1\rangle , \langle 2,1 \rangle,\langle 3,1 \rangle}\ .
\end{align}
\item Choosing an appropriate value for $\mu$, compute
\begin{align}
 R_{P_{\langle 1,1 \rangle}} = -\frac{\Gamma(-\frac14)\Gamma(\frac13)}{\Gamma(\frac14)\Gamma(-\frac13)} \ , \quad R_{P_{\langle 1,2 \rangle}} = -\frac{\Gamma(-\frac54)\Gamma(\frac53)}{\Gamma(\frac54)\Gamma(-\frac53)} \ , \quad R_{P_{\langle 2,1 \rangle}} = -\frac{\Gamma(\frac12)\Gamma(-\frac23)}{\Gamma(-\frac12)\Gamma(\frac23)}\ ,
\end{align}
and deduce the explicit formula \eqref{csse} for $\check{C}_{\sigma\sigma\epsilon}$.
\item Rederive that structure constant using the results of Exercise \ref{exosssc}.
 \end{enumerate}
\end{exo}

\chapter{Affine symmetry \label{secaff}}

Affine Lie algebras are infinite-dimensional Lie algebras whose universal enveloping algebras contain the Virasoro algebra.
In order to study conformal field theories based on these algebras, we will follow the same steps as in the case of the Virasoro algebra: study the algebras and their representations, deduce Ward identities and differential equations for correlation functions, and write spectrums and correlation functions for particular models. 

\section{Free bosons}\label{secfb}

\subsection{Symmetry algebra and representations \label{secaua}}

\subsubsection{The \boldmath $\hat{\mathfrak{u}}_1$ current and the energy-momentum tensor}

In Section \ref{secem}, we showed how the Virasoro symmetry algebra is encoded in the energy-momentum tensor $T(y)$.
Similarly, we now introduce the holomorphic \textbf{\boldmath $\hat{\mathfrak{u}}_1$ current}\index{current!u1@$\hat{\mathfrak{u}}_1$---} $J(y)$, the symmetry field that encodes the symmetry algebra of free bosonic theories.
This field is characterized by the OPE
\begin{align}
 \boxed{J(y)J(z) = \frac{-\frac12}{(y-z)^2} + O(1)}\ .
\label{jj}
\end{align}
Using such a $\hat{\mathfrak{u}}_1$ current, it is possible to build a one-parameter family of Virasoro fields,
\begin{align}
 \boxed{T(z) = -(JJ)(z) - Q\partial J(z)}\ ,
\label{tqz}
\end{align}
where the parameter $Q$ will shortly be found to coincide with the background charge. 
Here $(JJ)(z)$ is an instance of the \textbf{\boldmath normal-ordered product}\index{normal-ordered product}, 
\begin{align}
 (AB)(z) = \frac{1}{2\pi i} \oint_z \frac{dy}{y-z} A(y)B(z)\ .
\label{abz}
\end{align}
Equivalently, if $\cunderbracket{A}{(y)}{B}(z)$ is the singular part of the OPE $A(y)B(z)$,
we have
\begin{align}
(AB)(z) &= \underset{y\to z}{\lim} \left(A(y)B(z)-\cunderbracket{A}{(y)}{B}(z)\right)\ ,
\\
 A(y)B(z) &= \cunderbracket{A}{(y)}{B}(z) + (AB)(z) + O(y-z)\ .
 \label{abope}
\end{align}
In our example, this amounts to
\begin{align}
 (JJ)(z) = \underset{y\to z}{\lim} \left( J(y)J(z) + \frac{\frac12}{(y-z)^2}\right)\ .
\end{align}
In general $\cunderbracket{A}{(y)}{B}(z) \neq \cunderbracket{B}{(z)}{A}(y)$ and $(AB)\neq (BA)$ although $A(y)B(z)=B(z)A(y)$.
(See Exercise \ref{exoabba} for the computation of $(AB)-(BA)$.)
OPEs of normal-ordered products can be computed using \textbf{\boldmath Wick's theorem}\index{Wick's theorem},
\begin{align}
 \cunderbracket{A}{(z)}{(BC)}(y) = \frac{1}{2\pi i}\oint_y \frac{dx}{x-y}\left(\cunderbracket{A}{(z)}{B}(x)C(y) + B(x)\cunderbracket{A}{(z)}{C}(y)\right)\ .
\label{wick}
\end{align}
For example, we can compute 
\begin{align}
 (JJ)(y)J(z) = -\frac{J(y)}{(y-z)^2} +O(1) =  -{\frac{\partial}{\partial z}}\frac{J(z)}{y-z} + O(1)\ ,
\end{align}
which leads to 
\begin{align}
 \boxed{T(y)J(z) = \frac{-Q}{(y-z)^3} +{\frac{\partial}{\partial z}}\frac{1}{y-z}J(z) + O(1)}\ .
\label{tqj}
\end{align}
Then we can compute the OPE $T(y)T(z)$, and we find nothing but the Virasoro field OPE \eqref{tt}, where the central charge is given in terms of the background charge $Q$ by $c = 1+6 Q^2$ (repeating Eq. \eqref{cqb}). 
So $T(z)$ is a Virasoro field as announced. 
In order to build conformal field theories from the current $J(z)$, we choose a value of $Q$ and identify the corresponding Virasoro field with the energy-momentum tensor -- the Virasoro fields with $Q'\neq Q$ will henceforth play no role.

\subsubsection{The affine Lie algebra \boldmath $\hat{\mathfrak{u}}_1$}

Let us introduce the modes $J_n^{(z_0)}$ of the current $J(y)$, 
\begin{align}
 J_n^{(z_0)} = \frac{1}{2\pi i}\oint_{z_0} dy\ (y-z_0)^n J(y)\ .
\end{align}
Equivalently, the OPE of $J(y)$ with a generic field $V_\sigma(z_0)$ is 
\begin{align}
 \boxed{J(y) V_\sigma(z_0) = \sum_{n\in {\mathbb{Z}}} \frac{J_n V_\sigma(z_0)}{(y-z_0)^{n+1}}}\ ,
\label{jvn}
\end{align}
where we use the simplified notation $J_n=J_n^{(z_0)}$ when working at fixed $z_0$.
This OPE is consistent with dimensional analysis, if in addition to the known dimensions \eqref{zaz} we assume $[J]=1$ and $[J_n]=-n$.
The $JJ$ OPE \eqref{jj} is equivalent to the commutation relations
\begin{align}
 \boxed{ [J_m,J_n] =  \frac12 n \delta_{m+n,0}}\ ,
\label{jmjn}
\end{align}
which define the affine Lie algebra \textbf{\boldmath $\hat{\mathfrak{u}}_1$}\index{u1@$\hat{\mathfrak{u}}_1$ (affine Lie algebra)}.
This algebra can be written as a direct sum of commuting finite-dimensional subalgebras as 
\begin{align}
 \hat{\mathfrak{u}}_1 = \text{Span}(J_0) \oplus \bigoplus_{n=1}^\infty \text{Span}(J_n,J_{-n}) \ .
\end{align}
The construction \eqref{tqz} of $T(z)$ from the $\hat{\mathfrak{u}}_1$ current $J(z)$ is equivalent to the following formula for the modes $L_n$ \eqref{lit} of $T(z)$:
\begin{align}
 L_n &= -\sum_{m\in{\mathbb{Z}}} J_{n-m}J_m + Q(n+1)J_n\ , \qquad (n\neq 0)\ ,
\label{lnj}
\\
L_0 &=-2\sum_{m=1}^\infty J_{-m}J_m -J_0^2+QJ_0 \ ,
\label{lzj}
\end{align}
and the $TJ$ OPE \eqref{tqj} is equivalent to 
\begin{align}
 [L_m,J_n] = -nJ_{m+n} -\frac{Q}{2}m(m+1) \delta_{m+n,0} \ . 
\end{align}

Let us briefly discuss how $\hat{\mathfrak{u}}_1$ generators can behave under Hermitian conjugation.
It turns out that a conjugation of $\hat{\mathfrak{u}}_1$ that is compatible with the conjugation \eqref{ldn} of the Virasoro algebra  can exist only if $Q^2\in\mathbb{R}$ (see Exercise \ref{exocuo}). 
The conjugation is then 
\begin{align}
 J_n^\dagger = \operatorname{sign}(Q^2)\big( Q\delta_{n,0}-J_{-n} \big)\ ,\qquad (Q^2\in\mathbb{R}) \ .
\label{jdq}
\end{align}

\subsubsection{Primary fields and highest-weight representations}

We define a \textbf{\boldmath $\hat{\mathfrak{u}}_1$-primary field}\index{primary field!u1@$\hat{\mathfrak{u}}_1$---} $V_{\alpha}(z_0)$ with the \textbf{\boldmath $\hat{\mathfrak{u}}_1$ momentum}\index{momentum!u1@$\hat{\mathfrak{u}}_1$---} $\alpha$ by its OPE with $J(y)$,
\begin{align}
 \boxed{J(y) V_\alpha(z_0) = \frac{\alpha}{y-z_0} V_\alpha(z_0) + O(1)}\ .
\label{jva}
\end{align}
To a $\hat{\mathfrak{u}}_1$-primary field $V_\alpha(z_0)$ we associate a highest-weight representation $\mathcal{U}_\alpha$ of the  affine Lie algebra $\hat{\mathfrak{u}}_1$.
The states in $\mathcal{U}_\alpha$ correspond to $V_\alpha(z_0)$ and its $\hat{\mathfrak{u}}_1$-descendant fields $V_\sigma(z_0)=\left(\prod_i J_{-n_i}\right) V_\alpha(z_0)$ with $n_i\geq 1$, which all obey 
\begin{align}
 \left(J_0 - \alpha\right) V_\sigma(z_0) = 0
\label{jma}
\end{align}
For $Q^2\in\mathbb{R}$, 
a natural scalar product can be defined in $\mathcal{U}_\alpha$ using Eq. \eqref{jdq}.
Then, if $|\alpha\rangle$ is the $\hat{\mathfrak{u}}_1$-primary state of $\mathcal{U}_\alpha$, we have $\left< J_{-1}\alpha\middle|J_{-1}\alpha\right> = \frac12 \operatorname{sign}(Q^2) \langle \alpha|\alpha\rangle$.
So, for the scalar product to be positive definite, we need $Q\in \mathbb{R}$.
Moreover, since $J_0$ obeys $J_0^\dagger = Q-J_0$ and has the eigenvalue $\alpha$, we must have $\alpha \in \frac{Q}{2} + i{\mathbb{R}}$.

Any $\hat{\mathfrak{u}}_1$-primary field is also a primary field, whose OPE $T(y)V_\alpha(z_0)$ is given by Eq. \eqref{tvp}, with the conformal dimension 
\begin{align}
 \Delta(\alpha) = \alpha(Q-\alpha)\ .
 \label{daqu}
\end{align}
Comparing with
Eq. \eqref{daq}, we obtain the relation $\alpha = \frac{Q}{2}+P$ (up to reflection) between $\hat{\mathfrak{u}}_1$ and Virasoro momentums. 
Under the action of the Virasoro algebra Eqs. \eqref{lnj}-\eqref{lzj}, the $\hat{\mathfrak{u}}_1$ highest-weight representation $\mathcal{U}_\alpha$ coincides with a Verma module $\mathcal{V}_P$ of momentum $P=\pm(\alpha-\frac{Q}{2})$, provided this Verma module is irreducible. (For the case of reducible Verma modules, see Exercise \ref{exoazq}.)
To summarize,
\begin{align}
\renewcommand{\arraystretch}{1.3}
 \begin{tabular}{|l|c|c|}
  \hline
 Symmetry algebra  & $\mathfrak{V}$ & $\hat{\mathfrak{u}}_1 $
\\
\hline\hline
 Existence of scalar product & $Q\in {\mathbb{R}} \cup i{\mathbb{R}} $ & $Q\in {\mathbb{R}} \cup i{\mathbb{R}}$ 
\\
\hline
 Unitarity & $Q\in\mathbb{R}$ and $P\in i{\mathbb{R}} \cup ]-\frac{Q}{2},\frac{Q}{2}[ $ &  $Q\in\mathbb{R}$ and $\alpha \in \frac{Q}{2}+i{\mathbb{R}}$
\\
\hline 
 Reflection relation  & $\mathcal{V}_P = \mathcal{V}_{-P} $ & $\mathcal{U}_\alpha \neq \mathcal{U}_{Q-\alpha} $ 
\\
\hline
 \end{tabular}
\end{align}
Even in theories with no $\hat{\mathfrak{u}}_1$ symmetry (and therefore no $\hat{\mathfrak{u}}_1$ Ward identities), 
it can be useful to think of Virasoro representations as $\hat{\mathfrak{u}}_1$ representations. 
For example,
whenever Liouville theory is unitary i.e. $Q\in\mathbb{R}^*$, its spectrum \eqref{sad} contains only representations that are unitary as $\hat{\mathfrak{u}}_1$ representations, leaving out the representations with $P\in ]-\frac{Q}{2}, \frac{Q}{2}[$ although they are unitary as Virasoro representations. 
For example, the AGT relation leads to an expression for Virasoro conformal blocks as sums over particular descendant states, and these states simplify when written in terms of $\hat{\mathfrak{u}}_1$ creation modes \cite{aflt10}.

\subsection{Ward identities and Seiberg--Witten equations}

\subsubsection{Ward identities}

In order to derive $\hat{\mathfrak{u}}_1$ Ward identities, we need to know how the current $J(y)$ behaves as $y\to \infty$.
The relation \eqref{tqz} with the energy-momentum tensor $T(y)$, and the behaviour \eqref{tyi} of $T(y)$, suggest
\begin{align}
 \boxed{J(y) \underset{y\to \infty}{=} \frac{Q}{y} + O\left(\frac{1}{y^2}\right)}\ .
\label{jyi}
\end{align}
For any meromorphic function $\epsilon(y)$, with no poles outside $\{z_1,\cdots z_N\}$, we have 
\begin{align}
 \oint_\infty dy\ \epsilon(y) \left\langle \left(J(y)-\frac{Q}{y}\right)\prod_{j=1}^N V_{\sigma_j}(z_j)\right\rangle = 0   \quad \text{provided} \quad \epsilon(y) \underset{y\to\infty}{=} O(1)\ .
\end{align}
In the case $\epsilon(y)=1$, we obtain the $\hat{\mathfrak{u}}_1$ global Ward identity.
Assuming all fields are primary or descendant fields, and therefore obey Eq. \eqref{jma}, the global Ward identity reads
\begin{align}
 \boxed{\sum_{i=1}^N \alpha_i = Q} \ ,
\label{saq}
\end{align}
In the case $\epsilon(z) = \frac{1}{(z-z_i)^{n}}$ with $n\geq 1$, we obtain the $\hat{\mathfrak{u}}_1$ local Ward identity
\begin{align}
\left\langle \left(J_{-n}^{(z_i)}+ (-1)^{n}\sum_{j\neq i}\sum_{p=0}^\infty \frac{\binom{p+n-1}{p}}{z_{ij}^{n+p}}  J_p^{(z_j)}\right)\prod_{j=1}^N V_{\sigma_j}(z_j) \right\rangle = 0\ .
\label{jnjp}
\end{align}
This simplifies when the fields with indices $j\neq i$ are primary, 
\begin{align}
 \left\langle \left(J_{-n}^{(z_i)}  + \sum_{j\neq i} \frac{\alpha_j}{z_{ji}^n}\right) V_{\sigma_i}(z_i)\prod_{j\neq i} V_{\alpha_j}(z_j)\right\rangle= 0\ .
\label{jnz}
\end{align}
If all fields are primary, we actually have 
\begin{align}
 \boxed{\left\langle J(y) \prod_{i=1}^N V_{\alpha_i}(z_i) \right\rangle = \sum_{i=1}^N \frac{\alpha_i}{y-z_i} \left\langle \prod_{i=1}^N V_{\alpha_i}(z_i) \right\rangle  }\ .
\label{jsa}
\end{align}

\subsubsection{OPEs and fusion rule}

Since $\hat{\mathfrak{u}}_1$-primary fields are also Virasoro-primary, the OPE of two left and right 
$\hat{\mathfrak{u}}_1$-primary fields must be of the type 
\begin{align}
 V_{\alpha_1,\bar\alpha_1}(z_1) V_{\alpha_2,\bar\alpha_2}(z_2) = \sum_{\alpha_3,\bar\alpha_3} C_{12}^3 \left| z_{12}^{\Delta(\alpha_3)-\Delta(\alpha_1)-\Delta(\alpha_2)}\right|^2 \Big(V_{\alpha_3,\bar\alpha_3}(z_2) + O(z_{12}) \Big)\ .
\end{align}
Inserting $\oint_\infty J(z)dz$ on both sides of this OPE, we obtain a factor $\alpha_1+\alpha_2$ on the left, and a factor $\alpha_3$ on the right.
This leads to $(\alpha_3-\alpha_1-\alpha_2)C_{12}^3=0$, so that $V_{\alpha_3,\bar\alpha_3}$ can appear only if $\alpha_3=\alpha_1+\alpha_2$. 
In other words, we have the $\hat{\mathfrak{u}}_1$ fusion rule
\begin{align}
 \boxed{\mathcal{U}_{\alpha_1}\times \mathcal{U}_{\alpha_2} = \mathcal{U}_{\alpha_1+\alpha_2}}\ ,
\label{vvp}
\end{align}
i.e. the $\hat{\mathfrak{u}}_1$ momentum $\alpha$ is conserved.
The OPE then reads 
\begin{align}
\boxed{V_{\alpha_1,\bar\alpha_1}(z_1)V_{\alpha_2,\bar\alpha_2}(z_2) 
= 
\left| z_{12}^{-2\alpha_1\alpha_2}\right|^2 \big( V_{\alpha_1+\alpha_2,\bar\alpha_1+\bar\alpha_2}(z_2) + O(z_{12})\big)}\ ,
\label{vvoo}
\end{align}
where we have normalized the fields so that the OPE coefficient is one. (See Exercise \ref{exoone}.)
It is actually possible to write all the descendants explicitly,
\begin{align}
 V_{\alpha_1,\bar\alpha_1}(z_1)V_{\alpha_2,\bar\alpha_2}(z_2) 
= 
\left| z_{12}^{-2\alpha_1\alpha_2} \exp\left(-\sum_{p=1}^\infty \frac{2}{p}\alpha_1 z_{12}^pJ_{-p} \right)\right|^2 V_{\alpha_1+\alpha_2,\bar\alpha_1+\bar\alpha_2}(z_2) \ .
\end{align}

\subsubsection{Differential equations}

The conformal symmetry equations \eqref{spz} of course still hold in theories with the $\hat{\mathfrak{u}}_1$ symmetry algebra.
But the construction \eqref{tqz} of $T(z)$ from the $\hat{\mathfrak{u}}_1$ current $J(z)$ leads to additional differential equations.
In order to derive them, let us consider 
\begin{align}
 {\frac{\partial}{\partial z}} V_{\alpha}(z) = L_{-1}V_\alpha(z) = -2J_{-1}J_0 V_\alpha(z)\ ,
\end{align}
where we used Eq. \eqref{lnj} for expressing $L_{-1}$ in terms of $J_{-1}$ and $J_0$.
We therefore obtain
\begin{align}
{\frac{\partial}{\partial z}} V_{\alpha}(z)&= -2\alpha J_{-1}V_\alpha(z) = -2\alpha(JV_\alpha)(z)\ .
\label{pvaj}
\end{align}
Applying this to a field $V_{\alpha_i}(z_i)$ in an 
$N$-point function of $\hat{\mathfrak{u}}_1$-primary fields, and using 
the local Ward identity \eqref{jnz} for $J_{-1}^{(z_i)}$, we obtain 
\begin{align}
\left( {\frac{\partial}{\partial z_i}} +\sum_{j\neq i} \frac{2\alpha_i\alpha_j}{z_i-z_j} \right) \left\langle \prod_{i=1}^N V_{\alpha_i}(z_i) \right\rangle = 0 \ .
\label{kzl}
\end{align}
This implies 
\begin{align}
 \boxed{\left\langle \prod_{i=1}^N V_{\alpha_i}(z_i)\right\rangle  \propto \prod_{i<j} (z_i-z_j)^{-2\alpha_i\alpha_j}}\ ,
\label{pzz}
\end{align}
where the proportionality factor is an arbitrary antiholomorphic function of the positions.
In particular, the left and right $\hat{\mathfrak{u}}_1$-primary field $V_{0,0}(z)$ must be $z$-independent.
In the normalization of Eq. \eqref{vvoo}, $V_{0,0}(z)$ is actually the identity field.

\subsubsection{Seiberg--Witten equations}

Let us use Eq. \eqref{pvaj} to find out how primary fields depend on $\hat{\mathfrak{u}}_1$ momentums. 
Since conformal spins take integer values, we cannot vary left and right momentums independently, and we will consider a diagonal primary field $V_{\alpha,\alpha}(z)$. 
If $\nabla_z={\frac{\partial}{\partial z}} +2\alpha J(z)$ and $\overline{\nabla}_{z}={\frac{\partial}{\partial \bar z}} +2\alpha \bar J(z)$, then $\nabla_z V_{\alpha,\alpha}(z)=\overline{\nabla}_z V_{\alpha,\alpha}(z)=0$, neglecting the issue of regularizing operator products.
If we then define $\nabla_\alpha = {\frac{\partial}{\partial \alpha}}+2\int^z (J(y)dy +\bar J(y)d\bar y)$ so that $[\nabla_z,\nabla_\alpha]=[\overline{\nabla}_z,\nabla_\alpha]=0$, then $\nabla_\alpha V_{\alpha,\alpha}(z)$ is annihilated by $\nabla_z,\overline{\nabla}_z$, and must be proportional to $V_{\alpha,\alpha}(z)$.
The proportionality constant can be set to zero by renormalizing the field $V_{\alpha,\alpha}(z)$, and this normalization coincides with the normalization that we adopted in Eq. \eqref{vvoo}.
We thus obtain
\begin{align}
 {\frac{\partial}{\partial \alpha}} V_{\alpha,\alpha}(z) 
 = -2 \int^z V_{\alpha,\alpha}(z)\Big(J(y)dy+\bar J(y)d\bar y \Big)\ ,
\label{swp}
\end{align}
where the integral denotes a regularized value of a primitive of the integrand. (The integrand is a closed one-form because $\bar\partial J = \partial \bar J=0$.)
It can be checked that this equation holds when inserted into correlation functions of primary fields, using Eqs. \eqref{pzz} and \eqref{jsa}, provided we use the primitive $\int^z \frac{dy}{y-z_i} = \log(z-z_i)$.
From the OPE \eqref{jva}, we also have the equation
\begin{align}
 \alpha V_{\alpha,\alpha}(z) 
 = \frac{1}{2\pi i} \oint_{z} V_{\alpha,\alpha}(z)J(y)dy
 = \frac{1}{2\pi i} \oint_{z} V_{\alpha,\alpha}(z)\bar J(y)d\bar y\ .
\label{swa}
\end{align}
Equations \eqref{swp} and \eqref{swa} show that the actions of ${\frac{\partial}{\partial \alpha}}$ and $\alpha$ on $V_{\alpha,\alpha}(z)$ can be obtained by integrating field-valued one-forms along particular contours.
By analogy with supersymmetric gauge theory, these equations may be called \textbf{\boldmath Seiberg--Witten equations}\index{Seiberg--Witten equation}.

\subsection{Spectrum and correlation functions \label{secsacf}}

We define a \textbf{\boldmath free bosonic theory}\index{free bosonic theory} as a conformal field theory with $\hat{\mathfrak{u}}_1$ symmetry, such that each representation of the $\hat{\mathfrak{u}}_1 \times \bar{\hat{\mathfrak{u}}}_1$ symmetry algebra appears at most once in the spectrum. Let us parametrize highest-weight representations of $\hat{\mathfrak{u}}_1 \times \bar{\hat{\mathfrak{u}}}_1$ by their momentum vectors $\vec\alpha = (\alpha,\bar\alpha)\in\mathbb{C}^2$: by closure under fusion \eqref{vvp}, the set of the momentum vectors must close under addition. Moreover, we assume that the corresponding spins are integer,
\begin{align}
 S(\vec\alpha) = (\alpha-\bar\alpha)(\alpha+\bar\alpha-Q)\in\mathbb{Z}\ . 
\end{align}
Let us analyze the interplay between this integer spin condition and closure under fusion, depending on whether the background charge vanishes or not.

\subsubsection{Discrete spectrums with $Q=0$}

For $Q=0$, arbitrary integer linear combinations of two momentum vectors have integer spins if and only if the vectors and their sums have integer spins,
\begin{align}
 S\left(\mathbb{Z}\vec\alpha_1 + \mathbb{Z}\vec\alpha_2\right)\subset \mathbb{Z} \quad \iff \quad S\left(\left\{\vec\alpha_1,\vec\alpha_2,\vec\alpha_1+\vec\alpha_2\right\}\right)\subset \mathbb{Z}\ .
\end{align}
More generally, the elements of the spectrum $\sum_{i=1}^N \mathbb{Z}\vec \alpha_i$ have integer spins if and only if the vectors $\vec\alpha_i$ and $\vec\alpha_i+\vec\alpha_j$ (with $i\neq j$) have integer spins. This gives us $\frac12 N(N+1)$ constraints for the $2N$ complex components of the vectors $\vec\alpha_i$. We focus on spectrums where these constraints are underdetermined and leave us with at least one continuous parameter, i.e. $N=1,2$: we call these spectrums generic. We further focus on spectrums that cannot be extended by adding an extra vector, i.e. $N=2$: we call these spectrums generic and maximal.

A generic maximal spectrum may or may not contain diagonal states, see Exercise \ref{exofbd}. We focus on generic maximal spectrums that do, and we parametrize a diagonal generator as  $\vec\alpha_1=(\frac{i}{2R},\frac{i}{2R})$. For $\vec\alpha_2$ another generator, we have 
\begin{align}
 S(\vec\alpha_1 + \vec\alpha_2) - S(\vec\alpha_1)-S(\vec\alpha_2) = \frac{i}{R}(\alpha_2-\bar\alpha_2)\ .
\end{align}
This combination of spins must be integer, and since the spectrum is maximal we have $\alpha_2-\bar\alpha_2=iR$ (or equivalently $\alpha_2-\bar\alpha_2=-iR$). This implies $\alpha_2+\bar\alpha_2 \in \frac{i}{R}\mathbb{Z}$,
and by doing a redefinition of the type $\vec\alpha_2\to \vec\alpha_2+k\vec\alpha_1$ with $k\in\mathbb{Z}$ we set $\alpha_2+\bar\alpha_2=0$ and therefore $\vec\alpha_2=(\frac{iR}{2},-\frac{iR}{2})$. This leads to the spectrum
\begin{align}
 \boxed{\mathcal{S}_R = \bigoplus_{(n,w)\in {\mathbb{Z}^2}} \mathcal{U}_{\frac{i}{2}\left(\frac{n}{R} + Rw\right)} \otimes \bar{\mathcal{U}}_{\frac{i}{2}\left(\frac{n}{R} - Rw\right)} }\ .
\label{sr}
\end{align}
The corresponding theory is called the \textbf{\boldmath compactified free boson}\index{free boson!compactified---} theory, whose parameter $R$ is  the compactification radius.
The integers $n$ and $w$ are respectively called the momentum and winding number.
Free bosons are essential building blocks of string theory, which is at the origin of the names for $R$, $n$ and $w$.

The  algebra $\hat{\mathfrak{u}}_1$ has the automorphism
\begin{align}
 \omega(J_n) = -J_n \ ,
\end{align}
which preserves the Virasoro generators \eqref{lnj}-\eqref{lzj}. (For $Q\neq 0$ we would have to assume $\omega(Q)=-Q$.)
This automorphism acts on representations of $\hat{\mathfrak{u}}_1 \times \bar{\hat{\mathfrak{u}}}_1$ as $\omega(\mathcal{U}_\alpha \otimes \bar{\mathcal{U}}_{\bar{\alpha}}) =  \mathcal{U}_{-\alpha} \otimes \bar{\mathcal{U}}_{\bar{\alpha}}$, and on the spectrum of a compactified free boson as 
\begin{align}
 \omega(\mathcal{S}_R) = \mathcal{S}_{\frac{1}{R}}\ .
\end{align}
This shows that two compactified free bosons are equivalent if their radiuses are inverses of one another. In string theory, this is called T-duality.

\subsubsection{Discrete spectrums with $Q\neq 0$}

For $Q\neq 0$, $S(\vec\alpha)\in\mathbb{Z}\centernot\implies S(2\vec\alpha)\in\mathbb{Z}$, due to the identity
\begin{align}
 S(2\vec\alpha) - 4 S(\vec\alpha) = 2Q(\alpha-\bar \alpha)\ .
\end{align}
Looking again for a generic maximal spectrum of the type $\mathbb{Z}\vec\alpha_1 + \mathbb{Z}\vec\alpha_2$, we must have 
\begin{align}
 \alpha_i-\bar\alpha_i\in \frac{1}{2Q}\mathbb{Z}\ .
 \label{amba}
\end{align}
This implies that the spectrum contains diagonal states. And actually, the compactified free boson spectrum $\mathcal{S}_R$ \eqref{sr} is made of states with integer spins provided
\begin{align}
 R \in \frac{1}{iQ}\mathbb{Z}\ .
 \label{riqz}
\end{align}
In this spectrum however, we have $\alpha-\bar\alpha\in\frac{1}{Q}\mathbb{Z}$. 
If we insisted that the constraint \eqref{amba} be saturated, then we would find the alternative spectrum
\begin{align}
 \mathcal{S}'_R = \bigoplus_{n\in \mathbb{Z}} \bigoplus_{w\in \frac{n}{2}+\mathbb{Z}} \mathcal{U}_{\frac{i}{2}\left(\frac{n}{R} + Rw\right)} \otimes \bar{\mathcal{U}}_{\frac{i}{2}\left(\frac{n}{R} - Rw\right)} \qquad \text{with}\qquad R \in \frac{1}{iQ}(2\mathbb{Z}+1)\ .
\end{align}

\subsubsection{Continuous spectrums}

The large $R$ limit of the compactified free boson spectrum \eqref{sr} is a continuous, diagonal spectrum.
If $Q\neq 0$, the compactification radius $R$ must obey the condition \eqref{riqz}, and we can send it to infinity in a particular, $Q$-dependent direction. We find the spectrum 
\begin{align}
 \boxed{ \mathcal{S} = \int_{Q\mathbb{R}} d\alpha\ \mathcal{U}_\alpha\otimes \bar{\mathcal{U}}_\alpha}\ .
 \label{sc}
\end{align}
For $Q=0$ we can take $Q\mathbb{R}$ to be a straight line in an arbitrary direction, although the direction $\alpha\in i\mathbb{R}$ is singled out by unitarity.   
The resulting theory is called the (uncompactified) \textbf{\boldmath free boson}\index{free boson} theory.
In the context of string theory, this theory is also called the linear dilaton theory if $Q\neq 0$.

\subsubsection{Unitarity}

The only unitary free bosonic theories are the compactified free boson at $Q=0$ with a real compactification radius, and the uncompactified free boson that is obtained as its infinite radius limit. For $Q\neq 0$, free bosonic theories cannot be unitary, because the set $U=\frac{Q}{2}+i\mathbb{R}$ of unitary $\hat{\mathfrak{u}}_1$ momentums is not closed under fusion, i.e. $(U+U)\cap U = \emptyset$ if $Q\in \mathbb{R}^*$.

\subsubsection{Correlation functions}

The correlation functions of our free bosonic theories with discrete spectrums have the form
\begin{align}
 \left\langle \prod_{i=1}^N V_{\alpha_i,\bar\alpha_i}(z_i) \right\rangle = 
 \delta_{\sum_i \alpha_i-Q} \delta_{\sum_i\bar\alpha_i-Q} 
 \prod_{i<j} (z_i-z_j)^{-2\alpha_i\alpha_j}(\bar{z}_i-\bar{z}_j)^{-2\bar{\alpha}_i\bar{\alpha}_j}\ , 
\end{align}
where the two Kronecker deltas constrain the two discrete coordinates of $(\sum_i\alpha_i,\sum_i\bar\alpha_i)$.
In theories with continuous, diagonal spectrums, $\hat{\mathfrak{u}}_1$ momentum conservation is enforced by a Dirac delta function,
\begin{align}
 \left\langle \prod_{i=1}^N V_{\alpha_i,\alpha_i}(z_i)\right\rangle = \delta\left({\textstyle \sum}_{i=1}^N\alpha_i-Q\right)\ \prod_{i<j} |z_i-z_j|^{-4\alpha_i\alpha_j} \ .
\label{dpzz}
\end{align}
In the application to Liouville theory (Section \ref{seclld}), we will actually use these correlation functions for $\hat{\mathfrak{u}}_1$ momentums that do not necessarily belong to the spectrum \eqref{sc}. 

While the structure constants are trivial if the spectrum only involves even spins, they are nontrivial functions with values in $\{-1,1\}$ in the presence of odd spins, due to the permutation rule \eqref{css}. In the case of a compactified free boson, the two- and three-point structure constants are 
\begin{align}
B_{(n_1,w_1)} = (-1)^{n_1w_1} \quad , \quad 
 C_{(n_1,w_1),(n_2,w_2),(n_3,w_3)} = (-1)^{n_1w_2 + n_2w_3+n_3w_1}\ .
 \label{csigns}
\end{align}
(The Kronecker delta functions that appear in the two- and three-point functions are universal quantities, so we do not include them in the structure constants.)

\subsection{Free bosons and Liouville theory \label{seclld}}

Let us invert the 
construction \eqref{tqz} of a Virasoro field from a $\hat{\mathfrak{u}}_1$ current, and construct fields $J,\bar J$ from the energy-momentum tensors $T,\bar T$ of an arbitrary conformal field theory. 
These fields are in general not meromorphic, so they are not symmetry fields.
Nevertheless, we will use them for computing limits and particular values of correlation functions in Liouville theory.

\subsubsection{The Liouville equation}

Equation \eqref{tqz} is not enough for determining $J$ from $T$. We could add the condition that $J$ be holomorphic, but we will add the milder condition $\partial \bar J = \bar\partial J$, 
% This condition has to be compatible with the relations of J, \bar J with T, \bar T. Two reasons why it is compatible: it is milder than J being holomorphic, which is itself compatible because T is holomorphic. And the relations with T, \bar T both lead to the same Liouville equation.
which is sufficient for the Seiberg--Witten equation \eqref{swp} to make sense. This implies that locally there is a field $\phi$ such that 
\begin{align}
 J = -\partial \phi \qquad , \qquad \bar J = -\bar\partial \phi \ .
\end{align}
This field is called the \textbf{Liouville field}\index{Liouville!---field}. By formally solving the Seiberg--Witten equation, we can express primary fields in terms of the Liouville field as 
\begin{align}
 V_\alpha = e^{2\alpha \phi} \ .
\end{align}
We are now labelling primary fields using the $\hat{\mathfrak{u}}_1$ momentum $\alpha$, which is related to the conformal dimension by Eq. \eqref{daqu}, and to the momentum $P$ by $\alpha =\frac{Q}{2}-P$.

Given an $N$-point function $\left\langle \prod_{i=1}^N V_{\alpha_i}(z_i) \right\rangle$ in Liouville theory, we define
\begin{align}
 \boxed{F= -\log \left\langle \prod_{i=1}^N V_{\alpha_i}(z_i) \right\rangle} \quad , \quad \boxed{W(y) = \frac{\left\langle J(y)\prod_{i=1}^N V_{\alpha_i}(z_i) \right\rangle}{\left\langle \prod_{i=1}^N V_{\alpha_i}(z_i) \right\rangle}} \ .
 \label{fwy}
\end{align}
In terms of these objects, the Seiberg--Witten equation \eqref{swp} becomes 
\begin{align}
 {\frac{\partial}{\partial \alpha_i}} F = 2 \int^{z_i} \Big(W(y)dy+\bar W(y)d\bar y\Big)\ . 
\label{daf}
\end{align}
We will use these objects for studying Liouville theory in the limit $c\to \infty$ -- equivalently, $Q\to \infty$.
Equations such as \eqref{jyi} suggest $J=O(Q)$ which leads to $W=O(Q)$.
Using Eq. \eqref{daf}, we deduce $F=O(Q^2)$.
The natural expansion parameter is $Q^2$ (as in $c=1+6Q^2$) and we write the expansions
\begin{align}
 F &= Q^2 F^{(0)} + F^{(1)} + O(Q^{-2})\ ,
\\
W & = QW^{(0)} + O(Q^{-1})\ .
\end{align}
Now, in the $Q\to \infty$ limit of the $JJ$ OPE \eqref{jj} (at fixed positions), the singular term becomes negligible. So, in this limit, the correlation function $ \left\langle J(y_1)J(y_2)\prod_{i=1}^N V_{\alpha_i}(z_i) \right\rangle$ has no singularity at $y_1=y_2$. If this correlation function was determined by the behaviour of $J$ near its singularities, it would have to factorize,
\begin{align}
 \frac{\left\langle J(y_1)J(y_2)\prod_{i=1}^N V_{\alpha_i}(z_i)\right\rangle}{\left\langle \prod_{i=1}^N V_{\alpha_i}(z_i) \right\rangle} \underset{Q\to \infty}{=} Q^2W^{(0)}(y_1)W^{(0)}(y_2) + O(1)\ .
 \label{lqfac}
\end{align}
However, since $J(y)$ is not a symmetry field, it is not necessarily determined by its singular behaviour. For the moment, we assume that the factorization holds.
Inserting the definition \eqref{tqz} of $J(z)$ in $\left\langle \prod_{i=1}^N V_{\alpha_i}(z_i) \right\rangle$ then yields a nonlinear differential equation for $W^{(0)}$,
\begin{align}
 \boxed{\left(W^{(0)}\right)^2 + \partial W^{(0)} = - t}\ ,
\label{wwwt}
\end{align}
where we introduced
\begin{align}
 t(z) = \underset{Q\to \infty}{\lim} Q^{-2} \frac{\left\langle T(z) \prod_{i=1}^N V_{\alpha_i}(z_i) \right\rangle}{\left\langle \prod_{i=1}^N V_{\alpha_i}(z_i) \right\rangle}\ .
\end{align}
This can be computed using Eq. \eqref{dtz}, 
\begin{align}
 t(z) = \sum_{i=1}^N \left( \frac{\underset{Q\to\infty}{\lim} Q^{-2}\Delta(\alpha_i)}{(z-z_i)^2} + \frac{\beta_i}{z-z_i} \right)\ ,
\label{tzs}
\end{align}
where we defined the accessory parameters
\begin{align}
 \beta_i = -{\frac{\partial}{\partial z_i}} F^{(0)}\ .
\end{align}
The condition $t(z) \underset{z\to \infty}{=} O(\frac{1}{z^4})$, which follows from Eq. \eqref{tyi}, amounts to three constraints on the accessory parameters, which are thereby uniquely determined in the case $N= 3$. Studying the behaviour of $t(z)\prod_{i=1}^3(z-z_i)$ near $z=z_i$ and $z=\infty$, we find in this case
\begin{align}
 t(z) \underset{N=3}{=} \frac{1}{\prod_{i=1}^3(z-z_i)} \sum_{i=1}^3 \frac{\prod_{j\neq i} z_{ij}}{z-z_i}\underset{Q\to\infty}{\lim} Q^{-2}\Delta(\alpha_i)\ .
\end{align}
Let us reformulate our nonlinear equation for $W^{(0)}$ in terms of the functions $\psi$ and $\phi^\text{cl}$ such that
\begin{align}
 W^{(0)} = \frac{\partial\psi}{\psi} = -\partial \phi^\text{cl} \qquad , \qquad \bar W^{(0)} = \frac{\bar\partial\psi}{\psi}= -\bar\partial \phi^\text{cl}\ .
\end{align}
In terms of $\psi$, we obtain the linear differential equation
\begin{align}
 \boxed{ (\partial^2 + t ) \psi = 0 } \ .
 \label{dtp}
\end{align}
In terms of $\phi^\text{cl}$, the $\bar{z}$-derivative of Eq. \eqref{wwwt} reads $\partial\left(\log \partial\bar{\partial}\phi^\text{cl} -2\phi^\text{cl}\right)=0$. Together with $\bar{\partial} \left(\log \partial\bar{\partial}\phi^\text{cl} -2\phi^\text{cl}\right)=0$, this leads to 
\begin{align}
 \partial\bar{\partial}\phi^\text{cl} = \mu e^{2\phi^\text{cl}}\ ,
 \label{phieq}
\end{align}
where the integration constant $\mu$ coincides with the cosmological constant up to a normalization factor.
This equation for $\phi^\text{cl}$ is the \textbf{\boldmath Liouville equation}\index{Liouville!---equation}, which gives its name to Liouville theory.
This equation is equivalent to the condition that the two-dimensional metric
\begin{align}
 ds^2 = e^{2\phi^\text{cl}} dz d\bar{z}\ ,
\end{align}
has a constant scalar curvature.
Since all two-dimensional metrics can be brought to this form by changes of coordinates, Liouville theory can be interpreted as a quantum theory of two-dimensional gravity, whose classical equation of motion is the Liouville equation.

\subsubsection{Heavy asymptotic limit}

In order to define and compute large $c$ limits of correlation functions, we should specify how the momentums behave.
Since $J=O(Q)$, it is natural to assume $\alpha_i=O(Q)$, i.e. to consider the 
\begin{align}
 \textbf{\boldmath heavy\ asymptotic\ limit}\index{heavy\ asymptotic\ limit}: \quad \left\{\begin{array}{l}  Q \to \infty \ ,\\ \eta_i=\frac{\alpha_i}{Q}\ \text{fixed}\ .\end{array}\right.  
\end{align}
We then have $\underset{Q\to\infty}{\lim} Q^{-2}\Delta(\alpha_i)=\eta_i(1-\eta_i)$, and Eq. \eqref{wwwt} allows two possible asymptotic behaviours $W^{(0)}(y)\underset{y\to z_1}{=} \frac{\eta_1}{y-z_1} + O(1)$ and 
$W^{(0)}(y)\underset{y\to z_1}{=} \frac{1-\eta_1}{y-z_1} + O(1)$.
In the case $N=3$, the linear differential equation \eqref{dtp} for $\psi$ is equivalent to the hypergeometric equation, which has a unique single-valued solution (up to an overall constant factor).
The uniqueness of the solution provides an a posteriori justification for our factorization hypothesis \eqref{lqfac}.
We can then deduce the leading term $F^{(0)}$ of $F$, with the help of the heavy asymptotic limit of the Seiberg--Witten equation \eqref{daf}, 
\begin{align}
 {\frac{\partial}{\partial \eta_i}} F^{(0)} = 2\int^{z_i}\Big( W^{(0)}(y)dy + \bar W^{(0)}(y)d\bar y\Big) = -2\phi^\text{cl}(z_i)\ .
\end{align}
We do not carry out these calculations here, and instead refer the reader to \cite{zz95}, where the result for $F^{(0)}$ is shown to agree with the leading behaviour of the three-point function \eqref{caaa} in the heavy asymptotic limit.
Actually, the agreement does not stop at the leading order: similar techniques can be used for computing  $F$ order by order in $Q^2$ \cite{cer12}.

\subsubsection{Light asymptotic limit}

Let us now assume that conformal dimensions are fixed as the central charge goes to infinity.
This defines the 
\begin{align}
 \textbf{\boldmath light\ asymptotic\ limit}\index{light\ asymptotic\ limit}: \quad \left\{\begin{array}{l}  Q \to \infty \ ,\\ \eta_i=Q\alpha_i\ \text{fixed}\ .\end{array}\right.  
\end{align}
 The Seiberg--Witten equation \eqref{daf} now implies 
\begin{align}
 {\frac{\partial}{\partial \eta_i}} F^{(0)}  = 0 \qquad ,\qquad 
{\frac{\partial}{\partial \eta_i}} F^{(1)} & = -2\phi^\text{cl}(z_i) \ .
\label{pefo}
\end{align}
So $F^{(0)}$ is $\eta_i$-independent.
It is actually also $z_i$-independent, according to the following argument: since $\Delta(\alpha_i)=\eta_i + O(Q^{-2})$, its is natural to assume $t(z)=0$ where $t(z)$ is given by Eq. \eqref{tzs}, and this implies $\beta_i=-{\frac{\partial}{\partial z_i}} F^{(0)}=0$.
So $F^{(0)}$ is not affected by the presence of the fields $V_{\alpha_i}(z_i)$, which are then called light fields.
The first interesting term of $F$ is thus $F^{(1)}$.
To compute $F^{(1)}$, let us solve the linear differential equation \eqref{dtp}, which is now simply $\partial^2 \psi=0$.
Together with $\bar\partial^2\psi=0$, this leads to the solutions 
$\psi_h(z) = \left[\begin{smallmatrix} z \\ 1 \end{smallmatrix}\right]^\dagger h \left[\begin{smallmatrix} z \\ 1 \end{smallmatrix}\right]$, for $h$ a constant matrix such that $h=h^\dagger$ and $\det h = -\mu$, where $\mu$ is the parameter of the Liouville equation.

So the solution for $\psi$ is not unique. And individual solutions are not covariant under global conformal transformations, although of course the space of solutions is invariant.
Different solutions $\psi_h$ correspond to different linear forms on tuples of fields $(V_{\alpha_i}(z_i)) \mapsto e^{-F_h} = \langle \prod_i V_{\alpha_i}(z_i)\rangle_h$. 
Liouville correlation functions correspond to a linear form that is covariant under global conformal transformations, and does not obey the factorization assumption \eqref{lqfac}. 
This form can be obtained as a linear combination of the linear forms $\langle \rangle_h$, so that $e^{-F^{(1)}} = \int dh\, e^{-F^{(1)}_h} = \int dh \prod_{i=1}^N \psi_h(z_i)^{-2\eta_i}$. 
More explicitly, in the case where the integration constant in Eq. \eqref{phieq} is $\mu=-1$, we have
\begin{align}
\underset{\text{light asymptotic}}{\lim}\ \left\langle\prod_{i=1}^N V_{\alpha_i}(z_i)\right\rangle= \int_{H^+_3} dh\ \prod_{i=1}^N \left( \left[\begin{smallmatrix} z_i \\ 1 \end{smallmatrix}\right]^\dagger h \left[\begin{smallmatrix} z_i \\ 1 \end{smallmatrix}\right] \right)^{-2\eta_i}\ ,
\label{zih}
\end{align}
where the integral is over the space \textbf{\boldmath $H_3^+$}\index{H3+@$H_3^+$ (space)} of Hermitian matrices of size two and determinant one, with a measure $dh$ that we assume to be invariant under $h\mapsto g^\dagger hg$ for $g\in SL_2({\mathbb{C}})$.
In the case $N=3$, this agrees with the DOZZ formula for the three-point function \eqref{caaa}. (See \cite{zz95}.) In the light asymptotic limit, the spectrum of Liouville theory reduces to the space of functions on $H_3^+$, and the symmetry algebra reduces to the algebra of the global conformal transformations.
So the light asymptotic limit of Liouville theory is a two-dimensional global conformal field theory, which also describes harmonic analysis on $H_3^+$. (See Exercise \ref{exolight} for more details.)

This relation between harmonic analysis and conformal field theory can be generalized to higher-dimensional CFT \cite{kks18} and to the light asymptotic limit of 2d CFTs with larger symmetry algebras \cite{fr11}. Moreover, using this relation, the conformal bootstrap method can be applied to geometrical problems. In particular, in harmonic analysis on compact cosets of $PSL_2(\mathbb{R})$, numerical bootstrap techniques lead to bounds on eigenvalues of the Laplace operator \cite{kmp21}.

\subsubsection{Coulomb gas integrals}

To conclude our discussion of the use of free boson techniques in Liouville theory, let us review
the relation between certain Liouville correlation functions, and free bosonic correlation functions \cite{zz95}.
The relevant Liouville correlation functions are $N$-point functions $\left\langle \prod_{i=1}^N V_{\alpha_i}(z_i)\right\rangle^{\text{Liouville}}$ at $\sum\alpha_i = Q-bm$ where $m\in \mathbb{N}$.  
For the Liouville three-point structure constants $C$ and $\hat C$, we have in this case
\begin{align}
 \lim_{\sum\alpha_i \to Q-bm} \hat C_{\alpha_1,\alpha_2,\alpha_3} 
 = \underset{\sum \alpha_i = Q-bm}{\operatorname{Res}}  C_{\alpha_1,\alpha_2,\alpha_3}\ .
\end{align}
This follows from Eq. \eqref{chco} in the case $m=0$, from which the cases $m\neq 0$ are obtained using the shift equations.
For $c\notin ]-\infty,1]$, our Liouville $N$-point function has a simple pole at $\sum\alpha_i = Q-bm$, with the residue
\begin{align}
 \underset{\sum \alpha_i = Q-bm}{\operatorname{Res}} \left\langle \prod_{i=1}^N V_{\alpha_i}(z_i)\right\rangle^{\text{Liouville}} = \underset{\sum\alpha_i = Q-bm}{\operatorname{Res}} \left\langle e^{-\frac{\mu^b}{\pi\gamma(b^2)}\int_{\mathbb{C}} V_b(z)d^2z}\prod_{i=1}^N V_{\alpha_i}(z_i)\right\rangle^{\text{Free boson}}\ ,
\label{lild}
\end{align}
where $\mu$ is the cosmological constant.
On the free boson side, taking the residue of a correlation function \eqref{dpzz} means stripping out a delta function, and 
\begin{multline}
 \underset{\sum\alpha_i = Q-bm}{\operatorname{Res}} \left\langle e^{-\frac{\mu^b}{\pi\gamma(b^2)}\int_{\mathbb{C}} V_b(z)d^2z}\prod_{i=1}^N V_{\alpha_i}(z_i)\right\rangle^{\text{Free boson}}
\\
= \frac{1}{m!}\left[-\frac{\mu^b}{\pi\gamma(b^2)}\right]^m \prod_{i<j} |z_{ij}|^{-4\alpha_i\alpha_j} \int_{{\mathbb{C}}^m} \prod_{k=1}^m d^2y_k\ \prod_{k,i} |y_k-z_i|^{-4b\alpha_i}\prod_{k<\ell} |y_{k\ell}|^{-4b^2}\ .
\label{mint}
\end{multline}
The relation \eqref{lild} implies that Liouville theory can be viewed as a perturbation of a free bosonic theory.
This is an example of \textbf{\boldmath conformal perturbation theory}\index{conformal perturbation theory} -- the definition and study of two-dimensional quantum field theories (conformal or not) as perturbed conformal field theories.
The perturbing operator, here $V_b$, is chosen so that its conformal dimension is one, a necessary but not sufficient condition for the perturbed theory to still have conformal symmetry.
This example is rather pathological, as Liouville theory does not have a smooth $\mu\to 0$ limit where we could recover a free bosonic theory, and we do not really compute Liouville theory correlation functions, but only their residues at certain poles. 

The integral \eqref{mint} is called a \textbf{\boldmath Coulomb gas integral}\index{Coulomb gas integral} or Dotsenko--Fateev integral. Such integrals can be used for computing correlation functions not only in Liouville theory, but also in other CFTs, including minimal models. 
Since the number of integrals to be performed depends on $\alpha_i$ and not just on $N$, Coulomb gas integrals  
are less appealing for numerical computations than using decompositions into conformal blocks, together with Zamolodchikov's recursion.
Coulomb gas integrals
coincide with partition functions of random matrix models \cite{ekr15}, and we have the following matrix model interpretation of certain CFT quantities:
\begin{align}
\renewcommand{\arraystretch}{1.3}
 \begin{tabular}{|l|l|}
  \hline
CFT & Random matrix model
\\
\hline \hline
$N$-point function & partition function 
\\
\hline
$F, W(y)$ in Eq. \eqref{fwy}  & free energy, resolvent
\\
\hline
Eq. \eqref{wwwt} for $W^{(0)}(y)$ & loop equation 
\\
\hline $m$ in Eq. \eqref{lild} & matrix size
\\
\hline
$y_k$ in Eq. \eqref{mint} & matrix eigenvalue
\\
\hline
 \end{tabular}
\end{align}

\section{Nonabelian affine symmetry}

We will now build an affine Lie algebra $\hat{\mathfrak{g}}$ from a simple Lie algebra $\mathfrak{g}$, in the same way as we built the  affine Lie algebra $\hat{\mathfrak{u}}_1$ from the abelian Lie algebra $\mathfrak{u}_1$.
We will use the smallest simple Lie algebra $\mathfrak{g}=\mathfrak{sl}_2$ as the main example. 

Affine Lie algebras are special cases of Kac--Moody algebras, which are themselves special cases of Lie algebras. For more details on these algebras, their representations, and their classification using Cartan matrices, see \cite{fuc97}.

\subsection{Symmetry algebra}

\subsubsection{Reminders on Lie algebras}

A Lie algebra $\mathfrak{g}$ is defined by generators $t^a$ and commutation relations 
\begin{align}
 [t^a,t^b] = f^{ab}_c t^c \ ,
\label{ttft}
\end{align}
where the structure constants $f^{ab}_c$ are numbers.
The commutation relations are assumed to obey two axioms:
\begin{align}
 \text{antisymmetry} & : \quad [t^a,t^b] + [t^b,t^a] = 0\ ,
\\
\text{Jacobi identities} & : \quad  [t^a,[t^b,t^c]] + [t^b,[t^c,t^a]] + [t^c,[t^a,t^b]] = 0 \ .
\end{align}
For example, we already encountered the Lie algebra $\mathfrak{sl}_2$ \eqref{ttpm} in our study of global conformal symmetry.
Let us define the \textbf{\boldmath Killing form}\index{Killing form} 
\begin{align}
 K^{ab} = \frac{1}{2g} \operatorname{Tr} \left(\operatorname{ad}_{t^a}\operatorname{ad}_{t^b}\right) =\frac{1}{2g} f^{ac}_d f^{bd}_c \overset{\mathfrak{sl}_n}{=}\operatorname{Tr}_f t^at^b\ ,
\end{align}
where  $\operatorname{ad}_t(t') = [t,t']$ is the adjoint action, $f$ is the $n$-dimensional fundamental representation of $\mathfrak{sl}_n$, and $g$ is the dual Coxeter number of the Lie algebra $\mathfrak{g}$, in particular $g\overset{\mathfrak{sl}_n}{=}n$.
The normalization factor $\frac{1}{2g}$ is included so that the level, which we will shortly define in Eq. \eqref{jajb}, takes integer values in rational models.
We assume that the Killing form is non-degenerate, which is equivalent to $\mathfrak{g}$ being semi-simple. 
Using the Killing form to raise indices, we define
\begin{align}
 f^{abc} = K^{ad}f_d^{bc} = \frac{1}{2g}\operatorname{Tr}\left( \operatorname{ad}_{t^a} [\operatorname{ad}_{t^b}, \operatorname{ad}_{t^c}] \right)\ ,
\end{align}
whose second expression, which is manifestly totally antisymmetric, follows from the Jacobi identities.
The total antisymmetry of $f^{abc}$ implies that the \textbf{\boldmath quadratic Casimir operator}\index{quadratic Casimir operator}
\begin{align}
 C_2 = K_{ab} t^a t^b \overset{\mathfrak{sl}_2}{=} 2t^0t^0+ t^+t^-+t^-t^+\ ,
\label{ctk}
\end{align}
is a central element of the universal enveloping algebra $U(\mathfrak{g})$, that is $[C_2,t^a]=0$.

\subsubsection{Currents and Sugawara construction}

Let us build conformal field theories based on a Lie algebra $\mathfrak{g}$. 
We introduce a number $\dim \mathfrak{g}$ of holomorphic \textbf{\boldmath $\hat{\mathfrak{g}}$ currents}\index{current!g@$\hat{\mathfrak{g}}$---} $J^a(y)$ characterized by the OPEs 
\begin{align}
 \boxed{ J^a(y) J^b(z) = \frac{k K^{ab}}{(y-z)^2} +  \frac{ f^{ab}_c J^c(z)}{y-z}  + O(1)} \ ,
\label{jajb}
\end{align}
where the parameter $k$ is called the \textbf{level}\index{level (of an affine Lie algebra)}. (This parameter did not appear in the OPE \eqref{jj} of the $\hat{\mathfrak{u}}_1$ current $J$ with itself, as it could be absorbed in a rescaling of $J$.) For example, the OPEs of $\widehat{\mathfrak{sl}}_2$ currents are
\begin{align}
\begin{array}{ll}
  J^0(y)J^0(z) = \frac{\frac{k}{2}}{(y-z)^2} + O(1)\ ,  & J^0(y)J^\pm(z) = \frac{\pm J^\pm(z)}{y-z} + O(1)\ ,
\\
 J^\pm(y)J^\pm(z) = O(1) \ , & J^+(y)J^-(z) = \frac{k}{(y-z)^2} + \frac{2J^0(z)}{y-z} + O(1)\ .
\end{array}
\label{jjjj}
\end{align}
Let us now introduce the \textbf{\boldmath Sugawara construction}\index{Sugawara construction} of a field $T$ as a quadratic combination of $\hat{\mathfrak{g}}$ currents,
\begin{align}
\boxed{ T =  \frac{ K_{ab} (J^aJ^b)}{2(k+g)} } \ .
\label{tjj} 
\end{align}
Then $T$ is a Virasoro field with the central charge 
\begin{align}
 \boxed{ c = \frac{ k \dim \mathfrak{g}}{k+g} }\ ,
\label{ckg}
\end{align}
such that $J^a$ is a primary field of conformal dimension one, 
\begin{align}
\boxed{ T(y)J^a(z) = {\frac{\partial}{\partial z}} \frac{1}{y-z} J^a(z) + O(1)} \ .
\label{tja}
\end{align}
(See Exercise \ref{exosug}.)
We identify $T$ with the energy-momentum tensor.
(See Exercise \ref{exotqpj} for the generalized energy-momentum tensor $\hat{T} = T + Q_a\partial J^a$.)

\subsubsection{Affine Lie algebra}

We do not repeat the definition \eqref{jvn} of the modes $J^a_n$ of $J^a$.
These modes obey the commutation relations
\begin{align}
 \boxed{[J^a_m,J^b_n] =   f^{ab}_c J^c_{m+n} +kmK^{ab}\ \delta_{m+n,0}} \ , 
\label{jam}
\end{align}
which define the \textbf{\boldmath affine Lie algebra}\index{affine Lie algebra} $\hat{\mathfrak{g}}$. 
The OPE \eqref{tja} is equivalent to
\begin{align}
 [L_m,J^a_n] = -nJ^a_{m+n}\ ,
\end{align}
and the Sugawara construction \eqref{tjj} is equivalent to 
\begin{align}
 L_n &= \frac{K_{ab}}{2(k+g)} \sum_{m\in{\mathbb{Z}}} J^a_{n-m}J^b_m\ , \qquad (n\neq 0)\ ,
\label{ljj}
\\
L_0 & = \frac{K_{ab}}{2(k+g)}\left(2\sum_{m=1}^\infty J^a_{-m}J^b_m + J^a_0J^b_0\right)\ .
\label{lzjj}
\end{align}

\subsection{Fields, representations and fusion rules}
 
\subsubsection{Affine primary fields} 
 
Given a representation $\mathcal{R}$ of $\mathfrak{g}$, we define an \textbf{\boldmath affine primary field}\index{primary field!affine---} $\Phi^{\mathcal{R}}(z_0)$ by its OPE with $J^a(y)$,
\begin{align}
 \boxed{ J^a(y) \Phi^{\mathcal{R}}(z_0) = \frac{-t^a\Phi^{\mathcal{R}}(z_0)}{y-z_0} + O(1) } \ .
\label{jpr}
\end{align}
This assumes that for any $z_0$, the field $\Phi^{\mathcal{R}}$ is a vector in the representation $\mathcal{R}$. Given a basis $(e^i)$ of $\mathcal{R}$, the field $\Phi^{\mathcal{R}} = \Phi^{\mathcal{R}}_ie^i$ is a collection of scalar fields $\Phi^{\mathcal{R}}_i$, such that $J^a(y)\Phi^{\mathcal{R}}_i(z_0) = \frac{-(t^a)^j_i\Phi^{\mathcal{R}}(z_0)_j}{y-z_0}+O(1)$. This OPE involves the transpose $(t^a)^T$ of a Lie algebra generator. Transposition flips the sign of commutation relations, which is why we have a minus sign in Eq. \eqref{jpr}. This minus sign is needed for 
the associativity of the $J^aJ^b\Phi^\mathcal{R}$ OPE. (See Exercise \ref{exojjp}.)

Let us
compute the OPE $T(y)\Phi^{\mathcal{R}}(z_0)$, by applying Wick's theorem to $\cunderbracket{\Phi^\mathcal{R}}{(z_0)}{(J^aJ_a)}(y)$, using Eq. \eqref{jpr} in the form $\cunderbracket{\Phi^\mathcal{R}}{(z_0)}{J^a}(x) = \frac{-t^a\Phi^\mathcal{R}(x)}{x-z_0}$.
We find 
\begin{align}
 T(y)\Phi^{\mathcal{R}}(z_0) = \frac{K_{ab}t^a}{k+g}\left(\frac{\frac12 t^b\Phi^{\mathcal{R}}(z_0)}{(y-z_0)^2} -\frac{(J^b\Phi^{\mathcal{R}})(z_0)}{y-z_0}\right) + O(1)\ . 
\end{align}
Because $T$ is the energy-momentum tensor and obeys Axiom \ref{ax:dvz}, this determines the derivative of $\Phi^\mathcal{R}$ in terms of the action of currents,
\begin{align}
\partial\Phi^\mathcal{R} = -\frac{K_{ab}t^a(J^b\Phi^\mathcal{R})}{k+g} \ .
\label{lmp}
\end{align}
Then, we see that our $T\Phi^{\mathcal{R}}$ OPE is 
of the form \eqref{tvp}, which shows that $\Phi^{\mathcal{R}}$ is a primary field with the conformal dimension
\begin{align}
 \boxed{\Delta_\mathcal{R}  = \frac{C_2(\mathcal{R})}{2(k+g)}}\ ,
\label{dr}
\end{align}
where $C_2(\mathcal{R})$ is the eigenvalue of the quadratic Casimir operator $C_2$ \eqref{ctk} when acting in the representation $\mathcal{R}$. (We are assuming that $\mathcal{R}$ is indecomposable, so that it contains only one eigenspace of $C_2$.) 

Let us now consider the states $|v^\mathcal{R}\rangle$ that correspond to the $\mathcal{R}$-valued field $\Phi^\mathcal{R}(z_0)$ by the state-field correspondence.
The definition \eqref{jpr} of $\Phi^\mathcal{R}(z_0)$ is equivalent to 
\begin{align}
\renewcommand{\arraystretch}{1.3}
 \left\{\begin{array}{l}  J^a_{n>0}|v^\mathcal{R}\rangle = 0 \ ,  \\ J^a_0|v^\mathcal{R}\rangle = -t^a |v^\mathcal{R}\rangle\ . \end{array}\right. 
\end{align}
The states $|v^\mathcal{R}\rangle$ are killed by the annihilation operators, and are called \textbf{\boldmath affine primary states}\index{primary state!affine---}.
They transform in the representation $\mathcal{R}$ of the \textbf{\boldmath horizontal subalgebra}\index{horizontal subalgebra} $\mathfrak{g}\subset \hat{\mathfrak{g}}$ generated by $\{J^a_0\}$.
Acting on the affine primary states with creation operators $J^a_{n<0}$ generates affine descendant states, which form an \textbf{\boldmath affine highest-weight representation}\index{highest-weight representation!affine---} $\hat{\mathcal{R}}$ of the affine Lie algebra $\hat{\mathfrak{g}}$.

\subsubsection{Isospin variables}

Rather than introducing a discrete basis of the representation $\mathcal{R}$, it is often convenient to represent states in $\mathcal{R}$ 
as functions of an \textbf{\boldmath isospin variable}\index{isospin variable} $X$. 
Affine primary fields become functions $\Phi^\mathcal{R}_X(z_0)$ of $X$, on which $t^a$ acts as a differential operator $D_X^\mathcal{R}(t^a)$, 
and we have
\begin{align}
 J^a(y)\Phi^\mathcal{R}_X(z_0) = \frac{-D^\mathcal{R}_X(t^a)\Phi^\mathcal{R}_X(z_0)}{y-z_0}+O(1)\quad \text{with} \quad t^a \Phi^\mathcal{R}_X(z_0) = D_X^\mathcal{R}(t^a) \Phi^\mathcal{R}_X(z_0)\ .
\label{jprx}
\end{align}
The differential operators $D_X^\mathcal{R}(t^a)$ obey the commutation relations of $\mathfrak{g}$, and their quadratic invariant combination is $C_2(\mathcal{R})$,
\begin{align}
[D_X^\mathcal{R}(t^a),D_X^\mathcal{R}(t^b)] = f^{ab}_c D_X^\mathcal{R}(t^c)\ , \qquad K_{ab} D_X^\mathcal{R}(t^a)D_X^\mathcal{R}(t^b) = C_2(\mathcal{R})\ .
\label{dta}
\end{align}
The isospin variable $X$ is a collection of as many complex variables as $\mathfrak{g}$ has positive roots.
In particular, an $\mathfrak{sl}_n$ isospin variable has $\frac{n(n-1)}{2}$ components, and an $\mathfrak{sl}_2$ isospin variable has one component.
We already know differential operators $D^j_x(t^a)$ \eqref{ddz} that 
obey the $\mathfrak{sl}_2$ commutation relations \eqref{ttpm} for any choice of the 
\textbf{\boldmath spin}\index{spin (sl2)@spin ($\mathfrak{sl}_2$)} $j$.
The eigenvalue of the quadratic Casimir operator in the corresponding representation of $\mathfrak{sl}_2$ is
\begin{align}
 C_2(j) =  K_{ab}D_x^j(t^a)D_x^j(t^b) = 2j(j+1)\ ,
\end{align}
and the conformal dimension of the corresponding field is
\begin{align}
 \Delta_j =\frac{j(j+1)}{k+2}\ .
\label{dj}
\end{align}
Another triple of $\mathfrak{sl}_2$ differential operators is given by 
\begin{align}
\renewcommand{\arraystretch}{1.3}
\left\{\begin{array}{l}  
 D_\mu^j(t^-) = -\mu \ , \\  D_\mu^j(t^0) = -\mu{\frac{\partial}{\partial \mu}} \ , \\ D_\mu^j(t^+) = \mu\frac{\partial^2}{\partial\mu^2} -\frac{j(j+1)}{\mu}\ . \end{array}\right. 
\label{mub}
\end{align}
The \textbf{\boldmath $x$-basis field}\index{x-basis field@$x$-basis field} $\Phi^j_x(z_0)$ and \textbf{\boldmath $\mu$-basis field}\index{mu-basis field@$\mu$-basis field} $\Phi^j_\mu(z_0)$ are related by the formal Fourier transform
\begin{align}
 \Phi_x^j(z_0) = \int d\mu\ \mu^{-j-1}e^{\mu x} \Phi_\mu^j(z_0)\ .
\label{emx}
\end{align}
The $\widehat{\mathfrak{sl}}_2$ currents and $\mu$-basis fields can be represented in terms of free fields.
This \textbf{Wakimoto free-field representation}\index{Wakimoto free-field representation} of $\widehat{\mathfrak{sl}}_2$ is described in Exercise \ref{exowaki}.

\subsubsection{OPEs and fusion rules}

Let us study the fusion rules of affine highest-weight representations, equivalently the OPEs of affine primary fields.
To write such OPEs, we will omit the dependences on the positions $z_i$ of the fields, which are dictated by conformal symmetry since affine primary fields are also primary fields.
We will also omit the affine descendant fields because, as in the cases of the Virasoro and $\hat{\mathfrak{u}}_1$ algebras, the contributions of affine descendant fields are determined by the contributions of the affine primary fields. (This would however not be true in the case of a $W$-algebra.) So we write a generic OPE as 
\begin{align}
 \Phi^{\mathcal{R}_1}_{X_1}\Phi^{\mathcal{R}_2}_{X_2} \sim \sum_{\mathcal{R}_3} \int dX_3\ C^{\mathcal{R}_1,\mathcal{R}_2}_{\mathcal{R}_3}(X_1,X_2|X_3) \Phi^{\mathcal{R}_3}_{X_3}\ .
\end{align}
Inserting $\oint dz\, J^a(z)$ on both sides, and using the linear independence of the operators $\Phi^{\mathcal{R}_3}_{X_3}$, we obtain an equation for the structure function $C^{\mathcal{R}_1,\mathcal{R}_2}_{\mathcal{R}_3}(X_1,X_2|X_3)$,
\begin{align}
 \left(D_{X_1}^{\mathcal{R}_1}(t^a)+D_{X_2}^{\mathcal{R}_2}(t^a)-\left(D_{X_3}^{\mathcal{R}_3}(t^a)\right)^\dagger\right) C^{\mathcal{R}_1,\mathcal{R}_2}_{\mathcal{R}_3}(X_1,X_2|X_3) = 0\ ,
\label{dddc}
\end{align}
where the dagger denotes the Hermitian conjugate for $X_3$-differential operators, such that for any functions $f,g$ we have 
$\int fDg =\int g D^\dagger f$.
This equation characterizes $C^{\mathcal{R}_1,\mathcal{R}_2}_{\mathcal{R}_3}(X_1,X_2|X_3)$ as an intertwiner between the representations $\mathcal{R}_1\otimes \mathcal{R}_2$ and $\mathcal{R}_3$ of the Lie algebra $\mathfrak{g}$.
This shows that the fusion multiplicities of affine highest-weight representations of $\hat{\mathfrak{g}}$ are bounded by the tensor product multiplicities of the underlying representations of $\mathfrak{g}$, 
\begin{align}
 m_{\hat{\mathcal{R}}_1,\hat{\mathcal{R}}_2}^{\hat{\mathcal{R}}_3} \leq m_{\mathcal{R}_1,\mathcal{R}_2}^{\mathcal{R}_3}\ .
\end{align}
The presence of null vectors in the representations $\hat{\mathcal{R}}_i$ can lead to extra conditions on the structure function $C^{\mathcal{R}_1,\mathcal{R}_2}_{\mathcal{R}_3}(X_1,X_2|X_3)$, in which case $m_{\hat{\mathcal{R}}_1,\hat{\mathcal{R}}_2}^{\hat{\mathcal{R}}_3} < m_{\mathcal{R}_1,\mathcal{R}_2}^{\mathcal{R}_3}$. 
And nothing guarantees that only highest-weight representations appear in the fusion product of two highest-weight representations. 
In the case of the $\widetilde{SL}_2(\mathbb{R})$ WZW model, other types of representations do appear. (See Section \ref{secslr}.) 

For generic representations $\mathcal{R}_i$, the multiplicity $m_{\mathcal{R}_1,\mathcal{R}_2}^{\mathcal{R}_3}$ is the number of linearly independent solutions of Eq. \eqref{dddc} in the absence of further constraints, namely
\begin{align}
 m_{\mathrm{max}} = \left\{\begin{array}{l}  2 \quad \text{if}\ \mathfrak{g}=\mathfrak{sl}_2\ , \\ \infty \quad \text{if}\ \mathfrak{g}=\mathfrak{sl}_{n\geq 3}\ . \end{array}\right. 
\end{align}
We indeed have a set of $\dim \mathfrak{sl}_n=n^2-1$ equations, for a function $C^{\mathcal{R}_1,\mathcal{R}_2}_{\mathcal{R}_3}(X_1,X_2|X_3)$ of $3\frac{n(n-1)}{2}$ variables -- the components of $X_1,X_2$ and $X_3$.
If $n=2$ there are three equations and three variables, and this can actually be reduced to one second-order differential equation for a function of one variable. (See Eq. \eqref{pmf}.)
If $n\geq 3$ there are more variables than equations, so that $C^{\mathcal{R}_1,\mathcal{R}_2}_{\mathcal{R}_3}(X_1,X_2|X_3)$ has an arbitrary dependence on a number of variables.
In the case of finite-dimensional representations of $\mathfrak{sl}_{n\geq 3}$, the multiplicities are of course finite, but they can take arbitrarily high values, depending on the involved representations.

\subsection{Ward identities and Knizhnik--Zamolodchikov equations \label{secwikz}}

\subsubsection{Ward identities}

The Sugawara construction of the energy-momentum tensor $T(y)$ from the currents $J^a(y)$, and the behaviour \eqref{tyi} of $T(y)$ near $y=\infty$, suggest
\begin{align}
 \boxed{J^a(y) \underset{y\to \infty}{=} O\left(\frac{1}{y^2}\right)}\ .
\label{jayi}
\end{align}
Given $N$ fields $\Phi^{\sigma_i}(z_i)$, and a meromorphic function $\epsilon(z)$, with no poles outside $\{z_1,\cdots z_N\}$, we therefore have 
\begin{align}
 \oint_\infty dy\ \epsilon(y) \left\langle J^a(y)  \prod_{i=1}^N \Phi^{\sigma_i}(z_i)\right\rangle = 0 \quad \text{provided} \quad \epsilon(y) \underset{y\to\infty}{=} O(1)\ ,
\end{align}
In the case $\epsilon(y)=\frac{1}{(y-z_i)^n}$ with $n\geq 1$, we obtain the $\hat{\mathfrak{g}}$ local Ward identities, which are formally identical to the $\hat{\mathfrak{u}}_1$ local Ward identities \eqref{jnjp}. 
In particular, if all fields are affine primaries except possibly the field at $z_i$, we find 
\begin{align}
\left\langle J^a_{-n}\Phi^{\sigma_i}(z_i)\prod_{j\neq i} \Phi^{\mathcal{R}_j}_{X_j}(z_j)\right\rangle &=\sum_{j\neq i} \frac{D_{X_j}^{\mathcal{R}_j}(t^a)}{(z_j-z_i)^n} \left\langle \Phi^{\sigma_i}(z_i)\prod_{j\neq i} \Phi^{\mathcal{R}_j}_{X_j}(z_j)\right\rangle\ . 
\label{jmnz}
\end{align}
In the case $\epsilon(y)=1$, we obtain the $\hat{\mathfrak{g}}$ global Ward identities,
\begin{align}
 \left\langle \sum_{i=1}^N (J_0^a)^{(z_i)} \prod_{i=1}^N \Phi^{\sigma_i}(z_i)\right\rangle=0\ .
\end{align}
Let us specialize to correlation functions involving only affine primary fields.
Knowing the poles \eqref{jpr} of $J^a(y)$ and its behaviour near $y=\infty$, we have
\begin{align}
 \left\langle J^a(y) \prod_{i=1}^N \Phi^{\mathcal{R}_i}_{X_i}(z_i)\right\rangle = - \sum_{i=1}^N \frac{D^{\mathcal{R}_i}_{X_i}(t^a)}{y-z_i}\left\langle \prod_{i=1}^N \Phi^{\mathcal{R}_i}_{X_i}(z_i)\right\rangle\ ,
\label{dja}
\end{align}
and the global Ward identities become
\begin{align}
 \sum_{i=1}^N D_{X_i}^{\mathcal{R}_i}(t^a) \left\langle \prod_{i=1}^N \Phi^{\mathcal{R}_i}_{X_i}(z_i)\right\rangle   = 0 \ .
\label{drxt}
\end{align}

\subsubsection{$\widehat{\mathfrak{sl}}_2$ global Ward identities}

In the $x$-basis, the $\widehat{\mathfrak{sl}}_2$ global Ward identities are formally identical to the Virasoro global Ward identities \eqref{spz}. In the case $N=3$, the solution is Eq. \eqref{fzzz}, so that 
\begin{align}
 \left\langle \prod_{i=1}^3 \Phi^{j_i}_{x_i} \right\rangle \propto\ x_{12}^{j_1+j_2-j_3} x_{23}^{j_2+j_3-j_1} x_{31}^{j_3+j_1-j_2}\ ,
\label{xxx}
\end{align}
where we omitted the dependence on $z_i$.
In the $\mu$-basis, we find 
\begin{align}
 \left\langle \prod_{i=1}^3\Phi^{j_i}_{\mu_i}\right\rangle \propto\ \mu_2\delta(\mu_1+\mu_2+\mu_3) \mathcal{H}\left(-\frac{\mu_1}{\mu_2}\right)\ ,
\label{pmf}
\end{align}
where the function $\mathcal{H}(x)$, which parametrizes the general solution of the $t^-$ and $t^0$ equations, is constrained by the $t^+$ equation to obey the twisted hypergeometric differential equation \eqref{hj}.
It may seem strange that in the $\mu$-basis we obtain a second-order differential equation, whereas in the $x$-basis the space of solutions appeared to be one-dimensional.
The number of linearly independent solutions of the global Ward identities has the algebraic interpretation of a tensor product multiplicity for $\mathfrak{sl}_2$ representations, and this should not depend on our choice of isospin variable.
Actually it is the $x$-basis calculation that is misleading: analyticity conditions on the $x_i$-dependence of $\left\langle \prod_{i=1}^3 \Phi^{j_i}_{x_i} \right\rangle$ in general allow the existence of two solutions that differ globally, although they are locally identical \cite{rib09}.
The tensor product multiplicity for generic $\mathfrak{sl}_2$ representations is two, as correctly suggested by the $\mu$-basis calculation. 

\subsubsection{Knizhnik--Zamolodchikov equations}

Inserting the equation \eqref{lmp} in an $N$-point function of affine primary fields, and applying the local Ward identity \eqref{jmnz} to $J^b_{-1}\Phi^\mathcal{R}_X = (J^b\Phi^\mathcal{R}_X)$, we obtain the Knizhnik--Zamolodchikov equations or \textbf{\boldmath KZ equations}\index{KZ equations},
\begin{align}
 \boxed{\left\{(k+g){\frac{\partial}{\partial z_i}} + \sum_{j\neq i} \frac{K_{ab}D_{X_i}^{\mathcal{R}_i}(t^a)D_{X_j}^{\mathcal{R}_j}(t^b)}{z_i-z_j}\right\}\left\langle \prod_{i=1}^N \Phi^{\mathcal{R}_i}_{X_i}(z_i)\right\rangle  = 0}\ .
\label{kz} 
\end{align}
These are first-order differential equations in $z_i$, and like the analogous equations \eqref{kzl} for free boson correlation functions, they determine the dependence on $z_i$ of correlation functions of primary fields.
However, unlike the free boson equations, the KZ equations do not have simple solutions in general. 
It can be checked that the KZ equations imply the conformal global Ward identities. (See Exercise \ref{exokz}.)

The KZ equations can be rewritten as 
\begin{align}
 \left\{(k+g){\frac{\partial}{\partial z_i}} + H_i \right\}\left\langle \prod_{i=1}^N \Phi^{\mathcal{R}_i}_{X_i}(z_i)\right\rangle   = 0 \ ,
\label{phz}
\end{align}
where $H_i$ are the mutually commuting Gaudin Hamiltonians, the Hamiltonians of the \textbf{\boldmath Gaudin model}\index{Gaudin model} -- an integrable model associated with the Lie algebra $\mathfrak{g}$.
Techniques developed for studying the Gaudin model can be useful for solving the KZ equations, as we will now see in the case $\mathfrak{g} = \mathfrak{sl}_2$.

\subsection{\texorpdfstring{$\mathfrak{sl}_2$}{sl2} case: the KZ-BPZ relation} \label{seckzbpz}

We will now show that the $\mathfrak{sl}_2$ KZ equations are equivalent to certain BPZ equations via Sklyanin's separation of variables for the $\mathfrak{sl}_2$ Gaudin model. 

\subsubsection{Reformulation of the KZ equations}

Our derivation of the KZ equations amounted to inserting the Sugawara construction \eqref{tjj}  at the points $z_i$ in the $N$-point function $\left\langle \prod_{i=1}^N \Phi^{j_i}_{X_i}(z_i)\right\rangle $ of affine primary fields.
Inserting the Sugawara construction at an arbitrary point $y$ instead, we obtain the identity
\begin{align}
 \left\langle \left(T(y)  - \frac{1}{2(k+2)}\left[ 2(J^0J^0)(y)+(J^+J^-)(y)+(J^-J^+)(y)\right]\right) \prod_{i=1}^N \Phi^{j_i}_{X_i}(z_i)\right\rangle = 0\ .
\label{tmjj}
\end{align}
According to Eqs. \eqref{dtz} and \eqref{dja}, inserting the fields $T(y)$ and $J^a(y)$ amounts to acting with the differential operators 
\begin{align}
 \hat{T}(y) &= \sum_{i=1}^N \left(\frac{\Delta_{j_i}}{(z-z_i)^2} + \frac{1}{z-z_i}{\frac{\partial}{\partial z_i}}\right)\ ,
\label{tcy}
\\
 \hat{J}^a(y) &= - \sum_{i=1}^N \frac{D^{\mathcal{R}_i}_{X_i}(t^a)}{y-z_i}\ ,
\label{jay}
\end{align}
where $\Delta_j$ is defined in Eq. \eqref{dj}.
These differential operators obey the commutation relations
\begin{align}
 \left[\hat{T}(y),\hat{J}^a(z)\right] &= {\frac{\partial}{\partial z}} \frac{\hat{J}^a(y)-\hat{J}^a(z)}{y-z}\ ,
\label{dtd}
\\
 \left[ \hat{J}^a(y),\hat{J}^b(z)\right] &= f^{ab}_c \frac{\hat{J}^c(y)-\hat{J}^c(z)}{y-z}\ ,
\label{ddd}
\end{align}
where $f^{ab}_c$ are the structure constants of the Lie algebra $\mathfrak{sl}_2$, as encoded in the commutation relations \eqref{ttpm}.
And one can easily show that inserting a normal-ordered product, for instance $(J^-J^+)(y)$, amounts to acting with the product of the corresponding differential operators.
Therefore, the identity \eqref{tmjj} amounts to the differential equation
\begin{align}
 \left(\hat{T}(y) -\frac{1}{2(k+2)}\left[ 2\hat{J}^0(y)\hat{J}^0(y) +\hat{J}^+(y)\hat{J}^-(y)+\hat{J}^-(y)\hat{J}^+(y)\right]\right) \left\langle \prod_{i=1}^N \Phi^{j_i}_{X_i}(z_i)\right\rangle=0\ .
 \label{refkz}
\end{align}

\subsubsection{Insertion of the zeros of $\hat{J}^-(y)$}

Let us simplify Eq. \eqref{refkz} by taking $y$ to be one of the zeros $\hat{y}_j$ of $ \hat{J}^-(y)$.
These zeros are differential operators, and the value of a $y$-dependent differential operator $\hat{\mathcal{O}}(y)$ at $y=\hat{y}_j$ is defined by inserting $\hat{y}_j$ from the left, that is $ \hat{\mathcal{O}}(\hat{y}_j) = \frac{1}{2\pi i}\oint_{\hat{y}_j} \frac{dy}{y-\hat{y}_j} \hat{\mathcal{O}}(y) $.
Using Eq. \eqref{ddd} for bringing the $\hat{J}^-(y)$ factors to the left, we obtain
\begin{align}
 \left( \hat{T}(\hat{y}_j) -\frac{1}{k+2}\left[(\hat{J}^0)^2(\hat{y}_j) + \partial \hat{J}^0(\hat{y}_j)\right] \right)  \left\langle \prod_{i=1}^N \Phi^{j_i}_{X_i}(z_i)\right\rangle= 0 \ .
\label{jjpj}
\end{align}
Let us further study the differential operators $\hat{y}_j$ and $\hat{J}^0(\hat{y}_j)$.
According to Eq. \eqref{ddd}, we have $[\hat{J}^-(y),\hat{J}^-(z)]=0$.
Therefore $[\hat{y}_j,\hat{y}_k]=0$, and
\begin{align}
 \boxed{\hat{J}^-(y)  = \hat{Y}_2 \frac{\prod_{j}(y-\hat{y}_j)}{\prod_i(y-z_i)}}\ ,
\label{djm}
\end{align}
where $\hat{Y}_2$ is a differential operator such that $[\hat{Y}_2,\hat{y}_j]=0$.
Using Eq. \eqref{ddd} we find $[\hat{J}^0(\hat{y}_j),\hat{J}^-(z)] = \frac{\hat{J}^-(z)}{\hat{y}_j-z}$, and deduce
\begin{align}
 [\hat{p}_j,\hat{Y}_2]=0 \quad \text{and} \quad [\hat{p}_j,\hat{y}_k]=\delta_{j,k} \quad \text{where} \quad \hat{p}_j = \hat{J}^0(\hat{y}_j)\ .
\label{pyd}
\end{align}
We can now simplify the second term of Eq. \eqref{jjpj},
\begin{multline}
 (\hat{J}^0)^2(\hat{y}_j) + \partial \hat{J}^0(\hat{y}_j) = \frac{1}{2\pi i} \oint_{\hat{y}_j} \frac{dy}{y-\hat{y}_j}\left(\hat{J}^0(y) +{\frac{\partial}{\partial y}} \right) \hat{J}^0(y) \\
 = \frac{1}{2\pi i} \oint_{\hat{y}_j} \frac{dy}{y-\hat{y}_j}\left(\hat{p}_j +{\frac{\partial}{\partial y}} \right) \hat{J}^0(y)
 = \hat{p}_j \frac{1}{2\pi i} \oint_{\hat{y}_j} \frac{dy}{y-\hat{y}_j} \hat{J}^0(y) = \hat{p}_j^2\ .
\end{multline}
So Eq. \eqref{jjpj} becomes 
\begin{align}
 \boxed{\left(\frac{1}{k+2}\hat{p}_j^2 -\hat{T}(\hat{y}_j)\right) \left\langle \prod_{i=1}^N \Phi^{j_i}_{X_i}(z_i)\right\rangle= 0 }\ .
\label{ppdz}
\end{align}

\subsubsection{Sklyanin's separation of variables}

Let us define the Sklyanin variable $y_j$ as the eigenvalues of the mutually commuting operators $\hat{y}_j$. 
The $\widehat{\mathfrak{sl}}_2$ global Ward identity \eqref{drxt} associated with the current $J^a$ can be written as 
\begin{align}
 \underset{y\to \infty}{\lim} y \hat{J}^a(y) \left\langle \prod_{i=1}^N \Phi^{j_i}_{X_i}(z_i)\right\rangle = 0\ .
\end{align}
When combined with the definition \eqref{djm} of $\hat{y}_j$, the Ward identity associated with $J^-$ suggests that we have $N-2$ Sklyanin variables $y_j$.
Let us define two more variables as the eigenvalues $Y_1$ and $Y_2$ of the operators $\hat{Y}_1$ and $\hat{Y}_2$ \eqref{djm}, where we define
\begin{align}
 \hat{Y}_1 = \underset{y\to \infty}{\lim} y \hat{J}^-(y)\ .
\end{align}
We can now define  \textbf{\boldmath Sklyanin's separation of variables}\index{Sklyanin's separation of variables} for the $\mathfrak{sl}_2$ Gaudin model as the linear 
map $\mathcal{K}$ from functions of $(X_1,\cdots X_N)$ to functions of $(Y_1,Y_2,y_1,\cdots y_{N-2})$ that diagonalizes the operators $(\hat{Y}_1,\hat{Y}_2,\hat{y}_1,\cdots \hat{y}_{N-2})$, so that in particular $\mathcal{K} \hat{y}_j f(X_1,\cdots X_N) = y_j \mathcal{K} f(X_1,\cdots X_N)$.
We actually only define this map on 
functions that obey the global Ward identity associated with $J^-$ i.e.
that are killed by $\hat{Y}_1$, whose images therefore have a $\delta(Y_1)$ prefactor.
Using in addition the global Ward identity associated with $J^0$, we find that the combined dependence of $\mathcal{K}\left\langle \prod_{i=1}^N \Phi^{j_i}_{X_i}(z_i)\right\rangle$ on $Y_1$ and $Y_2$ is a $\delta(Y_1)Y_2$ prefactor.
(This is done with the help of Eq. \eqref{ddd}, which implies $\left[\underset{y\to \infty}{\lim} y \hat{J}^0(y),\hat{J}^-(z)\right]=-\hat{J}^-(z)$.)

Let us rewrite our equation \eqref{ppdz} in terms of Sklyanin variables.
Using Eq. \eqref{pyd}, we find
$\mathcal{K} \hat{p}_j \mathcal{K}^{-1}= {\frac{\partial}{\partial y_j}} $.
Therefore,  
Eq. \eqref{ppdz} is equivalent to the following equation for $\mathcal{K}\left\langle \prod_{i=1}^N \Phi^{j_i}_{X_i}(z_i)\right\rangle$,
\begin{align}
 \left(\frac{1}{k+2}\frac{\partial ^2}{\partial y_j^2} - \mathcal{K} \hat{T}(\hat{y}_j) \mathcal{K}^{-1}\right)\mathcal{K}\left\langle \prod_{i=1}^N \Phi^{j_i}_{X_i}(z_i)\right\rangle  = 0\ .
\label{pkkz}
\end{align}
This equation apparently only involves $y_j$ for a given index $j$, whereas each KZ equation involves $(X_1,\cdots X_N)$. 
So it may seem that using Sklyanin variables leads to a separation of variables not only in the Gaudin model, but also
in the KZ equations. 
This is however not true, because the differential operator $\hat{T}(y)$ \eqref{tcy} involves $z_i$-derivatives at fixed isospin variables $X_i$.
When writing $\mathcal{K} \hat{T}(\hat{y}_j) \mathcal{K}^{-1}$, we have to use $z_i$-derivatives at fixed Sklyanin variables $y_k$, and we find
\begin{align}
\mathcal{K} \hat{T}(\hat{y}_j) \mathcal{K}^{-1} = \sum_i\left[\frac{\Delta_{j_i}}{(y_j-z_i)^2}+ \frac{1}{y_j-z_i}\left({\frac{\partial}{\partial z_i}}+{\frac{\partial}{\partial y_j}}\right)\right]-\sum_{k\neq j}\frac{1}{y_{jk}}\left({\frac{\partial}{\partial y_j}}-{\frac{\partial}{\partial y_k}}\right)\ .
\label{dtyj}
\end{align}
(See Exercise \ref{exoktk}.) 
Introducing the function
\begin{align}
 \Theta_N = \frac{\prod_{i<i'\leq N}(z_i-z_{i'})\prod_{j<j'\leq N-2}(y_j-y_{j'})}{\prod_{i=1}^N\prod_{j=1}^{N-2}(z_i-y_j)}\ ,
\end{align}
the equation \eqref{pkkz} is equivalent to 
\begin{multline}
 \left\{\frac{1}{k+2}  \frac{\partial^2}{\partial y_j^2} - \sum_i\frac{1}{y_j-z_i}{\frac{\partial}{\partial z_i}} -\sum_{k\neq j}\frac{1}{y_j-y_k} {\frac{\partial}{\partial y_k}}
\right. \\ \left.
 -\sum_i\frac{\Delta_{j_i}-\frac{k}{4}}{(y_j-z_i)^2}  -\sum_{k\neq j}\frac{\frac{3k}{4}+1}{(y_j-y_k)^2}   \right\} \Theta_N^{\frac{k+2}{2}}\mathcal{K}\left\langle \prod_{i=1}^N \Phi^{j_i}_{X_i}(z_i)\right\rangle = 0\ .
 \label{vskz}
\end{multline}

\subsubsection{KZ-BPZ relation for conformal blocks}

As an equation for $\Theta_N^{\frac{k+2}{2}}\mathcal{K}\left\langle \prod_{i=1}^N \Phi^{j_i}_{X_i}(z_i)\right\rangle$,
Eq. \eqref{vskz} coincides with the BPZ equation \eqref{pvot} for the correlation function of $\left\langle \prod_{i=1}^N V_{P_i}(z_i) \prod_{j=1}^{N-2}V_{\langle 2,1\rangle}(y_j)\right\rangle$, provided the parameter $b$ of the Virasoro algebra is given by 
\begin{align}
 \boxed{ b^2 = -k-2}\ ,
\label{bk}
\end{align}
and the momentums $P_i$ are given in terms of the spins $j_i$ by 
\begin{align}
 \boxed{P = b^{-1}\left(j+\tfrac12\right)}\quad \implies \quad \boxed{\Delta(P) = \Delta_j-\frac{k}{4}}\ ,
\label{aj}
\end{align}
where $\Delta(P)$ is given by Eq. \eqref{daq}.
The BPZ equations only constrain the dependence on $y_j$, but we have already determined the dependence of  
$\mathcal{K}\left\langle \prod_{i=1}^N \Phi^{j_i}_{X_i}(z_i)\right\rangle$ on the variables $Y_1$ and $Y_2$.
This leads to Feigin, Frenkel and Stoyanovsky's \textbf{\boldmath KZ-BPZ relation}\index{KZ-BPZ relation},
\begin{align}
 \boxed{ \mathcal{K}\left\langle \prod_{i=1}^N \Phi^{j_i}_{X_i}(z_i)\right\rangle \sim \delta(Y_1) Y_2 \Theta_N^{\frac12 b^2} \left\langle \prod_{i=1}^N V_{P_i}(z_i)\prod_{j=1}^{N-2}V_{\langle 2,1\rangle}(y_j)\right\rangle}\ .
\label{dyy}
\end{align}
This equation means that the differential equations obeyed by both sides coincide.
By definition, \textbf{\boldmath $\widehat{\mathfrak{sl}}_2$ conformal blocks}\index{conformal block!sl2@$\widehat{\mathfrak{sl}}_2$---} are elements of bases of solutions of the KZ equations: we can therefore define $N$-point $\widehat{\mathfrak{sl}}_2$ conformal blocks from Virasoro $(2N-2)$-point conformal blocks. 
In the diagrammatic representation of conformal blocks as trees, the relevant Virasoro block is obtained from an $\widehat{\mathfrak{sl}}_2$ conformal block by adding a degenerate field near each node, and identifying the two fusion channels of that degenerate field with $\widehat{\mathfrak{sl}}_2$ fusion multiplicities.
For example, we can build a basis of $s$-channel four-point $\widehat{\mathfrak{sl}}_2$ conformal blocks, parametrized by an $s$-channel spin $j_s$ and two fusion multiplicities $\epsilon \in \{+,-\}$, as follows:
\begin{align}
\begin{tikzpicture}[baseline={(0,0)}, very thick, scale = .6]
\draw (-1,2) node [left] {$j_1$} -- (0,0) node[below right] {$\epsilon$} -- node [above] {$j_s$} (4,0) node[below left] {$\sigma$} -- (5,2) node [right] {$j_4$};
\draw (-1,-2) node [left] {$j_2$} -- (0,0);
\draw (4,0) -- (5,-2) node [right] {$j_3$};
\end{tikzpicture}
\quad \sim \quad  
\begin{tikzpicture}[baseline={(0,0)}, very thick, scale = .6]
\draw (-1,2) node [left] {$P_1$} -- (0,0) -- (1,0) -- node [above] {$P_s$} (5,0) -- (6.3,2.6) node [right] {$P_4$};
\draw (-1,-2) node [left] {$P_2$} -- (0,0);
\draw (5,0) -- (6,-2) node [right] {$P_3$};
\draw[dashed] (1,0) -- (1,2.2) node[above] {$\left<2,1\right>$};
\draw[dashed] (5.3, .6) -- (4.3, 2.6) node[above] {$\left<2,1\right>$};
\node[right] at (5.15, .2) {$P_4+\sigma \frac{b}{2}$};
\draw[thin, latex-] (.5, -.2) to[out = -90, in = 180] (1,-1) node [right] {$P_s + \epsilon \frac{b}{2}$};
\end{tikzpicture}
\end{align}
The separation of variables $\mathcal{K}$ is in general an integral transformation, but in the case of the $\mu$-basis \eqref{mub} of isospin variables, we have 
\begin{align}
\renewcommand{\arraystretch}{1.5}
 \hat{J}^-(y) = \sum_{i=1}^N \frac{\mu_i}{y-z_i} \quad \implies \quad \left\{\begin{array}{l}  Y_1 =\sum_{i=1}^N \mu_i\ , \\ Y_2 = \sum_{i=1}^N \mu_i z_i\ , \\ \mu_i = Y_2\frac{\prod_{j=1}^{N-2}(z_i-y_j)}{\prod_{i'\neq i}(z_i-z_{i'})}\ ,\end{array}\right. 
\label{my}
\end{align}
and the variables on both sides of the relation are functions of one another. 

The KZ-BPZ relation is useful for studying $\widehat{\mathfrak{sl}}_2$-symmetric theories such as the $H_3^+$ model, and one may wonder whether the $\mathfrak{g}$ KZ equations are involved in similar relations for more general choices of the Lie algebra $\mathfrak{g}$.
While Sklyanin's separation of variables for the $\mathfrak{sl}_3$ Gaudin model exists, writing the $\mathfrak{sl}_3$ KZ equations in Sklyanin variables however does not lead to the expected generalizations of the BPZ equations \cite{rib08b}, and the reason for this discrepancy is not known. 
Another tentative generalization of the KZ-BPZ relation \eqref{dyy} is to replace the field $V_{\langle 2,1\rangle}$ on the right-hand side with another primary field (degenerate or not).
Then the resulting expression can be interpreted as an $N$-point function in a theory whose symmetry algebra is a generalization of $\widehat{\mathfrak{sl}}_2$ \cite{rib08}.

\subsubsection{\texorpdfstring{$\widehat{\mathfrak{sl}}_2$}{sl2} degenerate representations and fusion rules}

Let us use the KZ-BPZ relation for deriving $\widehat{\mathfrak{sl}}_2$ degenerate representations and fusion rules. 
The idea is that an OPE of two $\widehat{\mathfrak{sl}}_2$-primary fields $\Phi^{j_1}\Phi^{j_2}$ is equivalent an OPE of three Virasoro-primary fields $V_{\langle 2, 1\rangle}V_{P_1}V_{P_2}$, with momentums given by Eq. \eqref{aj}.
We call the field 
$\Phi^{j_1}$ degenerate if the OPEs $\Phi^{j_1}\Phi^{j_2}$ contain finitely many $\widehat{\mathfrak{sl}}_2$-primary fields.
In order to achieve this, the corresponding Virasoro-primary field $V_{P_1}$ must be degenerate, $V_{P_1}=V_{\langle r, s\rangle}$ with $r,s\geq 1$. 
The relevant Virasoro OPE becomes 
\begin{align}
 V_{\langle 2, 1\rangle}V_{\langle r, s\rangle}V_{P_2} \sim V_{\langle r+1, s\rangle}V_{P_2}\ .
\end{align}
(If $r>1$, the OPE $V_{\langle 2, 1\rangle}V_{\langle r, s\rangle}\sim V_{\langle r-1, s\rangle}+ V_{\langle r+1, s\rangle}$ actually has two terms, but the first term's contributions are included in the second term's.)
We actually only assume $r\geq 0,s\geq 1$ rather than $r,s\geq 1$, as this is enough for $V_{\langle r+1, s\rangle}$ to be degenerate.
This suggests that there are degenerate fields $\Phi^{\langle r,s\rangle}$ with spins
\begin{align}
 j_{\langle r,s\rangle} = -\frac12 +\frac12 s -\frac{k+2}{2} r \ , \quad \text{for} \quad \left\{\begin{array}{l} r\geq 0\ , \\ s\geq 1\ , \end{array}\right.
 \label{jrs}
\end{align}
obtained by applying the relation \eqref{aj} to the degenerate momentums \eqref{ars}. And the resulting OPEs are
\begin{align}
 \Phi^{\langle r, s\rangle}\Phi^j \sim 
 \sum_{i=-\frac{r}{2}}^\frac{r}{2} \sum_{\ell=-\frac{s-1}{2}}^{\frac{s-1}{2}} \Phi^{j+(k+2)i+\ell}
 = \Phi^{j-j_{\langle r,s\rangle}} + \cdots + \Phi^{j+j_{\langle r,s\rangle}}\ .
 \label{prspj}
\end{align}
These OPEs agree with the fusion rules that can be derived from analyzing singular vectors in affine highest-weight representations \cite{ay92}, with $\Phi^{\langle r, s\rangle}$ having a vanishing descendant at the level $N=rs$. (See Exercise \ref{exolos} for the case $N=1$.) In contrast to the case of Virasoro algebra, there are nontrivial representations with singular vectors at the level $N=0$.

It is tempting to speculate that there exist $\widehat{\mathfrak{sl}}_2$ generalized minimal models.
For any given irrational value of the level $k$, this would be a diagonal model whose spectrum would be made of the representations that correspond to the fields $\Phi^{\langle r, s\rangle}$. For a rational value of the level $k$, there should exist a diagonal $\widehat{\mathfrak{sl}}_2$ minimal model, whose spectrum would contain finitely many such representations.
If we restrict ourselves to the set of half-integer spin fields $\Phi^{\langle 0, s\rangle}$, which is closed under OPEs, we actually obtain the (generalized) $SU_2$ WZW models of Section \ref{secsu}.

\section{The \texorpdfstring{$H_3^+$}{H3+} model\label{sechtp}}

The \textbf{\boldmath $H_3^+$ model}\index{h3+ model@$H_3^+$ model} is to the $\widehat{\mathfrak{sl}}_2$ algebra what Liouville theory is to the Virasoro algebra: the simplest model with the given symmetry and a continuous spectrum. 

In the $H_3^+$ model as in Liouville theory, it is possible to deduce the three-point function from the associativity of OPEs involving degenerate fields \cite{tes97a}.
We will take a shortcut, and build the $H_3^+$ model from Liouville theory, by extending the KZ-BPZ relation for conformal blocks \eqref{dyy} to a relation between correlation functions, called the \textbf{\boldmath $H_3^+$-Liouville relation}\index{h3+-Liouville relation@$H_3^+$-Liouville relation}.

Since Liouville theory exists for all complex values of the central charge, we expect that the $H_3^+$ model exists for all values of the level except $k=-2$, when the Sugawara construction breaks down and the model is no longer a conformal field theory.
According to the relation \eqref{bk}, the central charges of the $H_3^+$ model \eqref{ckg} and of the corresponding Liouville theory then take the values
\begin{align}
 \boxed{c^{H_3^+} = \frac{3k}{k+2} \in \mathbb{C}-\{3\}} \ , \quad c^{\text{Liouville}} = 1-6k-\frac{6}{k+2}\in \mathbb{C}\ .
\end{align} 

\subsection{Spectrum and correlation functions}

Since Liouville theory is diagonal, the $H_3^+$ model is diagonal.
The values of the spin $j$,
\begin{align}
 \boxed{j\in -\frac12 + ib\mathbb{R}_+}\ , 
\end{align}
are deduced from Liouville theory via the relation \eqref{aj}. The relation \eqref{my} between $\widehat{\mathfrak{sl}}_2$ isospin variables and positions of Liouville fields suggests that the left- and right-moving isospins $\mu$ and $\bar\mu$ are complex conjugates. Let us write $\Phi^j_\mu(z)$ an affine primary field of the $H_3^+$ model with isospins $\mu$ and $\bar\mu$. In terms of such fields, the $H_3^+$-Liouville relation reads 
\begin{align}
 \boxed{ \left\langle \prod_{i=1}^N \Phi^{j_i}_{\mu_i}(z_i)\right\rangle 
 = \delta^{(2)}\left({\textstyle\sum_{i=1}^N \mu_i}\right) 
 \left|{\textstyle\sum_{i=1}^N\mu_iz_i}\right|^2 |\Theta_N|^{-k-2} 
 \left\langle \prod_{i=1}^N V_{P_i}(z_i)\prod_{j=1}^{N-2}V_{\langle 2,1\rangle}(y_j)\right\rangle}\ .
\label{dyym} 
\end{align}
We will now unpack the $H_3^+$-Liouville relation, and in particular extract the three-point structure constant of the $H_3^+$ model. 
To begin with, let us produce a natural definition for this structure constant, using the $x$-basis.

\subsubsection{$x$-basis fields}

While using the isospin variable $\mu$ simplifies the $H_3^+$-Liouville relation, using the isospin variable $x$ simplifies the action of the global $\mathfrak{sl}_2$ symmetry. 
In particular, the dependence of three-point functions on this variable takes a simple form \eqref{xxx}.
We assume that the left- and right-moving isospins $x$ and $\bar x$ are complex conjugates just like $\mu$ and $\bar\mu$, and define $x$-basis fields by 
\begin{align}
 \Phi^j_{x} = \gamma(2j+1)\int_{{\mathbb{C}}} d^2\mu\ |\mu|^{-2j-2} e^{\mu x -\bar{\mu}\bar{x}}\Phi^j_{\mu}\ .
\end{align}
In this single-valued version of the Fourier transformation \eqref{emx}, 
we introduced a prefactor $\gamma(2j+1)$, which will ensure that the $H_3^+$ OPE has a finite limit when the spin becomes half-integer.
Moreover, we flipped the sign of the second term of the exponent of $e^{\mu x -\bar{\mu}\bar{x}}$, as compared to the expected $e^{\mu x +\bar{\mu}\bar{x}}$, in order to ensure the convergence of the integral when $x$ and $\bar{x}$ are complex conjugates.
At the level of the $J^a\Phi^j_\mu$ OPE \eqref{jprx}, this sign flipping manifests itself in the relation between the left- and right-moving differential operators,
\begin{align}
 \bar{D}^j_{\bar{x}}(t^a) = D^j_{\bar{x}}(t^a)  \quad , \quad \bar{D}^j_{\bar{\mu}}(t^a) = D^j_{-\bar{\mu}}(t^a)\ . 
\end{align}
Notice that the minus sign in the operators $D^j_{-\bar{\mu}}(t^a)$ modifies neither their commutation relations, nor the KZ equations.

In the $x$-basis, the dependence of the three-point function on isospin variables is given by Eq. \eqref{xxx}.
Assuming correlation functions are single-valued as functions of the isospin variables, we must have
\begin{align}
  \left\langle \prod_{i=1}^3 \Phi^{j_i}_{x_i} \right\rangle = C^{H_3^+}_{j_1,j_2,j_3}\ |x_{12}|^{2(j_1+j_2-j_3)} |x_{23}|^{2(j_2+j_3-j_1)} |x_{31}|^{2(j_3+j_1-j_2)}\ ,
\label{ch}
\end{align}
which provides a natural definition of the three-point structure constant $C^{H_3^+}_{j_1,j_2,j_3}$.

\subsubsection{Three-point structure constant}

In the case $N=3$, the $H_3^+$-Liouville relation involves one Sklyanin variables $y_1$, and we have
\begin{align}
 N=3 \quad \implies \quad  
 \renewcommand{\arraystretch}{1.6}
 \left\{\begin{array}{l} y_1 = -\frac{\mu_1z_2z_3+\mu_2z_3z_1+\mu_3z_1z_2}{\sum_{i=1}^3\mu_iz_i}\ , \\ \frac{(y_1-z_1)(z_2-z_3)}{(y_1-z_2)(z_1-z_3)} = -\frac{\mu_1}{\mu_2}\ . \end{array}\right.
\end{align}
The relation \eqref{dyym} then reads 
\begin{align}
 \left\langle \prod_{i=1}^3\Phi^{j_i}_{\mu_i} \right\rangle = \delta^{(2)}(\textstyle{\sum}_{i=1}^3\mu_i) |\mu_2|^2 \sum_\pm C_\pm(P_1)C_{P_1\pm\frac{b}{2},P_2,P_3} \left|\mathcal{H}^{(s)}_\pm(-\tfrac{\mu_1}{\mu_2})  \right|^2\ , 
\label{sfpm}
\end{align}
where we used the momentums $P_i$ \eqref{aj}, the Liouville theory structure constants $C_\pm(P)$ and $C_{P_1,P_2,P_3}$, and the functions $\mathcal{H}^{(s)}_\pm(x)$ that are related to the 
$s$-channel Virasoro conformal blocks  
$
 \mathcal{F}^{(s)}_\pm(x)  =  
\begin{tikzpicture}[baseline=(current  bounding  box.center), very thick, scale = .3]
\draw (-1,2) node [left] {$P_1$} -- (0,0) -- node [above] {$P_1\pm \frac{b}{2}$} (4,0) -- (5,2) node [right] {$P_2$};
\draw[dashed] (-1,-2) node [left] {$\langle 2,1 \rangle$} -- (0,0);
\draw (4,0) -- (5,-2) node [right] {$P_3$};
\end{tikzpicture}
$
by Eq. \eqref{fgs}. 

In order to extract the three-point structure constant, let us compute the Fourier transform of Eq. \eqref{ch}, using the formula \cite{rt05}
\begin{multline}
 \prod_{i=1}^3\left(\frac{|\mu_i|^{2j_i+2}}{\pi^2}\int_{{\mathbb{C}}}d^2x_i\ e^{-\mu_ix_i+\bar{\mu}_i\bar{x}_i}\right)
|x_{12}|^{2(j_1+j_2-j_3)} |x_{23}|^{2(j_2+j_3-j_1)} |x_{31}|^{2(j_3+j_1-j_2)} 
\\
= \frac{1}{\pi^2}\delta^{(2)}(\textstyle{\sum}_{i=1}^3\mu_i)|\mu_2|^2 \sum_\pm d_\pm \left|\mathcal{H}^{(s)}_\pm(-\tfrac{\mu_1}{\mu_2})\right|^2 \ ,
\label{iii}
\end{multline}
where we define 
\begin{align}
 d_+ = \frac{\gamma(-j_1+j_2+j_3+1)}{\gamma(-2j_1)} \quad , \quad d_- = \frac{\gamma(j_1+j_2-j_3+1)\gamma(j_1-j_2+j_3+1)}{\gamma(-j_1-j_2-j_3-1)\gamma(2j_1+2)}\ .
\end{align}
The right-hand side of Eq. \eqref{iii} is a combination of solutions \eqref{pmf} of the $\mu$-basis global Ward identities, written using the functions $\mathcal{H}^{(s)}_\pm(x)$ of Eq. \eqref{fpm}.
Comparing with Eq. \eqref{sfpm}, we obtain the three-point structure constant in terms of Liouville theory structure constants,
\begin{align}
 C^{H_3^+}_{j_1,j_2,j_3} = \frac{\pi^2\prod_{i=1}^3\gamma(2j_i+1)}{d_\pm}C_{\pm}(P_1)C_{P_1\pm \frac{b}{2},P_2,P_3}\ .
\end{align}
The single-valuedness of $\left\langle \prod_{i=1}^3\Phi^{j_i}_{\mu_i} \right\rangle $ as a function of $\mu_i$ guarantees that this does not depend on the sign $\pm$, and this can be checked explicitly using Eq. \eqref{eq:shift}. 
Using the DOZZ formula \eqref{caaa} (with $\mu=1$) for $C_{P_1\pm \frac{b}{2},P_2,P_3}$, and $C_+(P)=1$ from Exercise \ref{exodoc}, we obtain
\begin{align}
\boxed{C^{H_3^+}_{j_1,j_2,j_3} = \frac{\pi^2b^{-1}b^{\frac{2}{b^2}(j_1+j_2+j_3+1)} \Upsilon'_b(0)\Upsilon_b(-\frac{2j_1}{b})\Upsilon_b(-\frac{2j_2}{b})\Upsilon_b(-\frac{2j_3}{b})}
{\Upsilon_b(-\frac{j_1+j_2+j_3+1}{b})\Upsilon_b(-\frac{j_1+j_2-j_3}{b}) \Upsilon_b(-\frac{j_1-j_2+j_3}{b})\Upsilon_b(-\frac{-j_1+j_2+j_3}{b})}}\ ,
\label{chp}
\end{align}
where the parameter $b$ is given by Eq. \eqref{bk}.
(We would have a different formula in the case $b\in i\mathbb{R}$ i.e. $k>-2$, based on the alternative Liouville structure constant \eqref{hc}.)

\subsubsection{Reflection relation, two-point function and OPE}

The reflection relation for Liouville primary fields leads to a reflection relation for $H_3^+$ primary fields,
\begin{align}
 \Phi^j_{\mu} = R_{b^{-1}(j+\frac12)} \Phi^{-j-1}_{\mu} \ ,
\end{align}
which involves the Liouville reflection coefficient $R_P$ \eqref{ram}. The two-point function can be deduced from the two-point function \eqref{vvrdd} of Liouville theory, 
\begin{align}
 \left\langle \Phi^{j_1}_{\mu_1} \Phi^{j_2}_{\mu_2}\right\rangle = \delta^{(2)}(\mu_1+\mu_2) |\mu_1|^2 b\Big[\delta(j_1+j_2+1) + R_{b^{-1}(j_1+\frac12)} \delta(j_2-j_1)\Big]\ .
\end{align}
Let us deduce the reflection relation and two-point function for $x$-basis fields, using the formula  
\begin{align}
 \int_{{\mathbb{C}}}d^2\mu\ e^{\mu x-\bar{\mu}\bar{x}} |\mu|^{-4j-2} = |x|^{4j}\pi \gamma(-2j) \ .
\label{icmx}
\end{align}
We find the reflection relation 
\begin{align}
 \Phi^j_{x} = \frac{b^2}{\pi}\gamma\left(1-\tfrac{2j+1}{b^2}\right)\int_{{\mathbb{C}}}d^2x'\ |x-x'|^{4j}\Phi^{-j-1}_{x'}\ ,
\end{align}
and the two-point function 
\begin{align}
 \left\langle \Phi^{j_1}_{x_1} \Phi^{j_2}_{x_2} \right\rangle = \frac{-\pi^2 b}{(2j_1+1)^2}\, \delta(j_1+j_2+1)\delta^{(2)}(x_{12}) 
+ \pi b^{-1} \gamma\left(-\tfrac{2j_1+1}{b^2}\right)\, \delta(j_1-j_2)|x_{12}|^{4j_1}\, .
\label{pjpj}
\end{align}
Knowing the two-point function, we can easily deduce the OPE from the three-point function. 
The simplest expression is obtained by focussing on the $\delta(j_1+j_2+1)$ term in the two-point function, and we find the following $\mu$-basis and $x$-basis OPEs,
\begin{align}
\Phi^{j_1}_{\mu_1}\Phi^{j_2}_{\mu_2} &\sim  \int_{-\frac12 + ib\mathbb{R}_+} dj\int_{{\mathbb{C}}}\frac{d^2\mu}{b|\mu|^2} \left\langle \Phi^{j_1}_{\mu_1}\Phi^{j_2}_{\mu_2} \Phi^{-j-1}_{-\mu} \right\rangle \Phi^j_{\mu}\ , 
\label{mope}
\\
\Phi^{j_1}_{x_1}\Phi^{j_2}_{x_2} &\sim \int_{-\frac12 + ib\mathbb{R}_+} \frac{(2j+1)^2 dj}{-\pi^2b}\int_{{\mathbb{C}}}d^2x \left\langle \Phi^{j_1}_{x_1}\Phi^{j_2}_{x_2} \Phi^{-j-1}_{x} \right\rangle \Phi^j_{x}\ ,
\label{xope}
\end{align}
In the $\mu$-basis OPE, the integral over $\mu$ reduces to the value $\mu=\mu_1+\mu_2$, due to the delta-function prefactor of the three-point function. 
For a study of the analytic properties of the $x$-basis OPE, and its analytic continuation to degenerate fields, see Exercise \ref{exodrfrh}.

\subsection{Large level limit and geometrical interpretation}

Having defined the $H_3^+$ model by its spectrum and correlation functions, we will now propose a geometrical interpretation of the model, based on the manifold $H_3^+$.

\subsubsection{$SL_2({\mathbb{C}})$ symmetry group and functions on $H_3^+$}

Let us consider the \textbf{\boldmath large level limit}\index{large level limit} $k\to \infty$, which is sometimes called the minisuperspace limit. 
The KZ equations \eqref{kz} imply that correlation functions do not depend on the positions $z_i$ in this limit, and are functions of the sole isospin variables. 
Moreover, given a generator $J^a_m$, the commutator $[J^a_m,J^b_{-m}]$ \eqref{jam}  tends to infinity for some index $b$, unless $m=0$. 
So the generators $J^a_{m\neq 0}$, and the descendant states they create, disappear from the theory, and only the horizontal subalgebra of $\widehat{\mathfrak{sl}}_2$ survives.
Since the left and right isospin variables are complex conjugates, the corresponding symmetry group is $SL_2({\mathbb{C}})$, and the action of this group on a field of spin $j$ is 
\begin{align}
 U_g\Phi^j_{x} = |cx+d|^{4j}\Phi^j_{\frac{ax+b}{cx+d}} \quad \text{with} \quad g = \left(\begin{array}{cc} a & b \\ c & d \end{array}\right) \in SL_2({\mathbb{C}})\ .
\label{ugp}
\end{align}
(This is formally identical to the action \eqref{tgv} of global conformal transformations on quasi-primary fields.)

Let us consider the following functions on the space $H_3^+$ of Hermitian matrices of size two and determinant one,
\begin{align}
 \phi^j_{x}(h) = \left(\begin{bmatrix}
                                x \\ 1
                               \end{bmatrix}^\dagger 
h \begin{bmatrix}
   x \\ 1 
  \end{bmatrix}
 \right)^{2j}\ .
 \label{pjxh}
\end{align}
The natural action of $g\in SL_2({\mathbb{C}})$ on the space $\mathcal{F}(H_3^+)$ of functions on $H_3^+$ is
\begin{align}
 g\cdot f(h) = f(g^\dagger h g)\ ,
\end{align}
and we have 
\begin{align}
 g\cdot \phi^j_{x}(h) = U_g \phi^j_{x}(h)\ ,
\end{align}
with the same action $U_g$ as in Eq. \eqref{ugp}.
This shows that both $\Phi^j_x$ and $\phi^j_x$ transform in the \textbf{\boldmath principal series representation of $\mathfrak{sl}_2({\mathbb{C}})$}\index{principal series representation!of sl2c@---of $\mathfrak{sl}_2({\mathbb{C}})$} with spin $j$, which we denote as $\mathcal{C}^j$. 
As a representation of $\mathfrak{sl}_2({\mathbb{C}})$, the large level limit of the spectrum is 
\begin{align}
 \underset{k\to\infty}{\lim} \mathcal{S}^{H_3^+} =  \mathcal{F}(H_3^+) = \bigoplus_{j\in -\frac12+i{\mathbb{R}}_+} \mathcal{C}^j\ .
\end{align}
(For a more rigorous definition of $\mathcal{F}(H_3^+)$, and proof of its decomposition into irreducible representations of $\mathfrak{sl}_2({\mathbb{C}})$, see \cite{tes97b}.)

\subsubsection{Correlation functions}

In the large level limit, correlation functions of the $H_3^+$ model are given in terms of functions on $H_3^+$ by 
\begin{align}
 \underset{k\to \infty}{\lim} \left\langle \prod_{i=1}^N \Phi^{j_i}_{x_i}(z_i)\right\rangle \propto \int_{H_3^+} dh\ \prod_{i=1}^N \phi^{j_i}_{x_i}(h)\ , 
\end{align}
where the unknown proportionality factor is an $x$-independent field normalization.
This is because both sides of this equation obey the same symmetry equations $\left<\prod_{i=1}^N U_g \phi^{j_i}_{x_i} \right> = \left<\prod_{i=1}^N \phi^{j_i}_{x_i} \right>$, and the same axioms of associativity and commutativity of the operator product, if we define the operator product in $\mathcal{F}(H_3^+)$ to be the product of functions. 
An explicit check of this equation can be done in the case $N=3$ \cite{tes97b}.

So the large level limit of our two-dimensional conformal field theory is the quantum mechanics of a point particle on the space $H_3^+$, which justifies naming the theory the $H_3^+$ model.
Other names for the same theory include the $H_3^+$ WZW model and the $SL_2({\mathbb{C}})/SU_2$ WZW model; the latter name comes from the realization of the space $H_3^+$ as a quotient.
The large level limit of the $H_3^+$ $N$-point function is formally identical to the light asymptotic limit of the Liouville $N$-point function \eqref{zih}, whose interpretation is however quite different as it depends on positions $z$ instead of isospin variables $x$.

\subsubsection{Scalar product and unitarity}

There is a natural, positive definite scalar product on $\mathcal{F}(H_3^+)$, 
\begin{align}
 \left\langle f \middle| f' \right\rangle = \int_{H_3^+} dh\ \overline{f(h)} f'(h)\ ,
\end{align}
where $dh$ is the $SL_2({\mathbb{C}})$-invariant measure on $H_3^+$.
For this scalar product, the action of $g\in SL_2({\mathbb{C}})$ is a unitary tranformation, i.e. $\left\langle g\cdot f \middle| g\cdot f'\right\rangle = \left\langle f \middle| f'\right\rangle $.
Let us interpret this at the level of the symmetry algebra.
The Lie algebra $\mathfrak{sl}_2({\mathbb{C}})$ of the symmetry group $SL_2({\mathbb{C}})$ can be viewed as a six-dimensional real vector space, whose generators $t^a,it^a$ are related to $J^a_0, \bar{J}^a_0$ by the ${\mathbb{R}}$-linear map
\begin{align}
 \left\{\begin{array}{lcl} t^a & \mapsto & J_0^a + \bar{J}_0^a \ ,  \\ it^a & \mapsto & i(J^a_0 - \bar{J}^a_0)\ . \end{array}\right. 
\end{align}
(The minus sign in the image of $it^a$ comes from the complex conjugation of the matrix elements of $g$ in $U_g\Phi^j_{x}$ \eqref{ugp}.)
Our scalar product is such that $t^a$ and $it^a$ are antihermitian, which is equivalent to
\begin{align}
 (J^a_0)^\dagger = -\bar{J}^a_0\ .
\label{jzd}
\end{align}
This conjugation rule can be extended to the entire affine Lie algebra as 
\begin{align}
 \boxed{(J^a_n)^\dagger = -\bar{J}^a_{-n}}\ .
\label{jdj}
\end{align}
This is compatible with the structure \eqref{jam} of the affine Lie algebra $\hat{\mathfrak{sl}}_2$ provided 
\begin{align}
 k\in\mathbb{R}\ .
\label{kir}
\end{align}
Via the Sugawara construction \eqref{ljj}, the conjugation rule for $J^a_n$ implies $L_n^\dagger = \bar{L}_{-n}$, which differs from the conjugation rule \eqref{ldn} that we previously assumed. While the dilation generator $L_0+\bar L_0$ is still self-ajoint, the generator $L_0-\bar L_0$ is now antihermitian. However, the eigenvalues of this generator are conformal spins, which must be half-integer and therefore real. This shows that the $H_3^+$ model cannot be unitary -- except of course in the large level limit, where descendant modes disappear, correlation functions are $z$-independent, and conformal spins all vanish. For a more pedestrian proof that the model is non-unitary, see Exercise \ref{exonu}.

\section{WZW models}

\subsection{Definition and general properties}

\subsubsection{Elements of a definition}

Given a simple Lie group $G$, the $G$ Wess--Zumino--Witten model or $G$ \textbf{\boldmath WZW model}\index{WZW model} is usually defined by a Lagrangian, which depends on a parameter $k$.
This model can then be shown to be a conformal field theory with a $\hat{\mathfrak{g}}$ symmetry algebra, where $\mathfrak{g}$ is the Lie algebra of $G$.
The parameter $k$ coincides with the level of $\hat{\mathfrak{g}}$, and may have to be quantized for the model to be consistent, depending on the group $G$.

We will not use the Lagrangian definition of WZW models.
This raises the question of characterizing these models in the conformal bootstrap approach.
The fundamental axiom is the presence of the symmetry algebra $\hat{\mathfrak{g}}$, and some  authors call all models with this symmetry algebra WZW models.
Here we will insist that, among models with this symmetry, 
a WZW model can be associated with a particular Lie group $G$, such that its spectrum $\mathcal{S}_k$ obeys
\begin{align}
 \boxed{\underset{k\to \infty}{\lim} \mathcal{S}_k = \mathcal{F}(G)}\ ,
\label{lsfg}
\end{align}
where $\mathcal{F}(G)$ is the space of functions on $G$.
This property still does not fully characterize WZW models: in particular, nothing forces the level $k$ to be quantized whenever the Lagrangian definition dictates it.
It is plausible that the spectrum of WZW models can be characterized in terms of functions on a manifold related to the loop group of $G$, but such a characterization is not known.
Instead of a proper definition of WZW models, we will limit ourselves to giving a few known properties of the spectrum of the $G$ WZW model:
\begin{enumerate}
 \item If $G$ is compact, then the $G$ WZW model is rational.
\item If $G$ is compact, then the level $k$ takes positive integer values.
\item The $G$ WZW model is diagonal if and only if $G$ is simply connected. 
\end{enumerate}
The relation between the simple connectedness of $G$ and the diagonality of the associated model 
has analogs in the case of free bosonic theories. Such theories are not strictly speaking WZW models, and they have no parameter that would be analogous to the level. Nevertheless, the uncompactified free boson is analogous to a WZW model with $G=\mathbb{R}$, and it is diagonal. The compactified free boson is analogous to a WZW model with $G=U_1$, and its spectrum \eqref{sr} is not diagonal. 

\subsubsection{Features of the spectrum}

Let us start with the large level limit \eqref{lsfg}.
The space $\mathcal{F}(G)$ of functions on $G$ has a natural action of 
$G\times \bar{G}$, where the bar is here for distinguishing two copies of $G$, such that for $f\in \mathcal{F}(G)$ we have 
\begin{align}
\left( (g,\bar{g})\cdot f\right)(h) = f(g^{-1}h\bar{g})\ .
\end{align}
We identify the corresponding infinitesimal symmetry algebra $\mathfrak{g}\times \bar{\mathfrak{g}}$ with the large level limit of the symmetry algebra $\hat{\mathfrak{g}}\times \bar{\hat{\mathfrak{g}}}$ of the WZW model.
That is, $\mathfrak{g}$ and its generators $t^a$ are identified with the horizontal subalgebra of $\hat{\mathfrak{g}}$ and its generators $J^a_0$.

We now assume that $G$ is compact.
Then $\mathcal{F}(G)$ can be decomposed into irreducible representations of the symmetry group $G\times \bar{G}$ using the \textbf{\boldmath Peter--Weyl theorem}\index{Peter--Weyl theorem}, 
\begin{align}
 \mathcal{F}(G) = \bigoplus_{\mathcal{R}\in \text{Rep}(G)} \mathcal{R}\otimes \bar{\mathcal{R}}\ ,
\end{align}
% NB: Problem with notations here.
where $\text{Rep}(G)$ is the set of irreducible unitary representations of $G$, which coincides with the set of irreducible finite-dimensional representations of $G$.
If moreover $G$ is simply connected, then the spectrum of the WZW model is diagonal for all positive integer values of $k$ \cite{fms97},
\begin{align}
 \mathcal{S}_k = \bigoplus_{\mathcal{R}\in \text{Rep}_k(G)} \hat{\mathcal{R}}\otimes \bar{\hat{\mathcal{R}}}\ .
\end{align}
Here $\hat{\mathcal{R}}$ is the affine highest-weight representation of $\hat{\mathfrak{g}}$ that is built from $\mathcal{R}$ by acting with the creation modes and removing the null vectors, and the finite subset $\text{Rep}_k(G)$ of $\text{Rep}(G)$ is defined by certain $k$-dependent conditions.
The resulting representations $\hat{\mathcal{R}}$ are called the integrable highest-weight representations of $\hat{\mathfrak{g}}$. 
We have $\underset{k\to\infty}{\lim} \text{Rep}_k(G) =\text{Rep}(G)$ and $\underset{k\to \infty}{\lim} \hat{\mathcal{R}} = \mathcal{R}$, which imply that Eq. \eqref{lsfg} holds in the case of compact groups.

\subsubsection{Scalar product}

The natural, positive definite scalar product on $\mathcal{F}(G)$ is 
\begin{align}
 \langle f|f'\rangle = \int_G dh\ \overline{f(h)} f'(h)\ ,
\label{gbg}
\end{align}
where $dh$ is the Haar measure, which is invariant under the left and right actions of $G$ on itself.
As in the case of the $H_3^+$ model, the generators of the symmetry algebra are antihermitian for this scalar product, which now implies
\begin{align}
 (J^a_0)^\dagger = -J^a_0\  . 
\label{jzdj}
\end{align}
This conjugation rule is naturally extended to the following conjugation rule on the affine Lie algebra,
\begin{align}
 \boxed{(J^a_n)^\dagger = -J^a_{-n}}\ .
\end{align}
This is compatible with the commutation relations \eqref{jam} of the affine Lie algebra $\hat{\mathfrak{g}}$, provided the structure constants $f^{ab}_c$ and level $k$ are real.
This is also compatible with the conjugation rule \eqref{ldn} for the generators of the Virasoro algebra, via the Sugawara construction \eqref{ljj}.

\subsection{The \texorpdfstring{$SU_2$}{SU(2)} WZW model \label{secsu}}

Let us build conformal field theories from the finite-dimensional representations of $SU_2$, which have spins $j\in\{0,\frac12,1,\frac32,\dots\}$.

\subsubsection{Generalized $SU_2$ WZW model}

Since the spins of the finite-dimensional representations of $SU_2$ are of the type $j=j_{\langle 0,s\rangle}$ \eqref{jrs}, the 
corresponding affine highest-weight representations $\hat{\mathcal{R}}^{\langle 0,s\rangle}$ of $\widehat{\mathfrak{sl}}_2$ are degenerate, with level zero null vectors.
Their fusion rules are 
\begin{align}
 \hat{\mathcal{R}}^{\langle 0, s_1\rangle}\times\hat{\mathcal{R}}^{\langle 0, s_2\rangle} = \sum_{s\overset{2}{=}|s_1-s_2|+1}^{s_1+s_2-1} \hat{\mathcal{R}}^{\langle 0, s\rangle}\ .
\end{align}
For any value of the level $k\in\mathbb{C}-\{-2\}$, let us build a diagonal spectrum from these representations,
\begin{align}
 \mathcal{S}_k = \bigoplus_{s=1}^\infty \hat{\mathcal{R}}^{\langle 0, s\rangle}\otimes \bar{\hat{\mathcal{R}}}^{\langle 0, s\rangle}\ ,
\end{align}
thereby defining the \textbf{\boldmath generalized $SU_2$ WZW model}\index{WZW model!generalized $SU_2$---}. 
We will now look for finite sets of such representations that are closed under fusion.

\subsubsection{Spectrum of the $SU_2$ WZW model}

In the same way as we derived the spectrums of Virasoro minimal models by studying doubly-degenerate representations of the Virasoro algebra, let us study doubly degenerate representations of the affine Lie algebra $\widehat{\mathfrak{sl}}_2$. 

We restrict our attention to representations that have level zero null vectors, and are therefore of the type $\hat{\mathcal{R}}^{\langle 0, s\rangle}$. In order to have another null vector, the representation must also be of the type $\hat{\mathcal{R}}^{\langle r', s'\rangle}$ with $(r',s')\neq (0,s)$. It is easy to see that this is possible only if $r'\neq 0$, and for simplicity we assume $r'=1$. This implies $j_{\langle 0, s\rangle} = j_{\langle 1, s'\rangle} $ or $j_{\langle 0, s\rangle} + j_{\langle 1, s'\rangle} +1=0$ for some $s'\in\mathbb{N}^*$. We focus on the second possibility, which implies $s+s'=k+2$. The level $k$ must therefore be integer, and for a given value of the level there are only $k+1$ doubly degenerate representations of this type. The resulting spectrum is
\begin{align}
 \mathcal{S}_k = \bigoplus_{s=1}^{k+1} \hat{\mathcal{R}}^{\langle 0, s\rangle}\otimes \bar{\hat{\mathcal{R}}}^{\langle 0, s\rangle}\ .
\end{align}
Fusion is constrained by both null vectors, and becomes 
\begin{align}
 \hat{\mathcal{R}}^{\langle 0, s_1\rangle}\times\hat{\mathcal{R}}^{\langle 0, s_2\rangle} = \sum_{s\overset{2}{=}|s_1-s_2|+1}^{\min(s_1+s_2-1,2k+3-s_1-s_2)} \hat{\mathcal{R}}^{\langle 0, s\rangle}\ .
\end{align}

\subsubsection{Correlation functions}

Like the $H_3^+$ model, the (generalized) $SU_2$ WZW model has a diagonal spectrum made of highest-weight representations.
Therefore, it is natural to conjecture that correlation functions of the (generalized) $SU_2$ WZW model can be deduced from $H_3^+$ correlation functions by analytically continuing the spins and the level to discrete values, in the same way as two- and three-point functions of (generalized) minimal models  are deduced from Liouville theory. 
In particular, degenerate fusion rules can be deduced from the $H_3^+$ OPE, see Exercise \ref{exodrfrh}.

\subsubsection{Further rational models with an $\widehat{\mathfrak{sl}}_2$ symmetry algebra}

Among rational models with an $\widehat{\mathfrak{sl}}_2$ symmetry algebra,
the $SU_2$ WZW model has the peculiarities of being diagonal, and of having level zero null vectors. 
Relaxing these assumptions, further rational models can be found.
These include the $SO_3$ WZW model, which is non-diagonal as $SO_3 = \frac{SU_2}{\mathbb{Z}_2}$ is not simply connected.

\subsection{The \texorpdfstring{$\widetilde{SL}_2(\mathbb{R})$}{SL2(R)} WZW model \label{secslr}}

Work on the $\widetilde{SL}_2(\mathbb{R})$ WZW model has been motivated by its relevance to string theory in $AdS_3$.
The model has not been fully solved: the three-point function is known only partially, and crossing symmetry has not been proved.
We will limit ourselves to working out the fusion rules, and deriving Maldacena and Ooguri's widely believed and well-tested conjecture for the spectrum \cite{mo00a}.
This spectrum is more complicated than the spectrum of the $H_3^+$ model, which is why solving the $\widetilde{SL}_2(\mathbb{R})$ WZW model is more difficult.
In order to compute correlation functions, it is tempting to use a Wick rotation from the $H_3^+$ model \cite{mo01}: the representation-theoretic interpretation of the Wick rotation is however subtle, as is already apparent in the large level limit \cite{rib09}. 

\subsubsection{Large level limit}

The Lie group \textbf{\boldmath $\widetilde{SL}_2(\mathbb{R})$}\index{sl2tilder@$\widetilde{SL}_2(\mathbb{R})$ (group)} is defined as the universal covering group of the group $SL_2({\mathbb{R}})$ of matrices of size two with real coefficients and determinant one.
The group $SL_2({\mathbb{R}})$ is not simply connected, as the matrices of the type $\left(\begin{smallmatrix} \cos \tau & \sin\tau \\ -\sin\tau & \cos \tau \end{smallmatrix}\right)$ form a non-contractible loop, and the first homotopy group of $SL_2({\mathbb{R}})$ is ${\mathbb{Z}}$.
So $\widetilde{SL}_2(\mathbb{R})$ is obtained from $SL_2({\mathbb{R}})$ by decompactifying the $\tau$ direction, and we have $SL_2({\mathbb{R}}) = \widetilde{SL}_2(\mathbb{R})/{\mathbb{Z}}$, where the denominator subgroup $\mathbb{Z}$ is generated by $\exp 2\pi\left(\begin{smallmatrix} 0 & 1 \\ -1 & 0 \end{smallmatrix}\right)$. The center of $\widetilde{SL}_2(\mathbb{R})$ is the (twice larger) $\mathbb{Z}$ subgroup generated by $\exp\pi\left(\begin{smallmatrix} 0 & 1 \\ -1 & 0 \end{smallmatrix}\right)$.

Let us adopt a basis of the Lie algebra whose Cartan generator is proportional to the matrix $\left(\begin{smallmatrix} 0 & 1 \\ -1 & 0 \end{smallmatrix}\right)$. We could choose another timelike generator, i.e. another matrix of the type $g\left(\begin{smallmatrix} 0 & 1 \\ -1 & 0 \end{smallmatrix}\right) g^{-1}$ with $g\in SL_2({\mathbb{R}})$, such that the corresponding Cartan subgroup contains the center of $\widetilde{SL}_2(\mathbb{R})$. 
The following matrix representation of the Lie algebra obeys the commutation relations \eqref{ttpm},
\begin{align}
\renewcommand{\arraystretch}{1.3}
\left\{ \begin{array}{rl}
 M'(t^+)& = \frac12\left(\begin{smallmatrix} 1 & i \\ i& -1 \end{smallmatrix} \right) \ , 
\\
 M'(t^0) &= \frac{1}{2}\left(\begin{smallmatrix} 0 & -i \\ i & 0 \end{smallmatrix}\right)\ ,
\\
M'(t^-) & = \frac12\left(\begin{smallmatrix} 1 & -i \\ -i & 1 \end{smallmatrix} \right) \ .
\end{array} \right.
\label{mpta}
\end{align}
This is actually a basis of $\mathfrak{sl}_2(\mathbb{C})$, which is equivalent to $\mathfrak{sl}_2(\mathbb{R})$ when it comes to considering representations over complex vector spaces. The conjugation rule \eqref{jzdj} states that elements of $\mathfrak{sl}_2(\mathbb{R})$ are antihermitian, and we deduce the conjugation rule for the generators in our new basis, 
\begin{align}
 \left\{\begin{array}{l} (J_0^0)^\dagger = J_0^0\ , \\
         (J_0^\pm)^\dagger = -J_0^\mp \ .
        \end{array}
\right.
\label{jtdj}
\end{align}
In particular, $J_0^0$ is now Hermitian, and should have real eigenvalues in unitary representations -- and half-integer values in all representations of $SL_2(\mathbb{R})$.

Let us study the space $\mathcal{F}(\widetilde{SL}_2(\mathbb{R}))$ of functions on $\widetilde{SL}_2(\mathbb{R})$, which is assumed to be the large level limit of the spectrum of our model. 
The decomposition of $\mathcal{F}(\widetilde{SL}_2(\mathbb{R}))$ into irreducible representations of the global symmetry algebra $\mathfrak{sl}_2\times \overline{\mathfrak{sl}}_2$ involves two types of unitary representations of $\mathfrak{sl}_2$: the \textbf{\boldmath principal series representation}\index{principal series representation!of sl2@---of $\mathfrak{sl}_2$} $\mathcal{C}^j_\alpha$ and the \textbf{\boldmath discrete series representation}\index{discrete series representation (of $\mathfrak{sl}_2$)} $\mathcal{D}^{j,\pm}$, where $j$ is the spin. 
These representations can be characterized by the eigenvalues of the generator $J_0^0$,
\begin{align} 
\renewcommand{\arraystretch}{1.3}
\begin{tabular}{|l|l|l|}
  \hline
Representation & Parameter values & Eigenvalues of $J_0^0$
\\
\hline 
$\mathcal{C}^j_\alpha$  & $j\in -\tfrac12+i{\mathbb{R}}_+,\ \alpha\in{\mathbb{R}}\bmod {\mathbb{Z}}$ &  $\alpha + {\mathbb{Z}}$ 
\\
$\mathcal{D}^{j,+}$ & $j\in]-\infty,-\tfrac12[$ & $-j+{\mathbb{N}}$
\\
$\mathcal{D}^{j,-}$ & $j\in]-\infty,-\tfrac12[$ & $j-{\mathbb{N}}$
\\
\hline 
 \end{tabular}
\label{rpe}
\end{align}
According to \cite{rib09} and references therein, we have
\begin{align}
 \mathcal{F}(\widetilde{SL}_2(\mathbb{R})) &= \int^\oplus_{-\frac12+i{\mathbb{R}}_+} dj \int^\oplus_{]0,1[} d\alpha\ \mathcal{C}^j_\alpha \otimes \bar{\mathcal{C}}^j_{\alpha} \oplus \bigoplus_\pm \int^\oplus_{]-\infty,-\frac12[} dj\ \mathcal{D}^{j,\pm}\otimes \bar{\mathcal{D}}^{j,\pm} \ ,
\label{fst}
\\
 \mathcal{F}(SL_2(\mathbb{R})) &= \int^\oplus_{-\frac12+i{\mathbb{R}}_+} dj \bigoplus_{\alpha\in\{0,\frac12\}} \mathcal{C}^j_\alpha \otimes \bar{\mathcal{C}}^j_{\alpha} \oplus \bigoplus_\pm \bigoplus_{j=-\frac12, -1,-\frac32 \cdots} \mathcal{D}^{j,\pm}\otimes \bar{\mathcal{D}}^{j,\pm} \ .
\end{align}
The space $\mathcal{F}(SL_2(\mathbb{R}))$ is the subspace of $\mathcal{F}(\widetilde{SL}_2(\mathbb{R}))$ where the Hermitian generator $J_0^0$ has half-integer eigenvalues.

The groups $SL_2(\mathbb{R})$ and $\widetilde{SL}_2(\mathbb{R})$ have an outer automorphism: in $SL_2(\mathbb{R})$ it can be realized as the conjugation by the matrix $\left(\begin{smallmatrix} 0 & 1 \\ 1 & 0\end{smallmatrix}\right)$, which does not belong to the group because its determinant is $-1$. The action of this outer automorphism of the Lie algebra is 
\begin{align}
 (J_0^0)^* = -J_0^0 \quad , \quad (J_0^\pm)^* = - J_0^\mp\ .
\end{align}
Composing a representation with the outer automorphism, we obtain another representation which we call the conjugate representation. The conjugate representations of a discrete or continuous representation is the representation with the same spin and opposite $J_0^0$ eigenvalues:
\begin{align}
 (\mathcal{C}^j_\alpha)^* &= \mathcal{C}^j_{-\alpha} \ , 
\\
 (\mathcal{D}^{j,\pm})^* &= \mathcal{D}^{j,\mp}\ .
\end{align}
Tensor products of principal and discrete series representations can be written as 
\begin{align}
 \mathcal{R}_1\otimes \mathcal{R}_2 = \bigoplus_{\mathcal{R}_3} N_{\mathcal{R}_1,\mathcal{R}_2,\mathcal{R}_3^*} \mathcal{R}_3\ ,
\label{ror}
\end{align}
where $N_{\mathcal{R}_1,\mathcal{R}_2,\mathcal{R}_3}$ is an integer-valued function that obeys $N_{\mathcal{R}_1^*,\mathcal{R}_2^*,\mathcal{R}_3^*}=N_{\mathcal{R}_1,\mathcal{R}_2,\mathcal{R}_3}$ and is symmetric under permutations.
Modulo these symmetries, the only nonzero values of $N_{\mathcal{R}_1,\mathcal{R}_2,\mathcal{R}_3}$ are
\begin{align}
N_{\mathcal{D}^{j_1,+},\mathcal{D}^{j_2,+},\mathcal{D}^{j_3,-}} &= 1 \quad \text{if}\ j_3\in j_1+j_2-{\mathbb{N}}\ ,
\label{nddd}
\\
N_{\mathcal{C}^{j_1}_{\alpha_1},\mathcal{D}^{j_2,+},\mathcal{D}^{j_3,-}} &= 1 \quad \text{if}\ \alpha_1+j_2-j_3\in{\mathbb{Z}}\ ,
\\
 N_{\mathcal{C}^{j_1}_{\alpha_1},\mathcal{C}^{j_2}_{\alpha_2},\mathcal{D}^{j_3,+}} &= 1 \quad \text{if}\ \alpha_1+\alpha_2+j_3\in {\mathbb{Z}}\ ,
\\
N_{\mathcal{C}^{j_1}_{\alpha_1},\mathcal{C}^{j_2}_{\alpha_2},\mathcal{C}^{j_3}_{\alpha_3}} &= 2 \quad \text{if}\ \alpha_1+\alpha_2+\alpha_3\in {\mathbb{Z}}\ .
\end{align}
For example, the tensor product of two principal series representations is
\begin{align}
 \mathcal{C}^{j_1}_{\alpha_1}\otimes \mathcal{C}^{j_2}_{\alpha_2} &= 2\int^\oplus_{-\frac12+i{\mathbb{R}}_+} dj\ \mathcal{C}^j_{\alpha_1+\alpha_2} 
\oplus \bigoplus_\pm \bigoplus_{\substack{j\in \pm\alpha_1\pm\alpha_2+{\mathbb{Z}}\\ j\in]-\infty,-\frac12[}} \mathcal{D}^{j,\pm} \ ,
\label{coc}
\end{align}
where the factor of $2$ is the tensor product multiplicity that we already encountered in Section \ref{secwikz}.

There is a heuristic reasoning for deducing all the values of $N_{\mathcal{R}_1,\mathcal{R}_2,\mathcal{R}_3}$ from the values of $N_{\mathcal{D}^{j_1,+},\mathcal{D}^{j_2,+},\mathcal{D}^{j_3,-}}$.
The idea is to use the identification
\begin{align}
 \mathcal{C}^j_\alpha \sim \mathcal{D}^{\alpha-1,+} \oplus \mathcal{D}^{-\alpha,-}\ ,
\label{cjdd}
\end{align}
which holds as far as the eigenvalues \eqref{rpe} of $J^0_0$ are concerned.
This reasoning apparently predicts a nonzero value for $N_{\mathcal{C}^j_\alpha,\mathcal{D}^{j_2,+},\mathcal{D}^{j_3,+}}$.
That value should be discarded, because it depends on $\alpha$ whereas $\mathcal{C}^j_\alpha$ only depends on $\alpha\ \text{mod}\ {\mathbb{Z}}$.

\subsubsection{Spectral flow}

After these reminders on $\widetilde{SL}_2(\mathbb{R})$ and its representations, we are ready to consider the spectrum of the associated WZW model.
The principal and discrete series representations of $\mathfrak{sl}_2$ are naturally extended to affine highest-weight representations of $\widehat{\mathfrak{sl}}_2$: the principal series representations $\hat{\mathcal{C}}^j_\alpha$ and discrete series representations $\hat{\mathcal{D}}^{j,\pm}$.
However, it turns out that we cannot build this spectrum from these representations.
This can be understood in terms of an algebraic feature of $\widehat{\mathfrak{sl}}_2$ called the \textbf{\boldmath spectral flow}\index{spectral flow} \cite{mo00a}.
The spectral flow is a family $(\rho_w)_{w\in{\mathbb{Z}}}$ of automorphisms of $\widehat{\mathfrak{sl}}_2$ that obey $\rho_w\rho_{w'}=\rho_{w+w'}$ and act as 
\begin{align}
 \rho_w(J^0_n)&=  J^0_n + \frac12 kw \delta_{n,0}   \ ,
\\
 \rho_w(J^\pm_n) &= J^\pm_{n\pm w}  \ .
\end{align}
According to Eqs. \eqref{ljj} and \eqref{lzjj}, this implies
\begin{align}
 \rho_w(L_n)  = L_n + wJ^0_n +\frac14 kw^2 \delta_{n,0}\ .
\end{align}
Given a representation $\mathcal{R}$ of $\widehat{\mathfrak{sl}}_2$, i.e. an action of the generators $J^a_n$ on some vectors $|v\rangle$, we define the \textbf{\boldmath spectrally flowed representation}\index{spectrally flowed representation} $\rho_w(\mathcal{R})$ by the action $\rho_{-w}(J^a_n)|v\rangle$.  
Together with the extension of $\mathfrak{sl}_2$'s outer automorphism to $\widehat{\mathfrak{sl}}_2$, spectral flow generates a $\mathbb{Z}\rtimes \mathbb{Z}_2$ group of automorphisms, such that
\begin{align}
 \rho_w(\mathcal{R})^* = \rho_{-w}(\mathcal{R}^*)\ .
\end{align}
Moreover, it is believed that the action of spectral flow commutes with fusion, in the sense that \cite{gab01b}
\begin{align}
 \rho_{w}(\mathcal{R})\times \rho_{w'}(\mathcal{R}') = \rho_{w+w'}(\mathcal{R}\times \mathcal{R}')\ .
\label{rwr}
\end{align}
We assume that fusion products of representations of $\widehat{\mathfrak{sl}}_2$ have the form 
\begin{align}
 \mathcal{R}_1\times \mathcal{R}_2 = \bigoplus_{\mathcal{R}_3} N_{\mathcal{R}_1,\mathcal{R}_2,\mathcal{R}_3^*} \mathcal{R}_3\ ,
\end{align}
where the coefficients $N_{\mathcal{R}_1,\mathcal{R}_2,\mathcal{R}_3}$ are a permutation-symmetric, integer-valued fusion multiplicities such that $N_{\mathcal{R}_1^*,\mathcal{R}_2^*,\mathcal{R}_3^*}=N_{\mathcal{R}_1,\mathcal{R}_2,\mathcal{R}_3}$.
From Eq. \eqref{rwr}, we must then have 
\begin{align}
w_1+w_2+w_3=0 \quad \implies \quad N_{\rho_{w_1}(\mathcal{R}_1),\rho_{w_2}(\mathcal{R}_2),\rho_{w_3}(\mathcal{R}_3)}=N_{\mathcal{R}_1,\mathcal{R}_2,\mathcal{R}_3}\ .
\label{nrrr} 
\end{align}
Let us consider the action of spectral flow on our affine highest-weight representations. 
We introduce the notations 
\begin{align}
 \hat{\mathcal{C}}^{j,w}_\alpha &= \rho_w(\hat{\mathcal{C}}^j_\alpha) \quad , \quad (w\in{\mathbb{Z}})\ ,
\\
\hat{\mathcal{D}}^{j,w} &= \rho_{w-\frac12}(\hat{\mathcal{D}}^{j,+})\quad , \quad (w\in \tfrac12+{\mathbb{Z}})\ .
\end{align}
% NB: We could change notations and have \alpha (in C series) and j (in D series) depend on w, so that they correspond
% to J_0^0 eigenvalues and are conserved modulo integers. This would simplify fusion rules, but make spectral flow more complicated. 
If $w\neq 0$, then $\hat{\mathcal{C}}^{j,w}_\alpha$ cannot be an affine highest-weight representation, because the eigenvalues of $\rho_{-w}(L_0) = L_0-w J^0_0 +\frac14 kw^2$ in $\hat{\mathcal{C}}^j_\alpha$ are not bounded from below -- and actually, the representations $(\hat{\mathcal{C}}^{j,w}_\alpha)_{w\in{\mathbb{Z}}}$ all differ from one another.
Let us now focus on discrete series representations.
The representation $\hat{\mathcal{D}}^{j,\pm}$ can be characterized by the existence of a state $|v^{j,\pm}\rangle$ such that 
\begin{align}
J^\mp_{n\geq 0}|v^{j,\pm}\rangle = J^0_{n>0}|v^{j,\pm}\rangle =  J^\pm_{n>0}|v^{j,\pm}\rangle = (J^0_0\pm j)|v^{j,\pm}\rangle = 0\ .
\end{align}
So we can characterize $\rho_w(\hat{\mathcal{D}}^{j,\pm})$ by the action of $\rho_{-w}(J^a_n)$ on $|v^{j,\pm}\rangle$.
In particular, we notice that $\rho_1(J^a_n)|v^{j,+}\rangle = 0 \ \iff \ J^a_n|v^{\frac{k}{2}-j,-}\rangle=0$, which leads to 
\begin{align}
 \rho_{-1}(\hat{\mathcal{D}}^{j,+}) = \hat{\mathcal{D}}^{\frac{k}{2}-j,-} \ .
\label{rdd}
\end{align}
So the spectral flow orbit of $\hat{\mathcal{D}}^{j,+}$ contains another affine highest-weight representations.
(The rest of the orbit is made of representations where the eigenvalues of $L_0$ are not bounded from below.)
And discrete representations of both series can be written in our new notations as 
\begin{align}
 \hat{\mathcal{D}}^{j,+} = \hat{\mathcal{D}}^{j,\frac12} \quad , \quad \hat{\mathcal{D}}^{j,-} = \hat{\mathcal{D}}^{\frac{k}{2}-j,-\frac12}\ .
 \label{dpdo}
\end{align}
The dual representations of our spectrally flowed representations are
\begin{align}
 (\hat{\mathcal{C}}^{j,w}_{\alpha})^* &= \hat{\mathcal{C}}^{j,-w}_{-\alpha}\ ,
\label{cjwd}
\\
(\hat{\mathcal{D}}^{j,w})^* &= \hat{\mathcal{D}}^{\frac{k}{2}-j,-w}\ .
\label{djwd}
\end{align}
This concludes our discussion of spectral flow.
We will now argue that spectrally flowed representations must appear in the spectrum. 

\subsubsection{Spectrum}

By our definition of WZW models, the large level limit of the spectrum $\tilde{\mathcal{S}}_k$ of the $\widetilde{SL}_2(\mathbb{R})$ WZW model is $\mathcal{F}(\widetilde{SL}_2(\mathbb{R}))$ \eqref{fst}, and we therefore expect $\tilde{\mathcal{S}}_k$ to contain affine discrete representations of both series $\hat{\mathcal{D}}^{j,\pm}$.
Let us show that $\tilde{\mathcal{S}}_k$ cannot contain only affine highest-weight representations.
Consider spins $j_1,j_2,j_3$ such that $N_{\mathcal{D}^{j_1,+},\mathcal{D}^{j_2,-},\mathcal{D}^{j_3,+}}=1$.
If the level $k$ is large enough, we must then have $N_{\hat{\mathcal{D}}^{j_1,\frac12},\hat{\mathcal{D}}^{j_2,-\frac12},\hat{\mathcal{D}}^{j_3,\frac12}}=1$.
Using the behaviour \eqref{nrrr} of fusion multiplicities under spectral flow, this implies $N_{\hat{\mathcal{D}}^{j_1,-\frac12},\hat{\mathcal{D}}^{j_2,-\frac12},\hat{\mathcal{D}}^{j_3,\frac32}}=1$.
So the fusion product $\hat{\mathcal{D}}^{j_1,-\frac12}\times \hat{\mathcal{D}}^{j_2,-\frac12}$ contains the representation $(\hat{\mathcal{D}}^{j_3,\frac32})^*$, which is not an affine highest-weight representation.
Generalizing this argument, the spectrum must in fact contain 
representations of the type $\hat{\mathcal{D}}^{j,w}$ for all values of $w\in \frac12+{\mathbb{Z}}$.
Then it is natural to assume that the spectral flow actually leaves the spectrum invariant.
But what are the allowed values of the spin $j$ of $\hat{\mathcal{D}}^{j,w}$? We still impose the constraint $j<-\frac12$ that comes from the representation $\mathcal{D}^{j,\pm}$ of $\mathfrak{sl}_2$.
If this applies to both spins in the relation \eqref{rdd}, we must then have 
\begin{align}
 \frac{k+1}{2} <j < -\frac12\ .
\label{jimm}
\end{align}
This defines a non-empty interval provided
\begin{align}
 \boxed{k\in]-\infty,-2[}\ .
\end{align}
(Nevertheless, the model surely exists for $k\in \mathbb{C}-\{2\}$ by analytic continuation.)
The natural conjectures for the spectrums of the $\widetilde{SL}_2(\mathbb{R})$ and $SL_2(\mathbb{R})$ WZW models are then
\begin{align}
 \tilde{\mathcal{S}}_k &= \bigoplus_{w\in{\mathbb{Z}}}\int^\oplus_{-\frac12+i{\mathbb{R}}_+} dj \int^\oplus_{]0,1[} d\alpha\ \hat{\mathcal{C}}^{j,w}_\alpha \otimes \bar{\hat{\mathcal{C}}}^{j,w}_{\alpha} 
\oplus \bigoplus_{w\in\frac12+{\mathbb{Z}}}\int^\oplus_{]\frac{k+1}{2},-\frac12[} dj\ \hat{\mathcal{D}}^{j,w}\otimes \bar{\hat{\mathcal{D}}}^{j,w} \ ,
\\
 \mathcal{S}_k &= \bigoplus_{\substack{w_L,w_R\in{\mathbb{Z}}\\ w_L-w_R\in 2{\mathbb{Z}}}}\int^\oplus_{-\frac12+i{\mathbb{R}}_+} dj \bigoplus_{\alpha\in\{0,\frac12\}} \hat{\mathcal{C}}^{j,w_L}_\alpha \otimes \bar{\hat{\mathcal{C}}}^{j,w_R}_{\alpha} \oplus  \bigoplus_{\substack{w_L,w_R\in\frac12+{\mathbb{Z}}\\ w_L-w_R\in 2{\mathbb{Z}}}}\ \bigoplus_{\substack{j= -\frac12, -1,-\frac32,\cdots \\ j>\frac{k+1}{2}}} \hat{\mathcal{D}}^{j,w_L}\otimes \bar{\hat{\mathcal{D}}}^{j,w_R} \ .
\end{align}
Notice that the left and right spectral flow numbers $w_L$ and $w_R$ are independent in the case of $SL_2(\mathbb{R})$, and equal in the case of $\widetilde{SL}_2(\mathbb{R})$, so that the spectrum $\tilde{\mathcal{S}}_k$ is diagonal.
The rule, which can only be heuristic in the absence of a definition of WZW models based on the corresponding loop groups, is:
\begin{quote}
 In the $G$ WZW model the spectral flow takes values in the first homotopy group of the global symmetry group, that is $\pi_1(\frac{G\times \bar{G}}{Z(G)})$ where $Z(G)$ is the center of $G$.
\end{quote}
For our WZW models, the relevant homotopy groups are 
$ \pi_1(\frac{\widetilde{SL}_2(\mathbb{R})\times \overline{\widetilde{SL}_2(\mathbb{R})}}{{\mathbb{Z}}}) = {\mathbb{Z}}$ and $\pi_1(\frac{SL_2({\mathbb{R}})\times \overline{SL_2({\mathbb{R}})}}{{\mathbb{Z}}_2}) = \frac{{\mathbb{Z}}\times \overline{{\mathbb{Z}}}}{{\mathbb{Z}}_2}$.
The rule also applies to the case $G=U_1$ of the compactified free boson, if we consider the winding number as a spectral flow number. 

\subsubsection{Fusion rules}

Let us check that we can find fusion rules for the representations of $\widehat{\mathfrak{sl}}_2$, such that
\begin{enumerate}
 \item the rule \eqref{nrrr} is obeyed,
\item in the large level limit $k\to -\infty$, the fusion rules reduce to the tensor product rules for representations of $\mathfrak{sl}_2$,
\item the conjectured spectrums of the $\widetilde{SL}_2(\mathbb{R})$ and $SL_2(\mathbb{R})$ WZW models are closed under fusion.
\end{enumerate}
We first obtain the fusion multiplicities $N_{\hat{\mathcal{D}}^{j_1,\pm},\hat{\mathcal{D}}^{j_2,\pm},\hat{\mathcal{D}}^{j_3,\pm}}$ for affine highest-weight representations of the discrete series  by assuming that it can be nonzero only when the corresponding tensor product multiplicity \eqref{nddd} is nonzero, and when all spins obey the condition \eqref{jimm}.
The rest of the nonzero fusion multiplicities of the type    
$N_{\hat{\mathcal{D}}^{j_1,w_1},\hat{\mathcal{D}}^{j_2,w_2},\hat{\mathcal{D}}^{j_3,w_3}}$ are obtained by the rule \eqref{nrrr}.
Then we generalize the relation \eqref{cjdd} between $\mathfrak{sl}_2$ representations of the principal and discrete series, and obtain $\hat{\mathcal{C}}^j_\alpha \sim \hat{\mathcal{D}}^{\alpha-1,+} \oplus \hat{\mathcal{D}}^{-\alpha,-}$. Using the notations \eqref{dpdo} and applying spectral flow, this implies
\begin{align}
 \hat{\mathcal{C}}^{j,w}_\alpha \sim \hat{\mathcal{D}}^{\alpha-1,w+\frac12} \oplus \hat{\mathcal{D}}^{\alpha+\frac{k}{2},w-\frac12} \ .
\end{align}
This enables us to compute fusion multiplicities involving representations of the type $\hat{\mathcal{C}}^{j,w}_\alpha$.
The only subtlety is that we obtain terms in $N_{\hat{\mathcal{C}}^{j_1,w_1}_{\alpha_1},\hat{\mathcal{D}}^{j_2,w_2},\hat{\mathcal{D}}^{j_3,w_3}}$ that depend on $\alpha_1$ instead of $\alpha_1\ \text{mod}\ {\mathbb{Z}}$, and must be discarded.
Keeping the condition \eqref{jimm} on spins of discrete series representations implicit, the results are 
\begin{align}
 N_{\hat{\mathcal{D}}^{j_1,w_1},\hat{\mathcal{D}}^{j_2,w_2},\hat{\mathcal{D}}^{j_3,w_3}} 
&= \delta_{\sum w_i,\frac12} \delta_{\sum j_i-\frac{k}{2},{\mathbb{N}}} + \delta_{\sum w_i,-\frac12}\delta_{\sum j_i-k,-{\mathbb{N}}} \ ,
\\
 N_{\hat{\mathcal{C}}^{j_1,w_1}_{\alpha_1},\hat{\mathcal{D}}^{j_2,w_2},\hat{\mathcal{D}}^{j_3,w_3}} &= \delta_{\sum w_i,0} \delta_{\alpha_1-j_2-j_3+\frac{k}{2},{\mathbb{Z}}}\ ,
\\
 N_{\hat{\mathcal{C}}^{j_1,w_1}_{\alpha_1},\hat{\mathcal{C}}^{j_2,w_2}_{\alpha_2},\hat{\mathcal{D}}^{j_3,w_3}} & = \delta_{\sum w_i,\frac12} \delta_{\alpha_1+\alpha_2-j_3,{\mathbb{Z}}} + \delta_{\sum w_i,-\frac12}\delta_{\alpha_1+\alpha_2-j_3+\frac{k}{2},{\mathbb{Z}}}\ ,
\\
N_{\hat{\mathcal{C}}^{j_1,w_1}_{\alpha_1},\hat{\mathcal{C}}^{j_2,w_2}_{\alpha_2},\hat{\mathcal{C}}^{j_3,w_3}_{\alpha_3}} & = \delta_{\sum w_i,1}\delta_{\sum\alpha_i-\frac{k}{2},{\mathbb{Z}}} + 2\, \delta_{\sum w_i,0}\delta_{\sum \alpha_i,{\mathbb{Z}}} + \delta_{\sum w_i,-1}\delta_{\sum\alpha_i+\frac{k}{2},{\mathbb{Z}}}\ .
\end{align}
This leads to the following fusion rules:
\begin{multline}
\hat{\mathcal{D}}^{j_1,w_1}\times \hat{\mathcal{D}}^{j_2,w_2} = \int^\oplus_{-\frac12+i{\mathbb{R}}_+} dj\ \hat{\mathcal{C}}^{j,w_1+w_2}_{j_1+j_2+\frac{k}{2}} 
\\ \oplus 
\bigoplus_{\substack{j\in j_1+j_2-{\mathbb{N}} \\ j\in ]\frac{k+1}{2}, -\frac12[}} \hat{\mathcal{D}}^{j,w_1+w_2-\frac12} \oplus 
\bigoplus_{\substack{j\in j_1+j_2-\frac{k}{2}+{\mathbb{N}} \\ j\in ]\frac{k+1}{2}, -\frac12[}} \hat{\mathcal{D}}^{j,w_1+w_2+\frac12}\ ,
\end{multline}
\begin{align}
 \hat{\mathcal{C}}^{j_1,w_1}_{\alpha_1}\times \hat{\mathcal{D}}^{j_2,w_2} &= \int^\oplus_{-\frac12+i{\mathbb{R}}_+} dj\left( \hat{\mathcal{C}}^{j,w_1+w_2-\frac12}_{\alpha_1-j_2} \oplus \hat{\mathcal{C}}^{j,w_1+w_2+\frac12}_{\alpha_1-j_2+\frac{k}{2}}\right) \oplus \bigoplus_{\substack{j\in \alpha_1-j_2+{\mathbb{Z}} \\ j\in ]\frac{k+1}{2}, -\frac12[}} \hat{\mathcal{D}}^{j,w_1+w_2}\ ,
\end{align}
\begin{multline}
 \hat{\mathcal{C}}^{j_1,w_1}_{\alpha_1}\times \hat{\mathcal{C}}^{j_2,w_2}_{\alpha_2} = \int^\oplus_{-\frac12+i{\mathbb{R}}_+} dj \left(\hat{\mathcal{C}}^{j,w_1+w_2-1}_{\alpha_1+\alpha_2-\frac{k}{2}} \oplus 2\, \hat{\mathcal{C}}^{j,w_1+w_2}_{\alpha_1+\alpha_2} \oplus \hat{\mathcal{C}}^{j,w_1+w_2+1}_{\alpha_1+\alpha_2+\frac{k}{2}}\right) 
\\
\oplus \bigoplus_{\substack{j\in -\alpha_1-\alpha_2+\frac{k}{2}+{\mathbb{Z}} \\ j\in]\frac{k+1}{2}, -\frac12[}} \hat{\mathcal{D}}^{j,w_1+w_2-\frac12} 
\oplus \bigoplus_{\substack{j\in -\alpha_1-\alpha_2+{\mathbb{Z}} \\ j\in]\frac{k+1}{2}, -\frac12[}} \hat{\mathcal{D}}^{j,w_1+w_2+\frac12} \ .
\end{multline}
So, while the spectral flow number $w$ is not conserved, fusion violates it by at most one unit.
This can alternatively be shown at the level of correlation functions by studying how spectral flow affects the Ward identities and Knizhnik--Zamolodchikov equations \cite{rib05}.

\subsubsection{Unitarity}

With the conjugation rule \eqref{jtdj}, the $\mathfrak{sl}_2$ representations $\mathcal{C}^j_\alpha,\mathcal{D}^{j,\pm}$, and the large level spectrum $\mathcal{F}(\widetilde{SL}_2(\mathbb{R}))$, are unitary.
The natural extension of the conjugation rule to the affine algebra $\widehat{\mathfrak{sl}}_2$ is
\begin{align}
 \left\{\begin{array}{l} (J_n^0)^\dagger = J_{-n}^0\ , \\
         (J_n^\pm)^\dagger = -J_{-n}^\mp \ .
        \end{array}
\right.
\end{align}
By computing the norms of level one states, it is easy to see that the affine highest-weight representations $\hat{\mathcal{C}}^j_\alpha$  are not unitary. (See Exercise \ref{exonu}.)
But the $\widetilde{SL}_2(\mathbb{R})$ WZW model is not expected to be unitary, as the metric on the underlying group has a mixed signature. (What matters for applications to string theory is not unitarity, but another property called the no-ghost theorem.) 

\section{Exercises}

\begin{exo}[Normal-ordered commutator]
~\label{exoabba}
Let us consider holomorphic fields $A,B$ such that 
\begin{align}
 A(y)B(z) = \sum_{n=0}^N \frac{C_n(z)}{(y-z)^n} + O(y-z)\ ,
\end{align}
which is of the type of Eq. \eqref{abope} with $C_0 = (AB)$. 
Compute the expansion of $A(y)B(z)$ near $z=y$ (rather than $y=z$), and deduce the operator product expansion of $B(y)A(z)$. 
In particular, show that 
\begin{align}
 (AB)-(BA) = \sum_{n=1}^N \frac{(-1)^{n+1}}{n!} C_n^{(n)} = C_1'-\frac12 C_2'' + \frac16 C_3''' -\cdots \ .
\end{align}
(It is recommended to begin with the cases $N=0$ and $N=1$.)
\end{exo}

\begin{exo}[Hermitian conjugation in the affine Lie algebra $\hat{\mathfrak{u}}_1$]
~\label{exocuo}
Let us look for a Hermitian conjugation $J_n\mapsto J_n^\dagger$ in $\hat{\mathfrak{u}}_1$ that is compatible with the conjugation \eqref{ldn} of Virasoro algebra.
\begin{enumerate}
 \item Using the commutation relation $[L_0,J_n]=-nJ_n$, show that 
 \begin{align}
  J_n^\dagger =\lambda_n J_{-n} + \nu\delta_{n,0}\ .
 \end{align}
Using the more general commutator $[L_m,J_n]$, determine the coefficients $\lambda_n,\nu$ in terms of $Q$. 
\item 
Using the relation \eqref{lnj}-\eqref{lzj} of the algebra $\hat{\mathfrak{u}}_1$ with the Virasoro algebra, show that $Q^2\in\mathbb{R}$, and deduce that $J_n^\dagger$ is given by Eq. \eqref{jdq}. For $Q=0$, show that both signs $J_n^\dagger = \pm J_{-n}$ are possible.
\end{enumerate}

\end{exo}

\begin{exo}[Representations of $\hat{\mathfrak{u}}_1$  and reducible Verma modules]
 ~\label{exoazq}
Given a highest-weight representation $\mathcal{U}_\alpha$ of the affine Lie algebra $\hat{\mathfrak{u}}_1$, we would like to understand its properties as a representation of the Virasoro algebra. We already know that if the corresponding conformal dimension $\Delta(\alpha)$ \eqref{daqu} is not degenerate, then $\mathcal{U}_\alpha$ coincides with the Verma module $\mathcal{V}_{\Delta(\alpha)}$. Let us investigate the simplest degenerate case $\Delta(\alpha)=0$ i.e. $\alpha\in\{0, Q\}$. 
\begin{enumerate}
 \item For both values of $\alpha$, compute the actions of $J_{-1}$ and $L_{-1}$ on an $\hat{\mathfrak{u}}_1$-primary state, and the action of $L_1$ on its $J_{-1}$ descendant. 
 \item Conclude that $\mathcal{U}_Q$ coincides with a reducible Verma module, while $\mathcal{U}_0$ is another type of Virasoro representation: a reducible, indecomposable representation that contains the irreducible degenerate representation $\mathcal{R}_{\langle 1,1\rangle}$, but also has a level one state that is not a null vector.
\end{enumerate}

\end{exo}

\begin{exo}[Normalization of OPE coefficients in free bosonic theories]
 ~\label{exoone}
The aim of this exercise is to show that the OPE coefficient in the OPE of $\hat{\mathfrak{u}}_1$-primary fields \eqref{vvoo} can be set to one by renormalizing the fields.
We focus on diagonal fields for simplicity: in the presence of fields with odd spins, OPE coefficients take values in $\{-1,1\}$, see Eq. \eqref{csigns} for an example.
Calling the OPE coefficient $C_{\alpha_1,\alpha_2}$, we want to show that there exists a function $\lambda_{\alpha}$ such that 
\begin{align}
 C_{\alpha_1,\alpha_2} = \frac{\lambda_{\alpha_1+\alpha_2}}{\lambda_{\alpha_1}\lambda_{\alpha_2}}\ .
\label{clll}
\end{align}
\begin{enumerate}
 \item Use commutativity and associativity of the OPE, and show that 
\begin{align}
C_{\alpha_1,\alpha_2} &= C_{\alpha_2,\alpha_1}\ ,
\\
 C_{\alpha_1,\alpha_2}C_{\alpha_1+\alpha_2,\alpha_3} &= C_{\alpha_1,\alpha_2+\alpha_3}C_{\alpha_2,\alpha_3}\ .
\end{align}

\item Consider the ansatz
\begin{align}
  \lambda_{\alpha} = \frac{1}{C_{\alpha,0}} \exp \int_0^{\alpha} \varphi \ , \quad \text{where}\quad  
 \varphi(\alpha)&=\left.{\frac{\partial}{\partial \alpha_2}}\log C_{\alpha,\alpha_2}\right|_{\alpha_2=0}\ .
\end{align}
Show that $C_{\alpha,0}$ is actually an $\alpha$-independent constant, and that the function $\varphi(\alpha)$ is such that
\begin{align}
 {\frac{\partial}{\partial \alpha_2}} \log C_{\alpha_1,\alpha_2}  = \varphi(\alpha_1+\alpha_2)-\varphi(\alpha_2)\ . 
\end{align}

\item
Prove Eq. \eqref{clll} by showing that both sides have the same value at $\alpha_2=0$, and the same logarithmic derivative wrt $\alpha_2$. 
(If the spectrum is discrete, a similar proof can be done using finite differences instead of derivatives.) 
\end{enumerate}
\end{exo}

\begin{exo}[Diagonal states in free bosonic theories]
 ~\label{exofbd}
 Let $\mathbb{Z}\vec\alpha_1+\mathbb{Z}\vec\alpha_2$ be a generic maximal spectrum for a free bosonic theory with $Q=0$.
 Show that the following three statements are equivalent:
 \begin{itemize}
  \item The theory is a compactified free boson.
  \item The spectrum contains diagonal states.
  \item There are integer numbers $n_1,n_2,w_1,w_2$ such that 
  \begin{align}
  S(\vec\alpha_1)=n_1w_1 \ \ , \ \  S(\vec\alpha_2)=n_2w_2\ \ ,\ \  S(\vec\alpha_1+\vec\alpha_2) = (n_1+n_2)(w_1+w_2)\ .
  \end{align}
 \end{itemize}
To do this, show that the combination $x=\frac{\alpha_1-\bar{\alpha}_1}{\alpha_2-\bar{\alpha}_2}$ obeys the quadratic equation 
\begin{align}
S(\vec\alpha_2)x+ S(\vec\alpha_1)x^{-1} = S(\vec\alpha_1+\vec\alpha_2) -S(\vec\alpha_1)-  S(\vec\alpha_2)\ .
\end{align}
If the third statement holds, show that the two solutions are $x=\frac{n_1}{n_2}$ and $x=\frac{w_1}{w_2}$.
\end{exo}

\begin{exo}[The light asymptotic limit of Liouville theory as a global conformal field theory]
 ~\label{exolight}
 Let us admit that the functions $\phi^j_z(h)$ \eqref{pjxh} that appear in the $N$-point functions \eqref{zih} form a basis of functions on $H_3^+$, on which any function on $H_3^+$ can be decomposed via the Fourier-type formula \cite{tes97b}(Appendix A) 
 \begin{align}
 f(h)=  \int_{-\frac12+i\mathbb{R}} dj\ (2j+1)^2\int_\mathbb{C} d^2z\ \phi^j_z(h)\int_{H_3^+} dh'\ \phi^{-j-1}_z(h')f(h')\ .
 \end{align}
\begin{enumerate}
 \item Using the Fourier-type formula, reduce the four-point function to a product of three-point functions.
 \item Using Eq. \eqref{cff} for the three-point functions, write the four-point function in terms of structure constants and conformal blocks, with an expression for conformal blocks as integrals over $z$.
 \item Evaluate the integrals over $z$, and deduce the expression for global conformal blocks in terms of hypergeometric functions 
 \begin{align}
  \lim_{\mathrm{light\ asymptotic}} \mathcal{F}^{(s)}_{\Delta_s}(\Delta_i|x) = x^{\Delta_s-\Delta_1-\Delta_2} F(\Delta_s+\Delta_1-\Delta_2,\Delta_s+\Delta_4-\Delta_3,2\Delta_s,x)\ .
 \end{align}
 Check that this agrees with Eq. \eqref{eq:fsexp}.
\end{enumerate}
\end{exo}

\begin{exo}[Sugawara construction]
 ~\label{exosug}
 Let us check that the Sugawara construction does produce a Virasoro field.
 \begin{enumerate}
  \item 
Compute the $TJ^a$ OPE \eqref{tja}, by applying Wick's theorem to 
$\cunderbracket{J^a}{(z)K_{bc}}{(J^bJ^c)}(y)$, 
going through the following intermediate steps:
\begin{align}
 \cunderbracket{J^a}{(z)}{(J^bJ_b)}(y) \hspace{-2cm} &
 \nonumber
 \\
 &\ = \frac{1}{2\pi i}\oint_y\frac{dx}{x-y}\left(\frac{kJ^a(y)}{(x-z)^2} + \frac{f^{ab}_{c}J^c(x)J_b(y)}{z-x} + \frac{kJ^a(x)}{(y-z)^2} - \frac{f^{ab}_{c}J^c(x)J_b(y)}{z-y}\right)  ,
\\
 &\ = \frac{2kJ^a(y)}{(y-z)^2} + \frac{f^{ab}_c f^{dc}_b J_d(y)}{(y-z)^2} 
 = 2(k+g) {\frac{\partial}{\partial z}} \frac{J^a(z)}{y-z} + O(1)\ .
\end{align}
\item
Apply Wick's theorem to $\cunderbracket{T}{(y) K_{ab}}{(J^aJ^b)}(z)$, and check the following identities:
\begin{align}
 \cunderbracket{T}{(y)}{(J^aJ_a)}(z) & = \frac{1}{2\pi i} \oint_z \frac{dx}{x-z}\left({\frac{\partial}{\partial x}}\frac{J^a(x)J_a(z)}{y-x} + {\frac{\partial}{\partial z}} \frac{J^a(x)J_a(z)}{y-z}\right) \ ,
\\
&=  \frac{1}{2\pi i} \oint_zdx \frac{J^a(x)J_a(z)}{(x-z)(y-x)(y-z)} + {\frac{\partial}{\partial z}} \frac{1}{2\pi i} \oint_zdx \frac{J^a(x)J_a(z)}{(x-z)(y-z)} \ ,
\\
&= \frac{k\dim \mathfrak{g}}{(y-z)^4} + \frac{(J^aJ_a)(z)}{(y-z)^2} + {\frac{\partial}{\partial z}}\frac{(J^aJ_a)(z)}{y-z}\ .
\end{align}

\item
Conclude that the field $T$ satisfies the Virasoro field OPE \eqref{tt} with the central charge \eqref{ckg}.
\end{enumerate}
\end{exo}

\begin{exo}[Modified Sugawara construction]
 ~\label{exotqpj}
In analogy with the $\hat{\mathfrak u}_1$ case \eqref{tqz}, let us modify the Sugawara construction \eqref{tjj} and introduce the field $\hat{T} = T + Q_a\partial J^a$.
\begin{enumerate}
 \item Show that $\hat{T}$ is a Virasoro field, and compute the central charge. 
 \item Compute the OPE $\hat{T}J^a$: under which conditions are $J^a$ primary fields with integer dimensions?
 \item Discuss the asymptotic behaviour of $J^a$, and the resulting Ward identities.
\end{enumerate}
\end{exo}
% NB: See diary10.tex

\begin{exo}[Associativity of the $J^aJ^b\Phi^\mathcal{R}$ OPE]
 ~\label{exojjp}
Let us check the associativity of the $J^aJ^b\Phi^\mathcal{R}$ OPE, by performing two different computations of the behaviour near $y_2=z_0$ of 
\begin{align}
\mathcal{O}= \frac{1}{2\pi i} \oint_{y_2}dy_1\ J^a(y_1)J^b(y_2)\Phi^\mathcal{R}(z_0)\ .
\end{align}
\begin{enumerate}
 \item 
Firstly, use the $J^aJ^b$ OPE \eqref{jajb}, and check that
\begin{align}
 \mathcal{O}= f_c^{ab}J^c(y_2)\Phi^\mathcal{R}(z_0) = \frac{-f_c^{ab}t^c\Phi^\mathcal{R}(z_0)}{y_2-z_0} + O(1)\ .
 \label{oftp}
\end{align}

\item
Secondly, split the integration contour in two terms, $\oint_{y_2} = \oint_{y_2,z_0} - \oint_{z_0}$.
In the first term, use the $J^b(y_2)\Phi^\mathcal{R}(z_0)$ OPE, as no integration contour runs between these two operators.
In the second term, use the $J^a(y_1)\Phi^\mathcal{R}(z_0)$ OPE, and show that
\begin{align}
 \mathcal{O}& =\frac{1}{2\pi i} \oint_{y_2,z_0}dy_1 J^a(y_1)\frac{-t^b\Phi^\mathcal{R}(z_0)}{y_2-z_0} - \frac{1}{2\pi i} \oint_{z_0}dy_1 J^b(y_2)\frac{-t^a\Phi^\mathcal{R}(z_0)}{y_1-z_0} +O(1)\ .
\end{align}

\item
Use $J^a(y_1) t^b \Phi^\mathcal{R}(z_0) =t^b J^a(y_1)  \Phi^\mathcal{R}(z_0)$, and check that you recover Eq. \eqref{oftp}.
\end{enumerate}
\end{exo}

\begin{exo}[Wakimoto free-field representation of $\widehat{\mathfrak{sl}}_2$]
 ~\label{exowaki}
Consider fields $(J,\beta,\gamma)$ such that 
\begin{align}
 J(y)J(z) = \frac{-\frac12}{(y-z)^2} + O(1) \quad , \quad \beta(y)\gamma(z) = \frac{-1}{y-z} + O(1)\ ,
\end{align}
and the OPEs $J\beta,J\gamma,\beta\beta,\gamma\gamma$ have no singular terms.
Consider the fields 
\begin{align}
 J^- = -\beta \quad , \quad J^0 = -(\beta\gamma) - bJ \quad , \quad J^+ = (\beta\gamma^2)+2b(\gamma J)+k\partial\gamma\ ,
\end{align}
where the brackets are normal-ordered products, and the parameters $b$ and $k$ obey the relation \eqref{bk}.

\begin{enumerate}
 \item 
Show that the fields $(J^-,J^0,J^+)$ obey the OPEs \eqref{jjjj}, and are therefore $\widehat{\mathfrak{sl}}_2$ currents. 
\item
Show that the Sugawara construction yields the Virasoro field
\begin{align}
 T = -\beta \partial\gamma - J^2 +b^{-1}\partial J\ ,
\end{align}
and rederive the central charge from this formula.

\item
Defining the field $\Phi^j_\mu(z)$ by 
\begin{align}
 & \beta(y)\Phi^j_\mu(z) = \frac{-\mu}{y-z}\Phi^j_\mu(z)+ O(1) \quad , \quad \gamma(y)\Phi^j_\mu(z)=O(1)\ , 
\\
 & J(y)\Phi^j_\mu(z) = \frac{-b^{-1}(j+1)}{y-z}\Phi^j_\mu(z)+ O(1)\ ,
\end{align}
show that the field $\left({\frac{\partial}{\partial \mu}}-\gamma(z)\right)\Phi^j_\mu(z)$ satisfies the same relations, and is therefore proportional to $\Phi^j_\mu(z)$.

\item
Choosing the coefficient of proportionality such that
\begin{align}
 {\frac{\partial}{\partial \mu}}\Phi^j_\mu(z) =\left(\frac{j+1}{\mu}+\gamma(z)\right)\Phi^j_\mu(z)\ ,
\end{align}
write the OPEs $J^a(y)\Phi^j_\mu(z)$ in terms of differential operators as in Eq. \eqref{jprx}, and conclude that $\Phi^j_\mu(z)$ is a $\mu$-basis affine primary field.
\end{enumerate}
\end{exo}

\begin{exo}[Conformal global Ward identities from KZ equations]
 ~\label{exokz}
 Prove that the KZ equations imply the global Ward identities of conformal symmetry \eqref{spz}.
 Use the global Ward identities of the affine symmetry \eqref{drxt}, the Casimir relation for the isospin differential operators \eqref{dta}, and the conformal dimensions of affine primary fields \eqref{dr}. 
\end{exo}

\begin{exo}[Proof of the identity \eqref{dtyj}]
 ~\label{exoktk}
Check that the identity holds when applied to positions $z_i$, due to $\mathcal{K}^{-1}z_i = z_i$. 
To check that the identity also holds when applied to Sklyanin variables $y_j$, it is enough to check 
that both sides have the same commutator with $\mathcal{K}\hat{J}^-(y)\mathcal{K}^{-1}$. 
Compute the commutator of the left-hand side using $[\hat{T}(\hat{y}_j),\hat{J}^-(y)]={\frac{\partial}{\partial y}}\frac{1}{y-\hat{y}_j}\hat{J}^-(y)$, which is a consequence of Eq. \eqref{dtd}.
Compute the commutator of the 
right-hand side using the formula \eqref{djm} for $\hat{J}^-(y)$, and conclude.
\end{exo}

\begin{exo}[Level one null vectors in affine highest-weight representations]
 ~\label{exolos}
 Let $(|v_i\rangle)$ be a basis of a representation $\mathcal{R}_j$ of $\mathfrak{sl}_2$, with a spin $j\in\mathbb{C}$. For $c_a^i$ some coefficients, let $|\chi\rangle =  c_a^i J_{-1}^a|v_i\rangle$ be a level one vector in the corresponding highest-weight representation of $\widehat{\mathfrak{sl}}_2$.
 \begin{enumerate}
  \item Show that $|\chi \rangle $ is an affine null vector if an only if 
  \begin{align}
    c_a^i f^{ab}_c t^c|v_i\rangle + k K^{ab} c_a^i |v_i\rangle = 0\ .
  \end{align}
  \item Let $|w^a\rangle$ a basis of the adjoint representation $\mathcal{R}_1$ of $\mathfrak{sl}_2$, with $t^a|w^b\rangle = f^{ab}_c|w^c\rangle$. Show that our affine null vector condition is equivalent to $c_a^i|w^a\rangle\otimes |v_i\rangle\in \mathcal{R}_1\otimes \mathcal{R}_j$ being an eigenvector of the operator $Q=K_{ab}t^a\otimes t^b$, with the eigenvalue $k$.
  % NB: Actually I find the eigenvalue -k, there must be a sign mistake.
  \item Write the operator $Q$ in terms of the quadratic Casimir operators of the representations $\mathcal{R}_1, \mathcal{R}_j$ and $\mathcal{R}_1\otimes \mathcal{R}_j$, and show that its eigenvalues are $-2, 2j$ and $-2j-2$. 
  Compute these eigenvalues in the case $j=j_{\langle 1,1\rangle}$ \eqref{jrs}, and show that we have an affine null vector in this case. 
 \end{enumerate}
\end{exo}

\begin{exo}[$\widehat{\mathfrak{sl}}_2$ degenerate representations and fusion rules from the $H_3^+$ model]
 ~\label{exodrfrh}
Consider the OPE $\Phi^{j_1}_{x_1}\Phi^{j_2}_{x_2}$ \eqref{xope} where we initially assume $j_1,j_2\in -\frac12+ib{\mathbb{R}}$.
\begin{enumerate}
 \item 
Check that the line of integration $j\in -\frac12+ib{\mathbb{R}}$ and the eight cones of poles of the OPE coefficient look as follows:
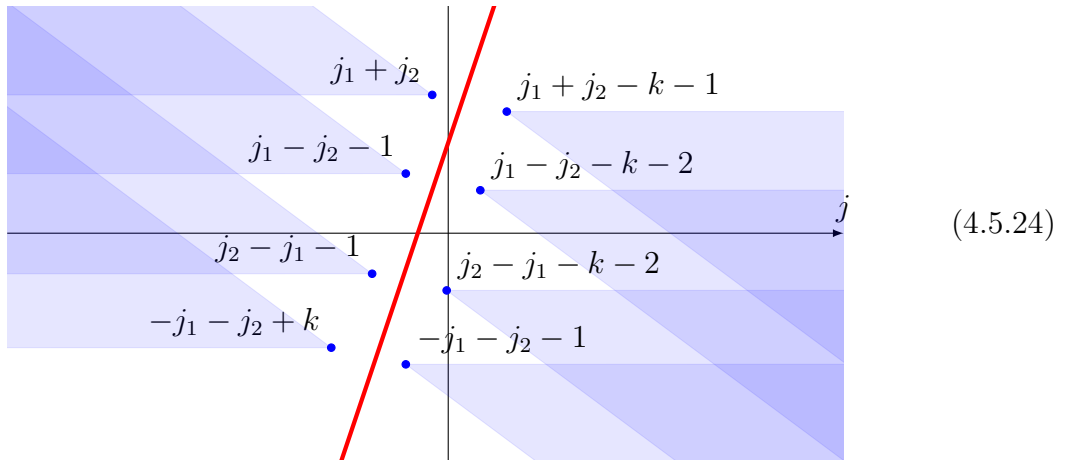
\begin{align}
\newcommand{\polewedge}[3]{
\begin{scope}[#1]
\node[blue, draw,circle,inner sep=1pt,fill] at (0, 0) {};
\node[#3] at (0,0) {#2};
\filldraw[opacity = .1, blue] (0,0) -- (6, -2) -- (6, 2) -- cycle;
\end{scope}
}
 \begin{tikzpicture}[baseline=(current  bounding  box.center), scale = 1]
  \draw[-latex] (-5,0) -- (.8, 0) -- (6,0) node [above] {$j$};
  \clip (-5, -3) -- (-5, 3) -- (6, 3) -- (6, -3) -- cycle;
  \draw (.8, -3.5) -- (.8, 3.5);
  \begin{scope}[rotate around = {-18.43:(.4,0)}]
  \draw[ultra thick, red] (.4, -4.5) -- (.4, 4.5);
  \polewedge{shift = {(0, .7)}, rotate = 180}{$j_1-j_2-1$}{above left};
  \polewedge{shift = {(0, -.7)}, rotate = 180}{$j_2-j_1-1$}{above left};
  \polewedge{shift = {(0, 1.8)}, rotate = 180}{$j_1+j_2$}{above left};
  \polewedge{shift = {(-.2, -1.8)}, rotate = 180}{$-j_1-j_2+k$}{above left};
  \polewedge{shift = {(1, .8)}}{$j_1-j_2-k-2$}{above right};
  \polewedge{shift = {(1, -.6)}}{$j_2-j_1-k-2$}{above right};
  \polewedge{shift = {(1, 1.9)}}{$j_1+j_2-k-1$}{above right};
  \polewedge{shift = {(.8, -1.7)}}{$-j_1-j_2-1$}{above right};
  \end{scope}
 \end{tikzpicture}
\end{align}

\item
We want to analytically continue our OPE to degenerate values of the spin $j_1$, i.e. values such that the integral over $j$ reduces to a finite sum. This happens whenever poles from the left of the line of integration, coincide with poles on the right of that line. (In principle we also need $j_1$ to be a zero of the OPE coefficient, but this can always be achieved by a field renormalization.) Compute the degenerate spins, and compare them with Eq. \eqref{jrs}.

\item 
Compute the limit of the OPE when $j_1$ takes degenerate values, and compare with the OPE Eq. \eqref{prspj}.
\end{enumerate}
\end{exo}

\begin{exo}[Non-unitarity of the $H_3^+$ and \texorpdfstring{$\widetilde{SL}_2(\mathbb{R})$}{SL2(R)} WZW models]
 ~\label{exonu}
We want to show that the $H_3^+$ and $\widetilde{SL}_2(\mathbb{R})$ WZW models are not unitary, either by a pedestrian calculation, or by a more algebraic reasoning. 
\begin{enumerate}
\item 
Let $|v\rangle$ be an affine primary state in an affine highest-weight representation. For $\lambda$ a complex number, compute the norm square of the level one descendant states  $|(J^0_{-1}+\lambda\bar{J}^0_{-1})v\rangle$ in both models. In the case of the $H_3^+$ model, show that this cannot be positive for all $\lambda\in\mathbb{C}$. In the case of the  $\widetilde{SL}_2(\mathbb{R})$ WZW model, find another family of level one descendant states whose norm square is not positive. 
\item
In an affine highest-weight representation, show that the level one subspace is a representation of the horizontal subalgebra, which is obtained by tensoring the level zero subspace with the adjoint representation. Decompose that representation into irreducibles, and discuss its unitarity. 
\end{enumerate}
\end{exo}

\bibliographystyle{cft}
\bibliography{cft}

\begin{thebibliography}{10}
\expandafter\ifx\csname url\endcsname\relax
  \def\url#1{\texttt{#1}}\fi
\expandafter\ifx\csname urlprefix\endcsname\relax\def\urlprefix{URL }\fi
\providecommand{\eprint}[2][]{\url{#2}}

\bibitem{rib14c}
\href{http://researchpracticesandtools.blogspot.fr/2014/09/modular-invariance-in-non-rational-cft.html}{S.~Ribault}
  (2014 blog post)\\ {\em Modular invariance in non-rational CFT\/}

\bibitem{zz90}
\href{http://libgen.io/book/index.php?md5=E51762AF50A22BA4441C55D902626C00}{A.~Zamolodchikov,
  A.~Zamolodchikov} (1990 book)\\ {\em {Conformal Field Theory and Critical
  Phenomena in Two-Dimensional Systems}\/}

\bibitem{tes17}
\href{http://arxiv.org/abs/1708.00680}{J.~Teschner} (2017 review)
  [arXiv:1708.00680]\\ {\em {A guide to two-dimensional conformal field
  theory}\/}

\bibitem{sch05}
\href{http://arxiv.org/abs/hep-th/0509155}{V.~Schomerus} (2006 review)
  [arXiv:hep-th/0509155]\\ {\em Non-compact string backgrounds and non-rational
  CFT\/}

\bibitem{fms97}
P.~Di~Francesco, P.~Mathieu, D.~S\'en\'echal (1997 book)\\ {\em Conformal field
  theory\/}

\bibitem{nak04}
\href{http://arxiv.org/abs/hep-th/0402009}{Y.~Nakayama} (2004 review)
  [arXiv:hep-th/0402009]\\ {\em Liouville field theory: A decade after the
  revolution\/}

\bibitem{gab99}
\href{http://arxiv.org/abs/hep-th/9910156}{M.~R. Gaberdiel} (2000 review)
  [arXiv:hep-th/9910156]\\ {\em {An Introduction to conformal field theory}\/}

\bibitem{car08}
\href{http://arxiv.org/abs/0807.3472}{J.~Cardy} (2008 review)
  [arXiv:0807.3472]\\ {\em {Conformal Field Theory and Statistical
  Mechanics}\/}

\bibitem{bs92}
\href{http://arxiv.org/abs/hep-th/9210010}{P.~Bouwknegt, K.~Schoutens} (1993
  review) [arXiv:hep-th/9210010]\\ {\em W symmetry in conformal field theory\/}

\bibitem{rib16}
\href{http://arxiv.org/abs/1609.09523}{S.~Ribault} (2016 review)
  [arXiv:1609.09523]\\ {\em {Minimal lectures on two-dimensional conformal
  field theory}\/}

\bibitem{kr18}
\href{http://arxiv.org/abs/1812.10713}{S.~Kanade, D.~Ridout} (2018 review)
  [arXiv:1812.10713]\\ {\em {NGK and HLZ: fusion for physicists and
  mathematicians}\/}

\bibitem{kqr21}
\href{http://arxiv.org/abs/2104.02090}{P.~Kravchuk, J.~Qiao, S.~Rychkov} (2021)
  [arXiv:2104.02090]\\ {\em {Distributions in CFT. Part II. Minkowski space}\/}

\bibitem{prv18}
\href{http://arxiv.org/abs/1805.04405}{D.~Poland, S.~Rychkov, A.~Vichi} (2018
  review) [arXiv:1805.04405]\\ {\em {The Conformal Bootstrap: Theory, Numerical
  Techniques, and Applications}\/}

\bibitem{ms89b}
G.~W. Moore, N.~Seiberg (1989)\\ {\em {Classical and Quantum Conformal Field
  Theory}\/}

\bibitem{car01}
\href{http://arxiv.org/abs/math-ph/0103018}{J.~L. Cardy} (2001 review)
  [arXiv:math-ph/0103018]\\ {\em {Lectures on conformal invariance and
  percolation}\/}

\bibitem{rw20}
\href{http://arxiv.org/abs/2001.05055}{I.~Runkel, G.~M.~T. Watts} (2020)
  [arXiv:2001.05055]\\ {\em {Fermionic CFTs and classifying algebras}\/}

\bibitem{flno09}
\href{http://arxiv.org/abs/0902.1331}{V.~A. Fateev, A.~V. Litvinov, A.~Neveu,
  E.~Onofri} (2009) [arXiv:0902.1331]\\ {\em {Differential equation for
  four-point correlation function in Liouville field theory and elliptic
  four-point conformal blocks}\/}

\bibitem{bhs17}
\href{http://arxiv.org/abs/1711.04361}{V.~Belavin, Y.~Haraoka, R.~Santachiara}
  (2017) [arXiv:1711.04361]\\ {\em {Rigid Fuchsian systems in 2-dimensional
  conformal field theories}\/}

\bibitem{mr17}
\href{http://arxiv.org/abs/1711.08916}{S.~Migliaccio, S.~Ribault} (2017)
  [arXiv:1711.08916]\\ {\em {The analytic bootstrap equations of non-diagonal
  two-dimensional CFT}\/}

\bibitem{fr11}
\href{http://arxiv.org/abs/1109.6764}{V.~Fateev, S.~Ribault} (2012)
  [arXiv:1109.6764]\\ {\em {The Large central charge limit of conformal
  blocks}\/}

\bibitem{ccy17}
\href{http://arxiv.org/abs/1703.09805}{M.~Cho, S.~Collier, X.~Yin} (2017)
  [arXiv:1703.09805]\\ {\em {Recursive Representations of Arbitrary Virasoro
  Conformal Blocks}\/}

\bibitem{rib18}
\href{http://arxiv.org/abs/1809.03722}{S.~Ribault} (2018) [arXiv:1809.03722]\\
  {\em {On 2d CFTs that interpolate between minimal models}\/}

\bibitem{aflt10}
\href{http://arxiv.org/abs/1012.1312}{V.~A. Alba, V.~A. Fateev, A.~V. Litvinov,
  G.~M. Tarnopolsky} (2011) [arXiv:1012.1312]\\ {\em {On combinatorial
  expansion of the conformal blocks arising from AGT conjecture}\/}

\bibitem{tv12}
\href{http://arxiv.org/abs/1202.4698}{J.~Teschner, G.~Vartanov} (2012)
  [arXiv:1202.4698]\\ {\em {6j symbols for the modular double, quantum
  hyperbolic geometry, and supersymmetric gauge theories}\/}

\bibitem{pt01}
\href{http://arxiv.org/abs/hep-th/0110244}{B.~Ponsot, J.~Teschner} (2002)
  [arXiv:hep-th/0110244]\\ {\em Boundary Liouville field theory: Boundary three
  point function\/}

\bibitem{dv95}
\href{http://arxiv.org/abs/hep-th/9508020}{K.~De~Vos, P.~Van~Driel} (1996)
  [arXiv:hep-th/9508020]\\ {\em {The Kazhdan-Lusztig conjecture for W
  algebras}\/}

\bibitem{rad13}
\href{http://arxiv.org/abs/1302.0801}{G.~Radobolja} (2013) [arXiv:1302.0801]\\
  {\em {Subsingular vectors in Verma modules, and tensor product modules over
  the twisted Heisenberg-Virasoro algebra and W(2,2) algebra}\/}

\bibitem{zam05}
\href{http://arxiv.org/abs/hep-th/0505063}{A.~B. Zamolodchikov} (2005)
  [arXiv:hep-th/0505063]\\ {\em On the three-point function in minimal
  Liouville gravity\/}

\bibitem{rib14b}
\href{http://researchpracticesandtools.blogspot.fr/2014/10/unitarity-and-reality-of-three-point.html}{S.~Ribault}
  (2014 blog post)\\ {\em Reality of three-point structure constants in CFT,
  unitary or not\/}

\bibitem{zz95}
\href{http://arxiv.org/abs/hep-th/9506136}{A.~B. Zamolodchikov, A.~B.
  Zamolodchikov} (1996) [arXiv:hep-th/9506136]\\ {\em Structure constants and
  conformal bootstrap in Liouville field theory\/}

\bibitem{rs15}
\href{http://arxiv.org/abs/1503.02067}{S.~Ribault, R.~Santachiara} (2015)
  [arXiv:1503.02067]\\ {\em {Liouville theory with a central charge less than
  one}\/}

\bibitem{tes03b}
\href{http://arxiv.org/abs/hep-th/0303150}{J.~Teschner} (2004)
  [arXiv:hep-th/0303150]\\ {\em A lecture on the Liouville vertex operators\/}

\bibitem{hjs09}
\href{http://arxiv.org/abs/0911.4296}{L.~Hadasz, Z.~Jaskolski, P.~Suchanek}
  (2010) [arXiv:0911.4296]\\ {\em {Modular bootstrap in Liouville field
  theory}\/}

\bibitem{zam03}
\href{http://arxiv.org/abs/hep-th/0312279}{A.~Zamolodchikov} (2004)
  [arXiv:hep-th/0312279]\\ {\em Higher equations of motion in Liouville field
  theory\/}

\bibitem{rw01}
\href{http://arxiv.org/abs/hep-th/0107118}{I.~Runkel, G.~M.~T. Watts} (2001)
  [arXiv:hep-th/0107118]\\ {\em A non-rational CFT with c = 1 as a limit of
  minimal models\/}

\bibitem{sch03}
\href{http://arxiv.org/abs/hep-th/0306026}{V.~Schomerus} (2003)
  [arXiv:hep-th/0306026]\\ {\em Rolling tachyons from Liouville theory\/}

\bibitem{mce07}
\href{http://arxiv.org/abs/0706.0365}{W.~McElgin} (2008) [arXiv:0706.0365]\\
  {\em Notes on Liouville Theory at $c\leq 1$\/}

\bibitem{clrs16}
\href{http://arxiv.org/abs/1611.02193}{X.~Cao, P.~Le~Doussal, A.~Rosso,
  R.~Santachiara} (2017) [arXiv:1611.02193]\\ {\em {Liouville field theory and
  log-correlated Random Energy Models}\/}

\bibitem{ilt13}
\href{http://arxiv.org/abs/1308.4092}{N.~Iorgov, O.~Lisovyy, Y.~Tykhyy} (2013)
  [arXiv:1308.4092]\\ {\em {Painlev\'e VI connection problem and monodromy of
  c=1 conformal blocks}\/}

\bibitem{cer12}
\href{http://arxiv.org/abs/1209.3984}{L.~Chekhov, B.~Eynard, S.~Ribault} (2013)
  [arXiv:1209.3984]\\ {\em {Seiberg-Witten equations and non-commutative
  spectral curves in Liouville theory}\/}

\bibitem{ekr15}
\href{http://arxiv.org/abs/1510.04430}{B.~Eynard, T.~Kimura, S.~Ribault} (2015)
  [arXiv:1510.04430]\\ {\em {Random matrices}\/}

\bibitem{fuc97}
\href{http://arxiv.org/abs/hep-th/9702194}{J.~Fuchs} (1997)
  [arXiv:hep-th/9702194]\\ {\em {Lectures on conformal field theory and
  Kac-Moody algebras}\/}

\bibitem{rib09}
\href{http://arxiv.org/abs/0912.4481}{S.~Ribault} (2010) [arXiv:0912.4481]\\
  {\em {Minisuperspace limit of the AdS3 WZNW model}\/}

\bibitem{rib08b}
\href{http://arxiv.org/abs/0811.4587}{S.~Ribault} (2009) [arXiv:0811.4587]\\
  {\em {On sl3 Knizhnik-Zamolodchikov equations and W3 null-vector
  equations}\/}

\bibitem{rib08}
\href{http://arxiv.org/abs/0803.2099}{S.~Ribault} (2008) [arXiv:0803.2099]\\
  {\em {A family of solvable non-rational conformal field theories}\/}

\bibitem{ay92}
H.~Awata, Y.~Yamada (1992)\\ {\em Fusion rules for the fractional level sl(2)
  algebra\/}

\bibitem{tes97a}
\href{http://arxiv.org/abs/hep-th/9712256}{J.~Teschner} (1999)
  [arXiv:hep-th/9712256]\\ {\em On structure constants and fusion rules in the
  $SL(2,\mathbb{C})/SU(2)$ {WZNW} model\/}

\bibitem{rt05}
\href{http://arxiv.org/abs/hep-th/0502048}{S.~Ribault, J.~Teschner} (2005)
  [arXiv:hep-th/0502048]\\ {\em $H_3^+$ correlators from Liouville theory\/}

\bibitem{tes97b}
\href{http://arxiv.org/abs/hep-th/9712258}{J.~Teschner} (1999)
  [arXiv:hep-th/9712258]\\ {\em The mini-superspace limit of the {SL(2,C)/SU(2)
  WZNW} model\/}

\bibitem{mo00a}
\href{http://arxiv.org/abs/hep-th/0001053}{J.~M. Maldacena, H.~Ooguri} (2001)
  [arXiv:hep-th/0001053]\\ {\em Strings in {$AdS_3$ and $SL(2,\mathbb{R})$ WZW
  model. I}\/}

\bibitem{mo01}
\href{http://arxiv.org/abs/hep-th/0111180}{J.~M. Maldacena, H.~Ooguri} (2002)
  [arXiv:hep-th/0111180]\\ {\em Strings in {AdS(3) and the SL(2,R) WZW model.
  III: Correlation functions}\/}

\bibitem{gab01b}
\href{http://arxiv.org/abs/hep-th/0105046}{M.~R. Gaberdiel} (2001)
  [arXiv:hep-th/0105046]\\ {\em {Fusion rules and logarithmic representations
  of a WZW model at fractional level}\/}

\bibitem{rib05}
\href{http://arxiv.org/abs/hep-th/0507114}{S.~Ribault} (2005)
  [arXiv:hep-th/0507114]\\ {\em Knizhnik-Zamolodchikov equations and spectral
  flow in $AdS_3$ string theory\/}

\bibitem{kks18}
\href{http://arxiv.org/abs/1809.05111}{D.~Karateev, P.~Kravchuk,
  D.~Simmons-Duffin} (2019) [arXiv:1809.05111]\\ {\em {Harmonic Analysis and
  Mean Field Theory}\/}

\bibitem{kmp21}
\href{http://arxiv.org/abs/2111.12716}{P.~Kravchuk, D.~Mazac, S.~Pal} (2021)
  [arXiv:2111.12716]\\ {\em {Automorphic Spectra and the Conformal
  Bootstrap}\/}

\end{thebibliography}

\printindex

\end{document}